%% file: main.tex
\documentclass[12pt]{report}
\usepackage[utf8]{inputenc}
\usepackage{physics}
\usepackage{amsmath}
\usepackage{subfiles}
\usepackage{amssymb}
\usepackage{mathtools}
\usepackage{hyperref}
\usepackage{geometry}
\usepackage{appendix}
\usepackage[usenames,dvipsnames]{color}
\usepackage{graphicx}
\usepackage{subcaption}
\usepackage{caption}
\usepackage{titling}
\usepackage{afterpage}

\newcommand{\ep}{{\epsilon}}
\newcommand{\Ea}{{E_\alpha}}
\newcommand{\PEa}{{ \hat{\mathcal{P}}_{\Ea} }}
\newcommand{\omL}{{\omega}}
\newcommand{\omc}{{\omega_c}}

\allowdisplaybreaks

\begin{document}

\setlength{\baselineskip}{22pt plus 2pt}
\normalsize

\subfile{Chapters/Oxford.tex}

\clearpage
\thispagestyle{plain}
\vspace{.35\textheight}

\pagebreak
\chapter*{Acknowledgements}
\thispagestyle{empty}
\setlength{\baselineskip}{22pt plus 2pt}

I would like to express my deepest gratitude to everyone who supported me throughout this journey.

To my supervisor, Dieter Jaksch, for constantly reminding me that critical thinking is the key. While many fascinating ideas can arise, it is through critical thinking that the initial sparks of inspiration are transformed into practical and meaningful achievements. You often encouraged me to start with the simplest possible problem and gradually incorporate complexities. In the last two years of my DPhil, I learned how to apply this approach, which allowed me to avoid unnecessary difficulties and focus on addressing the essential problems.

To Frank Schlawin, for your endless patience and guidance. You consistently emphasized the importance of mastering the finer details of conducting research, organizing thoughts, and reporting findings. Through our numerous discussions, I’ve gained a broader perspective on academia, and I've made discoveries that I never believed were possible.

To my colleagues in Oxford and Hamburg, and friends from around the world, thank you for your friendship, diverse perspectives, and the wisdom you’ve shared. Our connections transcend language barriers, and I deeply appreciate the inspiration you’ve all provided.

During the pandemic, I spent over a year in Beijing and Quanzhou, working on an idea that remains unresolved to this day. Yet, in those challenging times, I found comfort and encouragement from friends who helped me endure that unique period. I am forever grateful for their companionship.

To my parents and grandparents, your unwavering support gave me the foundation to pursue my own passion for physics. Even when I fell seriously ill several times during and after the pandemic, your care enabled me to persevere and ultimately achieve success. None of this would have been possible without you.

%\pagebreak

%\hspace{0pt}
%\vfill
%\begin{minipage}[h]{\textwidth}
%\vspace{6mm}
\chapter*{Abstract}
\thispagestyle{empty}

\setlength{\baselineskip}{22pt plus 2pt}

The development of future quantum devices requires understanding the dynamics of driven many-body systems, in which the Floquet-induced interactions play a central role. This understanding is crucial for coherently controlling quantum states, minimising errors, and benchmarking the performance of these devices. In this thesis, we analyse the enhancement on the Floquet-induced interactions by many-body correlations, and develop an advanced Floquet method to understand the Floquet-induced interactions relevant for future quantum devices. 

\vspace{2mm}
We first study Floquet-induced interactions that are generated in a two-dimensional two-band Hubbard model coupled to an optical cavity when it is driven in-gap by a strong laser. Starting from a Floquet description of the driven system, we derive effective low-energy Hamiltonians by projecting out the high-energy degrees of freedom and treating intrinsic interactions on a mean-field level. We then investigate how the virtual excitation of high-energy Frenkel excitons affects the Floquet-induced cavity-mediated interactions as well as the Floquet renormalisation of electron band dispersion. Floquet-induced interactions are enhanced strongly when the driving frequency approaches an exciton resonance. Additionally, the cavity-mediated interaction, as well as the Floquet band renormalisation, are strongly broadened in reciprocal space, which could further boost the impact of Floquet-induced interactions on the driven-dissipative steady state. 

\vspace{2mm}
To extend the above study to generic driven many-body systems, we develop a Floquet Schrieffer Wolff transform (FSWT) to obtain effective Floquet Hamiltonians and micro-motion operators of periodically driven many-body systems for any non-resonant driving frequency. FSWT is a more systematic and flexible Floquet method that generates the Floquet-induced interactions beyond mean-field approximations. Our FSWT perturbatively eliminates the oscillatory components in the driven Hamiltonian by solving operator-valued Sylvester equations with well-controlled approximations. It goes beyond various high-frequency expansion (HFE) methods commonly used in Floquet theory, which we demonstrate with the example of the driven Fermi-Hubbard model: Compared to the HFE results, the Floquet-induced interactions provided by FSWT offer a much more accurate prediction of the driven dynamics, over the regimes of non-resonant driving parameters. The Floquet-induced interactions provided by this method are useful for controlling correlated hopping in quantum simulations in optical lattices. 

\vspace{2mm}
Finally, we apply our FSWT to more generic driven cavity-QED setups. With the help of FSWT, we are able to study multi-orbital systems with long-range interactions driven by lasers with arbitrary polarisation. The corresponding Floquet Hamiltonian treats the cavity-independent Floquet-induced interaction on an equal footing with the cavity-mediated Floquet-induced interaction. The FSWT Hamiltonian offers a systematic way to predict driving-induced phase transitions in the cavity-QED setup.

\thispagestyle{empty}

%\end{minipage}

\vfill
%\hspace{0pt}
\pagebreak
\pagenumbering{roman}
\setcounter{page}{1}

\tableofcontents
\clearpage
\pagenumbering{arabic}
\setcounter{page}{1}

\chapter{Introduction}

\subfile{Chapters/Introduction}

\chapter{Sambe Space Floquet theory}\label{Chapter2}

\subfile{Chapters/SambeSpaceGaussianElimination}

\chapter{The driven cavity-material Hamiltonians}\label{Chapter3}

\subfile{Chapters/setup}

\chapter{The excitonic enhancement of Floquet-induced cavity-mediated interactions}\label{Chapter4}

\subfile{Chapters/exciton}

\chapter{Obtaining the self-consistent Floquet-induced interactions by solving Sylvester equations}\label{Chapter5}

\subfile{Chapters/FSWT}

\chapter{The self-consistent Floquet-induced interactions in Hubbard models}\label{Chapter6}

\subfile{Chapters/chain}

\chapter{The complete Floquet-induced two-particle interactions in cavity-semiconductor systems}\label{Chapter7}

\subfile{Chapters/exciton-revisited}

\chapter{Conclusions}
\subfile{Chapters/Conclusions}

\chapter{Appendix for Sambe Space Gaussian elimination}
\subfile{Chapters/appendix_A}

\chapter{Appendix for FSWT}
\subfile{Chapters/appendix_B}

\clearpage
\bibliographystyle{unsrt}
\bibliography{Bib}
\end{document}

%% file: Chapters/Oxford.tex
\graphicspath{ {images/Oxford/} }
\setlength{\unitlength}{1 cm} 
\thispagestyle{empty}

\text{ }
\vspace{3.5cm}

\begin{center}
\textbf{{\LARGE Floquet-induced interactions in many-body systems}}\\

\vspace{2.5cm}
\includegraphics[width=3cm]{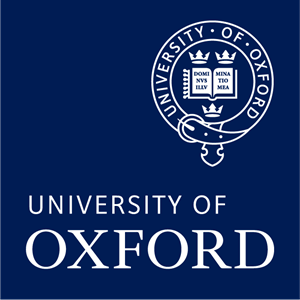}\\

\vspace{2.0cm}
{\Large Xiao Wang}\\
\vspace{0.5cm}
{\Large Keble College}\\
\vspace{0.5cm}
{\Large University of Oxford}\\

\vspace{2.5cm}
{\Large{A thesis submitted for the degree of}}\\
\vspace{0.5cm}
\centering
{\Large{\textit{Doctor of Philosophy}}}\\
\vspace{0.5cm}
{\Large{Michaelmas 2024}}

\end{center}

%% file: Chapters/Introduction.tex
\section{Coherent control of many-body systems by driving}
%

% Controlling the many-body dynamics is very useful, but challenging
Controlling the dynamics of quantum many-body systems is a pivotal frontier in contemporary physics, offering the tantalising prospect of harvesting complex quantum phenomena for technological applications \cite{schleich2016quantum}. This control allows us to simulate quantum dynamics that are intractable for classical computers \cite{RevModPhys.86.153,daley2022practical,bluvstein2021controlling} and to push matter into novel phases \cite{greiner2002quantum,ebadi2021quantum,oka2019floquet,weitenberg2021tailoring} that cannot appear in equilibrium. By steering the dynamical evolution of these systems, we can design functional quantum devices that exploit many-body entanglement \cite{RevModPhys.80.517}, enabling advanced technologies \cite{acin2018quantum} such as quantum computing, simulation, and secure communication. 

A common but striking feature of controlling interacting quantum many-body systems is the ability to modify the intrinsic interactions among numerous particles \cite{oka2019floquet,RevModPhys.82.1225,RevModPhys.80.885,RevModPhys.82.2313,Eckardt2017,bukov2015universal,goldman2014periodically}. This modification enables us to deliberately generate entanglement and other collective quantum behaviours. However, achieving this control is challenging for several reasons. First, manipulating quantum many-body systems often leads to decoherence \cite{schlosshauer2019quantum} before the desired control effect is realised. Second, precisely tracking the time evolution of a many-body wavefunction is generally unfeasible, as the memory required to store the wavefunction grows exponentially with the number of particles \cite{RevModPhys.71.1253,eisert2015quantum,schuch2008entropy}. Additionally, developing precise control protocols \cite{acin2018quantum} capable of operating within such intricate frameworks presents significant difficulties. Overcoming these obstacles is essential for advancing our fundamental understanding of quantum mechanics and unlocking the full potential of quantum technologies.

% laser is an ideal tool to realise coherent control 
Fortunately, in several scenarios, it remains feasible to control these many-body dynamics and harness their inherent potential. In this context, laser-based control of low-energy degrees of freedom in complex quantum many-body systems has emerged as a leading approach in fields such as condensed matter physics and cold atom physics. Utilizing a coherent light source \cite{scully1997quantum}, laser control enables precise temporal and spatial modulation of system parameters, stabilizing desired quantum phases and reducing decoherence. This control method is known as Floquet engineering~\cite{oka2019floquet,Eckardt2017} when the control relies on the coherent (i.e., reversible and phase-preserving) manipulation of the system's degrees of freedom. Alternatively, it is referred to as optical switching~\cite{RevModPhys.93.041002} when the control involves the deposition of energy, for instance, to melt competing orders in a transient nonthermal state.

%Thermal control, i.e., the control via temperature change, is widely used in areas such as the transition edge sensor in superconductor nanowires and the Curie points

% more examples of coherent Floquet engineering
Lasers have become an irreplaceable tool in the coherent control of many-body dynamics across various platforms, including cold atoms \cite{Eckardt2017,weitenberg2021tailoring}, correlated materials \cite{oka2019floquet,RevModPhys.93.041002}, multi-band electrons~\cite{rudner2020band,Lindner2010FloquetTI,PhysRevB.79.081406}, photons in wave-guides \cite{RevModPhys.91.015006}, and hybrid quantum systems \cite{mivehvar2021cavity,schlawin2022cavity}. Seminal experimental observations in cold atoms include, inter alia, light-induced gauge fields \cite{goldman2014light,RevModPhys.83.1523} and topology \cite{goldman2016topological,jotzu2014experimental}, Floquet-induced many-body localisation \cite{bordia2017periodically,singh2019quantifying}, discrete time crystals \cite{zhang2017observation,choi2017observation}, driving-induced superfluid-Mott transitions \cite{PhysRevLett.102.100403}, tuning of exchange interactions~\cite{gorg2018enhancement} and occupation-dependent tunneling \cite{meinert2016floquet}. In driven correlated solids, fascinating effects such as transient superconductivity \cite{doi:10.1126/science.1197294,mitrano2016possible,Rowe2023, doi:10.1080/00107514.2017.1406623, Buzzi2020, Buzzi2021,Tindall2020, Tindall2021,Tindall2021liebstheorem}, dressed surface states in topological insulators \cite{wang2013observation,mahmood2016selective}, and the light-induced anomalous Hall effect \cite{mciver2020light} have been demonstrated. Floquet topological insulators can be engineered in acoustic \cite{fleury2016floquet} and photonic \cite{rechtsman2013photonic} systems.
Additionally, laser excitation can stabilise coherent phases such as superconductivity above their equilibrium critical temperature~\cite{Fausti2011, Mitrano2016, Budden2021}, or transiently change the crystal structure to unlock new ground states~\cite{Foerst2011, Disa2021}. 

% off-resonant driving creates virtual excitations, which create Floquet-induced interactions
Off-resonant driving is a favoured choice to avoid Floquet heating \cite{PhysRevE.90.012110,PhysRevX.4.041048} in many-body systems. When the driving frequency is sufficiently detuned from any resonance of the many-body system, the laser cannot strongly excite the material but can still alter it by creating short-lifetime virtual excitations \cite{RevModPhys.93.041002}. These virtual excitations can induce non-intrinsic interactions in the many-body system, which we refer to as the \textit{Floquet-induced interactions}. These engineered interactions can be turned on and off with flexibility without heavily heating up the system. Thus, they have strong potential in the non-thermal coherent control of many-body systems. Through these Floquet-induced interactions, long-lived pre-thermal phases are created~\cite{RevModPhys.93.041002}, and new functionalities, such as Rydberg entangling gates~\cite{PhysRevLett.85.2208}, can be stabilized.

% cavity plays a similar control role as laser
Another approach to realise non-thermal coherent control of many-body systems is via the coupling to a cavity. A cavity is a condensed matter system that can trap the photons in continuous space to form bosonic quasiparticle excitations known as cavity photons. In the cavity approach to control quantum materials~\cite{schlawin2022cavity,bloch2022strongly, garcia2021manipulating, Mivehvar2021, Ruggenthaler2018}, the quantum fluctuations of light in an undriven cavity act as a tailored environment of the material, which alters the material's equilibrium behaviours at low temperatures. First experiments report, among other things, the polaritonic manipulation of chemical reaction rates~\cite{garcia2021manipulating}, cavity-modified carrier mobility in organic semiconductors~\cite{orgiu2015conductivity} and in Landau levels~\cite{Paravicini-Bagliani2019}, the change of superconducting critical temperatures~\cite{ThomasAnoop2019ESuS} and a metal-insulator transition~\cite{jarc2022cavity}, and the breakdown of topological protection of edge states under strong light-matter coupling~\cite{Appugliese2022}. Theoretically proposed effects include cavity-mediated long-range interactions~\cite{PhysRevLett.122.133602, andolina2022, Chakraborty2021, Ciuti2021, Schaefer2021}, the renormalisation of electronic bandwidths~\cite{PhysRevB.105.165121,sentef2020quantum} and magnetic interaction strengths~\cite{Kiffner2019b, Curtis2022}, the shift of phase transitions~\cite{PhysRevLett.122.167002, doi:10.1126/sciadv.aau6969, Li2020, PhysRevX.10.041027, Latini2021, Passetti2023}, or the opening of topological gaps~\cite{Xiao2019, Hubener2021,dmytruk2022controlling} when the cavity is coupled off-resonantly to a quantum material. Resonant coupling may generate exotic superconductor-polaritons~\cite{PhysRevB.99.020504} or Mott polaritons~\cite{Kiffner2019} and enable the control of exciton properties~\cite{latini2019cavity}. Transport properties of excitonic~\cite{schachenmayer2015cavity} or electronic systems~\cite{Bartolo2018, doi:10.1021/acs.jpcb.0c03227, Rokaj2019, Rokaj2022,Arwas2023} are predicted to be strongly influenced. In an ultrastrong coupling regime, where nonperturbative light-matter coupling may give rise to a superradiant phase transitions~\cite{Mazza2019, Andolina2019, Andolina2020, RomanRoche2021, SaezBlazquez2023}, exotic many-body phases of matter are predicted to emerge~\cite{Guerci2020, Rao2023, Rokaj2023, mercurio2023photon}. 
% carefully Explain the cavity-mediated interaction in the above bare-cavity control approach 
Among the various cavity-induced effects, cavity-mediated interaction stands out as particularly interesting. This interaction results from the virtual exchange of cavity photons between two charge carriers in the material. It emerges from quantum fluctuations of the electromagnetic field and has no classical counterpart—vanishing entirely if the cavity mode is replaced by an external laser drive.

% turn on/off the cavity effect still requires a laser
However, we still need to activate and deactivate the cavity-material coupling to switch the cavity-control effect on and off. The external laser drive is, again, a powerful tool for this task. When lasers drive a hybrid cavity-material system, more exotic Floquet engineering effects can emerge, which can be realised by neither bare-laser nor bare-cavity control. First pioneering applications include the generation of supersolid phases of matter through the use of cavity-mediated long-range interactions in cold atoms~\cite{LandigRenate2016Qpfc} or room-temperature exciton-polariton condensation in 2d semiconductor heterostructures~\cite{liu2015strong,schneider2016exciton}. Recent theory work further proposes important new directions for material control, which include photo-assisted tunable electron pairing in the Cooper channel~\cite{GaoHongmin2020Pepi}, laser-assisted cavity-mediated topological superfluidity~\cite{SchlawinFrank2019CUPi}, cavity-induced quantum spin liquids~\cite{chiocchetta2021cavity}, and a largely unexplored crossover between the quantum and a classical Floquet regime~\cite{sentef2020quantum}. 

% Floquet-induced cavity-mediated interactions in the driven cavity-material setup
In the cavity-material system, the external driving creates a Floquet-induced cavity-mediated interaction, whose strength can be enhanced by a small laser-cavity detuning and a large laser driving strength. It thereby can become much stronger than the cavity-mediated interaction in the undriven system, which requires ultrastrong coupling to become relevant~\cite{frisk2019ultrastrong}. This scheme is thus very similar to the established scheme in atomic cavity  QED~\cite{Mivehvar2021}. Since it arises from an off-resonant driving, this Floquet-induced cavity-mediated interaction separates from the highly active research on exciton-polaritons~\cite{kasprzak2006bose,doi:10.1126/science.aac9439,RevModPhys.82.1489,sanvitto2016road,dirnberger2022spin,keeling2020bose}, which is mainly concerned with a resonant coupling regime. This Floquet-induced interaction is experienced by the elementary charge carriers in the driven material. Thus, this interaction also separates from the Floquet engineering of excitons~\cite{Iorsh2022, Kobayashi2023, Conway2023}, which is concerned with the change of the excitonic states rather than the low-energy degrees of freedom. 

% Floquet-induced cavity-mediated interactions is accompanied by other driving-induced effects 
Based on these Floquet-induced cavity-mediated interactions, an interesting proposal is the competing long- and short-ranged interactions in driven correlated systems coupled to the cavity. In this case, the long-range interaction is given by the driving-induced cavity-mediated interaction, and the short-range interaction is the intrinsic interaction in the undriven material. Rich physics are predicted to emerge in these systems, such as Higgs mode stabilisation \cite{PhysRevB.104.L140503} and cavity-induced spin-liquid \cite{chiocchetta2021cavity}. Understanding the screening of the driving-induced interaction is vital for quantitatively predicting these phenomena. However, a standard method for analyzing this screening has yet to be developed. Besides, when these Floquet-induced cavity-mediated interactions are generated, they have to be treated on equal footing with other cavity-independent Floquet engineering effects, such as the AC Stark shift, the Bloch-Siegert shift~\cite{Sie2018}, as well as other unavoidable Floquet-induced interactions in the driven many-body systems. 

\section{Floquet perspective for many-body coherent control}\label{Floquet-pedalogical-intro}
% Floquet theory can describe these control effects
To accurately understand the coherent control effects on many-body systems, it is essential to develop powerful methods for predicting Floquet-induced interactions. Since the driving laser is in a highly populated coherent state \cite{scully1997quantum}, it can be approximated as a classical electromagnetic wave with perfect time periodicity and a stable phase. In this context, Floquet theory serves as the central theoretical framework for analyzing these interactions in driven many-body systems. When the driven Hamiltonian $\hat{H}_t$ is time-periodic, we can use Floquet Hamiltonian theory \cite{bukov2015universal,Giovannini_2020,oka2019floquet,rodriguez2021low,mori2023floquet} to describe the secular dynamics and the pre-thermal state of the driven system. 

% pedalogically introduce Floquet system:  What is quasi-energy, and why is stroboscopic evolution interesting/relevant? What is the role of micro-motion? Why these concepts can be understood from the Sambe space matrix?
In Floquet theory, the dynamics of a periodically driven quantum system are decoupled as stroboscopic evolution dressed by micro-motion. Stroboscopic evolution, where the system is observed at intervals matching the driving period, captures its long-term dynamics, while micro-motion refers to the fast oscillations within each driving cycle. In periodically driven systems where the time-translational invariance becomes a discrete symmetry, the dynamics conserve the quasi-energy, an analogue to energy in time-independent systems. These concepts are unified through the Sambe space matrix, which extends the Hilbert space to include the Fourier components of the time-periodic Hamiltonian. The eigenvalues of this Sambe space matrix represent the quasi-energies, and the corresponding eigenstates provide a comprehensive description of both stroboscopic and intra-period dynamics. 

% a review on the application of Floquet in few-body systems
The driven dynamics of various systems have been studied using Floquet theory through the diagonalisation of the Sambe space matrix.
These include for example, non-interacting lattice electrons~\cite{Rodriguez_Vega_2018,PhysRevB.93.144307}, one-band Hubbard dimmers~\cite{mentink2015ultrafast}, many-body spin systems~\cite{PhysRev.175.453,geier2021floquet}, off-resonantly driven one-band Mott Insulators~\cite{PhysRevB.96.014406, claassen2017dynamical,PhysRevLett.116.125301} or resonantly driven ones~\cite{okamoto2021floquet}, two-band Hubbard dimers~\cite{PhysRevX.11.011057}, kicked harmonic oscillators under RWA~\cite{liang2018floquet}, a two-body bi-exciton model~\cite{li2022excitonic}, driven disorder in a non-interacting lattice model~\cite{Martinez_2003}, cavity-magnons~\cite{yang2023theory} and cavity-Rydberg polaritons~\cite{clark2019interacting}. 
On the other hand, in driven many-body systems, the Sambe space matrix can no longer be fully diagonalised; nevertheless, Floquet theory can still be applied perturbatively.

% a review on many-body Floquet methods
A large number of expansion techniques have been derived within the framework of Floquet theory to describe driven many-body systems. These include the high-frequency expansion of Van-Vlek perturbation theory in Sambe space in Ref.~\cite{Eckardt_2015}, the Floquet-Magnus expansion \cite{casas2001floquet,mananga2011introduction}, Brillioun-Wigner theory~\cite{PhysRevB.93.144307}, the effective Hamiltonian method for multi-step driving sequences~\cite{goldman2014periodically}, the Sambe space flow equation approach \cite{PhysRevLett.111.175301} for driven Boson-Hubbard model, and the Schrieffer-Wolff transform of the driven Fermi-Hubbard model in the strongly correlated regime \cite{PhysRevLett.116.125301,PhysRevLett.120.197601,PhysRevB.98.035116}. All these methods are essentially high-frequency expansions (HFE) in orders of the inverse driving frequency $\omega$. When the energy scales in the undriven system are comparable to, or larger than, the driving frequency, the high-frequency expansion becomes inaccurate, necessitating the development of Floquet theories beyond HFE. So far, only a few Floquet methods beyond the HFE have been constructed ~\cite{Rodriguez_Vega_2018,Eckardt_2015} based on the perturbative (block-) diagonalisation of the Floquet Hamiltonian in Sambe space. Crucially, these approaches require knowledge of the eigenbasis of the undriven system and thus cannot be applied to interacting many-body systems. The Sambe-space Gaussian elimination method in Ref.~\cite{PhysRevB.101.024303} suffers from the same applicability issue. Recently, a Flow equation approach beyond HFE was developed~\cite{PhysRevX.9.021037}, which doesn't require knowledge of the eigenbasis. Instead, the Floquet Hamiltonian is constructed based on consecutive infinitesimal unitary transforms. However, the flow-truncation error becomes unpredictable when multiple fixed points \cite{claassen2021flow} exist, and the micro-motion is hard to obtain, as it requires evaluating the unitary flow.

\section{Research contribution: The Floquet-induced interactions in driven many-body systems}
In this thesis, we analyse the enhancement of the Floquet-induced interactions by many-body correlations and develop an advanced Floquet method to understand the Floquet-induced interactions relevant for the coherent control of many-body systems. 

\subsubsection{Enhancement of Floquet-induced cavity-mediated interactions by correlations}
 
%The Sambe space Hamiltonian becomes exponentially large when describing a many-body system. Vogl's Sambe space Gaussian elimination method becomes our starting point for breaking the exponential wall. 
In Chapter \ref{Chapter4}, we upgrade the Gaussian elimination Floquet method in Ref.~\cite{PhysRevB.101.024303} with a projector technique, which allows us to study the Floquet-induced interactions in a cavity-QED setup, where the single-mode cavity is coupled to two-band Hubbard models. Our Floquet method describes how the intrinsic interactions in the undriven system modify (i.e., screen) the Floquet-induced cavity-mediated interactions. 
This example shows that, in many body systems, to accurately describe the Floquet-induced interactions, a correlated picture becomes necessary to understand the virtual processes mediating these interactions. In our example, this correlated picture means that it is the virtual exciton, rather than an uncorrelated virtual band-excitation, that provides the cavity-mediated interaction. To work in this correlated picture, we include the electron correlation at the beginning of our Floquet treatment. In contrast, in the uncorrelated picture (used in previous adiabatic elimination methods), the cavity-mediated interaction is first derived in a non-interacting system and then added with the intrinsic electron interaction to form the overall Hamiltonian. In our multi-band model, our Floquet treatment in the correlated picture provides an excitonic enhancement of cavity-mediated interactions compared to the uncorrelated picture. Via forming virtual excitons, the Floquet-induced cavity-mediated interaction and the Floquet band renormalisation are strongly broadened in reciprocal space.

\subsubsection{Obtaining the Floquet-induced interactions by solving Sylvester equations}
The above projector-based Gaussian elimination method faces problems when going beyond mean-field approximations and finding higher-order driving effects, prompting us to develop a more advanced many-body Floquet method.
%Sylvester equations are a central concept in classical control theory, which turned out to be also central in deciding the Floquet Schrieffer Wolff transform in driven quantum systems. 
To solve these problems, In Chapter \ref{Chapter5}, we develop a Floquet Schrieffer Wolff transform (FSWT) to obtain effective Floquet Hamiltonians and micro-motion operators of periodically driven many-body systems for any non-resonant driving frequency. Our FSWT perturbatively eliminates the oscillatory components in the driven Hamiltonian in orders of driving strength. This elimination is achieved by solving the operator-valued Sylvester equations. Our FSWT goes beyond various high-frequency expansion methods commonly used in Floquet theory. 

In Chapter \ref{Chapter6}, using the driven Fermi-Hubbard model as an example, we show how to solve Sylvester equations for a driven many-body system without knowing the many-body eigenstates. We apply FSWT in driven Hubbard systems to obtain the Floquet Hamiltonian and micro-motions with well-controlled approximations. Compared with Floquet high-frequency expansion methods, we demonstrate that the Floquet-induced interactions provided by FSWT offer a much more accurate prediction of the driven dynamics over the regimes of non-resonant driving parameters. We find FSWT is also very powerful in the strong-driving frame. 
The Floquet-induced interactions provided by this FSWT method will be useful for designing Rydberg multi-qubit gates, controlling correlated hopping in quantum simulations in optical lattices, and describing multi-orbital and long-range interacting systems driven in-gap.

\subsubsection{The complete Floquet-induced interactions in cavity-semiconductor setups}

In Chapter \ref{Chapter7}, with the help of FSWT, we obtain the Floquet-induced interactions in a more generalised laser-driven cavity-QED setup, with long-range Coulomb interactions and arbitrary laser polarisation. In this many-body system, we demonstrate how to solve the Sylvester equations on a Bloch-electron basis. At the mean-field level, this FSWT provides the consistent result obtained by the previous Gaussian elimination method. FSWT furthermore allows us to go beyond mean-field description and find the Floquet-induced interaction in the Coulomb interacting semiconductor model in the absence of the cavity. This cavity-independent Floquet-induced interaction is unavoidable when generating cavity-mediated interactions by laser driving. Our FSWT method thus offers a systematic way to predict the driving-induced phase transitions in many-body systems coupled to a cavity, where the Floquet-induced cavity-mediated interactions compete with the cavity-independent Floquet-induced interactions.

%\cite{weinberg2017adiabatic}

%Understand the Floquet side-band in driven non-interacting bands. 
%Does this Sambe-space side-band originate from micro-motions or is it a gauge effect? If it is a gauge effect, why can we probe these side-bands in the ARPES photo-current? 
%The answer is that ARPES photo-current can be generated by absorbing non-zero photons, which creates the side-band in the photo-current spectrum. \cite{PhysRevB.94.155304}

\section{Publications}
%$~$

~~~ \textbullet{} Xiao Wang, Dieter Jaksch, and Frank Schlawin. ``Excitonic enhancement of cavity-mediated interactions in a two-band Hubbard model." Physical Review B 109.11 (2024): 115137.
Chapter \ref{Chapter4} is based on the results of this publication. XW performed all the calculations and simulations. XW wrote the initial version of the manuscript and made substantial contributions to the following editing. 

\vspace{5mm}

\textbullet{} Xiao Wang, Fabio Pablo Miguel Méndez-Córdoba, Dieter Jaksch, and Frank Schlawin. ``Floquet Schrieffer-Wolff transform based on Sylvester equations." Physical Review B 110.24 (2024): 245108.
Chapters \ref{Chapter5} and \ref{Chapter6} are based on the results of this publication. XW performed all the calculations and simulations. XW wrote the initial version of the manuscript and made substantial contributions to the following editing. 

\section{Thesis Structure}
$~$

In Chapter \ref{Chapter2}, we introduce the Sambe space Floquet theory. We first explain how to understand the driven dynamics with the Sambe space matrix. Then, we provide two Floquet Hamiltonian methods to simplify the Sambe space matrix, the Gaussian elimination method and the block-diagonalisation method, in many-body systems relevant to the thesis.

In Chapter \ref{Chapter3}, we describe the driven cavity-material setup used in the thesis. We first define a two-band square lattice Hubbard model coupled to a single-mode cavity driven by a linearly polarised laser. This minimal model will be studied using the Gaussian elimination method. Then, we define a generalised cavity-semiconductor model with long-ranged Coulomb interactions and arbitrary band dispersion driven by a laser with arbitrary polarization. This generalised model will be studied using FSWT.

In Chapter \ref{Chapter4}, we incorporate a many-body projector technique into the Gaussian elimination Floquet method. This updated method is then applied to obtain the Floquet Hamiltonian in the minimal cavity-material model. The resulting screened Floquet Hamiltonian describes the excitonic enhancement of the Floquet-induced cavity-mediated interaction in this system. 
The projector-based Gaussian elimination method used in Chapter \ref{Chapter4} only works under certain mean-field approximations. We need to go beyond this mean-field limitation to study more generalised driven many-body systems. 

For this purpose, in Chapter \ref{Chapter5}, we develop the FSWT method. We show how to obtain the micro-motion and the Floquet Hamiltonian from the solution of Sylvester equations. We then compare the FSWT to other block-diagonalisation Floquet methods and explain how our FSWT can be reduced to the Gaussian elimination result. We explain why FSWT no longer suffers from the spurious Floquet-induced interactions, which hinder the application of Gaussian elimination methods. 
In Chapter \ref{Chapter6}, as the first example, we apply FSWT to study the Floquet-induced interactions in a driven Fermi-Hubbard system. We show how to solve the Sylvester equation for this many-body system and compare the FSWT Hamiltonian with the result given by HFE. We demonstrate that the Floquet-induced interactions given by FSWT play an essential role in describing the Hubbard chain's driven dynamics and deciding the pre-thermal phases. We show that the applicability range of our FSWT method is much broader than the HFE methods over the non-resonant driving parameter regimes.

In Chapter \ref{Chapter7}, we apply FSWT to the generalised driven cavity-semiconductor model in Chapter \ref{Chapter3} and obtain the complete Floquet-induced interactions therein. 
By solving the Sylvester equations on the Bloch basis, we again reveal the exciton enhancement effect on the Floquet-induced interactions. The FSWT is no longer restricted by the mean-field approximations made in the previous Gaussian elimination method. This allows us to obtain the complete Floquet-induced interactions in the generalised cavity-material systems, where the cavity-mediated part competes with the unavoidable cavity-independent part.

%% file: Chapters/SambeSpaceGaussianElimination.tex
In this chapter, we provide a detailed review of the Sambe space Floquet theory, which was outlined pedagogically in Section \ref{Floquet-pedalogical-intro}. We first explain how to obtain the stroboscopic dynamics and micro-motion of the driven system using the Sambe space matrix. Then, we review two Floquet Hamiltonian methods to simplify the Sambe space matrix in many-body systems: the Gaussian elimination method and the block-diagonalisation method. This thesis will upgrade these methods to obtain the Floquet-induced interactions in the following chapters \ref{Chapter4}, \ref{Chapter6}, and \ref{Chapter7}.

%\section{Introduction to Flquet theory}

\section{The Sambe space eigen-value problem}
We begin with a generic time-periodic Hamiltonian, denoted by $\hat{H}_t$ where $t$ represents time, acting on the Hilbert space of the driven system. This Hamiltonian has a periodicity of $T=2\pi/\omL$, with $\omL$ denoting the basic driving frequency. This time-periodicity allows us to write $\hat{H}_t$ as
\begin{equation}\label{H_t-original}
\hat{H}_t = \sum\limits_{j=-\infty}^{\infty} \hat{H}_j e^{i j\omL t}
\end{equation}
where $j\in \mathbb{Z}$ is the Fourier index, and $\hat{H}_j$ represents the Fourier component of $\hat{H}_t$ oscillating at frequency $j\omL$. We want to study how a wavefunction $\vert \psi \rangle_{(t)}$ evolves under this Hamiltonian $\hat{H}_t$ according to the time-dependent Schr\"odinger equation 
\begin{equation}\label{SchEq}
    i \partial_t \vert \psi \rangle_{(t)} = \hat{H}_t \vert \psi \rangle_{(t)} 
\end{equation}
where we take $\hbar=1$ throughout this chapter. There are multiple solutions of Eq.~(\ref{SchEq}) corresponding to different initial conditions of $\vert \psi \rangle_{(0)}$. These solutions can be made orthogonal to each other, forming an orthonormal, time-evolving basis 
\footnote{The orthogonality between the basis states is maintained during the unitary evolution given by Eq.~(\ref{SchEq}). }. 
With such a basis, we can describe the evolution of an arbitrary initial state $\vert \psi \rangle_{(0)}$ by decoupling it into the basis states.
This basis is not unique, as we can rotate to another basis via a unitary transform. According to the Floquet theorem \cite{floquet1883equations,Giovannini_2020}, for the time-periodic Hamiltonian $\hat{H}_t$, there exists a particularly convenient basis which is formed by the \textit{Floquet states}, similar to the stationary states for a static Hamiltonian. These Floquet states will be denoted by $\vert \alpha \rangle_{(t)}$ with index $\alpha$. Apart from satisfying the Schr\"odinger equation (\ref{SchEq}), they also satisfy a unique property:
After evolving over one full driving cycle $T$, a Floquet state $\vert \alpha \rangle_{(t)}$ is changed only by an $\alpha$-dependent phase factor, i.e., $\vert \alpha \rangle_{(t+T)} = e^{-i E_{\alpha} T} \vert \alpha \rangle_{(t)} $. This means that $e^{i E_{\alpha} t} \vert \alpha \rangle_{(t)} $ has a periodicity of $T=2\pi/\omL$, and thus it can be Fourier transformed as $ e^{i E_{\alpha} t} \vert \alpha \rangle_{(t)} =  \sum\limits_{j=-\infty}^{\infty} e^{i j \omL t} \vert \alpha,j\rangle$, where $\vert \alpha,j\rangle$ represents the $j$-th Fourier component. Then, $\vert \alpha \rangle_{(t)}$ can be expressed as
\begin{equation}\label{eigen-oscillation}
    \vert \alpha \rangle_{(t)} = e^{-i E_{\alpha} t} \sum\limits_{j=-\infty}^{\infty} e^{i j \omL t} \vert \alpha,j\rangle.
\end{equation}
In Eq.~(\ref{eigen-oscillation}), the Floquet state $\vert \alpha \rangle_{(t)}$ contains a set of oscillatory components $\vert \alpha,j\rangle$ with oscillation frequency $\Ea - j \omL$. The central frequency $\Ea$ is known as the quasi-energy. Each oscillatory component $\vert \alpha,j\rangle$ is referred to as the ``$j$-th harmonic". We note that $\vert \alpha, j\rangle$ is not a normalised wavefunction, while its inverse Fourier transform $e^{i E_{\alpha} t} \vert \alpha \rangle_{(t)}$ is. Inserting Eq.~(\ref{eigen-oscillation}) back to the Schr\"odinger equation (\ref{SchEq}), we find the quasi-energy $\Ea$ and the harmonics $\vert \alpha,j\rangle$ of the Floquet state $\vert \alpha \rangle_{(t)}$ satisfy the following eigenvalue relation
\begin{equation}\label{Sambe-general}
    \left(\begin{array}{ccccccc}
  & \ddots & \vdots & \vdots & \vdots & \vdots & \\
\cdots & \hat{H}_{1} & \hat{H}_0-\omL & \hat{H}_{-1}& \hat{H}_{-2} & \hat{H}_{-3} & \cdots \\
\cdots & \hat{H}_{2} & \hat{H}_{1} & \hat{H}_0 & \hat{H}_{-1}& \hat{H}_{-2} & \cdots \\
\cdots & \hat{H}_{3} & \hat{H}_{2} & \hat{H}_{1} & \hat{H}_0+\omL & \hat{H}_{-1}& \cdots \\
& \vdots & \vdots & \vdots & \vdots & \ddots & 
\end{array}\right)\left(\begin{array}{c}
\vdots \\
\vert \alpha,-1 \rangle \\
\vert \alpha,0 \rangle \\
\vert \alpha,1 \rangle \\
\vdots
\end{array}\right)
= E_{\alpha} 
\left(\begin{array}{c}
\vdots \\
\vert \alpha,-1 \rangle \\
\vert \alpha,0 \rangle \\
\vert \alpha,1 \rangle \\
\vdots
\end{array}\right).
\end{equation}
Here, the matrix on the left-hand side is known as the \textit{Sambe space matrix}
\footnote{Strictly speaking, this Sambe space matrix is an operator. In Eq.~(\ref{Sambe-general}), this operator gains a matrix structure in the basis of Floquet photon space.}.
It acts on the Sambe space,
%, which comprises the direct sum of infinite copies of the original Hilbert space.
which can be understood as a tensor product space between the original Hilbert space and the Floquet photon space.
This photon space is spanned by the orthonormal basis states $ \vert j \rangle $ labelled 
\footnote{This Floquet photon space contains states like $ \vert j<0 \rangle $. Physically, its index $j$ represents the number of photons emitted into the coherent state of the laser. }
by an integer $j \in \mathbb{Z}$.
In this photon space basis, the Sambe space eigenvector in Eq.~(\ref{Sambe-general}) can be compactly written as
\begin{equation}\label{Sambe-eigen-compact}
\vert \alpha \rangle = \sum_{j=-\infty}^{\infty} \vert \alpha, j \rangle \otimes \vert j \rangle.
\end{equation}
The basis state $ \vert j \rangle $ locates a state in the Sambe space into its $j$-th \textit{Floquet sector}. For example, we find $\vert \alpha, j \rangle$ is the $j$-th Floquet sector of the Sambe space eigenvector $\vert \alpha \rangle$, according to the inner product $ \langle j \vert \alpha \rangle = \vert \alpha,j\rangle$. In other words, in Eq.~(\ref{Sambe-general}), the $j$-th Floquet sector of the eigenvector $\vert \alpha \rangle$ represents its $j$-th row.
The Sambe space matrix in Eq.~(\ref{Sambe-general}) has a compact notation
\begin{equation}\label{Sambe-compact}
    \hat{\mathcal{S}} = \sum\limits_{j,j'=-\infty}^{\infty} (\delta_{j,j'} j \omL + \hat{H}_{j-j'} ) \otimes \vert j \rangle \langle j' \vert.
\end{equation}
With Eqs.~(\ref{Sambe-eigen-compact}) and (\ref{Sambe-compact}), the Sambe space eigenvalue problem (\ref{Sambe-general}) can be compactly written as
\begin{equation}\label{Sambe-eigenproblem-compact}
    \hat{\mathcal{S}} \vert \alpha \rangle = \Ea \vert \alpha \rangle.
\end{equation}
Once the Sambe space matrix $\hat{\mathcal{S}}$ is diagonalised, the quasi-energy $\Ea$ is obtained as the $\alpha$-th eigenvalue. Then, in the $j$-th Floquet sector of the corresponding eigenstate $\vert \alpha \rangle$, we find the $j$-th harmonic $\vert \alpha,j\rangle$. This shows that the $\alpha$-th Floquet state $\vert \alpha \rangle_{(t)}$ in the Hilbert space is entirely decided by the $\alpha$-th Sambe space eigenvector $\vert \alpha \rangle$. In other words, the eigenstate $\vert \alpha \rangle$ is the Sambe space representation of the Floquet state $\vert \alpha \rangle_{(t)}$ defined in Eq.~(\ref{eigen-oscillation}).

After we obtain the time-evolving basis spanned by the Floquet states $\vert \alpha \rangle_{(t)}$, the wavefunction $\vert \psi \rangle_{(t)}$ can then be decomposed into this basis, such that $\vert \psi \rangle_{(t)} = \sum_{\alpha} c_\alpha \vert \alpha \rangle_{(t)}$. Once the initial condition $\vert \psi \rangle_{(0)}$ is fixed, the time-independent coefficients $c_\alpha$ can be directly obtained from its overlaps with $\vert \alpha \rangle_{(0)}$. This completely solves the Schr\"odinger equation (\ref{SchEq}).

There is a redundancy in the above Sambe space Floquet theory: For the $\alpha$-th eigenstate $\vert \alpha \rangle$ of the Sambe space matrix $\hat{\mathcal{S}}$ with eigenenergy $\Ea$, the following shift (on the index of the harmonics) 
\begin{equation}\label{gauge-freedom-explain}
    \begin{split}
        \vert \alpha , j \rangle &\to \vert \alpha , j-m \rangle 
    \end{split}
\end{equation}
transforms $\vert \alpha \rangle$ in Eq.~(\ref{Sambe-eigen-compact}) into a new Sambe space eigenstate 
\begin{equation}
    \vert \alpha' \rangle = \sum_{j=-\infty}^{\infty} \vert \alpha, j-m \rangle \otimes \vert j \rangle.
\end{equation}
We find this $\vert \alpha' \rangle$ still solves the eigenvalue problem (\ref{Sambe-general}) with a shifted eigenvalue $(\Ea + m \omL)$, i.e.,
\begin{equation}
    \hat{\mathcal{S}} \vert \alpha' \rangle = (\Ea + m \omL) \vert \alpha' \rangle.
\end{equation}
However, these two Sambe space eigenstates, $\vert \alpha \rangle$ and $\vert \alpha' \rangle$, describe the same Floquet state (\ref{eigen-oscillation}) in the Hilbert space, because
\begin{equation}
\vert \alpha' \rangle_{(t)} = e^{-i (E_{\alpha}+m\omL) t} \sum\limits_{j=-\infty}^{\infty} e^{i j \omL t} \vert \alpha,j-m\rangle = e^{-i (E_{\alpha}+m\omL) t} \sum\limits_{j=-\infty}^{\infty} e^{i (j+m) \omL t} \vert \alpha,j\rangle =\vert \alpha \rangle_{(t)} .
\end{equation}
In conclusion, by applying the simultaneous transform
\begin{equation}\label{gauge-freedom}
    \begin{split}
        \Ea&\to \Ea + m \omL \\
        \vert \alpha , j \rangle &\to \vert \alpha , j-m \rangle 
    \end{split}
\end{equation}
onto the Sambe space eigenstate $\vert \alpha \rangle$ and its eigenvalue $\Ea$, we can obtain a different Sambe space eigenstate $\vert \alpha' \rangle$, but they correspond to the same Floquet state in the Hilbert space.
This can be understood as a gauge degree of freedom in Sambe space Floquet methods, which must be fixed to avoid the double-counting of the same Floquet state.
In the weak-driving case, this gauge freedom can be conveniently fixed by requiring the 0-th harmonic $\vert \alpha,0 \rangle$ to have the largest norm~\cite{Giovannini_2020}, i.e.,
\begin{equation}\label{gauge_requirement}
    \langle \alpha,0 \vert \alpha,0 \rangle > \langle \alpha, j \vert \alpha,j\rangle  ~~~ \forall j\neq0 .
\end{equation}
Under this gauge choice, the 0-th harmonic $\vert \alpha , j=0 \rangle$ is identified as the \textit{stroboscopic} orbital describing the macro-motion of the $\alpha$-th Floquet state $\vert \alpha \rangle_{(t)}$ in Eq.~(\ref{eigen-oscillation}), while the Floquet \textit{micro-motion} is produced by $\vert \alpha , j \neq 0 \rangle$ in other Floquet sectors
\footnote{There is another gauge-fixing method which is commonly used in Floquet theory: for example, in Ref.~\cite{PhysRevB.101.024303}, the quasi-energy $\Ea$ is further folded to the first Floquet zone where $\Ea \in (-\omL/2,\omL/2)$ using the transform (\ref{gauge-freedom}), which creates Floquet bands. 
However, since we will mainly focus on driven many-body systems (whose static Hamiltonian $\hat{H}_0$ contains an unbounded energy spectrum), this folding does not offer an advantage in our case. 
%which will overlap infinite points in spectrum spaced by $\omL$, thus 
Therefore, we stick to the extended Floquet zone where $\Ea \in (-\infty,\infty)$.}.

This Sambe space diagonalisation method has been used to study various driven systems \cite{Rodriguez_Vega_2018,PhysRevB.93.144307,mentink2015ultrafast,PhysRev.175.453,geier2021floquet,PhysRevB.96.014406, claassen2017dynamical,PhysRevLett.116.125301,okamoto2021floquet,PhysRevX.11.011057,liang2018floquet,li2022excitonic,Martinez_2003,yang2023theory,clark2019interacting}. 
Among these previous works, when the driven system can no longer be described by a few-body Hamiltonian, the complete diagonalisation of the Sambe space matrix $\hat{\mathcal{S}}$ becomes impossible since the Sambe space's dimension grows exponentially with particle number. In this case, we can still obtain an effective static description of the driven system from the Sambe space matrix (\ref{Sambe-compact}) without fully diagonalising it. Below, we explain how this is achieved.
%To study correlated systems driven with lower frequencies, one usually resort to numerical simulations like Floquet-DMFT~\cite{PhysRevB.93.144307}.

\section{Effective Floquet Hamiltonian}
Although the Sambe space matrix (\ref{Sambe-compact}) can no longer be directly diagonalised when considering a driven many-body system, there are various ways to reduce this Sambe space matrix into an effective Floquet Hamiltonian which acts on the original Hilbert space and governs the stroboscopic dynamics of the driven system. Below, we review some of the representative approaches to finding this effective Floquet Hamiltonian, which will be upgraded and utilised in the thesis.

\subsection{Gaussian Elimination of the Sambe space Matrix}\label{Sec2.2.1}
Gaussian Elimination of the Sambe space matrix is the method we will use to find the Floquet Hamiltonian in Chapter \ref{Chapter4}. As developed by Vogl et al. in Ref.~\cite{PhysRevB.101.024303}, this method decouples one Floquet sector from other sectors, which provides an effective static Hamiltonian in the decoupled Floquet sector. This is a low-frequency Floquet method which allows the driving frequency $\omL$ to be smaller than other energy scales in the driven system. This low-frequency method applies only to many-body systems driven by a monochromatic laser. The full time-dependent Hamiltonian for this special case reads
\begin{equation}\label{H_Dip_newform}
    \hat{H}_t = \hat{H}_0 + \hat{H}_{-1} e^{-i \omL t} + \hat{H}_{1} e^{ i \omL t}
\end{equation}
where $\hat{H}_0$ denotes the static many-body Hamiltonian without laser driving, $\hat{H}_{\pm1}$ denotes the driving term. We assume $\hat{H}_{\pm1} \propto g$, where $g$ represents the \textit{driving strength} which is proportional to the laser amplitude. The laser frequency acts as the basic driving frequency $\omL$. 

According to the Floquet theory mentioned above~\cite{PhysRevA.91.033416,Giovannini_2020,PhysRevB.101.024303}, solving the dynamics of the time-periodic Hamiltonian (\ref{H_Dip_newform}) is equivalent to diagonalising the following Sambe space matrix
\begin{equation}\label{Sambe}
    \left(\begin{array}{ccccccc}
\ddots & \vdots & \vdots & \vdots & \vdots & \vdots & \\
\cdots & \hat{H}_{1} & \hat{H}_0-\omL & \hat{H}_{-1}& 0 & 0 & \cdots \\
\cdots & 0 & \hat{H}_{1} & \hat{H}_0 & \hat{H}_{-1}& 0 & \cdots \\
\cdots & 0 & 0 & \hat{H}_{1} & \hat{H}_0+\omL & \hat{H}_{-1}& \cdots \\
& \vdots & \vdots & \vdots & \vdots & \vdots & \ddots
\end{array}\right)\left(\begin{array}{c}
\vdots \\
\vert \alpha,-1 \rangle \\
\vert \alpha,0 \rangle \\
\vert \alpha,1 \rangle \\
\vdots
\end{array}\right)
= E_{\alpha} 
\left(\begin{array}{c}
\vdots \\
\vert \alpha,-1 \rangle \\
\vert \alpha,0 \rangle \\
\vert \alpha,1 \rangle \\
\vdots
\end{array}\right).
\end{equation}
With the gauge choice in Eq.~(\ref{gauge_requirement}), an embedding procedure (i.e., the Gaussian elimination) can be carried out~\cite{PhysRevB.84.235108,PhysRevA.91.033416,Giovannini_2020}, in which we represent all $\vert \alpha,j\neq0\rangle$ by $\vert \alpha,0\rangle$ using Eq.~(\ref{Sambe})
\footnote{More specifically, according to the 0-th sector in Eq.~(\ref{Sambe}), we have $ \hat{H}_1 \vert \alpha, -1 \rangle + \hat{H}_0 \vert \alpha, 0 \rangle + \hat{H}_{-1} \vert \alpha, 1 \rangle = \Ea \vert \alpha, 0 \rangle $. In this relation, to express $\vert \alpha, -1 \rangle$ by $\vert \alpha,0\rangle$, we can use the following relation obtained from the $-1$-th sector in Eq.~(\ref{Sambe}), which gives $ \hat{H}_1 \vert \alpha, -2 \rangle + (\hat{H}_0 -\omL) \vert \alpha, -1 \rangle + \hat{H}_{-1} \vert \alpha, 0 \rangle = \Ea \vert \alpha, -1 \rangle $. Thus we have $\vert \alpha, -1 \rangle = (\Ea+\omL-\hat{H}_0)^{-1}\hat{H}_{-1}\vert \alpha, 0 \rangle + (\Ea+\omL-\hat{H}_0)^{-1}\hat{H}_{1}\vert \alpha, -2 \rangle$. The same procedure can be conducted to obtain $\vert \alpha, 1 \rangle = (\Ea-\omL-\hat{H}_0)^{-1}\hat{H}_{1}\vert \alpha, 0 \rangle + (\Ea-\omL-\hat{H}_0)^{-1}\hat{H}_{-1}\vert \alpha, 2 \rangle$. If we ignore $\vert \alpha, \pm 2 \rangle$, we will arrive at the weak-driving result Eq.~(\ref{Heff_weak_drive}). If we keep on this elimination procedure which replaces $\vert \alpha, \pm 2 \rangle$ by $\vert \alpha, \pm 1 \rangle$ and $\vert \alpha, 0 \rangle$ using the $\pm2$-th sector in Eq.~(\ref{Sambe}), we can get higher orders of the Gaussian elimination result, as represented in Eq.~(\ref{Heff}). }.
The eigenvalue problem~(\ref{Sambe}) then becomes equivalent to a \textit{self-consistent} eigenvalue problem
\begin{equation}\label{0-HarmonicEigenProblem}
    \hat{H}^{\text{eff}}_{(\Ea)} \vert \alpha,0\rangle = \Ea \vert \alpha,0\rangle
\end{equation}
where the effective Floquet Hamiltonian $\hat{H}^{\text{eff}}_{(\Ea)}$ reads
\begin{equation}\label{Heff}
\begin{split}
    \hat{H}^{\text{eff}}_{(\Ea)} &= \hat{H}_0 + \hat{H}_{1} \frac{1}{\Ea-\hat{H}_0 +\omL-\hat{H}_{1} \frac{1}{\Ea-\hat{H}_0 +2 \omL-\ldots} \hat{H}_{-1}} \hat{H}_{-1} \\
    & ~~~~~~~~ +
    \hat{H}_{-1} \frac{1}{\Ea-\hat{H}_0 -\omL-\hat{H}_{-1} \frac{1}{\Ea-\hat{H}_0 -2 \omL-\ldots} \hat{H}_{1}} \hat{H}_{1}.
\end{split}
\end{equation}
Note that $\hat{H}^{\text{eff}}_{(\Ea)}$ acts on the 0-th Floquet sector, i.e., the physical Hilbert space of the many-body system, which is much smaller than the Sambe space matrix in (\ref{Sambe}). This simplification comes at a price that $\hat{H}^{\text{eff}}_{(\Ea)}$ has to be determined self-consistently with its eigenvalue $\Ea$, see Ref. \cite{PhysRevB.101.024303} for how this is achieved in a driven two-level system. In addition, the requirement (\ref{gauge_requirement}) is necessary for the expression (\ref{Heff}) to converge.

%In the weak-driving limit, 
In case we can treat the driving strength $g$ perturbatively, we can expand the effective Floquet Hamiltonian (\ref{Heff}) in orders of $\hat{H}_{-1}$ and $\hat{H}_{1}$.
To the lowest order, we obtain the following Hamiltonian \cite{PhysRevB.101.024303}
\begin{equation}\label{Heff_weak_drive}
    \hat{H}^{\text{eff}}_{(\Ea)} \approx  \hat{H}_0 + \hat{H}_{1}\hat{G}^{0}_{(\Ea +\omL)}\hat{H}_{-1} + \hat{H}_{-1}\hat{G}^{0}_{(\Ea -\omL)}\hat{H}_{1},
\end{equation}
where the Green operator (i.e. the resolvent) $\hat{G}^{0}$ reads
\begin{equation}\label{G0-chap2}
    \hat{G}^{0}_{(E)} = \frac{1}{ E - \hat{H}_0 }.
\end{equation}
Up to now we are following Refs.~\cite{PhysRevA.91.033416,Giovannini_2020,PhysRevB.101.024303}, whose main results~(\ref{Heff}) and (\ref{Heff_weak_drive}) can be equivalently derived by a Floquet-Green operator formalism~\cite{Martinez_2003}. 
However, when the static Hamiltonian $\hat{H}_0$ represents a correlated electron system coupled to bosonic modes (which we will encounter in the next chapter), the formula (\ref{Heff_weak_drive}) is still too complex to evaluate, requiring further simplifications.

The high-frequency expansion is frequently used to further simplify (\ref{Heff_weak_drive}), where $\omL$ is assumed to be much larger than any other energy scales in $\hat{H}_{0}$, such that $\vert\vert \Ea - \hat{H}_{0} \vert\vert \ll \omL$ where $\vert\vert .. \vert\vert$ represents the matrix norm. Then we have $\hat{G}^0_{(\Ea\pm\omL)}\approx\pm 1/\omL$ in Eq.~(\ref{Heff_weak_drive}), and thus the Floquet Hamiltonian is reduced to
$ \hat{H}^{\text{eff}}_{(\Ea)} \approx  \hat{H}_0 + \frac{1}{\omL}[\hat{H}_{1},\hat{H}_{-1}] + O[\frac{1}{\omL^2}]. $ In this limit, the self-consistent requirement disappears, i.e., $ \hat{H}^{\text{eff}}_{(\Ea)}$ no longer depends on $\Ea$, which greatly simplifies the eigenvalue problem (\ref{0-HarmonicEigenProblem}).
However, this high-frequency limit cannot accurately describe the in-gap driving of a multiband material, which is the scenario we will explore below. 
%inter-band red-detuned driving, which is the scenario to be explored below. 
Consequently, techniques going beyond the high-frequency expansion must be developed to simplify Eq.~(\ref{Heff_weak_drive}).

To achieve this, in Chapter \ref{Chapter4}, we incorporate the Sambe space Gaussian elimination method with a many-body projector technique. We find this combination enables us to eliminate the self-consistency requirement in Eq.~(\ref{Heff_weak_drive}) beyond the high-frequency limit, such that $\hat{H}^{\text{eff}}_{(\Ea)} $ reduces to $ \hat{H}_{\rm eff}$ which no longer depends on $\Ea$. We will see that this improvement allows the Gaussian elimination method to capture screening effects of the cavity-mediated interactions in the in-gap-driven many-body systems. 

However, this projector-based Gaussian elimination technique can only eliminate the self-consistency requirement in Eq.~(\ref{Heff_weak_drive}) after making a certain type of Hartree-type mean-field approximation. When trying to study the Floquet-induced interactions beyond this Hartree-type approximation, our projector technique finds enormous difficulties in eliminating the self-consistency requirement.
In Chapter \ref{Chapter5} of this thesis, to address the self-consistency-related issues faced by the projector-based Gaussian elimination method used in Chapter \ref{Chapter4}, we will develop and utilise a Floquet Hamiltonian method, which is named ``Floquet Schrieffer Wolff transform based on Sylvester equations" (FSWT). This FSWT is one of the main results of the thesis. 

\subsection{Block-diagonalisation of the Sambe space Matrix}\label{Sec_Block-diagonal-intro}

Our FSWT method can be categorised as a block-diagonalisation approach applied to the Sambe space matrix. In this approach, a unitary transformation $\hat{J}$ is used to eliminate the inter-sector components of the Sambe space matrix $\hat{\mathcal{S}}$ in the following way: The transformed matrix is given by $\hat{\mathcal{S}}' = \hat{J} \hat{\mathcal{S}} \hat{J}^\dagger$. This transformation ensures that $\langle j | \hat{\mathcal{S}}' | j' \rangle = \delta_{j,j'} \langle j | \hat{\mathcal{S}}' | j \rangle$, effectively making it block-diagonal. The Floquet Hamiltonian can then be extracted directly from this diagonal part.

Compared with the Gaussian elimination method described above, the block-diagonalisation methods have a distinct feature: The block-diagonalisation simultaneously decouples all Floquet sectors with each other, such that all sectors are simultaneously block-diagonalised, while in the Gaussian elimination method, only the 0-th sector is decoupled from others. This sector-independent feature indicates that the block-diagonalisation methods can be understood in a reduced way, where the gauge degree of freedom (\ref{gauge-freedom}) is absent. This alternative way of understanding the Sambe space block-diagonalisation is explained below.  

\subsubsection{Sambe space block-diagonalisation as a time-periodic unitary transform}

We first outline the general procedure for time-dependent unitary transform, which forms the basis of our FSWT and other block-diagonalisation methods \cite{casas2001floquet,PhysRevB.93.144307,Eckardt_2015}. The evolution operator from time $t_0$ to $t$ under any time-dependent Hamiltonian $\hat{H}_t$ is denoted by $\hat{\mathcal{U}}_{t,t_0}$. It satisfies the Schr\"odinger equation $i\partial_t \hat{\mathcal{U}}_{t,t_0} = \hat{H}_t \hat{\mathcal{U}}_{t,t_0}$. Using an arbitrary time-dependent unitary transform $\hat{U}_t$, the evolution operator can be decomposed as $\hat{\mathcal{U}}_{t,t_0} = \hat{U}_t^{\dag} \hat{\mathcal{U}}'_{t,t_0} \hat{U}_{t_0}$. The evolution operator $\hat{\mathcal{U}}'_{t,t_0}$ satisfies another Schrodinger equation $i\partial_t \hat{\mathcal{U}}'_{t,t_0} = \hat{H}'_t \hat{\mathcal{U}}'_{t,t_0}$ with the transformed Hamiltonian $\hat{H}'_t = \hat{U}_t \hat{H}_t \hat{U}_t^{\dag} + i (\partial_t \hat{U}_t)\hat{U}_t^{\dag}$. By properly choosing $\hat{U}_t$, the evolution can be greatly simplified, such that $\hat{H}'_t$ becomes (approximately) time-independent, $ \hat{H}'_t \approx \hat{H}' $, and then the evolution operator $\hat{\mathcal{U}}_{t,t_0}$ is reduced to
\begin{align} \label{eq.propagator-chap2}
\hat{\mathcal{U}}_{t,t_0} = \hat{U}_t^{\dag} e^{-i\hat{H}'(t-t_0)} \hat{U}_{t_0}.
\end{align}

\begin{figure}
\centering
\includegraphics[width=0.45\textwidth]{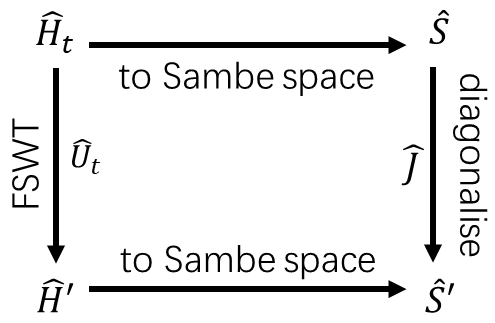}
\caption{
The relation between finding a unitary transform $\hat{U}_t$ that eliminates the oscillatory part in the driven Hamiltonian $\hat{H}_t$ (as achieved by our FSWT) and finding the matrix $\hat{J}$ which block-diagonalises the Sambe space matrix $\hat{\mathcal{S}}$.
%The relation between our FSWT and the Sambe space van Vleck block-diagonalisation in Ref.~\cite{Eckardt_2015}. The van Vleck method requires knowledge of the eigenbasis of the undriven system. This can be circumvented with the Sylvester equations in the FWST method.
}
\label{fig.relation-main}
\end{figure}

Next, following Refs.~\cite{goldman2014periodically,bukov2015universal}, we show that Sambe space block-diagonalisation is equivalent to finding a time-periodic unitary transform $\hat{U}_t$, which eliminates the time-dependence in the original Hamiltonian $\hat{H}_t$ in Eq.~(\ref{H_t-original}). We assume that $\hat{U}_t$ has the same period of $T=2\pi/\omL$ as $\hat{H}_t$, which transforms $\hat{H}_t$ into a time-independent Hamiltonian $\hat{H}'$ according to $\hat{H}' = \hat{U}_t \hat{H}_t \hat{U}_t^{\dag} + i (\partial_t \hat{U}_t)\hat{U}_t^{\dag}$. We expand $\hat{U}_t$ in terms of its Fourier coefficients $\hat{U}_j$, such that
\begin{equation}\label{Ut-Fourier}
    \hat{U}_t = \sum\limits_{j=-\infty}^{\infty} \hat{U}_j e^{i j \omL t}.
\end{equation}
We can directly check that the Sambe space matrix 
$\hat{\mathcal{S}} = \sum_{j,j'} (\delta_{j,j'} j \omL + \hat{H}_{j-j'} ) \vert j \rangle \langle j' \vert$, as given by Eq.~(\ref{Sambe-compact})
where $ \vert j \rangle $ is the basis state of the Floquet photon space,
becomes block-diagonalised in every Floquet sector after a unitary transform $ \hat{\mathcal{S}}' \to \hat{J} \hat{\mathcal{S}} \hat{J}^{\dag} $, where the matrix $\hat{J}$ is defined from $\hat{U}_j$ according to
\begin{equation}\label{Sambe-diag-J}
    \hat{J} = \sum\limits_{j,j'} \hat{U}_{j-j'} \otimes \vert j \rangle \langle j' \vert.  
\end{equation}
To see this, we consider the matrix element
\footnote{To the best of our knowledge, the construction of $\hat{J}$ and the derivation in Eq.~(\ref{check-bd}) are not provided in Refs.~\cite{goldman2014periodically,bukov2015universal} as well as other comprehensive reviews.}
%To check the transformed Sambe space matrix $\hat{S}'$ is indeed block-diagonalised, we look at its element 
\begin{equation}\label{check-bd}
\begin{split}
&\langle j \vert \hat{\mathcal{S}}' \vert j' \rangle = \langle j \vert \hat{J} \hat{\mathcal{S}} \hat{J}^{\dag} \vert j' \rangle \\
&= \sum_{j_1,j_2} \hat{U}_{j-j_1} (\delta_{j_1,j_2} j_1 \omL + \hat{H}_{j_1 - j_2} ) (\hat{U}_{j'-j_2})^{\dag} \\
&= \sum_{j_1,j_2,j_3} \int_{0}^{T} \frac{dt}{T} e^{i(j' - j_2 - j_3)\omL t} \hat{U}_{j-j_1} (\delta_{j_1,j_2} j_1 \omL + \hat{H}_{j_1 - j_2} ) (\hat{U}_{j_3})^{\dag} \\
&=  \int_{0}^{T} \frac{dt}{T} e^{i(j' - j )\omL t} \sum_{j_1} \hat{U}_{j-j_1} e^{i(j - j_1 )\omL t} \sum_{j_2} (\delta_{j_1,j_2} j_1 \omL + \hat{H}_{j_1 - j_2} ) e^{i(j_1 - j_2 )\omL t} (\sum_{j_3}\hat{U}_{j_3} e^{i j_3 \omL t})^{\dag} \\
&=  \int_{0}^{T} \frac{dt}{T} e^{i(j' - j )\omL t} \hat{U}_t \hat{H}_t \hat{U}_t^{\dag} + \sum_{j_1} \big( \hat{U}_{j-j_1} e^{i(j - j_1 )\omL t}  ( j_1 -j  ) \omL \hat{U}_t^{\dag}  + \hat{U}_{j-j_1} e^{i(j - j_1 )\omL t}  j   \omL \hat{U}_t^{\dag} \big)   \\
&= \int_{0}^{T} \frac{dt}{T} e^{i(j'-j)\omL t} \left( \hat{U}_t \hat{H}_t \hat{U}_t^{\dag} + i (\partial_t \hat{U}_t ) \hat{U}_t^{\dag} + j\omL \right) \\
&= \hat{H}'_{j-j'} + j\omL \delta_{j,j'}
\end{split}
\end{equation}
where $T=2\pi/\omL$ is the driving period. To derive the fifth and the sixth line, we have used the Fourier series expansion relations $ \sum_{j} \hat{H}_j e^{i j \omL t} =\hat{H}_t  $ and $ \sum_{j} \hat{U}_j e^{i j \omL t} = \hat{U}_t $. The last line identifies the Fourier components of the transformed Hamiltonian $\hat{H}'_t \equiv \hat{U}_t \hat{H}_t \hat{U}_t^{\dag} + i (\partial_t \hat{U}_t ) \hat{U}_t^{\dag}$. Since the $\hat{U}_t$ in our FSWT makes the transformed Hamiltonian $\hat{H}'_t$ time-independent, its Fourier component satisfies $\hat{H}'_{j-j'} \sim \delta_{j,j'}$, we see that $\hat{\mathcal{S}}'$ is indeed block-diagonalised. This shows that the properly chosen time-periodic unitary transform $\hat{U}_t$ equivalently block-diagonalises the Sambe space matrix, as schematically represented in Fig.\ref{fig.relation-main}.

The above relation means that the Floquet Hamiltonian given by the Sambe space block-diagonalisation can be obtained once we find the proper time-periodic unitary transform $\hat{U}_t$, which transforms Eq.~(\ref{H_t-original}) into a static Hamiltonian. The Floquet sector redundancy in Sambe space, Eq.~(\ref{gauge-freedom}), is completely absent in this unitary transform picture, and we can identify the transformed static Hamiltonian $\hat{H}'$ as the effective Floquet Hamiltonian governing the stroboscopic dynamics, whose eigenenergy provides the quasi-energy spectrum $\Ea$. Meanwhile, according to Eq.~(\ref{eq.propagator-chap2}), $\hat{U}_t^\dag$ transforms the stationary states of the stroboscopic Hamiltonian $\hat{H}'$ into the Floquet states $\vert \alpha \rangle_{(t)}$. This means that \cite{Eckardt_2015} $\hat{U}_t^\dag$ generates the sub-harmonic oscillations in the Floquet states from their stroboscopic dynamics. For this reason, the transform $\hat{U}_t$ is identified as the micro-motion operator.

Great efforts have been paid to find this time-periodic transform $\hat{U}_t$. Below, we review some commonly used block-diagonalisation methods. The ``Sylvester equation-based FSWT method" developed in Chapter \ref{Chapter5} belongs to the same block-diagonalisation category as these methods do, but we will demonstrate its advantage compared with these previous methods, in the application of understanding the Floquet-induced interactions. 

\subsubsection{High-Frequency expansion (HFE) in the Hilbert space}
HFE is a way to perturbatively block-diagonalize the Sambe space Floquet Hamiltonian in orders of the inverse driving frequency. The Magnus expansion \cite{casas2001floquet,PhysRevB.93.144307} is one of the most widely used HFE methods. However, we will review the HFE method proposed by Goldman and Dalibard \cite{goldman2014periodically}, which has a closer link to our FSWT method constructed in Chapter \ref{Chapter5}. In Ref. \cite{goldman2014periodically}, the form of the time-dependent unitary transform $\hat{U}_t$ in Eq.~(\ref{eq.propagator-chap2}) is chosen to be
\begin{equation}
    \hat{U}_t = e^{\hat{F}_t}
\end{equation}
where $\hat{F}_t$ is an anti-Hermitian operator. We assume $\hat{F}_t$ has a periodicity of $T=2\pi/\omL$, such that $\hat{F}_{t+T} = \hat{F}_t$. Using the derivative of the exponential map, the transformed Hamiltonian $\hat{H}'_t$ reads
\begin{equation}\label{H'-chap2}
\begin{split}
        \hat{H}'_t &= \hat{U}_t \hat{H}_{t} \hat{U}_t^{\dag} + i (\partial_t \hat{U}_{t}) \hat{U}_t^{\dag} \\
        &= \hat{H}_{t} + \hat{G}_{t} + \frac{1}{2!} [\hat{F}_t,\hat{G}_t] + \frac{1}{3!} [\hat{F}_t,[\hat{F}_t,\hat{G}_t]] + ...
\end{split}
\end{equation}
where
\footnote{The expansion in Eq.~(\ref{H'-chap2}) follows from the 
Baker–Campbell–Hausdorff formula and its infinitesimal case. Eq.~(\ref{H'-chap2}) can be understood as an expansion over orders of $\hat{F}_t$.}
the operator $\hat{G}_t$ is defined as
\begin{equation}\label{G_t-chap2}
    \hat{G}_{t} = [\hat{F}_{t},\hat{H}_{t}] + i \partial_t \hat{F}_{t}.
\end{equation}
%Eq.~(\ref{H'-chap2}) describes the relation between $\hat{H}'$, $\hat{H}_t$ and $\hat{F}_t$. In Magnus HFE, this relation is equivalently rewritten as the Magnus equation \cite{casas2001floquet,PhysRevB.93.144307}
%\begin{equation}\label{van-Vlek-eqn-chap2}
%\begin{split}
%i\partial_t \hat{F}_t &= \sum\limits_{k=0}^{\infty} \frac{B_k}{k!} (ad_{\hat{F}_t})^k \big[(-1)^{(k+1)} \hat{H}_t + \hat{H}' \big], 
%\end{split}
%\end{equation}
%where $ad_{\hat{F}_t}$ is an adjoint operator defined by $ad_{\hat{F}_t} \hat{X} = [\hat{F}_t,\hat{X}]$, and $B_k$ is the $k$-th Bernoulli number.
%In the high-frequency limit, we expand $\hat{F}_t$ and $\hat{H}'$ in Eq.~(\ref{van-Vlek-eqn-chap2}) in orders of $\omega^{-1}$ \cite{PhysRevB.93.144307}, 
In the high-frequency limit, we expand $\hat{F}_t$ and $\hat{H}'_t$ in orders of $\omega^{-1}$, such that
\begin{equation}\label{Dalibard_HFE}
\begin{split}
    \hat{H}'_t &= \sum_{n=0}^{\infty} \frac{1}{\omL^n} \hat{H}_t'^{(HFE,n)} ~~~~~~ \text{and}~~~~~~ 
    \hat{F}_t = \sum_{n=1}^{\infty} \frac{1}{\omL^n} \hat{F}_{t}^{(HFE,n)}.
\end{split}
\end{equation}
Inserting the original Hamiltonian $\hat{H}_t$ (\ref{H_t-original}) and the expansion (\ref{Dalibard_HFE}) into the transformed Hamiltonian $\hat{H}'_t$ in Eq.~(\ref{H'-chap2}), and requiring $\hat{H}'_t$ to be time-independent in each order of $\omL^{-1}$, we can perturbatively
\footnote{In the high-frequency expansion (\ref{Dalibard_HFE}), $\hat{F}_t$ starts with order $(1/\omL)^1$. However, since $\hat{F}_t$ is periodic in $T$, the derivative $i \partial_t \hat{F}_t$ starts with order $(1/\omL)^0$.}
determine $\hat{F}_{t}^{(HFE,n)}$, which eliminates the oscillatory part of $\hat{H}'_t$ in orders of $\omL^{-1}$.
The remaining static Hamiltonian, $\hat{H}'_t \approx \hat{H}'_{HFE}$, is identified as the HFE Floquet Hamiltonian, which reads
%\footnote{Here, we only show the HFE result for a monochromatically driven system described by Eq.~(\ref{H_Dip_newform}). See, e.g., Ref.~\cite{PhysRevB.93.144307} for the HFE Floquet Hamiltonian $\hat{H}'_{HFE}$ for the multi-frequency driven systems.}
\begin{equation}\label{van-Vlek-result-chap2}
\begin{split}
&\hat{H}'_{HFE} 
= \hat{H}_0  
+ \frac{1}{\omega} \sum_{j=1}^{\infty} \frac{1}{j} [ \hat{H}_j , \hat{H}_{-j} ] 
~~ +\frac{1}{2 \omega^2} \sum_{j=1}^{\infty} \frac{1}{j^2}\left(\left[\left[\hat{H}_j, \hat{H}_0\right], \hat{H}_{-j}\right]+\text { H.c. }\right) \\
&~~~~~ +\frac{1}{3 \omega^2} \sum_{j, l=1}^{\infty} \frac{1}{j l}\left(\left[\hat{H}_{j},\left[\hat{H}_{l}, \hat{H}_{-j-l}\right]\right]-2\left[\hat{H}_{j},\left[\hat{H}_{-l}, \hat{H}_{l-j}\right]\right]+\text { H.c. }\right)
+ \mathcal{O} (\omL^{-3}).
\end{split}
\end{equation}

The HFE result converges quickly only when the driving frequency $\omL$ is much larger than any other energy scales in the driven Hamiltonian $\hat{H}_t$. Thus, HFE suffers from a restricted application range. 
%To apply HFE in strongly correlated systems where the interaction strength is comparable to the driving frequency (see Ref.~\cite{PhysRevLett.116.125301} for example), we first have to make a frame rotation such that the interaction enters into the oscillation frequency. Then, we make the HFE in this rotating frame. However, this rotating frame HFE method still suffers the convergence issue when the driving frequency in this rotating frame becomes low, i.e., when the correlation strength becomes low enough to be comparable to the hopping.
For this reason, enormous efforts have been made to develop Floquet methods beyond HFE.

\subsubsection{The Van Vleck block-diagonalisation method in Sambe space }
Ref.~\cite{Eckardt_2015} shows that the Van Vleck method offers a systematic way to block-diagonalise the Sambe space matrix $\hat{\mathcal{S}}$ in Eq.~(\ref{Sambe-compact}). In this van Vleck method, the Sambe space block-diagonalisation transform $\hat{J}$ in Eq.~(\ref{Sambe-diag-J}) is written as a matrix exponential
\begin{equation}
    \hat{J} = e^{\hat{G}}
\end{equation}
where $\hat{G}$ is an anti-Hermitian matrix in Sambe space. Then $\hat{G}$ is obtained from the block-diagonalisation requirement, $\langle j \vert \hat{J} \hat{\mathcal{S}} \hat{J}^{\dag} \vert j' \rangle =0$ for $j\neq j'$, order by order in the driving strength $g$. 
We recall that $g$ measures the strength of the off-diagonal terms of the Sambe space matrix $\hat{\mathcal{S}}$ in Eq.~(\ref{Sambe-general}), and thus the expansion in $g$ goes beyond HFE. 
However, when used beyond the high-frequency limit, this Van Vleck method in Ref.~\cite{Eckardt_2015} is formulated using the eigenstates of the undriven Hamiltonian. Thus, it is hard to be applied to many-body systems. 

The FSWT method constructed in Chapter \ref{Chapter5} can be understood as achieving the same block-diagonalisation using the aforementioned time-periodic unitary transform picture of Floquet theory. FSWT achieves the block-diagonalisation in orders of $g$ (which the van Vleck method aims at) by solving the Sylvester equation, which doesn't suffer from the redundancies arising from the Floquet sectors in the Sambe space. In chapter \ref{Chapter6} and \ref{Chapter7}, we see that the Sylvester equation can be solved without knowing the eigenstates of the many-body system. Thus, our FSWT practically provides the many-body Floquet Hamiltonian, which is beyond the reach of the van Vleck method in Ref.~\cite{Eckardt_2015} (since the many-body eigenstates are unknown).

\section{Conclusion}

In this chapter, we have introduced the Floquet theory in Sambe space, which turns the time evolution problem in Eq.~(\ref{SchEq}) under the time-periodic Hamiltonian (\ref{H_t-original}) into the time-independent eigen-value problem (\ref{Sambe-general}) in Sambe space. By diagonalising the Sambe space matrix, the quasi-energy of a Floquet state can be obtained, together with its micro-motion and stroboscopic dynamics. In driven many-body systems where the complete diagonalisation of the Sambe space matrix becomes unfeasible, we can still reduce the Sambe space matrix to an effective Floquet Hamiltonian that governs the stroboscopic evolution of the system. We reviewed two types of methods for obtaining the many-body effective Floquet Hamiltonians. The first method is the Sambe space Gaussian elimination, which will be upgraded in Chapter \ref{Chapter4} to study the minimal cavity-material system described in Section \ref{subsec:setup&model}. The second method is the block-diagonalisation of the Sambe space matrix, which will be upgraded in Chapter \ref{Chapter5} to study the generalised cavity-material system described in Section \ref{sec:general-setup}.

\pagebreak

%% file: Chapters/setup.tex
In this thesis, the Floquet methods described in Chapter \ref{Chapter2} will be applied to obtain the driving-induced interactions in cavity-material setups, where a single-mode cavity is coupled to a 2d material. In this chapter, we construct the driving Hamiltonian $\hat{H}_t$ for these setups, where electrons interact with light via minimal coupling. In Section \ref{sec:dipolar-Hamiltonian-construct}, we will describe a minimal model for the driven cavity-material system, where the multi-band electron-electron interactions are on-site and the laser's polarisation is linear. We will use this minimal model in Chapter \ref{Chapter4} to analyse how electronic correlations can enhance the Floquet-induced interactions. In Section \ref{sec:general-setup}, we will extend this minimal setup into a generalised cavity-material setup, where the electron-electron interactions become long-ranged, and the laser can have an arbitrary polarisation. The complete Floquet-induced interactions in this generalised model will be obtained in Chapter \ref{Chapter7} using FSWT.

\section{The minimal model for driven Cavity-material setups}
\label{sec:dipolar-Hamiltonian-construct}

\begin{figure}[h]
\centering
\includegraphics[width=0.5\textwidth]{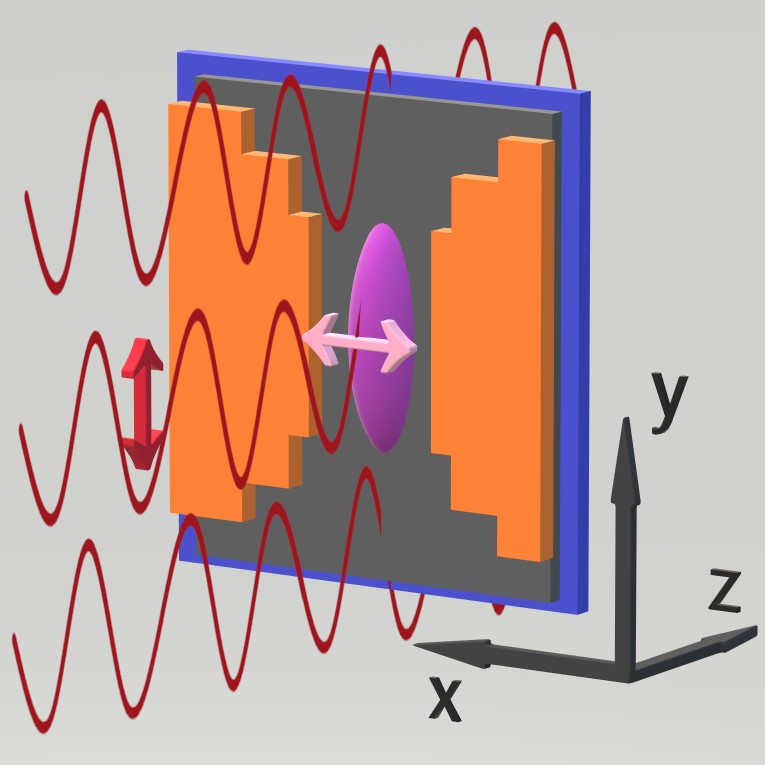} \captionsetup{justification=raggedright,singlelinecheck=false}
\caption{The laser-driven cavity-material system.  The red and the pink arrows denote the polarisation direction of the laser and the cavity mode, respectively. The laser field propagates along the $z$-axis. Here, we choose a square split-ring cavity structure similar to Ref.~\cite{butt2019multichannel} for illustration. The insulating substrate with relative permittivity $\ep_r$ is represented by the grey plane. }
\label{fig:setup}
\end{figure}

In Chapter \ref{Chapter4}, we will study the Floquet-induced interactions in a minimal laser-driven cavity-material system, whose setup is illustrated in Fig.~\ref{fig:setup}. It consists of a single-mode nanoplasmonic cavity with frequency $\omc$ (which we depict as an orange structure with a split) fabricated on a 2d electronic system (blue plane) and separated by a substrate (grey plane). A spatially uniform laser with frequency $\omL$ (red wavy line) is polarised along the y-axis, and propagates along the z-axis perpendicular to the material plane. The cavity mode, which is spatially confined in the purple region, is evanescently coupled to the material and polarised along the x-axis. Due to their perpendicular polarisation directions, the laser does not directly drive the cavity, so we only need to consider the laser-material coupling, but there is no direct laser-cavity coupling. 
We consider a material with only one electronic band crossing the Fermi level, and a large band gap to another empty band. The band gap is of a similar order of magnitude as the laser frequency $\omL$ and the cavity frequency $\omc$.% The 2d lattice contains $N$ sites.

%The reason why we consider evanescent cavity-material coupling is that, in our proposal, the driving laser must reach the material without pumping or being blocked by the cavity. 
Here we estimate the experimental detail of the exemplary setup in Fig.~\ref{fig:setup}: the nanoplasmonic cavity (orange structure) is about 50$\sim$100nm in diameter. Its split-gap is 5$\sim$20nm, supporting resonances in the infrared-to-optical regime ($\hbar\omc$ ranges from $ 1.5 \sim 2.6$eV)~\cite{butt2019multichannel}, close to the inter-band excitonic resonance of many molecular crystals~\cite{lof1992band,Cudazzo_2015,ni2022structural,PhysRevLett.110.216403}, perovskites~\cite{doi:10.1021/acs.jpclett.2c02436}, and TMD layers~\cite{chaves2020bandgap,rogalski20202d}. The grey plane represents an insulating substrate between the 2d material and the cavity. For monolayer molecular crystals, this substrate separation is realised by hBN in Ref. \cite{koo2021extraordinary}. 
%In real experiments, 
Apart from the split-ring cavity~\cite{ScalariG2013Ulca,maissen2014ultrastrong,rockstuhl2006resonances,KimNayeon2018CTFE,butt2019multichannel} considered here, the evanescent coupling between cavity and 2d-material can also be realised in other cavity-geometries, for example, in nanosphere plasmonic cavities~\cite{chikkaraddy2016single,kleemann2017strong}, in all-fiber Fabry–Perot cavities~\cite{PhysRevLett.115.093603,Wuttke:12}, and in photonic crystal cavities~\cite{gan2012strong}.  

% split-gap = 30nm, we find cavity resonance at 1.5eV in \ref{butt2019multichannel}.

%The Floquet Hamiltonian method described in Section \ref{Sec2.2.1} is most conveniently applied in the dipolar gauge, where the diamagnetic electron-laser coupling terms (oscillating at $e^{\pm 2 i \omL t}$) vanish. 
%We follow Ref.~\cite{PhysRevB.101.205140} and, starting from a minimal coupling Hamiltonian, carry out a multi-center Power-Zienau-Woolley (PZW) transformation (see Appendix \ref{appendix:dipolar}). We then truncate the continuum-space Hamiltonian to the two near-resonant electronic bands mentioned above, and invoke the electric dipole approximation. 

\subsubsection{The derivation of two-band Hamiltonian in Coulomb gauge}\label{subsec:setup&model}
We next derive a two-band Hamiltonian for this minimal setup, which will be analysed in Chapter \ref{Chapter4} by the Floquet method described in Section \ref{Sec2.2.1}. 
We start from the continuum-space Coulomb gauge minimal-coupling Hamiltonian \cite{cohen1998atom}, which describes a material driven by an off-resonant laser and coupled to a single-mode cavity, 
\begin{equation}\label{eq.H_Coulomb}
\begin{split}
    \hat{H}^{\text {Cont}}_t &= \sum\limits_{s} \int_{\mathbf{r}} \hat{\psi}_{s \mathbf{r} }^{\dag} \bigg( \frac{\big(\hat{\mathbf{p}}+ e \hat{\mathbf{A}}(\mathbf{r}, t) \big)^{2}}{2m} + V_{bg}(\mathbf{r}) -\mu \bigg) \hat{\psi}_{s \mathbf{r} } \\
    & + \frac{1}{2} \sum\limits_{s,s'} \int_{\mathbf{r}}\int_{\mathbf{r}'} \hat{\psi}_{s \mathbf{r} }^{\dag} \hat{\psi}_{s'\mathbf{r}'}^{\dag} \frac{e^2}{4\pi \epsilon_0 \epsilon_r \vert \mathbf{r}' - \mathbf{r} \vert}  \hat{\psi}_{s' \mathbf{r}' } \hat{\psi}_{s \mathbf{r} } \\
    & + \hbar \omega_c ~ \hat{a}^{\dag}\hat{a}
\end{split}
\end{equation}
where the bosonic operator $\hat{a}$ annihilates a cavity photon with frequency $\omega_c$, the fermionic operator $\hat{\psi}_{s \mathbf{r} }$ annihilates an electron with spin $s$ at position ${\bf r}$, $-e$ is the electron charge, $m$ is the electron mass, and $c$ is the speed of light, $\epsilon_0$ is the vacuum permittivity. $\hbar$ is the reduced Plank constant. $\mu$ is the electron chemical potential in the grand canonical ensemble. $V_{bg}(\mathbf{r})$ denotes the static lattice potential provided by the positively-charged ions in the 2d material, whose vibration (phonon) is omitted. The Coulomb interaction between electrons is screened by the background substrate with relative dielectric constant $\epsilon_r$. 
The  vector potential of the cavity mode and the laser obeys $\nabla \cdot \hat{\mathbf{A}}(\mathbf{r}, t) =0$ in the Coulomb gauge, it is given in Ref.~\cite{Bandrauk1995}, which reads 
\begin{equation}\label{A}
\begin{split}
    \hat{\mathbf{A}}(\mathbf{r}, t) &= \Omega_L  ~ \mathbf{u}_{L} \cos(\omL t - k_L  z) \\
    & ~ + \sqrt{\frac{\hbar}{ 2 \epsilon_{0} \epsilon_{r} \omega_{c} \mathcal{V}_{c}}}   \bigg( \mathbf{u}_{c}(\mathbf{r}) \varphi_{c}(\mathbf{r}) \hat{a} + \mathbf{u}_{c}^*(\mathbf{r}) \varphi_{c}^*(\mathbf{r})\hat{a}^{\dag} \bigg) .
\end{split}
\end{equation}
Here, the laser is uniform across the material, which has intensity $(\Omega_L)^2$ with polarisation unit vector $\mathbf{u}_{L}$ and driving frequency $\omL$. The 2d material plane is located at $z=0$, perpendicular to the wave-vector $\bf{k}_{L} \parallel \bf{z}$ of the laser field. We assume a linearly polarised laser with polarisation vector $\mathbf{u}_{L} \parallel \bf{y}$, aligned parallel to the material plane.
${\bf u}_c({\bf r})$ denotes the polarisation unit vector of the cavity mode. $\mathcal{V}_{c} = \Lambda (\frac{2\pi c}{\omega_c})^3$ is the cavity mode volume, and $\Lambda$ is the mode compression factor. The cavity mode wavefunction $\varphi_{c}(\mathbf{r})$ denotes the amplitude of finding the photon at position $\bf r$ which is normalised as $\int_{\mathbf{r}} \vert \varphi_{c}(\mathbf{r}) \vert^2 = \mathcal{V}_{c}$ . This mode wavefunction $\varphi_{c}({\bf r}) $ vanishes outside the purple region in Fig.~\ref{fig:setup}. For a linearly polarised cavity mode, we can take $\mathbf{u}_{c}$ and $\varphi_{c}$ to be real, satisfying the Helmholtz equation $\left(\nabla^{2}+ \epsilon_{r} \omega_{c}^{2} / c^{2}\right) \big(\mathbf{u}_{c}(\mathbf{r}) \varphi_{c}(\mathbf{r}) \big)=0$. In addition, the Coulomb gauge also requires $\nabla \cdot \big( \mathbf{u}_{c}(\mathbf{r}) \varphi_{c}(\mathbf{r}) \big) = 0$.
Importantly, we focus on a situation where the cavity mode is not directly driven by the laser. This is guaranteed by choosing the polarisation vectors of the cavity mode and the laser mode perpendicular to one another, i.e. we require $\mathbf{u}_{c}(\mathbf{r}) \perp \mathbf{u}_{L}$.

%We ignore the vector potential of all other environmental modes. Coupling to these environmental modes (a) makes the electrons interact with each other at light speed in Coulomb gauge, and (b) makes the life-time of the cavity mode finite.

Next, we derive an electronic band model from the above continuum space Hamiltonian by projecting the Hamiltonian~(\ref{eq.H_Coulomb}) on the Wannier orbitals of two electronic bands near the Fermi surface  
%This requires special care in the band-truncation process because the electrons are strongly coupled to a quantized photon mode~\cite{PhysRevB.101.205140}. To lower the band-truncation error, we follow Ref.~\cite{PhysRevB.101.205140} and first apply a multi-center PZW transformation, instead of directly projecting the Coulomb gauge Hamiltonian onto two Wannier bands. This gives us the dipolar Hamiltonian for a driven lattice system strongly coupled to the cavity. Then we apply the two-band-truncation, which results in  
\footnote{In this thesis, we do not consider the ultra-strong electron-cavity coupling regime \cite{frisk2019ultrastrong}, where the coupling strength reaches one-tenth of the cavity frequency, $\omega_c$. In such ultra-strong coupling cases, before truncating to low-energy bands, we may need to apply a multi-center Power-Zienau-Woolley (PZW) \cite{PhysRevB.101.205140} transformation to reduce the band truncation errors.}.
Specifically, we decompose the continuum-space electronic operator on an orthonormal Wannier basis, such that
\begin{equation}\label{Wannier-decompose}
    \hat{\psi}_{s{\bf r}} = \sum\limits_{{\bf R},b} w_{b({\bf r} - {\bf R})} \hat{c}_{{\bf R},b,s} ,
\end{equation}
where the operator $\hat{c}_{\mathbf{R} b s}$ annihilates an electron on a Wannier function $\langle {\bf r} \vert {\bf R} b \rangle = w_{b({\bf r} - {\bf R})} $ centered at the unit cell $\bf R$ with spin $s$ in band $b$. Here, the summation over $\bf R$ runs over all unit cells in the 2d material, whose total number is denoted by $N$.
The Wannier basis electron operator $\hat{c}_{\mathbf{R} b s}$ is obtained from the Bloch basis operator $\hat{c}_{{\bf k}bs}$ through the lattice Fourier transform $\hat{c}_{{\bf R}bs} = \frac{1}{\sqrt{N}} \sum_{\bf k} e^{i {\bf k} \cdot {\bf R}} \hat{c}_{{\bf k}bs} $, where the summation over $\bf k$ runs over all quasi-momenta 
\footnote{Here the quasi-momentum $\bf k$ is dimensionless as we take the lattice constant to be the unit length.}
in the first Brillouin zone.
The Bloch basis operator $\hat{c}_{{\bf k}bs}$ is chosen to diagonalise the single-electron part of the Hamiltonian (\ref{eq.H_Coulomb}), i.e.,
\begin{equation}\label{Bloch-choice}
    \sum\limits_{s} \int_{\mathbf{r}} \hat{\psi}_{s \mathbf{r} }^{\dag} \bigg( \frac{\hat{\bf p}^2 }{2m} + V_{bg}(\mathbf{r}) -\mu \bigg) \hat{\psi}_{s \mathbf{r} } = \sum_{{\bf k}, b, s} \big( \epsilon_{{\bf k}, b} -\mu \big) \hat{c}_{{\bf k} b s}^{\dag} \hat{c}_{{\bf k} b s},
\end{equation}
where $\epsilon_{{\bf k}, b}$ represents the bandstructure. We insert Eqs.~(\ref{A}), (\ref{Wannier-decompose}) and (\ref{Bloch-choice}) into the Hamiltonian (\ref{eq.H_Coulomb}), and truncate to the lowest two bands, where the lower band ($b=1$) crosses the Fermi level and the upper band ($b=2$) remains empty but is near-resonantly coupled to the lower band by the laser driving. This results in the following truncated two-band Hamiltonian
\begin{equation}\label{H-dip-full-terms}
\begin{split}
\hat{H}_{t} &= \sum_{{\bf k}, b, s} \big( \epsilon_{{\bf k}, b} -\mu \big) \hat{c}_{{\bf k} b s}^{\dag} \hat{c}_{{\bf k} b s} + \hbar ~ \omega_c ~ \hat{a}^{\dag}\hat{a} \\
& +    (e^{i \omL t} + e^{-i \omL t}) \sum_{\mathbf{R}, s} ( g  \hat{c}_{\mathbf{R} 2 s}^{\dag} \hat{c}_{\mathbf{R} 1 s}+\text { h.c. } ) \\
& +    (\hat{a}^{\dag} + \hat{a}) \sum_{\mathbf{R}, s} ( g_{c,{\bf R}} \hat{c}_{\mathbf{R} 2 s}^{\dag} \hat{c}_{\mathbf{R} 1 s} + \text { h.c. } ) \\
&+ \hat{U}_{11} + \hat{U}_{22} + \hat{U}_{12} ,
\end{split}
\end{equation}
where the interaction terms in the last line will be introduced later.
In Eq.~(\ref{H-dip-full-terms}), the second and third lines come from the paramagnetic electron-light coupling, i.e., the ${\bf p}\cdot {\bf A}$ term in Eq.~(\ref{eq.H_Coulomb}). Below, we explain how they are derived. In the Wannier basis given by Eq.~(\ref{Wannier-decompose}), this paramagnetic coupling term reads
\begin{equation}\label{H_paramagnetic_coup}
\begin{split}
\hat{H}_{{\bf p}\cdot {\bf A}}&=\frac{e}{m} \sum_{s} \sum_{{\bf R},{\bf R'}} \sum_{b,b'} \left( \int_{\mathbf{r}} w^*_{b({\bf r}-{\bf R})} ~ \hat{\bf A}({\bf r},t) \cdot \hat{\bf p} ~ w_{b'({\bf r}-{\bf R'})} \right)~  \hat{c}_{{\bf R}bs}^\dag \hat{c}_{{\bf R'}b's} \\
&\equiv \frac{e}{m} \sum_{s} \sum_{{\bf R},{\bf R'}} \sum_{b,b'} \langle {\bf R} b \vert \hat{\bf A}({\bf r},t) \cdot \hat{\bf p} \vert {\bf R'} b' \rangle~  \hat{c}_{{\bf R}bs}^\dag \hat{c}_{{\bf R'}b's} \\
&\approx \frac{e}{m} \sum_{s} \sum_{{\bf R},{\bf R'}} \sum_{b,b'} \hat{\bf A}({\bf R},t) \cdot \langle {\bf R} b \vert \hat{\bf p} \vert {\bf R'} b' \rangle~  \hat{c}_{{\bf R}bs}^\dag \hat{c}_{{\bf R'}b's}
\end{split}
\end{equation}
In the last line of Eq.~(\ref{H_paramagnetic_coup}), we apply the dipole approximation, where the vector potential is taken outside the integral. This dipole approximation is valid because the overlap between two Wannier orbitals, $ w_{b({\bf r}-{\bf R})}$ and $ w_{b'({\bf r}-{\bf R'})}$, rapidly vanishes as the distance between their centers, ${\bf R}-{\bf R'}$, increases. At this length scale (of few unit cells), the long-wavelength vector potential only shows negligible spatial variation, we can thus approximate it as $\hat{\bf A}({\bf r},t) \approx \hat{\bf A}({\bf R},t)$ and then take it outside the integral. 

In our minimal model, we also assume $ \langle {\bf R} b \vert \hat{{\bf p}} \vert {\bf R'} b' \rangle \propto \delta_{{\bf R},{\bf R'}}$, such that a Wannier electron cannot hop to another unit cell when interacting with light.
Furthermore, we assume the Wannier function $w_{b({\bf r})}$ has a definite parity, i.e., $w_{b(-{\bf r})} = \pm w_{b({\bf r})}$, such that the intra-band matrix element vanishes by symmetry, i.e, $\langle {\bf R} b \vert \hat{{\bf p}} \vert {\bf R} b \rangle =0$.
Under these approximations, we substitute the vector potential $\hat{\bf A}$ in Eq.~(\ref{A}) into the paramagnetic term (\ref{H_paramagnetic_coup}), which leads to the second and third lines of Eq.~(\ref{H-dip-full-terms}). The resulting site-independent electron-laser coupling strength $g$ and site-resolved electron-cavity coupling strength $g_{c,{\bf R}}$ in Eq.~(\ref{H-dip-full-terms}) are given by
\begin{equation}
    \begin{split}
        g &= \frac{e \Omega_L }{2m}  ~ \mathbf{u}_{L} \cdot \langle {\bf R} 2 \vert \hat{{\bf p}} \vert {\bf R} 1 \rangle ~~~~ \forall {\bf R} \\
        g_{c,{\bf R}} &= \frac{e}{m} \sqrt{\frac{\hbar}{2\varepsilon_{0} \varepsilon_{r} \omega_{c} \mathcal{V}_{c}}} \varphi_{c}(\mathbf{R}) \mathbf{u}_{c}(\mathbf{R}) \cdot \langle {\bf R} 2 \vert \hat{\bf p} \vert {\bf R} 1 \rangle,
    \end{split}
\end{equation}
%Due to the orthogonality of the Wannier function $\langle {\bf R} 2 \vert  {\bf R} 1 \rangle = 0$, 
where the site-independent inter-band matrix element is given by
\begin{equation}
\begin{split}
        &\langle {\bf R} 2 \vert \hat{{\bf p}} \vert {\bf R} 1 \rangle  = \int d{\mathbf{r}} \; w_{2({\bf r} - {\bf R})}^{*}  (-i \hbar \nabla_{\bf r})  w_{1({\bf r} - {\bf R})} 
        = \int d{\mathbf{r}} \; w_{2({\bf r})}^{*} (-i \hbar \nabla_{\bf r}) w_{1({\bf r})}.
\end{split}
\end{equation}
The Fourier transform of the electron-cavity coupling coefficient reads $ g_{c,{\bf q}} = \frac{1}{\sqrt{N}} \sum_{\bf R} e^{-i {\bf q} \cdot {\bf R} } g_{c,{\bf R}}$, which represents the amplitude of an inter-band electron-cavity scattering event in Eq.~(\ref{H-dip-full-terms}), during which a cavity photon is absorbed or emitted and the electron's quasi-momentum changes by $\bf q$. 
If the cavity mode encapsulates the full lattice but vanishes outside it, i.e., $\varphi_{c}(\mathbf{r})=1$ at position $\bf r$ inside the lattice, and $\varphi_{c}(\mathbf{r})=0$ at position $\bf r$ outside the lattice, then we have
\begin{equation}
    \vert g_{c,{\bf q}={\bf 0}} \vert^2 = N \vert g_{c,{\bf R}} \vert^2 = N \frac{e^2}{m^2} \frac{\hbar}{2\varepsilon_{0} \varepsilon_{r} \omega_{c} \mathcal{V}_{c}} 
    \big(  \mathbf{u}_{c} \cdot \langle {\bf R} 2 \vert \hat{\bf p} \vert {\bf R} 1 \rangle   \big)^2,
\end{equation}
which is finite in the thermodynamic limit $N\to \infty$, because $ \mathcal{V}_{c} \sim N a^2 h $ where $a$ is the lattice constant, and $h$ is the effective height of the cavity mode (perpendicular to the material plane). In nanoplasmonic cavities, $h$ can be much smaller than the vacuum wavelength of light at frequency $\omc$. For organic molecules coupled to cavity, $N \vert g_{c,{\bf R}} \vert\sim 1$ eV has been realised \cite{doi:10.1021/ph500266d,frisk2019ultrastrong}, much stronger than the electron-cavity coupling strength we will consider in this thesis ($\vert g_{c,{\bf R}} \vert \lesssim 0.1$ eV).

The diamagnetic electron-light coupling term, i.e., the $\hat{\bf A}^2$ term in Eq.~(\ref{eq.H_Coulomb}), has been ignored in the truncated Hamiltonian Eq.~(\ref{H-dip-full-terms}) for the following reasons. According to Eq.~(\ref{A}), the squared vector potential $\hat{\bf A}^2$ contains 3 terms: The laser-laser interaction, the cavity self-interaction, and the laser-cavity interaction. First, the laser-laser interaction is spatially homogeneous, which only contributes a term $(\Omega_L)^2 \cos^2(\omL t) \hat{N}_e$ in Eq.~(\ref{eq.H_Coulomb}) where $\hat{N}_e$ denotes the total electron number. Since $\hat{N}_e$ is a conserved quantity, we can ignore this laser-laser interaction in Eq.~(\ref{H-dip-full-terms}). Second, the cavity self-interaction in $\hat{\bf A}^2$ is time-independent, which only becomes relevant in the ultra-strong electron-cavity coupling regime \cite{frisk2019ultrastrong}, e.g., when $g_c \gtrsim 0.1 \omc$. For the parameters considered in this thesis, this self-interaction term can be ignored. Finally, the laser-cavity interaction in $\hat{\bf A}^2$ is proportional to the mode-overlap between laser and cavity. Since we require $\mathbf{u}_{c}(\mathbf{r}) \perp \mathbf{u}_{L}$, this final term completely vanishes. Hence, the entire diamagnetic electron-light coupling term can be excluded from the truncated Hamiltonian Eq.~(\ref{H-dip-full-terms}).

\subsubsection{The minimal driving model with Hubbard interaction}
We collect the truncated driving Hamiltonian in Eq.~(\ref{H-dip-full-terms}) in the form of Eq.~(\ref{H_Dip_newform}), i.e., 
\begin{equation}\label{Ht-chap3.1}
    \hat{H}_t = \hat{H}_0 + \hat{H}_{-1} e^{-i \omL t} + \hat{H}_{1} e^{ i \omL t}
\end{equation} 
where $\hat{H}_0$ denotes the static part and $\hat{H}_{-1}$ denotes driving term. Here, $\hat{H}_{0}$ reads in second quantisation
\begin{equation}\label{H0}
    \begin{split}
        \hat{H}_0 &= \hat{H}_b  + \hat{H}_c  \\
    \end{split}
\end{equation}
where $\hat{H}_b$ includes the bare (i.e., decoupled) material and cavity Hamiltonians and $\hat{H}_c$ describes the cavity-material coupling. We first describe $\hat{H}_b$, which reads
\begin{equation}\label{H1}
    \hat{H}_b = \hat{h} + \hat{U}, 
\end{equation}
which we further split into the single-particle Hamiltonian $\hat{h}$ and the inter-electron interaction $\hat{U}$. 
The single-particle Hamiltonian $\hat{h}$ reads
\begin{equation}\label{H1h}
    \begin{split}
        \hat{h}   &= \hbar \omc  \hat{a}^{\dag}\hat{a} + \sum_{{\bf{k}}, b, s} \big( \ep_{{\bf{k}}, b} -\mu \big) \hat{c}_{{\bf{k}} b s}^{\dag} \hat{c}_{{\bf{k}} b s}.  
    \end{split}
\end{equation}
%Here, the bosonic operator $\hat{a}$ annihilates a cavity photon with frequency $\omc$, the fermionic operator $\hat{c}_{{\bf q} b s}$ annihilates an electron with band index $b$, spin $s$ and quasi-momentum $\bf q$. Here $\bf q$ is dimensionless as we take the lattice constant to be the unit length. The number of sites of the 2d lattice is denoted by $N$.
%The various terms will be explained term by term in the following.
%The static Hamiltonian $\hat{H}_0$ is split into two parts, the "decoupled electron-cavity" part $\hat{H}_1$ and the electron-cavity coupling part $\hat{H}_c$. 
As introduced above, the chemical potential $\mu$ crosses only with the lower band with band-index $b=1$, so that the Fermi surface lies solely in the lower band, and the upper band is empty in the absence of external driving. For reasons explained later, we focus on a situation where the chemical potential is very close to the top of the lower-band, such that the Fermi surface is a tiny circle, e.g., with radius $k_F\sim\pi/30$ in units of inverse lattice constant.

For the interaction term $\hat{U}$ in Eq.~(\ref{H1}), in our minimal model, we only consider local electron density-density repulsion in the two-band model, such that 
\begin{equation}\label{H1U}
\hat{U}   = \hat{U}_{11} + \hat{U}_{22} + \hat{U}_{12}.
\end{equation}
Specifically, the intra-band repulsion terms read, for $b=1,2$,
\begin{equation}\label{U11-term}
    \begin{split}
        \hat{U}_{bb} = \frac{ U_{bb} }{2N}  \sum_{{\bf k},{\bf k'},{\bf q}, s \neq s'}  \hat{c}_{{\bf k}-{\bf q} + {\bf G} b s}^{\dag} \hat{c}_{{\bf k'}+{\bf q}+ {\bf G'} b s'}^{\dag} \hat{c}_{{\bf k'} b s'} \hat{c}_{{\bf k} b s},
    \end{split}
\end{equation}
where the quasi-momenta ${\bf k},{\bf k'},{\bf q}$ all belong to the first Brillouin zone, and the reciprocal lattice vectors ${\bf G}$ and ${\bf G}'$ are chosen such that ${\bf k}-{\bf q} + {\bf G}$ and ${\bf k'}+{\bf q}+{\bf G'}$ are again in the first Brillouin zone. 
Since we consider an on-site intra-band repulsion, only opposite spin states interact with one another
%Note that the ${\bf q}=0$ contribution is included in Eq.~(\ref{U11-term}), in contrast to the usual treatment in one-band jellium models \cite{bruus2004many}. 
%As we will see in section \ref{sec:correlated-model-k-GRPA}, it gives rise to the Hartree contribution in the screening calculation. 
%Here we exclude the $s=s'$ in the summation, because these nonphysical on-site inter-band same-spin self-interaction terms naturally vanish during Fourier transform.
% If we consider intra-band long-range repulsion, then the s=s' part will exist, giving rise to intra-band Fock self energy effect.
\footnote{Note that the ${\bf q}=0$ contribution is included in (\ref{U11-term}), in contrast to the usual treatment in one-band jellium models \cite{bruus2004many}. In two-band models, this ${\bf q}=0$ contribution can no longer be exactly cancelled by the positive background charges. Thus we keep the ${\bf q}=0$ term, meanwhile the influence of the background charges is accounted into the bandstructure (and the chemical potential).}.
Similarly, we write the inter-band repulsion $\hat{U}_{12}$ as
\begin{equation}\label{U12-term}
    \begin{split}
        \hat{U}_{12} = \frac{ U_{12} }{N}  \sum_{{\bf k},{\bf k'},{\bf q}, s, s'}  \hat{c}_{{\bf k}-{\bf q} + {\bf G} 1 s}^{\dag} \hat{c}_{{\bf k'}+{\bf q}+ {\bf G'} 2 s'}^{\dag} \hat{c}_{{\bf k'} 2 s'} \hat{c}_{{\bf k} 1 s}.
    \end{split}
\end{equation}
Note that the factor $\frac{1}{2}$ in Eq.~(\ref{U11-term}) is absent here because $\hat{U}_{12}$ comprises two equivalent terms (i.e., $\hat{c}_2^\dag \hat{c}_1^\dag \hat{c}_1 \hat{c}_2$ and $\hat{c}_1^\dag \hat{c}_2^\dag \hat{c}_2 \hat{c}_1$) arising from the band-truncation. With the exception of few counter-examples~\cite{nomura2015unified}, the intra-band repulsion is generally stronger than the inter-band one, i.e. $U_{11} > U_{12}$.

\begin{figure}[h]
\centering
\includegraphics[width=0.55\textwidth]{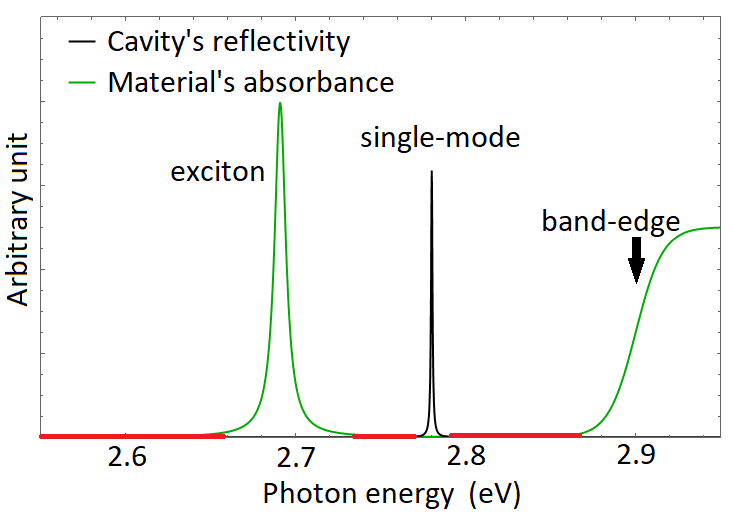} \captionsetup{justification=raggedright,singlelinecheck=false}
    \caption{Schematic illustration of the optical resonances in the cavity-material system. The black curve represents the reflectivity of the single-mode cavity. 
    The green curve represents the absorbance spectrum of our two-band Hubbard model.  The exciton and band-edge resonant frequency are respectively estimated from Eq.~(\ref{exciton-resonance}) and (\ref{band-resonance}), based on the parameters given in Eq.~(\ref{dispersion}). 
    When the driving frequency lies within the off-resonant regime (red region) with low absorbance, the laser-heating on the cavity-material system is suppressed.
    %This allows the driven system to be described by a Floquet Hamiltonian for a longer time.
    The lower-band is chosen to be near fully occupied with a tiny Fermi surface (e.g., $k_F\sim\pi/30$, corresponding to a hole-doping of $\sim5\times 10^{11}~\text{cm}^{-2}$ in tetracene), such that the exciton resonance is not evidently suppressed by the scattering with the free charge carriers.
    }
    \label{fig:Frenkel-resonance}
\end{figure}

These on-site repulsion terms generate Frenkel-type excitons in the material, as illustrated by the optical absorbance in Fig.~\ref{fig:Frenkel-resonance}. Frenkel excitons can be understood as tightly bound pairs of upper-band electrons and lower-band holes, formed due to inter-band electron-electron repulsion. Our interaction $\hat{U}$ gives a prototypical description of Frenkel excitons found in organic molecular crystals~\cite{lof1992band,PhysRevLett.110.216403}, perovskites~\cite{doi:10.1021/acs.jpclett.2c02436} and the bi-layer semiconductor structures listed in Ref.~\cite{Kuneš_2015,rademaker2013determinant}. 
In Chapter \ref{Chapter4}, the strength of the on-site interactions are chosen to match a tetracene-type molecular crystal~\cite{Cudazzo_2015}, which is known to host these Frenkel excitons. 
%To plot Fig.~(\ref{fig:Frenkel-resonance}), 
In particular, we take $U_{11}=1.6$ eV, $U_{12}=0.8$ eV. We further assume a simple square lattice bandstructure 
\begin{equation}\label{dispersion}
    \ep_{{\bf k},b} = \ep_{b} + 2 t_{b} \big(cos(k_x) + cos(k_y) \big)
\end{equation}
with band-index $b=1,2$. This dispersion gives the following momentum-dependent bandgap 
\begin{equation}\label{k-dependent-band-gap}
    \ep_{{\bf k},21 } = \ep_{21} + 2 t_{21} \big(cos(k_x) + cos(k_y) \big)
\end{equation}
where $t_{21}\equiv t_2 -t_1$, $\ep_{21}\equiv \ep_2 - \ep_1$. Unless specified otherwise, the band parameters are chosen as $\ep_{21}=3.7$ eV, $t_{21}=-0.2$ eV, with $t_{1}=0.05$ eV, $t_{2}=-0.15$ eV. These parameters are used in Fig.~\ref{fig:Frenkel-resonance}. We note, however, that all of the analytical results derived in Chapter \ref{Chapter4} remain valid for arbitrary bandstructures $\ep_{{\bf k},b}$. 
%, which can be more realistic than the one shown in Eq.~(\ref{dispersion}).

In this prototypical material model, we only consider the direct on-site repulsions, but ignore the local field effect of the Coulomb interaction~\cite{RevModPhys.74.601,PhysRevLett.34.155}, i.e., in (\ref{U11-term}) and (\ref{U12-term}) we will ignore the Umklapp (${\bf G}+{\bf G'}\neq0$) processes. This local field effect could be important, for instance, when there are several molecules in a unit cell (creating bands with exchange interactions) and the material shows strong Davydov splitting and anisotropic optical absorbance~\cite{Cudazzo_2015} \footnote{Specifically, the omitted local field terms include the inter-site dipolar interaction, which also contributes to the inter-site hopping of excitons~\cite{Cudazzo_2015}. We ignore this interaction in our prototypical Hubbard model, and focuses solely on the bandstructure's contribution to the exciton transport.}. The main result presented in Chapter \ref{Chapter4} will be independent of these material-specific details, thus they are not included in our prototypical model. 
%In future work, these additional terms could be included in the electron interaction $\hat{U}$ with more than 2 electron bands, to provide more accurate Floquet effective Hamiltonians for organic crystals. Moreover, by considering long-range repulsion, our Floquet method can also study materials which host Wannier-Mott excitons.

In Eq.~(\ref{H0}), the term $\hat{H}_c$ describes the electron-cavity coupling, which reads
\begin{equation}\label{H2}
    \begin{split}
        \hat{H}_c &=    (\hat{a}^{\dag} + \hat{a}) \frac{1}{\sqrt{N}} \sum_{{\bf{k}},{\bf q}, s} ( g_{c,{\bf q}} \hat{c}_{{\bf{k}+{\bf q}} 2 s}^{\dag} \hat{c}_{{\bf{k}} 1 s} + \text { h.c. } ).
    \end{split}
\end{equation}
Here, the coupling constant $g_{c,{\bf q}}$ depends only on the transferred momentum $\bf q$, due to the approximations employed on Eq.~(\ref{H_paramagnetic_coup}). %, but not on the absolute incoming momentum $\bf k$. 
%We assume that 
In our minimal model, the on-site inter-band dipolar transition dominates the coupling, while the inter-site electron-light coupling, for example given by the Peierls phase coupling \cite{peierls1996quantum}, is ignored in $\hat{H}_c$.
%As we work in the dipolar gauge, there is also a dipolar self-energy term from the cavity-electron coupling in the full Hamiltonian (see Appendix~\ref{appendix:dipolar}), which we neglect in Eq.~(\ref{H0}). This energy is proportional to $\vert g_c \vert^2 / \omc$, which is much weaker than the energy scales we will consider in the following (see Section~\ref{sec.in-gap-driving} below). 
Overall, our cavity-material system resembles the one proposed in Ref.~\cite{PhysRevB.106.205401}, while in our case, we assume the cavity resonance is detuned from the exciton resonance.

Finally, the electron-laser coupling term $\hat{H}_{-1}$ in Eq.~(\ref{Ht-chap3.1}), i.e., the driving term, reads
\begin{equation}\label{D}
    \hat{H}_{-1} = \left( \hat{H}_{1} \right)^\dag =  \sum_{{\bf{k}}, s} ( g \hat{c}_{{\bf{k}} 2 s}^{\dag} \hat{c}_{{\bf{k}} 1 s}+\text { h.c. } ),
\end{equation}
where we invoke the same approximations on Eq.~(\ref{H_paramagnetic_coup}) as in the cavity-electron interaction $\hat{H}_c$. As the laser propagates perpendicularly through the material with uniform strength at each lattice site, the electron-laser coupling constant $g$ only allows optical (de-)excitations with vanishing momentum transfer. The electron-laser coupling $g$ is proportional to the square root of laser intensity, and we consider in particular strong driving, such that $  g \gg g_c $. In this case, the Floquet-induced interaction becomes much stronger than the intrinsic cavity-mediated interaction in the undriven material.
%, so that $  g \gg g_c $ for "MW/cm2" lasers. 
%The direct laser-cavity coupling vanishes due to their perpendicular polarization vector.

%We have finished the setup description of our driven cavity-material system. 

\subsubsection{The RWA and non-RWA part of the Hamiltonian}
Under the rotating wave approximation (RWA), where fast oscillating light-matter interaction terms are neglected, the electron-cavity coupling $\hat{H}_c$ becomes
\begin{equation}\label{H2-RWA}
    \begin{split}
        \hat{H}_c^{\text{RWA}} &=    \hat{a} \frac{1}{\sqrt{N}} \sum_{{\bf{k}},{\bf q}, s}  g_{c,{\bf q}} \hat{c}_{{\bf{k}+{\bf q}} 2 s}^{\dag} \hat{c}_{{\bf{k}} 1 s} + \text { h.c. } 
    \end{split}
\end{equation}
and the driving term $\hat{H}_{-1}$ becomes
\begin{equation}\label{D-RWA}
    \hat{H}_{-1}^{\text{RWA}} =  \sum_{{\bf{k}}, s}  g \hat{c}_{{\bf{k}} 2 s}^{\dag} \hat{c}_{{\bf{k}} 1 s}.
\end{equation}
In Chapter \ref{Chapter4}, we will compare the effective Hamiltonians generated by $\hat{H}_{-1}^{\text{RWA}}$ with those obtained from $\hat{H}_{-1}$, where non-RWA driving effects are present. We will not apply this RWA throughout the derivations of the Floquet Hamiltonian in Chapter \ref{Chapter4}. This approximation will only be used in Appendix \ref{sec.GRPA}, where our Floquet result in Chapter \ref{Chapter4} is verified by an alternative rotating frame diagrammatic method.

\section{Generalised driven semiconductor-cavity setups}\label{sec:general-setup}
In Chapter \ref{Chapter7}, we will consider a similar but more generalised driven semiconductor-cavity model, where the electronic repulsion becomes long-ranged, and the laser contains an arbitrary polarisation. The driven Hamiltonian reads,
\begin{equation}\label{Ht-example-chap5}
\hat{H}_t = \hat{H}^{(0)} + \hat{H}^{(1)}_{-1}e^{-i\omL t} + \hat{H}^{(1)}_1e^{i\omL t}
\end{equation}
Here, the upper scripts denote the order of driving strength $g$. They are added to match the FSWT method used in Chapter \ref{Chapter7}. The undriven Hamiltonian, denoted as $\hat{H}^{(0)}$ here, is again given by
\begin{equation}
\hat{H}^{(0)} = \hat{h} + \hat{H}_c + \hat{U}.
\end{equation}
Here, the single-particle Hamiltonian $\hat{h}$ and the electron-cavity coupling $\hat{H}_c$ are unchanged, given by Eqs.~(\ref{H1h}) and (\ref{H2}) respectively. Without driving, the upper band is again empty. However, in Chapter \ref{Chapter7}, instead of using the trivial bandstructure in Eq.~(\ref{dispersion}), more generalised bandstructures $\ep_{{\bf k},b}$ will be considered in $\hat{h}$.  

In our generalised setup, the two-band interaction term $\hat{U}$ also deviates from Eqs.~(\ref{U11-term}) and (\ref{U12-term}). It reads
\begin{equation}\label{U-generalised}
\begin{split}
\hat{U} &= \sum\limits_{\bf q}  \frac{ V_{\bf q} }{2N} \sum\limits_{{\bf k},{\bf k'},s,s'} 
\left( \bigg( \sum_{b=1}^2 
\hat{c}_{{\bf k}bs}^{\dag} \hat{c}_{{\bf k'}bs'}^{\dag} \hat{c}_{{\bf k'}-{\bf q} bs'} \hat{c}_{{\bf k}+{\bf q} bs}   \bigg)
+ 2~  \hat{c}_{{\bf k}2s}^{\dag} \hat{c}_{{\bf k'}1s'}^{\dag} \hat{c}_{{\bf k'}-{\bf q} 1s'} \hat{c}_{{\bf k}+{\bf q} 2s}
\right)  \\
&= \sum\limits_{\bf q}  \frac{ V_{\bf q} }{2N} \sum\limits_{{\bf k},{\bf k'},s,s'} \sum_{b,b'}
\left(
\hat{c}_{{\bf k}bs}^{\dag} \hat{c}_{{\bf k'}b's'}^{\dag} \hat{c}_{{\bf k'}-{\bf q} b's'} \hat{c}_{{\bf k}+{\bf q} bs} 
\right)
\end{split}
\end{equation}
where $V_{\bf q}$ is the substrate-screened 2d Coulomb interaction~\cite{raja2017coulomb,chaves2020bandgap}. A series of Wannier exciton resonances has been observed in 2d semiconductors (such as TMDC \cite{chernikov2014exciton,he2014tightly}), which is attributed to this long-range interaction $V_{\bf q}$. Wannier excitons can be understood as hydrogen-like, loosely bound pairs of upper-band electrons and lower-band holes, formed due to long-range interband interactions. 

Compared to the dispersionless driving term in Eq.~(\ref{D}), the generalised model allows an electron to hop to another unit cell when interacting with the laser drive. This means in Eq.~(\ref{H_paramagnetic_coup}), we allow $ \langle {\bf R} b \vert \hat{{\bf p}} \vert {\bf R'} b' \rangle$ to be non-zero even when ${\bf R}\neq{\bf R'}$. In this more generic case, the electron-laser coupling strength depends on quasi-momentum $\bf k$, which means that
\begin{equation}\label{drive}
\hat{H}^{(1)}_{1} =
g  \sum\limits_{{\bf k},s} \sum\limits_{b,b'}  J^{bb'}_{{\bf k}s} \hat{c}_{{\bf k}bs}^{\dag} \hat{c}_{{\bf k}b's} 
\end{equation}
where the product, $g J^{bb'}_{{\bf k}s}$, represents the $\bf k$-dependent Rabi frequency. Here, $J^{bb'}_{{\bf k}s}$ is a dimensionless complex-valued coefficient whose maximum norm is 1, which depends on the bandstructure and the laser's polarisation. In Eq.~(\ref{drive}), the driving strength $g$ denotes the maximum Rabi frequency.

In the non-interacting model where $\hat{U}=0$, the phase property of $J^{12}_{{\bf k}s}$ has no physical consequences. Because in this case, a unitary transform $\hat{U}_{\varphi} = \exp (i \sum_{{\bf k},s} J^{12}_{{\bf k}s} ~ \hat{n}_{{\bf k},1,s} )$ can always absorb the phase of $J^{12}_{{\bf k}s}$ in Eq.~(\ref{drive}) into the definition of the lower-band operator $\hat{c}_{{\bf k}1s}$. With this redefinition given by $\hat{U}_{\varphi}$, the non-interacting term $\hat{h}$ is unchanged. This means that for a driven system described by the non-interacting Hamiltonian $\hat{h} + \hat{H}^{(1)}_{1} e^{i \omL t} + \hat{H}^{(1)}_{-1} e^{-i \omL t}$, the phase property of $J^{12}_{{\bf k}s}$ cannot create any physical effect. However, this redefinition of $\hat{c}_{{\bf k}1s}$ will change the form of the interaction $\hat{U}$ in Eq.~(\ref{U-generalised})
\footnote{This unitary transform $\hat{U}_{\varphi}$ on the Bloch basis $\hat{c}_{{\bf k}bs}$ will change the locality of the Wannier basis given by the lattice Fourier transform, $\hat{c}_{{\bf R}bs} = \frac{1}{\sqrt{N}} \sum_{\bf k} e^{i {\bf k} \cdot {\bf R}} \hat{c}_{{\bf k}bs} $. In Eq.~(\ref{U-generalised}), the interaction term $\hat{U}$ takes the form of density-density coupling in real space. This form of $\hat{U}$ is an approximation, which is valid only when the Wannier basis is localised. This means there is an optimal choice of the Bloch basis $\hat{c}_{{\bf k}bs}$ which makes the corresponding Wannier basis $\hat{c}_{{\bf R}bs}$ maximally localised \cite{marzari2012maximally}, and thus makes Eq.~(\ref{U-generalised}) a good approximation to the interaction term. 
%In the current model, we will assume that in such an optimised Bloch basis, the electron-light coupling coefficient $J^{12}_{{\bf k}s}$ shows a non-analytic behaviour when ${\bf k}\to{\bf K'}$. With numerical calculations that include the spin-orbital coupling effects, we could find the maximally localised Wannier functions \cite{marzari2012maximally} that give additional corrections to $J^{12}_{{\bf k}s}$.
To obtain the maximally localised Wannier functions for 2d materials, we need to resort to DFT numerical calculations \cite{marzari2012maximally}. However, this is beyond the purpose of the thesis. }
, such that $ \hat{U}_{\varphi} \hat{U} \hat{U}_{\varphi}^\dag \neq \hat{U}$.
This means that for the generalised driven setup described by Eq.~(\ref{Ht-example-chap5}), the phase property of $J^{12}_{{\bf k}s}$ can have physical consequences, because it can no-longer be eliminated by the transform $\hat{U}_{\varphi}$ without modifying other parts of Eq.~(\ref{Ht-example-chap5}). In Chapter \ref{Chapter7}, we indeed find that when the phase of $J^{12}_{{\bf k}s}$ exhibits non-analyticity around the high-symmetry points in the Brillouin zone, destructive interference occurs in the Floquet-induced interactions.

\section{Conclusion}
In this chapter, we have constructed the two-band driven cavity-semiconductor models that we will analyse in this thesis. In Chapter \ref{Chapter4}, the minimal model in Section \ref{sec:dipolar-Hamiltonian-construct} will be studied by the Sambe space Gaussian elimination method. From the resulting Floquet Hamiltonian, we will demonstrate how the multi-band Hubbard interaction enhances the Floquet-induced interactions in this minimal model. 

The study of the generalised cavity-semiconductor model in Section \ref{sec:general-setup} requires going beyond the mean-field limitations of the Gaussian elimination method. This is achieved by the FSWT method developed in Chapter \ref{Chapter5}. In Chapter \ref{Chapter7}, we will use FSWT to derive the Floquet-induced interaction in this generalised model.

%% file: Chapters/exciton.tex
%\title{Cavity-mediated interactions in a laser-driven two-band correlated material: \\virtual exciton effects in the Floquet many-body Hamiltonian}

We will use the Sambe space Gaussian elimination method formulated in section \ref{Sec2.2.1} to study the cavity-mediated interactions in a two-band Hubbard model described in section \ref{sec:dipolar-Hamiltonian-construct}. The Floquet Hamiltonian for this model shows that, when trying to engineer long-range interactions in the cavity-coupled material by laser drive, the intrinsic interactions in the material can qualitatively modify the Floquet-induced cavity-mediated long-range interaction. 

Using the Gaussian elimination Floquet Hamiltonian method, we show how the multi-band electronic on-site repulsion screens the driving-induced effects in this electron-cavity hybrid system via the scattering through virtual excitonic resonance. As the driving frequency approaches the exciton resonance, these virtual particles enhance the rotating-wave-approximation (RWA) driving effects like the optical Stark shift and cavity-mediated electron interactions. This excitonic enhancement strengthens the cavity-mediated interaction at an anomalously broad range of incoming electron momenta, and thus efficiently couples the Fermi surface with electrons far from it. 

%Compared to a simulation of the non-interacting electron system, we find this enhances the tendency of the cavity-mediated interaction to induce an s-wave Fermi-surface deformation (Pomeranchuk) instability. In a prototypical model for molecular crystals, near the driving-induced van Hove singularity, the enhanced cavity-mediated interaction triggers the Pomeranchuk instability at the temperature of $\sim$ 1 $K$. 

The chapter is organised as follows: In section \ref{sec:dipolar-Hamiltonian-construct}, we have already described the Hamiltonian of the minimal model of the driven cavity-material system. In section \ref{sec.in-gap-driving}, we discuss the energy and time scales in our driving scheme to generate the Floquet-induced interactions. In section \ref{floquet-summary}, we incorporate the projector technique into the above-mentioned Sambe space Gaussian elimination method, from which we obtain a Floquet Hamiltonian accounting for the screening effects arising from the driving-induced virtual excitons. 
%, which is detailed in Appendix \ref{sec:floquet-formalism}. 
In section \ref{sec:correlated-model-k-GRPA}, we discuss the screened Floquet Hamiltonian and analyse the main result, i.e. the excitonic-enhancement effects and the non-RWA effects. In section \ref{sec:conclude}, we analyse the validity of our Floquet Hamiltonian.
%We review the control calculations, which are included in Appendices~\ref{sec:zero-correlation-model}-\ref{appendix:GRPA-cavity-screening}. 
We compare our Floquet method to a GRPA screening calculation of the RWA effects, and we derive an estimate for the accuracy of our screening result. Finally, we conclude with a discussion of potential further applications of this Gaussian elimination Floquet method in driven many-body systems.

\section{In-gap driving and energy scales}
\label{sec.in-gap-driving}
We first explain the in-gap off-resonant driving scheme to generate the Floquet-induced cavity-mediated interactions in the minimal setup described in section \ref{sec:dipolar-Hamiltonian-construct}.

Laser heating presents a constant problem in any Floquet engineering application. 
In this regard, it is known that Floquet theory can only be applied if the condition $\omL \tau_{sc} > 1$ (where $\tau_{sc}$ is the electron scattering lifetime)\cite{PhysRevB.34.3625} is fulfilled. As we do not consider disorder explicitly in this thesis, we will assume that this is the case. 
In addition, we need to suppress heating due to the excitation of population in the upper electron band. 
To suppress this source of heating, we consider the red-detuned driving of the material, and refer to Ref.~\cite{aeschlimann2021survival} for an estimation of how red-detuning suppresses the heating. The laser frequency $\omL$ lies in the bandgap of the material (see the red region in Fig.~\ref{fig:Frenkel-resonance}), so that the driving-induced excitations in the material can only be created virtually, with a short lifetime inversely proportional to the laser's detuning to the optical resonances.

\begin{figure}
\centering
    \includegraphics[width=0.6\textwidth]{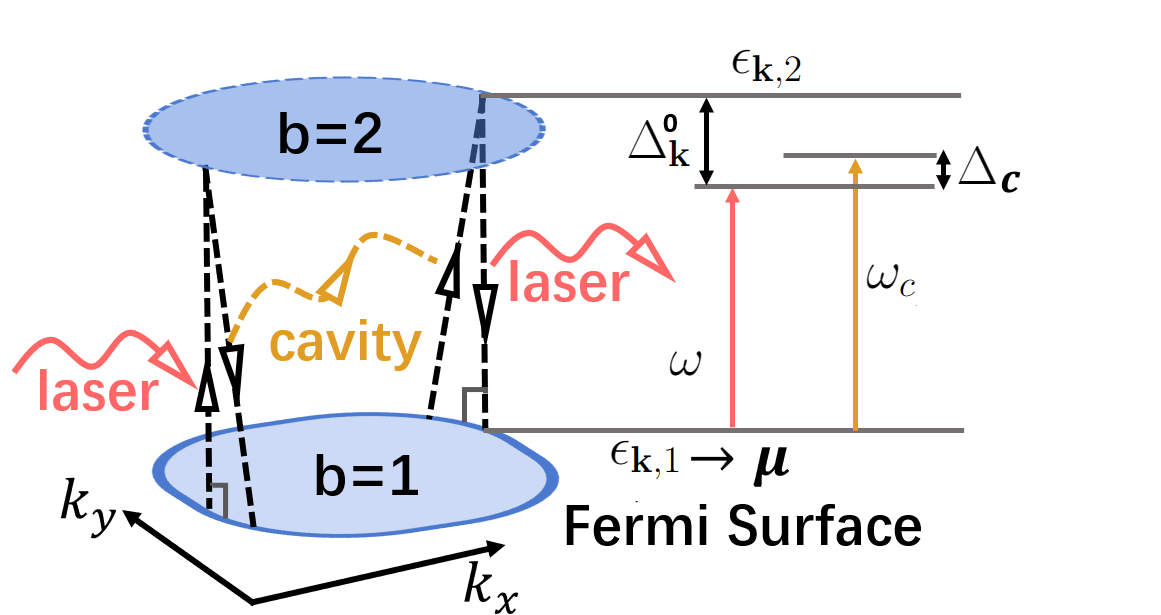} \captionsetup{justification=raggedright,singlelinecheck=false}
    \caption{Illustration of energy levels. In a two-band material, we focus on the Bloch electrons with quasi-momentum near the Fermi-surface in the lower band (thick blue curve). The Fermi surface we will consider is a tiny circle with radius $k_F\ll \pi$. In the non-interacting case, these electrons effectively interact with each other via the cavity-mediated Floquet-induced interaction, which involves two virtual inter-band optical processes, connected by the exchange of a virtual cavity photon. }
    \label{fig:band}
\end{figure}

Specifically, as illustrated in Fig.~\ref{fig:band}, in the non-interacting case ($\hat{U}=0$), we require the laser frequency $\omL$ to be smaller than the bandgap $\ep_{{\bf k},21}$ defined in Eq.~(\ref{k-dependent-band-gap}) at any quasi-momentum ${\bf k}$. To avoid power broadening, we further require the laser-bandgap detuning $\Delta_{\bf k}^0 \equiv \ep_{{\bf k}, 21} - \hbar \omL$ to be much larger than the laser-material coupling strength $g$, i.e.
\begin{equation}\label{band-detuning-requirement}
    \ep_{{\bf k}, 2} - \ep_{{\bf k}, 1} > \hbar \omL \gg \Delta_{\bf k}^0 \gg \vert  g \vert.
\end{equation} 
Furthermore, to avoid the creation of cavity photons, the laser-cavity detuning $\Delta_c \equiv \hbar( \omc - \omL)$ should also be large compared to the respective coupling $g_c$,
\begin{equation}\label{cavity-detuning-requirement}
    \hbar \omL \gg  \vert \Delta_c \vert \gg g_c. %\gg \frac{\vert  g \vert }{ \Delta_{\bf k}} \frac{1}{\sqrt{N}}  \vert g_{c,{\bf k}'} \vert_{\text{max}}.
\end{equation}
Under these off-resonant conditions, the Floquet-induced interactions arise from the consecutive virtual processes illustrated in Fig.~\ref{fig:band}: The red-detuned laser first creates a virtual band excitation, which transfers its energy to a virtual cavity photon. This photon then transfers the energy to create another virtual electron, which finally returns the energy to the laser. The overall process creates an effective long-ranged electronic interaction in the lower-band.

This non-interacting picture of the Floquet-induced cavity-mediated interaction will be qualitatively modified once the non-vanishing $\hat{U}$ is taken into account. In the following sections, we will analyse this correlation effect using the Gaussian elimination Floquet Hamiltonian method reviewed in Section \ref{Sec2.2.1}.

\section{Incorporating the many-body projector technique into the Sambe space Gaussian elimination}\label{floquet-summary}
For the minimal driven cavity-QED setup described by Eqs.~(\ref{H0} $\sim$ \ref{D}) in Section \ref{sec:dipolar-Hamiltonian-construct}, the Floquet Hamiltonian $\hat{H}^{\text{eff}}_{(\Ea)}$ in Eq.~(\ref{Heff_weak_drive}) contains many-body green operators $\hat{G}^{0}_{(\Ea \pm \omL)}$. To read out the Floquet-induced interaction from $\hat{H}^{\text{eff}}_{(\Ea)}$, in general we have to determine $\hat{H}^{\text{eff}}_{(\Ea)}$ self-consistently by its eigenenergy $\Ea$, which is not possible in our many-body system. However, the $\Ea$-dependence in $\hat{H}^{\text{eff}}_{(\Ea)}$ can be eliminated if the following energy filter of the undriven Hamiltonian 
\begin{equation}
    \PEa = \delta(\Ea-\hat{H}_0)
\end{equation}
is applied to the green operators in $\hat{H}^{\text{eff}}_{(\Ea)}$, such that $\hat{G}^{0}_{(\Ea + \Delta)} \PEa = \frac{1}{\Ea + \Delta - \hat{H}_0} \PEa = \frac{1}{\Delta} \PEa $. Once $\Ea$ is eliminated in $\hat{H}^{\text{eff}}_{(\Ea)}$ by $\PEa$, the self-consistency requirement (\ref{0-HarmonicEigenProblem}) is fulfilled. Below, we go through the details of this projector-based strategy for eliminating the self-consistency requirement.

\subsection{Projecting the Floquet Hamiltonian by the energy filter}
\label{appendix-filter}
According to Eq.~(\ref{Heff_weak_drive}), the Floquet Hamiltonian has the form $ \hat{H}^{{\rm Fl}}_{(\Ea)} = \hat{H}_0 + \vert  g \vert^2 \hat{V}_{(\Ea)} $, where the static Hamiltonian $\hat{H}_0$ represents the unperturbed system, and the perturbation operator $\hat{V}_{(\Ea)}$ becomes independent of driving strength $g$. The effective Hamiltonian of $\hat{H}^{{\rm Fl}}_{(\Ea)}$ near eigenenergy $\Ea$, according to the second order Kato's expansion in Refs.~\cite{essler2005one,klein1974degenerate}, is given by
\begin{equation}
\hat{H}_{\rm eff} (\Ea) = \PEa \big(   \hat{H}_0  + \vert  g \vert^2 \hat{V}_{(\Ea)}  +   \vert  g \vert^4  \hat{V}_{(\Ea)}  \sum\limits_{E_{\beta} \neq \Ea} \frac{\hat{\mathcal{P}}_{E_{\beta}}}{ \Ea - E_{\beta} }  \hat{V}_{(\Ea)}     \big) \PEa .
\end{equation}
where $\PEa = \delta(\Ea-\hat{H}_0)$ is the energy projector of the unperturbed system. 
Consistent with the weak-driving approximation made in Eq.~(\ref{Heff_weak_drive}), we only retain to the $\vert  g \vert^2$ order, thus we get
\begin{equation}\label{Heff_weak_drive-chap3}
    \hat{H}_{\rm eff} (\Ea) = \PEa \hat{H}^{{\rm Fl}}_{(\Ea)} \PEa 
    = \PEa \hat{H}_0 \PEa + \PEa \hat{H}_{1}\hat{G}^{0}_{(\Ea +\hbar\omL)}\hat{H}_{-1} \PEa + \PEa \hat{H}_{-1}\hat{G}^{0}_{(\Ea -\hbar\omL)}\hat{H}_{1} \PEa
\end{equation}
which is the starting point of the following procedures.

\subsection{Moving the Green operators}
In Eq.~(\ref{Heff_weak_drive-chap3}), the Green operator $\hat{G}^{0}_{(\Ea \pm \hbar\omL)}$ still cannot be eliminated by $\PEa$, since they are separated by a driving operator $\hat{H}_{\pm1}$. Thus, we need to move $\hat{G}^{0}_{(\Ea \pm \hbar\omL)}$ rightward (or equivalently, leftward) to make it adjacent to $\PEa$. This means we will achieve the following move-and-eliminate procedure
\begin{equation}\label{right-moving}
    \hat{G}^{0}_{(\Ea +\hbar\omL)}\hat{H}_{-1} \PEa \approx \sum\limits_{i} \hat{F}_{i} \hat{G}^{0}_{(\Ea +\Delta_i)} \PEa = \sum\limits_{i} \hat{F}_{i} \frac{1}{\Delta_i} \PEa 
\end{equation}
If $\hat{F}_{i}$ still contains the Green operator $\hat{G}^{0}$, we will repeat the above move-and-eliminate procedure. In this way, we are able to derive an operator product that is normal-ordered, which offers an approximation to the original term $\hat{G}^{0}_{(\Ea +\omL)}\hat{H}_{-1} \PEa$ in Eq.~(\ref{Heff_weak_drive-chap3}). This operator product takes the form of $\hat{c}_{i}^{\dag} \hat{c}_{j}^{\dag} \hat{c}_{k} \hat{c}_{l} \PEa$, which can be identified as the Floquet-induced interaction given by the Sambe space Gaussian elimination method.

In the right-moving procedure, as indicated by the $\approx$ sign in Eq.~(\ref{right-moving}), we have made several approximations, which we scrutinise in Appendix~\ref{appendix:useful-relations}. In the non-interacting limit where $\hat{U} \to 0$, we make two approximations in Eq.~(\ref{right-moving}). First, we assume the driven cavity-material system is in its low-energy state, such that the energy filter $\PEa$ projects to a subspace with an empty cavity and empty upper-band [see Eq.~(\ref{low-energy-approx})]. Second, we assume $g_c \ll \ep_{21}$, such that the electron-cavity coupling is too weak to evidently modify the undriven 2-band system. As described in Eq.~(\ref{G1-reduced-to-number}), this allows $\PEa$ to eliminate the material's cavity-decoupled Green operator $\hat{G}^b$ defined in Eq.~(\ref{definition-G1-g}).

With these approximations, in the non-interacting case, we apply the above-mentioned moving procedure and derive the Floquet Hamiltonian in the low-energy limit (see Appendix \ref{sec:zero-correlation-model}). The Floquet Hamiltonian, in this case, shows lowest-order Floquet engineering effects (correct to $\vert  g \vert^2$), including optical Stark shift (\ref{U=0_ACS}), Bloch-Siegert shift (\ref{BSS_U=0}), and cavity-mediated interactions (\ref{appendix:subsec:cav-med-int_U=0}).

\subsection{Multi-band Screening calculation during the moving procedure}
Next, we take a non-zero $\hat{U}$ into account (see Appendix \ref{appendix:correlated-model-k-GRPA} for the complete derivation). Here, the interaction $\hat{U}$ is weak compared to the inter-band excitation energy $\ep_{21}$. In the lower-band, its strength $U_{11}$ can be much larger than the hopping $t_1$. In this case, an additional approximation is made in the right-moving procedure in Eq.~(\ref{right-moving}). In particular, we use the following approximation to move $\hat{U}$ in the green operator $\hat{G}^{0}_{(\Ea +\hbar\omL)}$ to the right of $\hat{H}_{-1}$,
\begin{equation}\label{U-b-commutator-reduced-chap4}
    \begin{split}
        &\hat{U} \hat{b}_{{\bf q},s}^{\dag} \approx \sum\limits_{\bf k} \hat{b}_{{\bf k},s}^{\dag} \hat{f}_{{\bf k},{\bf q}}^{s}
    \end{split}
\end{equation}
where $\hat{b}_{{\bf k},s}^{\dag} \equiv \hat{c}_{{\bf k},2,s}^{\dag} \hat{c}_{{\bf k},1,s}$ denotes the inter-band polarization operator at momentum $\bf k$ and spin $s$, which occurs in the driving operator $\hat{H}_{-1}$ (and also in $\hat{H}_{c}$), and
\begin{align} \label{eq.f-def-chap4}
\hat{f}_{{\bf k},{\bf q}}^{s} &= \delta_{{\bf k},{\bf q}}  
        \big( \hat{U} 
        - U_{11} \hat{\nu}_{\Bar{s}}
        + U_{12}\sum\limits_{s'} \hat{\nu}_{s'}
        \big) - \frac{U_{12}}{N} \hat{n}_{{\bf q}, 1,s}.
\end{align}
We have also defined 
the spin-resolved filling operator in the lower band
\begin{equation}\label{filling-operator-chap4}
    \hat{\nu}_{s} \equiv \frac{1}{N}\sum\limits_{{\bf k'}}\hat{n}_{{\bf k'}, 1,s}.
\end{equation}
For this interacting case, in the right-moving procedure in Eq.~(\ref{right-moving}), the switching between $\hat{U}$ and $\hat{b}^\dag$ in Eq.~(\ref{U-b-commutator-reduced-chap4}) will appear for infinite times, as shown in Appendix \ref{appendixsec:Screened_ACS}. We then carry out the geometric infinite re-summation over the terms created during each switching between $\hat{U}$ and $\hat{b}^\dag$. The resulting re-summed operator is finally made adjacent to $\PEa$, and it has a similar form of the green operator $\hat{G}^0$ [see Eq.~(\ref{k-space-GRPA-Stark-middle})]. 

\subsection{Hartree-type Mean-field approximation}\label{sec:Hartree-type-mean-field-chap4}
However, an additional \textit{Hartree-type mean-field approximation} has to be made to turn this re-summed operator exactly into the green operator $\hat{G}^0$ (for the undriven Hamiltonian $\hat{H}_0$). This means we need to replace the filling operator by its expectation value, 
\begin{equation}\label{mean-field-global-chap4}
    \hat{\nu}_s \rightarrow \nu_s = \langle \hat{\nu}_s \rangle,
\end{equation}
which is evaluated at equilibrium. We assume that the driving-induced changes to this mean field value $\langle \hat{\nu}_s \rangle$ are negligible, so we do not need to determine it self-consistently. With this Hartree-type mean-field approximation, the re-summed operator in Eq.~(\ref{k-space-GRPA-Stark-middle}) can finally be eliminated by $\PEa$ according to Eq.~(\ref{right-moving}), as shown in Eqs.~(\ref{reduce-gHs-filling}) and (\ref{reduce-gHs-single-occupation}), and thus the self-consistency requirement on $\Ea$ is finally eliminated.

This re-summation followed by mean-field decoupling is linked closely to an inter-band generalised random phase approximation (GRPA) \cite{rowe1968equations,PhysRev.112.1900}, which we will further explore in Section \ref{Sec:exciton-resonance-structure}. With the additional treatments in Eqs.~( \ref{U-b-commutator-reduced-chap4} $\sim$ \ref{mean-field-global-chap4} ) in the interacting model, we derive the screened Floquet Hamiltonian. We note that, without using the final Hartree-type mean-field replacement (\ref{mean-field-global-chap4}), the projector-based method is no longer able to eliminate the self-consistency requirement in $\hat{H}^{\text{eff}}_{(\Ea)}$.

\section{The screened Floquet Hamiltonian}
\label{sec:correlated-model-k-GRPA}
%\textcolor{blue}{The main Focus is "the Floquet-induced cavity-mediated interaction is excitonically enhanced by the multi-band interactions" }

Collecting the calculations in Appendix \ref{appendix:correlated-model-k-GRPA}, we derive the screened Floquet low-energy Hamiltonian for electrons in the lower electronic band as 
\begin{align}\label{eq.main-result}
\hat{H}_{\rm eff} &= \hat{h}_{\rm eff} + \hat{U}_{\rm eff},
\end{align}
with 
\begin{align}\label{eq.h_eff}
\hat{h}_{\rm eff} &= \sum_{{\bf k},  s} \big( \ep_{{\bf k}} - \frac{\vert g \vert^2}{\Delta_{{\bf k},s} } - \frac{\vert  g \vert^2}{\Delta_{{\bf k},s}^{BS}} -\mu \big) \hat{c}_{{\bf k} s}^{\dag} \hat{c}_{{\bf k} s}
\end{align}
and
\begin{align}\label{eq.U_eff}
\hat{U}_{\rm eff} &= \hat{U}_{11}  ~ - ~ \frac{1}{N}  \sum_{\substack{{\bf k}, s \\ {\bf k'}, s'}} \frac{\vert  g g_c \vert^2 }{ \Delta_c  \Delta_{{\bf k'},s'}   \Delta_{{\bf k},s}  } \hat{c}_{{\bf k'}, s'} ^{\dag}  \hat{c}_{{\bf k'}, s'} \hat{c}_{{\bf k}, s}^{\dag}  \hat{c}_{{\bf k}, s}.
\end{align}
To reduce clutter, we neglected the electronic band index ($b=1$) here and in the following. 
In these equations we have introduced the laser-cavity detuning $\Delta_c = \hbar (\omc - \omL)$, as well as the screened denominators $\Delta_{{\bf k},s} $ and $\Delta_{{\bf k},s}^{BS}$, which are defined as
 
    \begin{equation}\label{renormalised-denominator}
        \begin{split}
            \Delta_{{\bf k},s} \equiv \Delta^{0}_{\bf k}- U_{11} \nu_{\Bar{s}} + U_{12} \sum\limits_{s'}\nu_{s'}
            - \frac{U_{12}}{N} \sum\limits_{ {\bf k}' }  \langle \hat{n}_{{\bf k}',s} \rangle
            \frac
            {  \Delta^{0}_{\bf k}- U_{11} \nu_{\Bar{s}}
            + U_{12} \sum\limits_{s'}\nu_{s'} }
            { \Delta^0_{\bf k'} - U_{11} \nu_{\Bar{s}}
            + U_{12} \sum\limits_{s'}\nu_{s'}},
        \end{split}
    \end{equation}
    and
    \begin{equation}\label{eq.BS-denominator}
    \begin{split}
        \Delta_{{\bf k},s}^{BS}&=
        \Delta^{0}_{\bf k}+2\omL - U_{11} \nu_{\Bar{s}}
        + U_{12} \sum\limits_{s'}\nu_{s'} - \frac{U_{12}}{N} \sum\limits_{ {\bf k}' }  \langle \hat{n}_{{\bf k}',s} \rangle
        \frac{ \Delta^{0}_{\bf k}+2\omL - U_{11} \nu_{\Bar{s}}
        + U_{12} \sum\limits_{s'}\nu_{s'} }
        { \Delta^0_{\bf k'} +2\omL - U_{11} \nu_{\Bar{s}}
        + U_{12} \sum\limits_{s'}\nu_{s'}}.
    \end{split}
\end{equation}
  
Here, $\Delta^{0}_{\bf k}= \ep_{{\bf k}, 2} - \ep_{{\bf k}, 1}  - \hbar \omL$ is the bare laser-bandgap detuning, with the bandstructure $\ep_{{\bf k}, b}$ defined in Eq.~(\ref{dispersion}). 
The detunings are renormalised by the interband interaction $U_{12}$ and the intraband interaction $U_{11}$ of the lower band. Interactions in the upper band, described by $U_{22}$, are not significant in the present weak driving limit, where no real population is created in the upper band.

Let us discuss the Hamiltonian structure: The single-particle sector $\hat{h}_{\rm eff}$ contains the band dispersion $\ep_{\bf k} - \mu$ of the bare material, which is renormalised by two terms. 
We identify the first term $- |  g |^2 / \Delta_{{\bf k},s}$ as the optical Stark shift, i.e. the laser-induced renormalization of the electronic band energy. 
In semiconductor quantum wells, this effect was first reported in Ref.~\cite{mysyrowicz1986dressed}.
The second term $- |  g |^2 / \Delta_{{\bf k},s}^{BS}$ is the Bloch-Siegert shift in materials~\cite{PhysRev.57.522, Sie2018} which further decreases the energy of electrons in the lower band. It stems from the non-RWA terms in the laser-electron coupling. The Bloch-Siegert shift in two-level systems has been previously derived by various Floquet methods~\cite{shirley1965solution,PhysRevA.81.022117,PhysRevA.79.032301,PhysRevLett.105.257003}, our Floquet Hamiltonian method generalizes this derivation to a many-body system.

In the interaction term $\hat{U}_{\rm eff}$, the local intra-band repulsion $\hat{U}_{11}$ is not screened by projecting out the high-energy degrees of freedom - as one would expect intuitively. However, the presence of the optical cavity gives rise to a cavity-mediated Floquet-induced interaction $\sim |  g  g_c|^2 / ( \Delta_c \Delta_{\bf k} \Delta_{\bf k'})$. 
It is induced by the scattering of a laser photon into and out of the cavity via the virtual excitation of two lower-band electrons into the upper band. A very similar effect was recently proposed theoretically in quantum spin systems~\cite{chiocchetta2021cavity}, and it is also closely connected to well-established methods in cold atoms to generate and control long-ranged interactions~\cite{Ritsch2013, Mivehvar2021}. More discussions on $\hat{U}_{\rm eff}$ will be given in section \ref{sec.cavity-interactions}.

\subsection{The screened denominator}
\label{sec.screened-denominator}
The different terms in the screened denominator~(\ref{renormalised-denominator}) can be readily interpreted physically: 
In the absence of electron interactions, the energy required to excite a Bloch electron with quasi-momentum ${\bf k}$ is given by the energy difference $\ep_{{\bf k},2} - \ep_{{\bf k},1}$, such that the energy mismatch to the laser drive is the bare detuning $\Delta_{\bf k}^{0}$. 
This detuning is reduced by $U_{11} \nu_{\Bar{s}}$, which originates from the absence of the intra-band repulsion $\hat{U}_{11}$, once a hole is created below the Fermi surface. 
It is counteracted by $U_{12} \nu_{\Bar{s}}$ and $U_{12} \nu_{s}$. This third term in Eq.~(\ref{renormalised-denominator}) originates from the opposite (same) spin part of the inter-band repulsion $\hat{U}_{12}$ after an electron is excited to the upper band. 
The final term has the most complex structure. It is the inter-band electron-hole excitonic attraction stemming from fermionic statistics: Once a fermion with spin-momentum quantum numbers $({\bf k},s)$ is excited to the upper band, it leaves a hole at the same momentum $({\bf k},s)$. This hole counteracts the third term, but since the electrons can hop between adjacent lattice sites, the energy shift cannot completely compensate the previous effect. 

The screened detuning~(\ref{renormalised-denominator}) is also contained in the low-temperature absorbance spectrum of the bare material, where it emerges from the Kubo formula (see Appendix~\ref{sec.absorbance}). Absorbance resonances are found when the detuning vanishes, i.e. $\Delta_{{\bf k},s} = 0$. This is the case when 
\begin{equation} \label{band-resonance}
\Delta_{\bf k}^0- U_{11} \nu_{\Bar{s}} + U_{12} \sum\limits_{s'}\nu_{s'} = 0,
\end{equation}
i.e. when the laser resonantly couples to the interband transition, which is shifted by the Hartree-type mean-field contribution $- U_{11} \nu_{\Bar{s}} + U_{12} \sum\limits_{s'}\nu_{s'}$,
or when 
\begin{equation}\label{exciton-resonance}
\frac{U_{12}}{N} \sum\limits_{ {\bf k}' } 
\frac{  \langle \hat{n}_{{\bf k}',s} \rangle } { \Delta_{{\bf k}'}^0 - U_{11} \nu_{\Bar{s}} + U_{12} \sum\limits_{s'}\nu_{s'}} = 1.
\end{equation}
This latter condition describes the formation of an exciton, it can also be derived from the Wannier equation (see Ref.~\cite{katsch2022excitonic,kira2006many}) for the case of on-site interactions: For a fully occupied lower-band where $\langle \hat{n}_{{\bf k},s} \rangle = \nu_{s}= \nu_{\Bar{s}}= 1$, Eq.~(\ref{exciton-resonance}) is equivalent to the semiconductor excitonic optical resonance derived in Refs.~\cite{katsch2022excitonic,kira2006many,PhysRevB.29.4401, davis1986photoemission}
\footnote{
The Wannier equation in Ref.~\cite{kira2006many}, in our on-site repulsion model with fully-occupied lower-band, becomes 
$$(\ep_{{\bf k}, 2} - \ep_{{\bf k}, 1} - U_{11}  + 2 U_{12} ) \phi({\bf k}) -\frac{U_{12}}{N}\sum\limits_{\bf k'} \phi({\bf k'}) =\hbar \omega_{ex} \phi({\bf k}) $$
which has a solution $\phi({\bf k})=(\ep_{{\bf k},2}-\ep_{{\bf k}, 1} -\hbar\omega_{ex} - U_{11}  + 2 U_{12} )^{-1}$ representing the exciton's momentum-distribution. The eigenvalue $\omega_{ex}$ satisfies
$$ \frac{U_{12}}{N}\sum\limits_{\bf k} (\ep_{{\bf k},2}-\ep_{{\bf k}, 1} -\hbar\omega_{ex} - U_{11}  + 2 U_{12} )^{-1} = 1 $$
which is identical to our result Eq.~(\ref{exciton-resonance}) when $\langle \hat{n}_{{\bf k},s} \rangle =\nu_s =1$. 
}. 
For the parameters chosen in Section \ref{sec:dipolar-Hamiltonian-construct}, Eq.~(\ref{exciton-resonance}) predicts an exciton resonance at $\hbar \omega_{\text{ex}}\approx 2.71$ eV, and we will denote the laser-exciton detuning as $\Delta_{ex}\equiv \hbar (\omega_{\text{ex}} - \omL)$.  
Crucially, the condition Eq.~(\ref{exciton-resonance}) is independent of the momentum index $\bf k$ in $\Delta_{{\bf k},s} = 0$. This will have profound consequences for the emergent Floquet physics, as we will explore in the remainder of this chapter.

\subsection{Bandstructure Floquet engineering}

\paragraph{Bandstructure renormalization}
\begin{figure}
\centering
\includegraphics[width=0.75\textwidth]{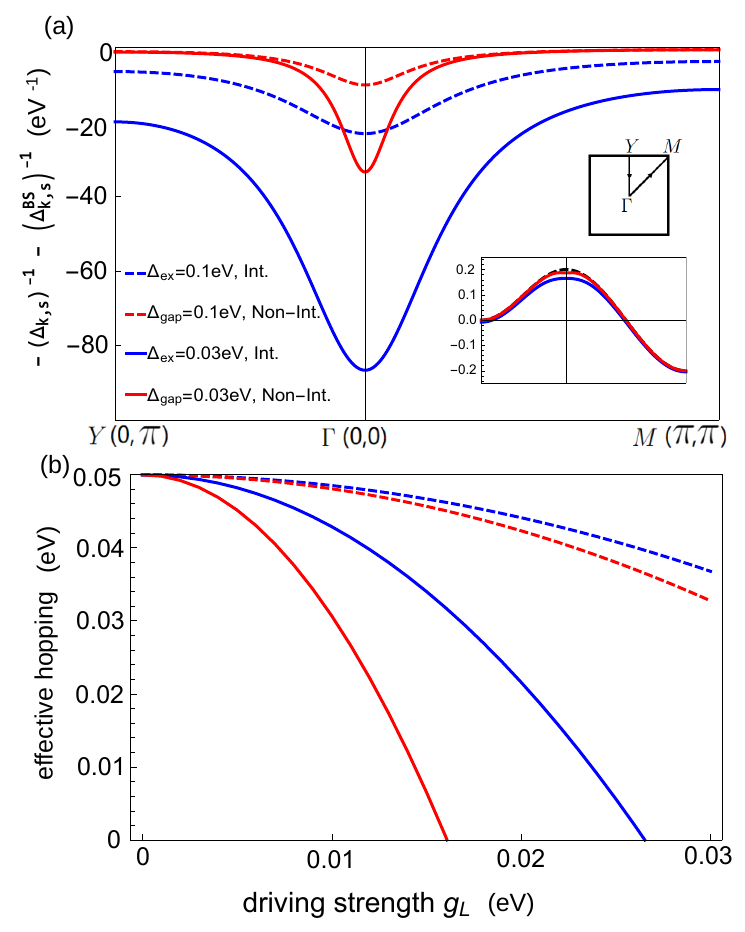}
\caption{
(a) The change of the single-electron energy of the lower-band according to Eq.~(\ref{eq.h_eff}) is shown 
along the path $Y \to \Gamma \to M$ in the first Brillouin zone. 
Since this change is proportional to $|  g |^2$, this constant is removed. The inset in the lower-right corner shows the engineered lower band for $g=0.02$eV.  
(b) The effective hopping rate $\Tilde{t}$, as extracted from the electronic dispersion at the $\Gamma$-point, is plotted vs. the laser driving strength $g$.
}
\label{fig.dispersion}
\end{figure}

We next explore how this peculiar resonance structure is reflected in the Floquet bandstructure in the single-particle sector $\hat{h}_{\rm eff}$ of Eq.~(\ref{eq.main-result}). 
The excitonic screening behaviour in our interacting model is revealed by comparing with an unscreened case in the non-interacting model where $U_{11}=U_{12}=0$. 
To allow for a fair comparison between the screened and the unscreened cases, we fix the detuning to the respective resonances, i.e. we chose the driving frequency such that we have the same detuning to the excitonic resonance (in the screened case) and to the $\Gamma$-point of upper electron band (unscreened case). 
That is, we let $\Delta_{\text{ex}}$ in the screened case to be equal to $\Delta_{\text{gap}} \equiv \Delta_{{\bf k}=(0,0)}^0$ in the unscreened case. 
Consequently, the Stokes term $|  g |^2 / \Delta_{{\bf k},s}$ in $\hat{h}_{\rm eff}$ has approximately the same size, and the major deviation stems from the distinct momentum dependence of the exciton resonance. 

The change of the bandstructure is shown along the high-symmetry points of the first Brillouin zone in Fig.~\ref{fig.dispersion}(a). 
Naturally, the largest renormalization effect can be achieved in the vicinity of the $\Gamma$-point for the unscreened case. The reason is obvious, as changes in different regions of reciprocal space are suppressed by a much larger detuning. 
In contrast, the screening causes changes of the bandstructure in a much larger region of reciprocal space.
This behaviour can again be explained easily: due to the lack of momentum dependence in Eq.~(\ref{exciton-resonance}), 
the detuning at momentum far from the $\Gamma$-point still vanishes as $\omL$ approaches $\omega_{\text{ex}}$. 
Hence, the screening enables the driving to renormalise the bandstructure at larger regions of reciprocal space. 
If we take $\Delta_{\text{ex}}\to0$ in Fig.~\ref{fig.dispersion}(a), the shape of the (blue) interacting curve will converge to $\phi({\bf k})$, which is the exciton's broadened momentum distribution. On the contrary, if we take $\Delta_{\text{gap}} \to 0$, the shape of the (red) non-interacting curve becomes a narrow delta peak at $\Gamma$-point. 
Naively, this would suggest an overall enhanced renormalisation effect. 

In Fig.~\ref{fig.dispersion}(b), we demonstrate that this is not necessarily true: 
We plot the renormalisation of the effective hopping rate $\Tilde{t}$ at the $\Gamma$-point, which we extract from the curvature of the Floquet bandstructure, versus the driving strength $g$.  
For a weakly hole-doped lower electron band, as we consider here, this hopping rate describes the effective kinetic energy at the Fermi surface, i.e. it is proportional to the electron mobility. 
In both screened and unscreened cases, the driving leads to a reduction of the hopping rate, indicating the dynamical localisation of the mobile charge carriers due to the driving~\cite{Kuwahara2016}. 
But we find that screening reduces the renormalization of this hopping rate - even though the overall change of the bandstructure is increased. 
If we reduce the detuning to the resonance to be 0.03 eV, the unscreened calculation predicts a vanishing of the effective hopping rate at $  g  \simeq 0.015$~eV. 
This would indicate a photo-induced van Hove singularity~\cite{vanHove1953}, where the density of states diverges (even though higher orders in $|  g |$ will likely shift or destroy this singularity).
The screening counteracts the reduction of the hopping rate, such that a singularity (to leading order in $g$) only occurs at very large driving strengths $|  g | \sim \Delta_{ex}$ where the weak driving approximation in Eq.~(\ref{Heff_weak_drive}) becomes very questionable and Floquet heating will become a fundamental problem.

\paragraph{Ratio between Stark and Bloch-Siegert shifts}
\begin{figure}
\centering
\includegraphics[width=0.7\textwidth]{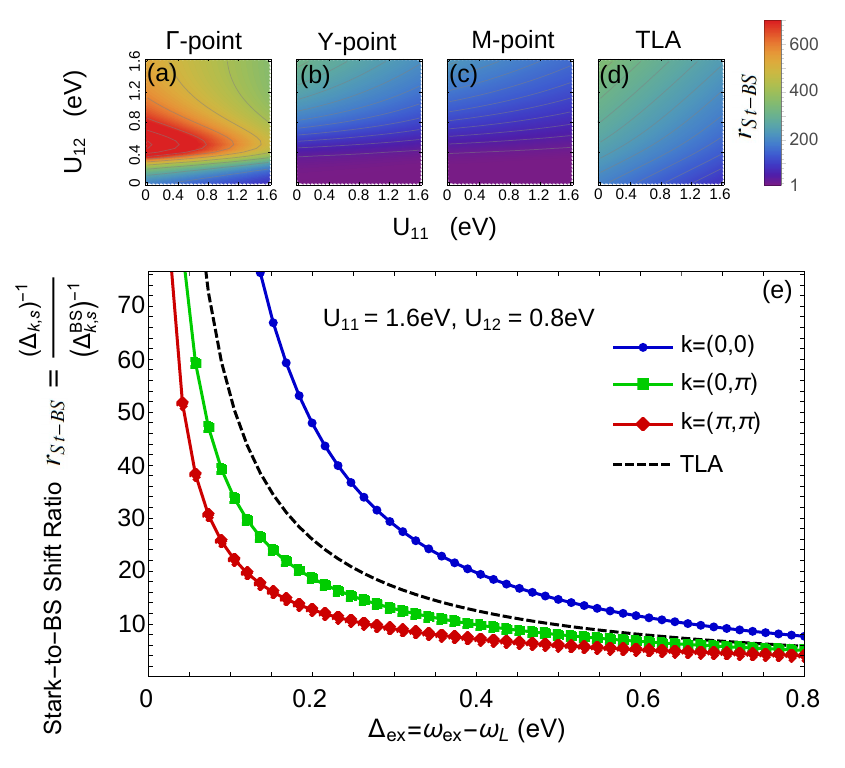}
\caption{
The ratio between the Stark shift~(\ref{renormalised-denominator}) and the BS shift~(\ref{eq.BS-denominator}) is plotted vs. the detuning between laser and exciton resonance for selected points in the Brillouin zone. For comparison, the same ratio for a two-level treatment (TLA), Eqs.~(\ref{eq.Stark-TLA}) and (\ref{eq.BS-TLA}), is shown in blue. 
The dependence of the ratio on the electronic interactions is shown at selected points in the top panels, where the laser-exciton detuning is fixed $\Delta_{\text{ex}}=0.03$eV. 
}
\label{fig.ratio}
\end{figure}
We next explore how the relative size of the Stark effect and the Bloch-Siegert shift are affected by electronic interactions. 
Recent experiments~\cite{Conway2023} on monolayer WS$_2$ revealed an enormous enhancement of the Bloch-Siegert shift compared to what one would expect from treating the exciton resonance as an effective two-level atom (TLA), where we would have
\begin{align} \label{eq.Stark-TLA}
\Delta\ep_{Stark}^{TLA} &= \frac{|  g |^2}{\hbar(\omL -\omega_{\text{ex}})},
\end{align}
and
\begin{align} \label{eq.BS-TLA}
    \Delta\ep_{BS}^{TLA} &= \frac{|  g |^2}{\hbar(\omL + \omega_{\text{ex}})},
\end{align}
where $\omega_{\text{ex}}$ is the excitonic resonance extracted from Eq.~(\ref{exciton-resonance}). 
In Fig.~\ref{fig.ratio}, we plot the ratio
\begin{align} \label{eq.r_St-BS}
r_{St-BS} &\equiv \frac{\Delta\ep_{Stark}}{\Delta\ep_{BS}}
\end{align}
vs. the detuning. We find that at the $\Gamma$-point the ratio according to Eqs.~(\ref{renormalised-denominator}) and (\ref{eq.BS-denominator}) is strongly enhanced compared to the TLA treatment in Eqs.~(\ref{eq.Stark-TLA}) and (\ref{eq.BS-TLA}). 
The excitonic enhancement benefits the Stark effect to a greater degree than the Bloch-Siegert shift, where the enhancement is suppressed by the large prefactor $2\omL$ [see Eq.~(\ref{eq.BS-denominator})]. 
However, at momenta (Y- and M-point) far from the $\Gamma$-point, our many-body approach indeed gives ratios smaller than TLA's prediction. 
If we scan the strength of the intra- and interband interactions [see the top panels of Fig.~\ref{fig.ratio}], we further find a remarkable non-monotonic dependence of this ratio at the $\Gamma$-point. 
Note that there is a weak dependence even in the TLA treatment, because we have fixed $\Delta_{\text{ex}}=0.03$eV and let $\omega_{\text{ex}}$ variate with different interaction strengths, then we cannot change $\omega_{\text{ex}}$ without changing the ratio~(\ref{eq.r_St-BS}). 
Remarkably, the ratio~(\ref{eq.r_St-BS}) displays a clear resonance-like feature, peaking at $U_{12} \simeq 0.5$~eV. 
This interesting phenomenon will be discussed in more detail in the following Section~\ref{sec.cavity-interactions}. 

A direct comparison of our results with Ref.~\cite{Conway2023} will require a simulation of the two-dimensional spectroscopy performed in this work. But as the optical signal will involve an integration over the Brillouin zone, our results already show that a many-body treatment of the excitonic resonance fundamentally can easily change the relative size of Stark and Bloch-Siegert shifts by an order of magnitude.

\subsection{Cavity-mediated interactions}
\label{sec.cavity-interactions}
\begin{figure}
\centering
\includegraphics[width=0.6\textwidth]{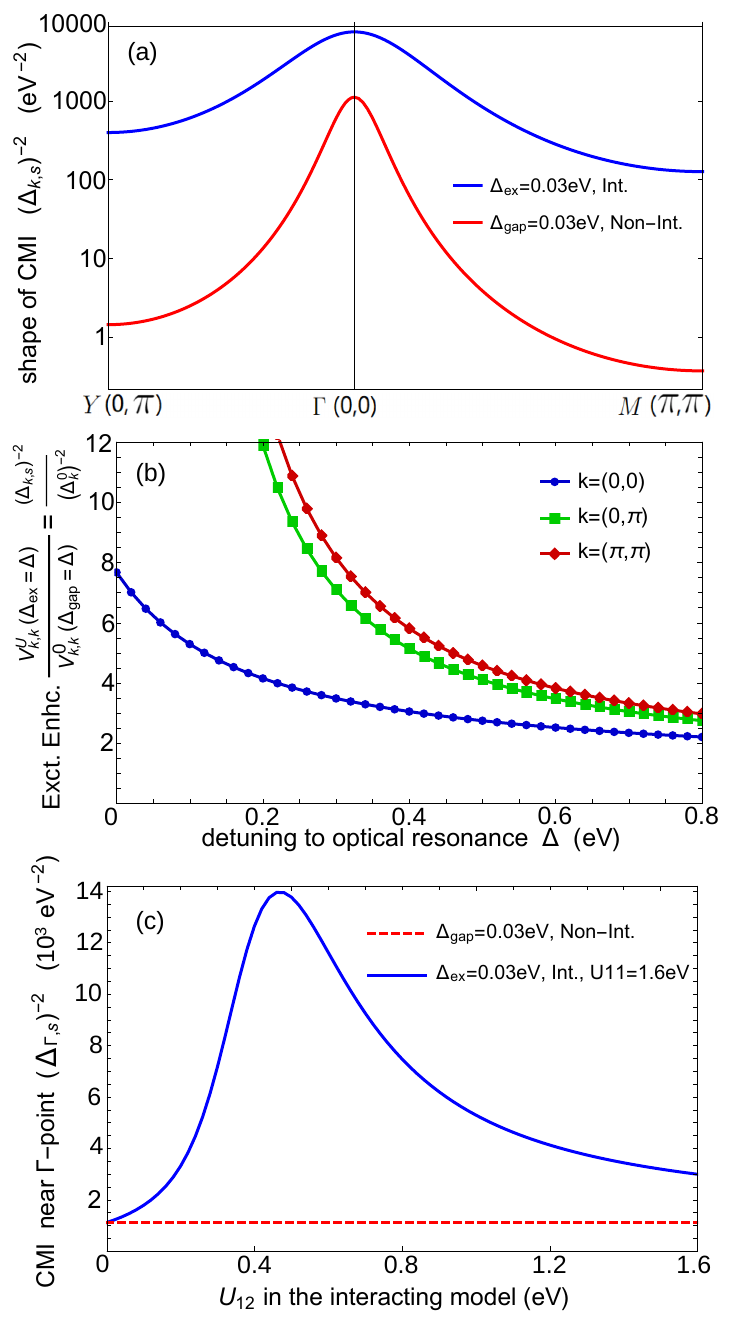}
\caption{
(a) Strength of cavity-mediated interaction in Eq.~(\ref{eq.U_eff}) in the forward scattering direction along a cut of the first Brillouin zone. The constant prefactor $|g_c   g |^2$, which appears in Eq.~(\ref{eq.U_eff}), is removed. 
(b) The excitonic enhancement, i.e. the ratio between the cavity-mediated interaction of the interacting and the non-interacting model is plotted vs. the detuning to the respective resonance at the $\Gamma$- (${\bf k} = (0,0)$), $Y$- (${\bf k} = (0,\pi)$) and $M$-points (${\bf k} = (\pi,\pi)$).  
(c) Strength of cavity-mediated interaction in Eq.~(\ref{eq.U_eff}) at the $\Gamma$-point $(k_x, k_y) = (0, 0)$ is plotted vs. the interband interaction strength $U_{12}$. The noninteracting model is plotted as a dashed line for comparison. 
}
\label{fig.interaction}
\end{figure}

We next investigate how electronic screening affects the cavity-mediated interaction in $\hat{U}_{\rm eff}$, Eq.~(\ref{eq.U_eff}), in forward-scattering direction when ${\bf k} = {\bf k}'$, where it is proportional to $\sim 1/ (\Delta_{{\bf k},s})^2$. 

In Fig.~\ref{fig.interaction}(a), we plot the inverse square of Eq.~(\ref{renormalised-denominator}) with the momentum ${\bf k}$ along the path $Y \to \Gamma \to M$. 
In addition, we provide the unscreened interaction for comparison. 
As before, in order to keep the comparison fair, we fix the detuning to the resonance (i.e. either to the exciton or to the upper electronic band). 
Remarkably, we find that the screening drastically \textit{enhances} the cavity-mediated interaction. 
This is clearly seen in Fig.~\ref{fig.interaction}(b), where we depict the ratio between the two cases as a function of the detuning to the respective resonances. 
The enhancement decreases with increasing detuning, but remains finite even for large detunings of half an electronvolt. 

This naturally leads us to investigate optimal conditions to maximise the excitonic enhancement, which we illustrate in Fig.~\ref{fig.interaction}(c). 
At a fixed detuning $\Delta = 0.05$~eV and fixed intraband repulsion $U_{11} = 1.6$~eV, we find a pronounced peak of the excitonic enhancement at $U_{12} \simeq 0.5$~eV. 
In this optimal regime, a massive enhancement of a factor $\gtrsim 10$ relative to the noninteracting model with the same detuning and driving strength is observed.
This enhancement is tied to the emergence of the exciton resonance in Eq.~(\ref{exciton-resonance}), and therefore the mixing of electronic momenta via the scattering of the virtual exciton. It has a non-trivial dependence on electron dispersion $\ep_{{\bf k},b}$: 
If we neglect the electronic band dispersion and replace $\Delta_{\bf k}^0$ in Eq.~(\ref{exciton-resonance}) by the constant detuning $\ep_{21}-\omL$, the exciton binding energy is simply given by $U_{12}$. 
%However, as this exciton does not have a momentum structure, which is distinct from the electrons, the enhancement vanishes in this case, and we recover the same interaction strength as the non-interacting model. 
In this dispersionless case, according to Eq.~(\ref{renormalised-denominator}), the excitonic enhancement vanishes, and we recover the same interaction strength as the non-interacting model. When the electrons become dispersive, it becomes difficult to find a simple analytical expression for $U_{12}$ to realise the largest enhancement effect in Fig.~\ref{fig.interaction}(c). 
Nevertheless, with the help of the effective Hamiltonian~(\ref{eq.main-result}), one can numerically estimate which parameter setup exhibits the greatest enhancement effect under a fixed laser-exciton detuning $\Delta_{ex}$.

\subsection{The excitonic resonance structure}\label{Sec:exciton-resonance-structure}
\begin{figure}[t]
\centering
\includegraphics[width=0.8\textwidth]{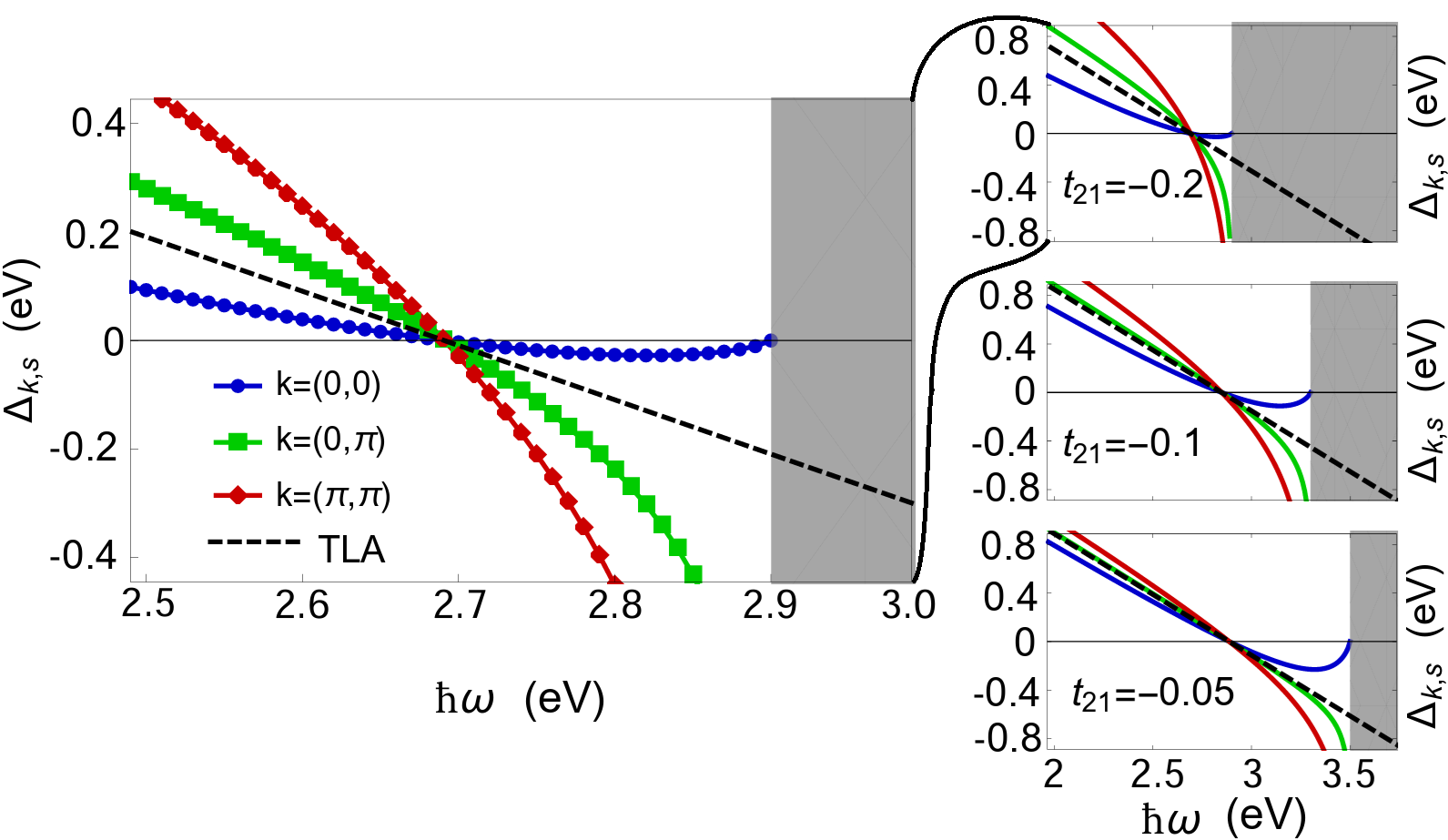}
\caption{The screened denominator $\Delta_{{\bf k},s}$~(\ref{renormalised-denominator}) is plotted vs. the driving frequency $\omL$. The model parameters are given in Section~\ref{subsec:setup&model}, with $U_{11}=2 U_{12}=1.6$eV, $\ep_{21}=\ep_2 -\ep_1=3.7$eV, $t_{2}=-0.15$eV, and with $t_1$ changed to obtain $t_{21}=t_2 -t_1$ as indicated in the panels. 
The black dashed line indicates the detuning in the non-interacting model at the $\Gamma$-point with $\Delta_{\text{gap}}=\Delta_{\text{ex}}$, which is equivalent to the two-level treatment of the Stark shift in Eq.~(\ref{eq.Stark-TLA}). 
    The shaded region represents the upper electronic band, where the effective Hamiltonian~(\ref{eq.main-result}) diverges due to resonant coupling. 
    }
\label{fig.denominator}
\end{figure}

We finish with a discussion of the excitonic resonance structure, which we already touched upon in Section~\ref{sec.screened-denominator}. 
Evidently, all our results can be traced back to the momentum structure of the detuning~(\ref{renormalised-denominator}). 
Thus, we plot the denominator~(\ref{renormalised-denominator}) in Fig.~\ref{fig.denominator} as a function of the driving frequency $\omL$ at several points in the Brillouin zone. 
In the vicinity of the $\Gamma$-point (blue line), the detuning is strongly reduced relative to the TLA treatment, which gives rise the observed enhancement of the cavity-mediated interaction. 
When the bandwidth of the lower band is reduced in the smaller panels on the right, this reduction is lost, and the detuning (at every point in the Brillouin zone) approaches the TLA.  
Moreover, Fig.~\ref{fig.denominator} also illustrates the origin of the broadening of the Floquet-induced interaction: the detuning vanishes at the excitonic resonances \textit{for all momenta}. 
In contrast, at the band edge, the detuning only ever vanishes at a single resonant momentum, while the remaining momenta are off-resonant. 
This vanishing-for-all-momenta behaviour leads to the broadening of the Floquet renormalisation of the band structure, which we observed in Fig.~\ref{fig.dispersion}, and the broadening of the cavity-mediated interaction in Fig.~\ref{fig.interaction}. 

In Appendix~\ref{sec.GRPA}, we show that the RWA component of the screened Floquet Hamiltonian Eq.~(\ref{eq.main-result}) can be derived by an alternative diagrammatic method in a rotating reference frame. The excitonic resonance structure discussed above can be reproduced by the particle-hole t-matrix which we evaluate in the generalised random phase approximation (GRPA).

\section{Discussion and Conclusion}\label{sec:conclude}

\subsection{Validity and Accuracy of the Floquet Hamiltonian}

%We leave the non-perturbative BCS-channel analysis~\cite{vsimkovic2021superfluid,Deng_2015} of the cavity-mediated interaction, and the corresponding instability competition/coexistence between Pomeranchuk and pairing, for future works. 

In our approach, to simplify the Floquet Hamiltonian, we start from a non-interacting band-model, and carry out a resummation over RPA-type inter-band interaction vertices. This approximation is reliable (see e.g. Refs.~\cite{PhysRevB.40.3802} and \cite{haug1984electron}) only if the undriven system can be described as a Fermi liquid, which is indeed the case in our weakly hole-doped semiconductor system: the strong on-site repulsion cannot create Fermi-surface instabilities as the charge carriers (holes in the lower-band in our case) are dilute \cite{arovas2022hubbard}. More precisely, one can define the diluteness parameter $1/\ln (1/n)$, which in our case (the hole-occupancy is $n\sim5\times10^{-4}$, corresponding to a hole-density of $\sim 10^{11} / \text{cm}^2$ in tetracene-type molecular semiconductors) evaluates to $\sim 1/8$, which is too small for Fermi surface instabilities to appear. Previous cluster perturbation simulations \cite{wang2020emergence} also agree that the Fermi liquid is a valid description for the parameters considered here up to a doping level of $n~\sim0.3$. However, we note that this diluteness is sensible to the interaction range: The diluteness parameter becomes no longer small when the interaction becomes long ranged \cite{engelbrecht1992landau}, which makes Fermi-liquid instability easier to trigger.

In the formula for the Floquet-induced cavity-mediated interaction shown in Eq.~(\ref{eq.U_eff}), we assume the cavity mode only carries 0 momentum, which allows us to focus on the forward scattering channel. This treatment is consistent with our assumption that the cavity mode is much larger than the material's lattice constant. On the other hand, if we consider a finite volume of the cavity mode, the electron density variation will appear near the boundary of the cavity mode. To observe this cavity-mediated density variation, we will need to include the ${\bf q}\neq0$ part of the cavity-mediated interaction which is given by Eq.~(\ref{cavity-meidated-interaction-k-GRPA}).

Finally, to study the non-linear driving effects in the $\vert   g \vert^4$ order, one needs to address two issues in this projector-based Gaussian elimination method: cancellation of infinities (see Appendix \ref{appendix:self-consistent-gL4-Hamiltonian}), as well as identifying the \textit{spurious Floquet-induced interactions} (see Appendix \ref{subsec:leading-gL4}). Both issues arise from the self-consistency requirement in $\hat{H}^{\text{eff}}_{(\Ea)}$. In the upgraded Floquet method developed in the next chapter, these self-consistency-related issues of the higher-order Floquet Hamiltonians disappear.

\subsection{Summary}

In this chapter, we show how the virtual excitons in an off-resonantly driven hole-doped semiconductor enhance the Floquet-induced interactions in a cavity QED setup. By an inter-band screening calculation, we find both the optical Stark shift and the cavity-mediated interaction are enhanced by the virtual exciton mechanism: 
In particular, in the presence of multi-band on-site repulsion, we find the excitonic screening results in a renormalisation of the detuning-denominator $ \Delta_{\bf q}^0 \to \Delta_{{\bf q},s} $ in the driving effects (compared with the non-interacting case). We find the frequencies for exciton- and band-resonance can both be obtained from this renormalised denominator $\Delta_{{\bf q},s}$. According to this screened denominator, near the exciton resonance, the Floquet-induced cavity-mediated interaction will be strongly enhanced at all incoming electron momenta. The physical reason for this enhancement is that, via the formation of a virtual exciton, the inter-band interaction allows an electron at momentum $\bf q$ to be influenced by the virtual band-excitation at every momentum $\bf k$, which we denote as the momentum-mixture mechanism.

Our current analysis is based on the screened Floquet Hamiltonian Eq.~(\ref{eq.main-result}), which is derived by a projector-based Gaussian elimination Floquet Hamiltonian method. This method is capable of describing the minimal cavity-material model with Hubbard interaction and linearly polarised laser. By including the long-range repulsion into $\hat{U}$ as described in Section \ref{sec:general-setup}, various effects will emerge during the screening calculation: bi-exciton states and their excitonic Stark shift\cite{combescot1988excitonic}, trion states~\cite{mak2013tightly},  charge-transfer excitons\cite{Cudazzo_2015} and Wannier-Mott excitons\cite{RevModPhys.90.021001,latini2019cavity}. 

However, the perturbative expansions and re-summations of the Green operator in our projector-based method become quite tedious in the generalised driven models with long-range electron interactions. This prompts us to develop a Floquet Hamiltonian method that allows us to derive the Floquet-induced interactions with the least possible effort. Besides, we also want this new Floquet method to transcend the Hartree-type mean-field limitations faced by the current projector method in Eq.~(\ref{right-moving}). As explained below, going beyond this mean-field limitation is vital for accurately understanding the driving-induced phase transitions in the cavity-material setups considered in this thesis.  

\subsubsection{Why do we need a better Floquet Hamiltonian method to study the driving-induced phase transition in this system?}
The Floquet Hamiltonian given by the Sambe space Gaussian elimination method reveals the in-gap exciton resonance. In our model, once we increase the driving frequency above this resonance, the renormalised denominator (representing the effective laser-material detuning) changes sign. This sign-changing property is qualitatively consistent with the two-level atom treatment \cite{Sie2018}, where the exciton resonance is treated as a single eigenstate coupled to the ground state by laser drive with a Coulomb-enhanced coupling strength. Our projector-based Gaussian elimination method reveals the relevance of this renormalised denominator in two important driving-induced effects: excitonically enhanced AC Stark shift and cavity-mediated interaction.

However, we note that the excitonically enhanced AC Stark shift is a mean-field effect which reflects the Floquet-induced interaction at the level of single-particle self-energy. When trying to predict the phase transition triggered by the Floquet drive, we need to derive the Floquet-induced interactions at the level of two-particle interaction (which gives rise to the renormalised Stark shift at the mean-field level). This Floquet-induced interaction exists without a cavity, and it will compete/collaborate with the excitonically enhanced cavity-mediated interaction. To derive this cavity-independent interaction, we need to go beyond the Hartree-type mean-field treatment in Eq.~(\ref{mean-field-global-chap4}). However, as we have seen in Section ~\ref{sec:Hartree-type-mean-field-chap4}, without using Eq.~(\ref{mean-field-global-chap4}), our projector-based Floquet method can no longer eliminate the self-consistency requirement in $\hat{H}^{\text{eff}}_{(\Ea)}$. Consequently, the current method cannot provide the self-consistent Floquet-induced cavity-independent interactions, which requires going beyond the mean-field approximations.

To predict the driving-induced phase transition in this cavity-material system, we need to find the Floquet Hamiltonian revealing the complete Floquet-induced interactions, including both the cavity-independent term and the cavity-mediated term. To achieve this, the corresponding Floquet method should be able to provide the exact self-consistent Floquet Hamiltonian: This Floquet Hamiltonian goes beyond the mean-field approximation while it still preserves self-consistency in $\Ea$. Besides, this Floquet method should also offer a systematic way to simplify its exact many-body Floquet Hamiltonian. Developing such a Floquet Hamiltonian method will be the main focus of the next chapter.

%\bibliography{Bib}

%\appendix

%% file: Chapters/FSWT.tex
In this chapter, we present a Floquet Schrieffer Wolff transform (FSWT) to obtain effective Floquet Hamiltonians and micro-motion operators of periodically driven many-body systems for any non-resonant driving frequency. The FSWT perturbatively eliminates the oscillatory components in the driven Hamiltonian by solving operator-valued Sylvester equations. We will show how to solve these Sylvester equations without knowledge of the eigenstates of the undriven many-body system. In the limit of high driving frequencies, these solutions reduce to the well-known high-frequency limit of the Floquet-Magnus expansion.

\begin{figure}
    \centering
    \includegraphics[width=0.45\textwidth]{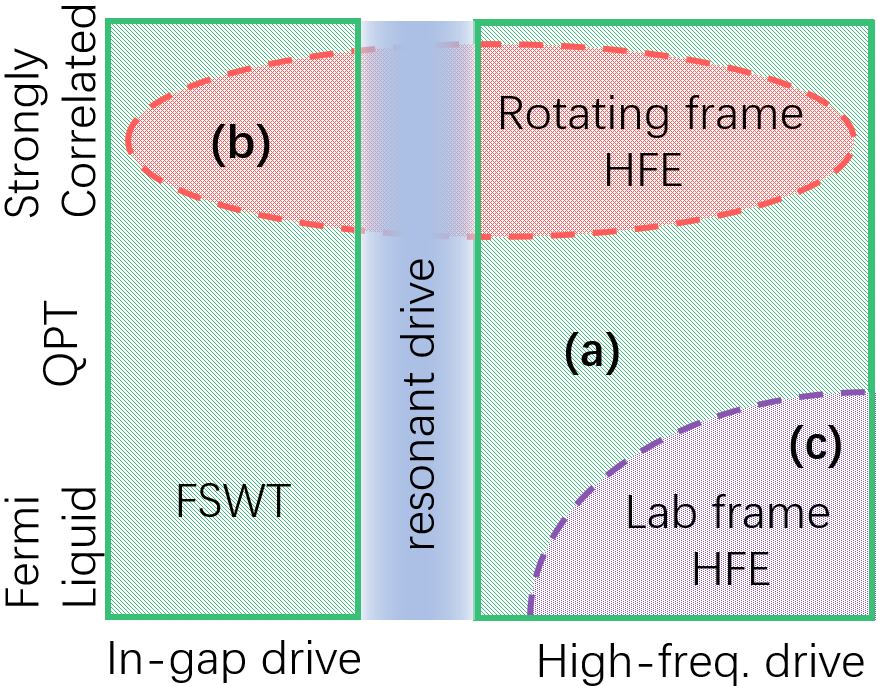}
    \caption{
    Sketch of the application regimes of different Floquet theories in terms of the driven system's interaction strength and the driving frequency relative to a system resonance: 
    %The convergence regime of different Floquet theories in the parameter space of a driven interacting system. 
    Our lab-frame FSWT, marked by (a), provides a unified result that converges for all non-resonant driving regimes. The high-frequency expansion 
    %, whichever the frame it is constructed in, 
    encounters convergence issues when the oscillation of the Hamiltonian %in that frame 
    becomes slow. The rotating-frame HFE in Ref.~\cite{PhysRevLett.116.125301}, labelled (b), diverges when the correlations are weak. The lab-frame HFE, labelled (c), diverges under the in-gap drive and converges slowly in the strongly correlated case when the interaction is comparable to the driving frequency. }
    \label{fig:illustration}
\end{figure}

In particular, this FSWT method constructs an effective Floquet Hamiltonian which is generated perturbatively in the driving strength $g$ rather than the inverse driving frequency $1/\omL$. This method is, therefore, particularly well suited to describe low-frequency or in-gap driving of multi-orbital systems, as illustrated in Fig.~\ref{fig:illustration}. The FSWT constructed in this thesis is conceptually equivalent to a time-dependent transform previously used in circuit QED systems \cite{blais2007quantum,ann2022two}, which perturbatively eliminates the time-dependence in the two-body Hamiltonian of the driven Jaynes-Cummings model. Here, we use the derivative of the exponential map to formulate this transform as a systematic many-body perturbation theory based on Sylvester equations, which we can solve with well-controlled approximations for driven many-body systems.

In the projector-based Gaussian elimination method used in chapter \ref{Chapter4}, the self-consistent Floquet Hamiltonian can only be obtained with the help of mean-field approximations. These approximations are no longer needed in our FSWT method, where the self-consistency of the Floquet Hamiltonian is guaranteed during the solution of the Sylvester equation. Therefore, the Floquet-induced interactions obtained by FSWT are not only self-consistent (their formula remains unchanged across all quasi-eigenenergies $\Ea$) but also free from mean-field limitations.

%We anticipate this method will be useful for describing multi-orbital and long-range interacting systems driven in-gap. 

The chapter \ref{Chapter5} is organised as follows: In Section \ref{Floq-Schrieffer-Wolff}, we construct our Floquet Schrieffer Wolff transform for general periodically driven systems. 
We perturbatively generate the underlying Sylvester equations and explain how to find the Floquet Hamiltonian and micro-motion operators in our FSWT. In Section \ref{compare}, we compare the FSWT with other Floquet methods, such as the HFE, 
%the methods with Floquet-induced interaction, 
the Sambe space van Vleck block-diagonalisation, and the Gaussian-elimination methods. 
We explain why FSWT gets rid of the spurious Floquet-induced interactions, which appear in other methods suitable for low-frequency driving.

\section{The Floquet Schrieffer-Wolff transform \\ for multi-frequency driving}\label{Floq-Schrieffer-Wolff}

%We first outline the general procedure for time-dependent unitary transform, which forms the basis of our FSWT. The evolution operator from time $t_0$ to $t$ under any time-dependent Hamiltonian $\hat{H}_t$ is denoted by $\hat{\mathcal{U}}_{t,t_0}$. It satisfies the Schr\"odinger equation $i\partial_t \hat{\mathcal{U}}_{t,t_0} = \hat{H}_t \hat{\mathcal{U}}_{t,t_0}$ where we set $\hbar=1$. Using an arbitrary time-dependent unitary transform $\hat{U}_t$, the evolution operator can be decomposed as $\hat{\mathcal{U}}_{t,t_0} = \hat{U}_t^{\dag} \hat{\mathcal{U}}'_{t,t_0} \hat{U}_{t_0}$. The evolution operator $\hat{\mathcal{U}}'_{t,t_0}$ satisfies another Schrodinger equation $i\partial_t \hat{\mathcal{U}}'_{t,t_0} = \hat{H}'_t \hat{\mathcal{U}}'_{t,t_0}$ with the transformed Hamiltonian $\hat{H}'_t = \hat{U}_t \hat{H}_t \hat{U}_t^{\dag} + i (\partial_t \hat{U}_t)\hat{U}_t^{\dag}$. By properly choosing $\hat{U}_t$, the evolution can be greatly simplified, such that $\hat{H}'_t$ becomes (approximately) time-independent, and the time-ordering becomes unnecessary when computing $\hat{\mathcal{U}}_{t,t_0}$. 

Based on the general procedure for time-dependent unitary transformations discussed in Section~\ref{Sec_Block-diagonal-intro}, we will construct the unitary transformation $\hat{U}_t$ that eliminates the time dependence of the Hamiltonian $\hat{H}_t$ order by order in the driving strength $g$. The transformed Hamiltonian, $\hat{H}'_t = \hat{U}_t \hat{H}_t \hat{U}_t^{\dagger} + i (\partial_t \hat{U}_t)\hat{U}_t^{\dagger}$, then becomes time-independent up to the truncated order in $g$. In Section~\ref{subsec:general-formalism}, we first construct our FSWT in time-domain for a general time-dependent Hamiltonian. Then, in Section~\ref{Sectiontime-periodic} we focus on the special case when the Hamiltonian is time-periodic, where FSWT is formulated in the frequency domain with the help of the Sylvester equations. We will set $\hbar =1$ for compactness in the FSWT formulas.

\subsection{General Formalism}\label{subsec:general-formalism}
We consider a general driven system described by the following time-dependent Hamiltonian
\begin{equation}\label{H_t}
    \hat{H}_{t} = \hat{H}^{(0)} + \sum\limits_{n=1}^{\infty} \hat{H}_t^{(n)} 
\end{equation}
Here, we assume the driving strength is characterized by a quantity $g$ (which can be proportional to the electric field's amplitude during the laser pulse), 
and $\hat{H}_t^{(n)} \sim g^n$. 
%is the driving Hamiltonian with strength at order $g^n$, 
Each $\hat{H}_t^{(n)}$ term may contain both time-dependent and static terms (which appear in, e.g. the co-rotating frame). 
%For electron-light dipolar coupling, $\hat{H}_t^n$ exists only for $n=1$. 
%While for Coulomb gauge minimal coupling and inter-lattice Peierls coupling, $\hat{H}_t^n$ also exists for $n>1$. 
$\hat{H}^{(0)}$ is the undriven Hamiltonian with $\mathcal{O}(g^0)$, which is time-independent. Thus the time-dependence starts at order $\mathcal{O}(g^1)$ in $\hat{H}_t$.

In the following, we aim to find a time-dependent unitary transform $\hat{U}_t$ which eliminates the time-dependency %(e.g. oscillation) 
in $\hat{H}_{t}$ order by order in $g$, such that the remaining time-dependence in the transformed Hamiltonian $\hat{H}'_t$ only contains higher orders of driving strength $g$. Therefore, the residual time-dependence is much weaker than its original strength $g^1$ and can be neglected, $H'_t \simeq H'$. 
The evolution operator of the driven system, 
\begin{align} \label{eq.propagator}
\hat{\mathcal{U}}_{t,t_0} = \hat{U}_t^{\dag} e^{-i\hat{H}'(t-t_0)} \hat{U}_{t_0}, 
\end{align}
is greatly simplified. 
For time-periodic driving, which we analyze in Section~\ref{Sectiontime-periodic}, this procedure is equivalent to block-diagonalising the Sambe space Hamiltonian.

Analogous to the Schrieffer-Wolff transform in a time-independent system, the form of the time-dependent unitary transform $\hat{U}_t$ is chosen to be
\begin{equation}
    \hat{U}_t = e^{\hat{F}_t}
\end{equation}
where $\hat{F}_t$ is an anti-Hermitian operator. Using the derivative of the exponential map, the transformed Hamiltonian $\hat{H}'_t$ reads
\begin{equation}\label{H'}
\begin{split}
        \hat{H}'_t &= \hat{U}_t \hat{H}_{t} \hat{U}_t^{\dag} + i (\partial_t \hat{U}_{t}) \hat{U}_t^{\dag} \\
        &= \hat{H}_{t} + \hat{G}_{t} + \frac{1}{2!} [\hat{F}_t,\hat{G}_t] + \frac{1}{3!} [\hat{F}_t,[\hat{F}_t,\hat{G}_t]] + ...
\end{split}
\end{equation}
where the operator $\hat{G}_t$ is defined as
\begin{equation}\label{G_t}
    \hat{G}_{t} = [\hat{F}_{t},\hat{H}_{t}] + i \partial_t \hat{F}_{t}.
\end{equation}
Eq.~(\ref{H'}) has been used in previous high-frequency expansions
\cite{goldman2014periodically,abanin2017effective}; however, below, we do not make use of the inverse-frequency expansion therein. 
Instead, we expand $\hat{F}_t$ in orders of driving strength $g$
%%%%%%%%%%%%%%%%%%%%%%
% Here, we do not define a dimensionless parameter since how to compare g is highly dependent on the specific problem to solve, as is commonly the case for other expansions such as the HFE \cite{...}. 
\begin{equation}\label{Ft-expand}
    \hat{F}_t = \sum\limits_{n=1}^{\infty}  \hat{F}_t^{(n)} ,
\end{equation}
where $\hat{F}_t^{(n)} \sim g^n$. 
%The anti-Hermiticity of $\hat{F}_t$ implies $\hat{F}_t^{(n)} = - (\hat{F}_t^{(n)})^{\dag}$. 
We require the unitary $\hat{U}_t$ generating the Floquet micro-motion to ($\bf{i}$) preserve the undriven dynamics and ($\bf{ii}$) preserve the macro-motion. Thus, we set
\begin{equation}\label{ansatz}
    \hat{F}^{(0)}_t = 0 ~\text{ and } \lim\limits_{\mathcal{T}\to\infty} \frac{1}{2\mathcal{T}} \int_{-\mathcal{T}}^{\mathcal{T}} dt ~ \hat{F}^{(n)}_t = 0 ,
\end{equation}
which guarantees that $\hat{F}_t$ vanishes if there is no drive, i.e. $g\to0$, and that $\hat{F}_t$ has no static part.
We next expand the transformed Hamiltonian (\ref{H'}) in powers of $g$, 
\begin{equation}
\hat{H}'_t = \sum\limits_{n=0}^{\infty} \hat{H}'^{(n)}_t    
\end{equation}
where $\hat{H}'^{(n)}_t \sim g^n$. Inserting Eqs.~(\ref{G_t}) and (\ref{Ft-expand}) into (\ref{H'}) we collect $\hat{H}'^{(n)}_t$ order by order

\begin{subequations}\label{H'-expansion}
\begin{align}
\hat{H}'^{(0)}_t &= \hat{H}^{(0)} \label{H'-expansion-0} \\
\hat{H}'^{(1)}_t &= \hat{H}^{(1)}_t + [\hat{F}^{(1)}_t,\hat{H}^{(0)}] + i \partial_t \hat{F}^{(1)}_t \label{H'-expansion-1} \\
\hat{H}'^{(2)}_t &= \hat{H}^{(2)}_t + \frac{1}{2}[\hat{F}^{(1)}_t,\hat{H}^{(1)}_t + \hat{H}'^{(1)}_t] + [\hat{F}^{(2)}_t,\hat{H}^{(0)}] + i \partial_t \hat{F}^{(2)}_t \label{H'-expansion-2} \\
\hat{H}'^{(3)}_t &= \hat{H}^{(3)}_t + \frac{1}{2}[\hat{F}^{(1)}_t,\hat{H}^{(2)}_t + \hat{H}'^{(2)}_t] + \frac{1}{2}[\hat{F}^{(2)}_t,\hat{H}^{(1)}_t + \hat{H}'^{(1)}_t]  
%\notag \\ &   
 \label{H'-expansion-3} \\
& 
+ \frac{1}{12} [\hat{F}^{(1)}_t,[\hat{F}^{(1)}_t, \hat{H}^{(1)}_t -\hat{H}'^{(1)}_t  ]] + [\hat{F}^{(3)}_t,\hat{H}^{(0)}] + i \partial_t \hat{F}^{(3)}_t  \notag  \\
\hat{H}'^{(4)}_t  &= \hat{H}^{(4)}_t + \frac{1}{2}[\hat{F}^{(1)}_t,\hat{H}^{(3)}_t + \hat{H}'^{(3)}_t]  
+ \frac{1}{12} [\hat{F}^{(1)}_t,[\hat{F}^{(1)}_t, \hat{H}^{(2)}_t -\hat{H}'^{(2)}_t  ]]
 \label{H'-expansion-4}
 \\
&+ \frac{1}{2}[\hat{F}^{(2)}_t,\hat{H}^{(2)}_t + \hat{H}'^{(2)}_t] + \frac{1}{2}  [\hat{F}^{(3)}_t  ,\hat{H}^{(1)}_t + \hat{H}'^{(1)}_t] 
+ \frac{1}{12} [\hat{F}^{(1)}_t,[\hat{F}^{(2)}_t, \hat{H}^{(1)}_t -\hat{H}'^{(1)}_t  ]]
\notag \\ &  
+ \frac{1}{12} [\hat{F}^{(2)}_t,[\hat{F}^{(1)}_t, \hat{H}^{(1)}_t -\hat{H}'^{(1)}_t  ]] + [\hat{F}^{(4)}_t,\hat{H}^{(0)}] + i \partial_t \hat{F}^{(4)}_t  \notag 
\end{align}
\end{subequations}

Now, we can determine the functions $\hat{F}_t^{(n)}$ for each order $n=1,2,3,...$, which eliminates the time-dependence at order $g^n$ in the transformed Hamiltonian $\hat{H}'$. This is achieved iteratively: the time-independency requirement $\partial_t \hat{H}'^{(n)}_t =  0$ determines $\hat{F}_t^{(n)}$, and $\hat{F}_t^{(n)}$ in turn determines the static part of $\hat{H}'^{(n+1)}_t$. To detail this iteration, we start at $n=1$ in Eq.~(\ref{H'-expansion}): The requirement $\partial_t \hat{H}'^{(1)}_t =  0$ on (\ref{H'-expansion-1}) eliminates the time-dependence to order $g^1$ in $\hat{H}'$, from which $\hat{F}_t^{(1)}$ is determined. Inserting $\hat{F}_t^{(1)}$ back to Eq.~(\ref{H'-expansion}) not only renders the entire $\hat{H}'^{(1)}_t$ static in (\ref{H'-expansion-1}), but also completely determines the static part of $\hat{H}'^{(2)}_t$ in (\ref{H'-expansion-2}). Then we proceed to $n=2$: Using the result for $\hat{F}_t^{(1)}$ derived above, the equation $\partial_t \hat{H}'^{(2)}_t =  0$ for (\ref{H'-expansion-2}) eliminates the time-dependence to order $g^2$ in $\hat{H}'$, determining $\hat{F}_t^{(2)}$. This elimination procedure can be repeated iteratively. Once the $n$-th order equation in Eq.~(\ref{H'-expansion}) is solved, the operator $\hat{F}_t^{(n)}$ will be determined which fixes the static part of $\hat{H}'^{(n+1)}_t$.

\subsection{Time-periodic driving}
\label{Sectiontime-periodic}
The above perturbative procedure works for arbitrary time-dependent Hamiltonians. Below, we assume the driven Hamiltonian $\hat{H}_t = \hat{H}_{t+2\pi/\omega}$ is time-periodic with angular frequency $\omega$. 
We can Fourier transform each order of its driving terms in Eq.~(\ref{H_t}), such that (for all $n\geq1$)
\begin{equation}\label{Ht-Fourier}
\hat{H}^{(n)}_t = \sum\limits_{j=-\infty}^{\infty} \hat{H}^{(n)}_j e^{i j\omega t}
\end{equation}
where we allow $\hat{H}^{(n)}_t$ to contain multi frequency components at $j \omega$, with $j$ an arbitrary integer. 
Here $\hat{H}^{(n)}_{j=0}$ denotes the time-independent driving terms (e.g. the resonant driving term in the rotating frame). 
%The Hermiticity of $\hat{H}_t$ means $\hat{H}_j^{(n)} =  (\hat{H}^{(n)}_{-j})^{\dag}$. 
In this case, we find $\hat{F}_t$ also becomes time-periodic, $\hat{F}_t = \hat{F}_{t+2\pi/\omega}$, and we expand its $n$-th order into a Fourier series
\begin{equation}\label{Ft-Fourier}
    \hat{F}^{(n)}_t = \sum\limits_{j=-\infty}^{\infty}  \hat{f}^{(n)}_j e^{i j \omega t},
\end{equation}
where $\hat{f}_{j}^{(n)} \propto (1-\delta_{n,0}) (1-\delta_{j,0})$ according to our ansatz (\ref{ansatz}), and the anti-Hermiticity of $\hat{F}_t$ means $\hat{f}_j^{(n)} = - (\hat{f}^{(n)}_{-j})^{\dag}$. %We remind that $n$ is the order of driving strength $g$, and $j$ is the Fourier index.

Inserting Eqs.~(\ref{Ht-Fourier}) and (\ref{Ft-Fourier}) into Eq.~(\ref{H'-expansion}), we can determine $\hat{f}_j^{(n)}$ for each order $n=1,2,3,...$, which eliminates the oscillatory component in $\hat{H}'^{(n)}_t$. 
%In the periodic driving case, these equations, i.e. $\partial_t \hat{H}'^{(n)}_t =0$, read: 

\subsubsection{Lowest order Floquet Hamiltonian}
For $n=1$, the equation $\partial_t \hat{H}'^{(1)}_t =0$ becomes, $\forall j\neq0$, 
\begin{equation}\label{formula-for-f_1^1}
    \hat{H}^{(1)}_{j} + [\hat{f}_j^{(1)},\hat{H}^{(0)}] - j\omega \hat{f}_j^{(1)} = 0,
\end{equation}
which eliminates the term oscillating at $e^{i j\omega t}$ in $\hat{H}'^{(1)}_t$. %This is the Sylvester equation. 
Equations of the form~(\ref{formula-for-f_1^1}) are known as \textit{Sylvester equations} in control theory \cite{bhatia1997and}. 

Eq.~(\ref{formula-for-f_1^1}) determines the solution of $\hat{f}_j^{(1)}$ for all $j$. 
Based on the unitary transform $\hat{U}_t = \exp\big( \hat{F}^{(1)}_t \big) = \exp\big( \sum_{j=1}^{\infty}  \hat{f}^{(1)}_j e^{i j \omega t} - H.c. \big)$, we find a transformed Hamiltonian $\hat{H}'_t$ with leading oscillatory contribution $\mathcal{O} (g^2)$.
Based on the ansatz~(\ref{ansatz}), we know the solution $\hat{f}_j^{(1)}$ completely determines the static part in $\hat{H}'^{(2)}_t$ in Eq.~(\ref{H'-expansion-2}). 
Thus, we can directly pick out the static part in the transformed Hamiltonian (\ref{H'-expansion}) up to order $g^2$, which gives 
\begin{equation}\label{H'-lowest}
    \hat{H}' = \hat{H}'^{(0)} + \hat{H}'^{(1)} + \hat{H}'^{(2)}
\end{equation}
with the zeroth order $\hat{H}'^{(0)}=\hat{H}^{(0)}$, the first order $\hat{H}'^{(1)} =\hat{H}^{(1)}_0$, and the second order contribution
\begin{equation}\label{H'2}
\begin{split}
\hat{H}'^{(2)}=\hat{H}^{(2)}_0 + \frac{1}{2}\sum\limits_{j\neq0} [\hat{f}_j^{(1)},\hat{H}^{(1)}_{-j}].
\end{split}
\end{equation}
This $\hat{H}'^{(2)}$ is identified as the lowest order Floquet Hamiltonian.

\subsubsection{Higher order Floquet Hamiltonians}
For $n=2$, using $\hat{f}_j^{(1)}$ and $\hat{H}'^{(1)}=\hat{H}^{(1)}_0$ determined above, the requirement $\partial_t \hat{H}'^{(2)}_t =0$ becomes, $\forall j\neq0$,

\begin{equation}\label{formula-for-f_1^2}
\begin{split}
\hat{H}^{(2)}_j + \frac{1}{2} \sum\limits_{j' \neq 0} [\hat{f}_{j'}^{(1)},\hat{H}_{j-j'}^{(1)}] + \frac{1}{2} [\hat{f}_j^{(1)},\hat{H}_0^{(1)}] + [\hat{f}_j^{(2)},\hat{H}^{(0)}]   - j \omega \hat{f}_j^{(2)} = 0,
\end{split}
\end{equation}
which eliminates the $\mathcal{O} (g^2)$ terms oscillating at $e^{i j \omega t}$ in $\hat{H}'^{(2)}_t$. These are again Sylvester equations, determining $\hat{f}_j^{(2)}$ 
and thus the static part of $\hat{H}'^{(3)}_t$. 
Collecting the static terms in Eq.~(\ref{H'-expansion-3}), we find this static part in $\mathcal{O} (g^3)$ to be
\begin{equation}\label{H'3}
\begin{split}
\hat{H}'^{(3)} &= \hat{H}^{(3)}_0 + \frac{1}{2}\sum\limits_{j\neq0} [\hat{f}_j^{(1)},\hat{H}^{(2)}_{-j}]
+ \frac{1}{2}\sum\limits_{j\neq0} [\hat{f}_j^{(2)},\hat{H}^{(1)}_{-j}] \\
&~~~~~~~~~~ 
+\frac{1}{12} \sum\limits_{j\neq0} \sum\limits_{j'\neq0} [\hat{f}^{(1)}_j,[\hat{f}^{(1)}_{j'}, \hat{H}^{(1)}_{-j-j'} ]]
-\frac{1}{12} \sum\limits_{j\neq0} [\hat{f}^{(1)}_j,[\hat{f}^{(1)}_{-j}, \hat{H}^{(1)}_0  ]]
.
\end{split}
\end{equation}

The corresponding transformed Hamiltonian $\hat{H}' = \hat{H}'^{(0)} + \hat{H}'^{(1)} + \hat{H}'^{(2)} + \hat{H}'^{(3)}$  is generated by the transform $\hat{U}_t = \exp\big( \hat{F}^{(1)}_t + \hat{F}^{(2)}_t \big) = \exp\big( \sum_{j=1}^{\infty}  (\hat{f}^{(1)}_j +\hat{f}^{(2)}_j) e^{i j \omega t} - H.c. \big)$. 
%is identified as the second lowest order Floquet Hamiltonian, which is correct to $g^3$ order. Again, the further perturbation in $n=3$ order will only correct this Floquet Hamiltonian $\hat{H}'$ by $g^4$ order.

This elimination can be extended to arbitrary orders of $g$. In the $n$-th order perturbation process, we first transform $\partial_t \hat{H}'^{(n)}_t =0$ into Sylvester equations, from which $\hat{f}_j^{(n)}$ are determined. 
Then we use the solution $\hat{f}_j^{(n)}$ to obtain the static term in $\hat{H}'^{(n+1)}_t$, which represents the correction to the Floquet Hamiltonian $\hat{H}'$ at $g^{n+1}$ order. The corresponding transform to get this $\hat{H}'$ is given by $\hat{U}_t = \exp \big( \sum_{j=1}^{\infty}   \sum_{k=1}^{n}\hat{f}^{(k)}_j  e^{i j \omega t} - H.c. \big)$.

\subsubsection{The Floquet micro-motion}\label{sec:micro-motion}
The time-periodic operator $\hat{F}_t$, found in FSWT by solving the Sylvester equations, is also known as the Floquet micro-motion operator \cite{Eckardt_2015}. 
%We show in Appendix \ref{linear-response} how this micro-motion plays an important role in describing the driven dynamics: 
%In Appendix \ref{linear-response}, we find the operator $\hat{F}_t^{(1)}$ directly links to the linear response of the driven system. This link stems from the construction of $\hat{F}_t$ in orders of $g$ in our FSWT. 
%To better explain the link, 
The formal solution to our Sylvester equations is given in Appendix \ref{linear-response}, which expresses the solution $\hat{f}^{(n)}_j$ as a Laplace-transformed Heisenberg operator. For example, the formal solution to the lowest order Sylvester equation (\ref{formula-for-f_1^1}) is, $\forall j \neq 0$,
\begin{equation}\label{formal-solution}
    \hat{f}_j^{(1)} = -i \int_0^{\infty} dt ~ e^{i j \omL t} ~ e^{- 0^+ t} ~ e^{i \hat{H}^{(0)} t} \hat{H}_j^{(1)}  e^{-i \hat{H}^{(0)} t}
\end{equation}
which directly relates the micro-motion $\hat{f}_j^{(1)}$ to the retarded Green function of the undriven system, and thus to the linear response. This link stems from the construction of $\hat{F}_t$ in orders of $g$, instead of $1/\omL$, in our FSWT. 

%In Appendix \ref{lab-vs-rot}, we compare the FSWT constructed in the lab frame and in the rotating frame, from which we find the micro-motion $\hat{F}_t$ could play a less significant role when constructing the FSWT in another frame. 

\section{Comparison to other Floquet methods}\label{compare}
\subsection{Comparison to High-frequency expansion}\label{appendix:compare_HFE}

For this comparison, we consider a system driven by a single frequency described by
\begin{equation}\label{Ht-chap5-2}
\hat{H}_t = \hat{H}^{(0)} + \hat{H}_0^{(1)} + \hat{H}^{(1)}_{-1}e^{-i\omega t} + \hat{H}^{(1)}_1e^{i\omega t} .
\end{equation}
We will show that the HFE expansion in Refs.~\cite{goldman2014periodically,casas2001floquet} can be recovered by solving the Sylvester equation in orders of inverse driving frequency $1/\omL$. This confirms that the low-order Floquet Hamiltonians given by our FSWT in Eqs.~ (\ref{H'2}) and (\ref{H'3}), in the high-frequency limit, reduces to the HFE results in Eq.~(\ref{van-Vlek-result}).

For the driven system described in Eq.~(\ref{Ht-chap5-2}), the HFE method described in chapter \ref{Chapter2} yields the following Floquet Hamiltonian [See Eq.~(\ref{van-Vlek-result-chap2})] 
\begin{equation}\label{van-Vlek-result}
\begin{split}
&\hat{H}'_{HFE} 
= \hat{H}^{(0)} + \hat{H}_{0}^{(1)} 
+ \frac{1}{\omega} [ \hat{H}^{(1)}_1 , \hat{H}^{(1)}_{-1} ] \\
&+ [\frac{[ \hat{H}^{(1)}_1 , \hat{H}^{(0)} + \hat{H}_{0}^{(1)} ]}{2\omega^2} , \hat{H}^{(1)}_{-1} ] 
+ [\frac{[ \hat{H}^{(1)}_{-1} , \hat{H}^{(0)} + \hat{H}_{0}^{(1)} ]}{2\omega^2} , \hat{H}^{(1)}_1 ]  \\
&+ \mathcal{O} (\omL^{-2}).
\end{split}
\end{equation}
In FSWT, if we assume the driving frequency $\omL$ to be the largest energy scale, and expand $\hat{f}_{j}^{(n)}$ in the Sylvester Eqs.~(\ref{formula-for-f_1^1}) and ({\ref{formula-for-f_1^2}}) in orders of $\omega^{-1}$, we find the solution
\begin{equation}
\begin{split}
\hat{f}_{1}^{(1)} &= \frac{1}{\omega} \hat{H}^{(1)}_1 + \frac{1}{\omega^2} [\hat{H}^{(1)}_1,\hat{H}^{(0)}] + O[\frac{1}{\omega^3}] \\
\hat{f}_{1}^{(2)} &=  \frac{1}{\omega^2} [\hat{H}^{(1)}_1,\hat{H}_{0}^{(1)}] + O[\frac{1}{\omega^3}] 
\end{split}
\end{equation}
Inserting this high-frequency solution back into Eqs.~(\ref{H'2}) and (\ref{H'3}), we find our Floquet Hamiltonian $\hat{H}'=\hat{H}'^{(0)} +\hat{H}'^{(1)} +\hat{H}'^{(2)} +\hat{H}'^{(3)} $ indeed reduces to $\hat{H}'_{HFE}$ in Eq.~(\ref{van-Vlek-result}). 
This confirms that the FSWT reduces to the HFE if the Sylvester equations are solved in orders of inverse driving frequency $1/\omL$.
%This confirms that the Floquet Magnus high-frequency method is equivalent to our FSWT if the Sylvester equations are solved in orders of inverse driving frequency $1/\omL$.

As an example, in the driven Hubbard system which will be studied in Chapter \ref{Chapter6}, we additionally have $\hat{H}_0^{(1)} =0$, $\hat{H}^{(1)}_{-1} = \hat{H}^{(1)}_{1}$ and $[ \hat{H}^{(1)}_{-1}, \hat{U} ] = 0$ where $\hat{U}$ is the interaction term in $\hat{H}^{(0)}$. For this specific system, we find that in the Floquet Hamiltonian given by the Magnus expansion [see e.g. Eq.~(33) in Ref.~\cite{PhysRevB.93.144307}], the leading order $\sim \omL^{-1}$ vanishes, the order $\mathcal{O}(\omL^{-2})$ only contains the driving-induced bandwidth renormalisation effect. 
The order $\mathcal{O}(\omL^{-3})$ vanishes, such that correlation effects only appear starting from order $\omL^{-4}$. This exactly agrees with our FSWT result Eq.~(\ref{Floquet-H'}) once we expand it in orders of $1/\omL$.

\subsection{Equivalence of FSWT to the Sambe space van Vleck block-diagonalisation}\label{sec:FSWT=vanVleck}

%Vogl et al. recently derived a low-frequency Floquet expansion~\cite{PhysRevB.101.024303}, which relies on the downfolding (i.e. Gaussian elimination) of the Sambe space Hamiltonian to the zeroth Floquet mode. The self-consistency requirement therein stems from the fact that the downfolding to the zeroth Floquet mode is different from the block-diagonalisation of each Floquet sector of the Sambe space Hamiltonian. This method becomes impractical in a many-body system as one has to solve the quasi-eigenenergy self-consistently. In our previous publication~\cite{PhysRevB.109.115137}, we used this Gaussian method without the self-consistency (to the lowest order driving strength) using a many-body projector. Proceeding to higher-order driving effects becomes unfeasible, as divergences appear in the self-consistent determination of the many-body quasi-eigenenergy, which have to be cancelled at each order. 

The Floquet Schrieffer-Wolff transform presented here is equivalent to block-diagonalising the Sambe space Floquet Hamiltonian
%, as shown in Fig.\ref{fig.relation} 
(as explained in Section \ref{Sec_Block-diagonal-intro}). 
We find the following connection between our FSWT approach and the previous Sambe space van Vleck block-diagonalisation method in Ref.~\cite{Eckardt_2015}: 
%van Vleck perturbation theory was used in Ref.~\cite{Eckardt_2015} to block-diagonalise the Sambe space Hamiltonian. 
The latter method results in an expression for the micro-motion operators in the eigenbasis of the undriven Sambe-space Hamiltonian (see Eq.C.40 therein). This expression can be obtained in our approach by expanding the Sylvester equations~ (\ref{formula-for-f_1^1}) and (\ref{formula-for-f_1^2}) in the eigenbasis of $\hat{H}^{(0)} = \sum_{l} \ep_l \vert l \rangle \langle l \vert$, which provides the formal solution to $\hat{f}^{(n)}_j$. For example, the expansion of Eq.~(\ref{formula-for-f_1^1}) yields
\begin{equation}\label{Formal-solu-eigen-expansion}
\langle l \vert \hat{f}^{(1)}_j \vert l' \rangle = \frac{ \langle l \vert \hat{H}^{(1)}_j \vert l' \rangle }{j \omega -(\ep_{l'}-\ep_{l})},
\end{equation}
which coincides with Eq. (C.40) of Ref.~\cite{Eckardt_2015}.
Although our FSWT provides the same formal solution given by the previous van Vleck method, in our case, the Sylvester equations can be solved (at least perturbatively, see Chapters \ref{Chapter6} and \ref{Chapter7}) without knowledge of the eigenstates $\vert l \rangle$ and eigenenergies $\ep_l$. This enables us %our Sylvester equation approach 
to describe many-body systems under in-gap drive, where the eigenbasis cannot be obtained, 
%$\hat{H}^{(0)} = \sum_{l} \ep_l \vert l \rangle \langle l \vert$ is unobtainable, 
as we will demonstrate in the following chapters.
%, and the whole concept of the Floquet zone is no longer used.

%Ref.~\cite{Rodriguez_Vega_2018} constructs a complete diagonalisation of the Sambe space Hamiltonian (\ref{Sambe}) based on a standard perturbation over driving strength. This complete diagonalisation requires the knowledge of the spectrum and the eigenstate of the undriven Hamiltonian, which hinders its application to driven many-body systems. This complete diagonalisation becomes even more tedious when the spectrum extends beyond the First Floquet zone in the low-frequency driving case. 

\subsection{Reduction of FSWT to the Gaussian elimination result}

Below, we show that FSWT reduces to the previous Gaussian elimination method in the weak-driving limit: the Sambe space Gaussian elimination result in Section \ref{Sec2.2.1} can be re-derived from the FSWT Hamiltonian, after making an approximation (energy projection) to the latter. From the eigenbasis of $\hat{H}^{(0)} = \sum_{l} \ep_l \vert l \rangle \langle l \vert$ and the corresponding formal solution of $\hat{f}^{(1)}_1$ in Eq.~(\ref{Formal-solu-eigen-expansion}), we find
\begin{equation}
\hat{f}^{(1)}_1 = \sum_{l} \frac{1}{\omL - \ep_l + \hat{H}^{(0)} } \hat{H}^{(1)}_1 \vert l \rangle \langle l \vert = \sum_{l'} \vert l' \rangle \langle l' \vert \hat{H}^{(1)}_1 \frac{1}{\omL + \ep_{l'} - \hat{H}^{(0)} }  
\end{equation}
Thus according to Eq.~(\ref{H'2}) the $\mathcal{O}(g^2)$ FSWT Hamiltonian is formally given by
\begin{equation}\label{FSWT-to-Sambe}
\begin{split}
&\frac{1}{2} \big( [\hat{f}^{(1)}_1, \hat{H}^{(1)}_{-1}] + H.c. \big) \\ 
&= \frac{1}{2} \left( \sum_{l'} \vert l' \rangle \langle l' \vert \hat{H}^{(1)}_1 \frac{1}{\omL + \ep_{l'} - \hat{H}^{(0)} } \hat{H}^{(1)}_{-1} 
+ \sum_{l} \hat{H}^{(1)}_{-1}  \frac{1}{- \omL + \ep_l - \hat{H}^{(0)} } \hat{H}^{(1)}_1 \vert l \rangle \langle l \vert
+ H.c. \right) \\
&\approx \frac{1}{2} \left( \sum_{l'} \vert l' \rangle \langle l' \vert \hat{H}^{(1)}_1 \frac{1}{\omL + \Ea - \hat{H}^{(0)} } \hat{H}^{(1)}_{-1} 
+ \sum_{l} \hat{H}^{(1)}_{-1}  \frac{1}{- \omL + \Ea - \hat{H}^{(0)} } \hat{H}^{(1)}_1 \vert l \rangle \langle l \vert
+ H.c. \right) \\
&= \hat{H}^{(1)}_1 \frac{1}{\Ea + \omL - \hat{H}^{(0)} } \hat{H}^{(1)}_{-1} 
+ \hat{H}^{(1)}_{-1}  \frac{1}{ \Ea - \omL  - \hat{H}^{(0)} } \hat{H}^{(1)}_1 
\end{split}
\end{equation}
where in the second line, we want to find the effective FSWT Hamiltonian near an eigenenergy $\Ea$ of the undriven system $\hat{H}^{(0)}$, in this case, we approximate $\ep_l \approx \Ea$. Then, using the identity $\sum_{l} \vert l \rangle \langle l \vert = 1$, we reduce our FSWT Hamiltonian back to the previous result Eq.~(\ref{Heff_weak_drive}) given by the Sambe-space Gaussian elimination.

This reduction shows that, by solving the Sylvester equation, our FSWT provides a Floquet Hamiltonian that works at all eigenenergies $\Ea$. Thus, FSWT gets rid of the self-consistency requirement on $\Ea$ in the previous Gaussian elimination method. This is achieved merely by solving the Sylvester equation, which doesn't rely on any mean-field approximations. We will also see that solving our Sylvester equations in many-body systems is, in general, much easier than simplifying the Green operators in the Gaussian elimination method.

\subsection{Absence of spurious Floquet-induced interactions in FSWT}
For a driven many-body system, even in the absence of intrinsic interactions, several Floquet methods find spurious Floquet-induced interactions, as reported in Ref.~\cite{PhysRevB.93.144307}. We also find them in the Gaussian elimination method \cite{PhysRevB.101.024303}, as shown in Appendix \ref{appendix:self-consistent-gL4-Hamiltonian}. These interactions are spurious because, in the absence of intrinsic interactions, we can first decouple the system in momentum space and then construct the Floquet Hamiltonian for each decoupled system using the same method, which can no longer provide any Floquet-induced interactions. 

The spurious Floquet-induced interactions account for the correlations found in a specific Fourier frequency component (e.g. the 0-th Floquet sector) of the driven many-body wavefunction, as the wavefunction remains a product state in the time domain. When this spurious Floquet-induced interaction appears in an interacting system, it is difficult to separate it from the higher-order (non-linear) driving effects. However, our FSWT method doesn't suffer from this issue at all.
The exponential structure $\exp (\hat{F}_t)$ in our FSWT properly allocates this fake correlation into the Floquet micro-motion: Given by the transform $\exp (\hat{F}_t)$, the Floquet Hamiltonian $\hat{H}'$ is expressed entirely in terms of commutators, see Eq.~(\ref{H'3}), which cannot create any spurious Floquet-induced interactions from a non-interacting system. Consequently, these correlations do not appear in the macro-motion described by our FSWT.

The spurious Floquet-induced interaction found in the Gaussian elimination method only occurs in the higher order driving effects, e.g., in the $\mathcal{O}(g^4)$ Floquet Hamiltonian. In the weak-driving limit such that $\mathcal{O}(g^2)$ becomes the only relevant order, the Sambe space Gaussian elimination result is well-behaved, which can be obtained from the FSWT result via the energy projection in Eq.~(\ref{FSWT-to-Sambe}). This is why we didn't encounter the issue of spurious Floquet-induced interaction in chapter \ref{Chapter4}. However, to go beyond the Hartree-type mean-field approximations or to understand the higher-order driving effects, we have to resort to the more powerful FSWT method developed in this Chapter.

%In conclusion, our transform transcends the high-frequency limit and can handle the many-body interactions in $\hat{H}_0$.

\section{Conclusion}
In this chapter, we have constructed the Floquet Schrieffer Wolff transform based on Sylvester equations. This FSWT perturbatively eliminates the oscillatory components in the driven Hamiltonian in orders of driving strength, which goes beyond various high-frequency expansion methods. By solving the Sylvester equation, our FSWT can obtain the effective Floquet Hamiltonians and micro-motion operators of periodically driven many-body systems for any non-resonant driving frequency. This FSWT is an optimal method for deriving the Floquet-induced interactions in driven many-body systems since it no longer suffers from the mean-field limitations, and meanwhile, it can provide the self-consistent Floquet-induced interactions which remain unchanged at all quasi-energies $\Ea$. Furthermore, the only computational effort in our FSWT is to solve the Sylvester equations. In the following chapters, we will show how to analytically solve these Sylvester equations using two representative driven many-body systems. 

In chapter \ref{Chapter6}, as a first example, we will show how to solve the Sylvester equations for driven many-body systems with on-site Hubbard interactions relevant to cold atom experiments. We will demonstrate the advantage of our FSWT in accurately predicting the Floquet-induced interactions in these driven cold-atom setups. Then, in chapter \ref{Chapter7}, we will apply FSWT to the generalised cavity-material system described in Section \ref{sec:general-setup} and obtain the complete Floquet-induced interactions in this cavity-material setup which transcends the previous Hartree-type mean-field approximations.

%% file: Chapters/chain.tex
In this chapter, we will illustrate our FSWT method with the example of a driven Fermi Hubbard chain. We show how to solve the Sylvester equation for this many-body system without knowing its eigenstates. The Floquet-induced interactions in this system are identified as correlated hopping terms, whose strength is underestimated in HFE. We demonstrate the importance of these Floquet-induced interactions. They can substantially change the Floquet-induced metal-insulator transition when the driving frequency is not too large. Using tensor network simulations \cite{orus2014practical} and exact diagonalisation, we simulate the return rates of the driven Hubbard chain. From these time evolution simulations, we benchmark the FSWT method and find substantial improvement compared to HFE Floquet approaches in large regions of parameter space (corresponding to the right part of Fig.~\ref{fig:illustration}). 

In Section \ref{example}, we apply the FSWT to a monochromatically driven Hubbard system and find a Floquet Hamiltonian which remains valid for all ratios of $U/J$, provided that the driving is off-resonant. In Section \ref{conclude}, we summarise our results and discuss the applications of FSWT in the strong driving frame. We will set $\hbar=1$.

\section{The Driven single-band Fermi Hubbard model}\label{example}

%Here, we consider the 
Below, we explore the applicability of the FWST with the example of the
one-band Hubbard model with on-site repulsion $U$ and nearest neighbour hopping $J$. It is driven monochromatically with a 
driving frequency $\omL$, which is much higher than the hopping energy $J$ (i.e. $\omL\gg J$) 
\footnote{We will not consider the $\omL\sim J$ case, as this would correspond to resonant driving of the band electrons.}.
The FWST provides an effective Floquet Hamiltonian that remains accurate for arbitrary ratios of $U/J$. 

The Hamiltonian for this driven system reads
\begin{equation}\label{Ht-example}
\hat{H}_t = \hat{H}^{(0)} + \hat{H}^{(1)}_{-1}e^{-i\omL t} + \hat{H}^{(1)}_1e^{i\omL t},
\end{equation}
where $\hat{H}_1^{(1)} =  (\hat{H}^{(1)}_{-1})^{\dag}$ due to the Hermiticity of $\hat{H}_t$. The undriven Hamiltonian is given by
\begin{equation}\label{H^0-example}
\hat{H}^{(0)} = \hat{h} + \hat{U} - \mu \hat{N},
\end{equation}
with kinetic energy %a one-particle term 
\begin{equation}
\hat{h} = -J \sum\limits_{s} \sum\limits_{j=1}^{L-1} (\hat{c}_{j+1,s}^{\dag} \hat{c}_{j,s} + \hat{c}_{j,s}^{\dag} \hat{c}_{j+1,s} ),
\end{equation}
local repulsion 
\begin{equation}\label{U}
\hat{U} = U \sum\limits_{j=1}^{L} \hat{n}_{j,\uparrow} \hat{n}_{j,\downarrow},
\end{equation} 
and chemical potential $- \mu \hat{N}$, where $\hat{N}= \sum_{j, s} \hat{n}_{j,s}$ is the total particle number. 
We consider the electric dipole driving in the long-wavelength limit, 
\begin{equation}\label{H_1-example}
\hat{H}^{(1)}_{1} = g \sum\limits_{j=1}^{L} 
j ~ \sum\limits_{s}  \hat{n}_{j,s}. 
\end{equation}
which can be realised, e.g., in cold atom platforms using shaken lattices ~\cite{PhysRevLett.99.220403} or in electronic platforms ~\cite{hensgens2017quantum}. 
The total particle number commutes with the driving term, and therefore, the chemical potential is not altered by the driving. In the following, we will set $\mu = 0$, unless specified otherwise. It can added to the effective Hamiltonian, if necessary. 
%Unless otherwise specified, we will set $\mu = 0$ (half-filling) in the following.
%The chemical potential $\mu$ cannot influence the driven dynamics, thus we set $\mu=0$ unless further specified. 

In Hamiltonian~(\ref{Ht-example}), we encounter the particularly simple situation $\hat{H}^{(n\geq1)}_j \propto \delta_{n,1}\delta_{j,\pm1}$, i.e. all driving terms other than $\hat{H}^{(1)}_{\pm1}$ vanish identically. This implies $\hat{H}'^{(1)}=0$, i.e. there is no %$\mathcal{O}(g^1)$
linear contribution to the Floquet Hamiltonian. To the lowest order of driving strength ($n=1$), in Eq.~(\ref{formula-for-f_1^1}), there is only one Sylvester equation for $\hat{f}^{(1)}_j$ which needs to be solved. It reads
\begin{equation}\label{FSWT-f^1_1}
\hat{H}^{(1)}_{1} + [\hat{f}^{(1)}_1,\hat{H}^{(0)}] - \omL \hat{f}^{(1)}_1 = 0,
\end{equation}
and the corresponding Floquet Hamiltonian~(\ref{H'-lowest}) 
to quadratic order in $g$ %to order $\mathcal{O} (g^2)$ 
simplifies to 
\begin{equation}\label{H'-1st}
\begin{split}
\hat{H}' &= \hat{H}'^{(0)} + \hat{H}'^{(2)} \\
&= \hat{H}^{(0)}  + \frac{1}{2} \big( [\hat{f}^{(1)}_1, \hat{H}^{(1)}_{-1}] + H.c. \big) .
\end{split}
\end{equation}
We next solve the %lowest order 
Sylvester equation~(\ref{FSWT-f^1_1}). % in the driven single-band Hubbard model. 
Without any approximation, %truncation, 
the exact solution to Eq.~(\ref{FSWT-f^1_1}) in a general driven many-body system with interactions will contain infinitely long operator-product terms, as does the Floquet Hamiltonian given by Eq.~(\ref{H'-1st}). This is shown in Appendix~\ref{appendix:first-order-example} and was also noticed in other Floquet methods~\cite{PhysRevX.9.021037,mori2023floquet,PhysRevX.4.041048}.
Below, we show how to find approximate solutions of the Sylvester equations for the driven Hubbard model, using neither the inverse-frequency expansion in HFE nor the eigenbasis of the undriven many-body Hamiltonian. 
Instead, since $J\ll \omL$, we perturbatively solve the lowest order Sylvester equation (\ref{FSWT-f^1_1}) by expanding $\hat{f}^{(1)}_1$ in orders of the hopping $J$. We will see that the result constructed in this way remains applicable regardless of the ratio of $U/J$, ranging from strongly correlated to weakly correlated cases. This expansion in $J$ still converges when $U>\omL$, %which is shown to be 
in sharp contrast with the HFE results in ~\ref{appendix:compare_HFE}.% where $\hat{f}^{(1)}_1$ is expanded in orders of $1/\omL$.

\subsection{Driven Hubbard dimer}\label{subsec:driven-dimer}
%We start from the simple case with only two lattice sites $j=1,2$.
We start our discussion with a Hubbard dimer, with site index $j=1,2$. 
We expand the function $\hat{f}^{(1)}_1$ in orders of $J$, such that $\hat{f}^{(1)}_1 = \sum_{n=0}^{\infty} \hat{y}_n$ where $\hat{y}_n \sim J^n$, and insert it into Eq.~(\ref{FSWT-f^1_1}). This results in a set of equations for different orders of $J$,
\begin{subequations}\label{y}
    \begin{align}
& \hat{H}^{(1)}_{1} + [\hat{y}_0,\hat{U}] - \omL \hat{y}_0 = 0  \label{y0} \\
& [\hat{y}_0,\hat{h}] + [\hat{y}_1,\hat{U}] - \omL \hat{y}_1 = 0 \label{y1} \\
& [\hat{y}_1,\hat{h}] + [\hat{y}_2,\hat{U}] - \omL \hat{y}_2 = 0, %~~~~ \text{and so on}  
\label{y2}
    \end{align}
\end{subequations} 
and so on. This system can be solved order by order.
The solution, including the $\hat{y}_0 \sim J^0$ and $\hat{y}_1 \sim J^1$ order, is 
\begin{equation}
\begin{split}
& \hat{f}^{(1)}_1 = \frac{g}{\omL} \sum\limits_{s} \sum\limits_{j=1,2} j ~ \hat{n}_{j,s} \\
&+ \frac{g J}{\omL^2} \sum\limits_{s}   
\hat{c}_{1,s}^{\dag} \hat{c}_{2,s} ( 1 + \beta' \hat{n}_{1,\bar{s}} + \gamma' \hat{n}_{2,\bar{s}} + \delta' \hat{n}_{1,\bar{s}} \hat{n}_{2,\bar{s}} ) \\
&- \frac{g J}{\omL^2} \sum\limits_{s}   
\hat{c}_{2,s}^{\dag} \hat{c}_{1,s} ( 1 + \beta' \hat{n}_{2,\bar{s}} + \gamma' \hat{n}_{1,\bar{s}} + \delta' \hat{n}_{2,\bar{s}} \hat{n}_{1,\bar{s}} ) \\
&+ \mathcal{O} \left( J^2 \right)
\end{split}
\end{equation}
\\
where the coefficients are given by
\begin{equation}\label{parameters'-main}
\begin{split}
\beta' &= \frac{-U}{\omL+U}    ~~~~~~~~ 
\gamma' = \frac{U}{\omL-U} ~~~~~~~
%\delta' = \frac{-2U^2}{(\omL+U)(\omL-U) } = - \beta' - \gamma'  
\delta' =  - \beta' - \gamma'  
\end{split}
\end{equation}
We stress that the form of $\hat{y}_1$ is not determined using the eigenbasis of the undriven Hamiltonian, but instead by symmetries: in Eq.~ (\ref{y1}), we find $[\hat{y}_0,\hat{h}]$ contains hopping between site 1 and 2, while $\hat{U}$ cannot create hopping. Thus, $\hat{y}_1$ must contain hopping. And since $\hat{U}$ correlates two spins, the hopping in $\hat{y}_1$ must be accompanied by local terms $\hat{n}_{j,\bar{s}}$ in the opposite spin. This symmetry argument completely determines the form of $\hat{y}_1$, and then the coefficients are obtained from Eq.~(\ref{y1}).

\subsection{Driven Hubbard chain}

In an $L$-site Hubbard chain, following the same perturbation procedure (in orders of $J$) and the same symmetry argument (for details, see Appendix \ref{appendix:first-order-example}), we find the solution of Eq.~(\ref{FSWT-f^1_1}), which reads

\begin{equation}\label{FSWT-H'-chain-g^2}
\begin{split}
& \hat{f}^{(1)}_1 = \frac{g}{\omL} \sum\limits_{s} \sum\limits_{j=1}^{L} j ~ \hat{n}_{j,s} 
+ \frac{g J}{\omL^2} \sum\limits_{s} \sum\limits_{i,j} \left( \delta_{i-j,1} - \delta_{j-i,1} \right) \hat{c}_{j,s}^{\dag} \hat{c}_{i,s} 
%\\&~~~~~~~~~~~~~~~~~ \times 
( 1 + \beta' \hat{n}_{j,\bar{s}} + \gamma' \hat{n}_{i,\bar{s}} + \delta' \hat{n}_{j,\bar{s}} \hat{n}_{i,\bar{s}} ) \\
&~~~~~~~~~~~~~~~~~~~~~~~~~~~~~   + \hat{y}_2 + \mathcal{O} ( J^3 )
\end{split}
\end{equation}
where the parameters $\beta'$, $\gamma'$ and $\delta'$ are given in Eq.~(\ref{parameters'-main}), and the $\mathcal{O} ( J^2 )$ contribution $\hat{y}_2$ is given in Eq.~(\ref{y2-full-result}).
According to Eq.~(\ref{H'-1st}), the corresponding lowest-order corrected Floquet Hamiltonian is given by%\begin{equation}\label{Floquet-H'2}
%\begin{split}
%&\hat{H}'^{(2)} = \frac{g^2 J}{\omL^2} \sum\limits_{s} \sum\limits_{j=1}^{L-1}  ( \hat{c}_{j,s}^{\dag} \hat{c}_{j+1,s} + \hat{c}_{j+1,s}^{\dag} \hat{c}_{j,s}   )  
%\left( 1 + \frac{\beta' + \gamma'}{2} (\hat{n}_{j,\bar{s}} +  \hat{n}_{j+1,\bar{s}} ) +  \delta' \hat{n}_{j,\bar{s}} \hat{n}_{j+1,\bar{s}} \right) 
%~~ + \frac{1}{2}([\hat{y}_2,\hat{H}^{(1)}_{-1}]+ H.c.)  + \mathcal{O} ( J^3 ).
%\end{split}
%\end{equation}
%Thus, the whole $\mathcal{O} ( g^2 )$ Floquet Hamiltonian $\hat{H}'^{(0)} + \hat{H}'^{(2)}$ reads
\begin{equation}\label{Floquet-H'}
\begin{split}
\hat{H}'^{(0)} &+ \hat{H}'^{(2)}= -J(1-\frac{g^2}{\omL^2}) \sum\limits_{s} \sum\limits_{j=1}^{L-1} ( \hat{c}_{j+1,s}^{\dag} \hat{c}_{j,s} + \hat{c}_{j,s}^{\dag} \hat{c}_{j+1,s} )  ~~~~~ + ~~~~ U \sum\limits_{j=1}^{L} \hat{n}_{j,\uparrow} \hat{n}_{j,\downarrow} \\
&~~~~~~  - J ~ \frac{g^2 U}{\omL^2} \left( \frac{1}{U-\omL} + \frac{1}{\omL+U} \right) \sum\limits_{s} \sum\limits_{j=1}^{L-1}  ( \hat{c}_{j,s}^{\dag} \hat{c}_{j+1,s} + \hat{c}_{j+1,s}^{\dag} \hat{c}_{j,s}   )
\left( \frac{ \hat{n}_{j,\bar{s}} +  \hat{n}_{j+1,\bar{s}} }{2} - \hat{n}_{j,\bar{s}} \hat{n}_{j+1,\bar{s}} \right) \\
&~~~~~~  + \frac{1}{2}([\hat{y}_2,\hat{H}^{(1)}_{-1}]+ H.c.)  + \mathcal{O} ( J^3 ).
\end{split}
\end{equation}

The $\mathcal{O} ( J^2 )$ term, $\frac{1}{2}([\hat{y}_2,\hat{H}^{(1)}_{-1}]+ H.c.)$ in the third line, is given explicitly in Eq.~(\ref{Floquet-H'2-t2}).
The non-interacting part of the Floquet Hamiltonian~(\ref{Floquet-H'}) shows the lowest order driving-induced bandwidth renormalisation (dynamical localisation), as predicted theoretically in, e.g.  Refs.~\cite{PhysRevB.34.3625,PhysRevLett.95.260404,PhysRevB.78.235124} and realised in e.g. Refs.~\cite{PhysRevLett.99.220403,PhysRevLett.81.5093}. The interacting part of Eq.~(\ref{Floquet-H'}) %is composed of the
contains a driving-induced interaction in the second line of Eq.~(\ref{Floquet-H'}), which modifies the interaction term $\hat{U}$ in the undriven system and renders the overall interaction non-local. This Floquet-induced interaction in Eq.~(\ref{Floquet-H'}) has the form of correlated hopping \cite{PhysRevB.88.115115}.
%, with a coefficient differing by a factor of $\beta'+\gamma'$ compared to the bandwidth renormalisation.
Similar correlated hopping was previously identified in the effective Hamiltonians of undriven systems (e.g. in cold atoms near Feshbach resonances \cite{Duan_2008} and in cavity-coupled quantum materials \cite{PhysRevB.99.085116}). 
The correlated hopping may lead to eta-pairing scars in one dimension~\cite{PhysRevB.102.075132} and superconductivity in two dimensions~\cite{PhysRevB.39.11653,Duan_2008}. It may also modify the Kibble-Zurek mechanism \cite{PhysRevB.95.104306} when the driving quenches the system across a quantum phase transition (QPT) \cite{PhysRevLett.120.127601}.
%, where the effective Hamiltonian for cold atoms near Feshbach resonance was derived based on a configuration argument.
%in dressed pictures. 
%Unlike Ref.~\cite{Duan_2008}, 
In our driven system, as given by FSWT,
%our Floquet method doesn't require the rotating wave approximation (as done for $\hat{H}_{ij}$ in Ref.~\cite{Duan_2008})
both the near-resonant contribution $1 / ( U - \omL )$ and the off-resonant contribution $1 / (\omL + U)$ can be found in the coefficient of the correlated hopping in Eq.~(\ref{Floquet-H'}). 
%where the accurate coefficient for interaction strength given by our FSWT becomes relevant to explain the observed driving-induced phase transitions.

The HFE result (see \ref{appendix:compare_HFE}) is equivalent to expanding our FSWT result Eq.~(\ref{Floquet-H'}) in orders of $1/\omL$. In HFE, this correlated hopping term only occurs to order $\omL^{-2} (U^2 - \omL^2)^{-1} \sim \mathcal{O}(\omL^{-4})$, and then we might expect it to be much weaker than the bandwidth renormalisation effect which is of order $\omL^{-2} $. Thus, in the lowest order HFE, one typically ignores the entire correlated hopping term. 
However, in our FSWT result Eq.~(\ref{Floquet-H'}), we find the correlated hopping may be of similar strength as the bandwidth renormalisation. 
%However, in our FSWT result Eq.~(\ref{Floquet-H'}), we find the correlated hopping may become stronger than the bandwidth renormalisation (in terms of the coefficient) when $\omL \lesssim 1.7 U$.
%\omL \lesssim 1.73 U$. %In this case, HFE must be corrected by our FSWT. 
%For this lowest order HFE to be accurate, we need $\omL \gtrsim 4 U$, which makes the coefficient of the correlated hopping evidently smaller (i.e. by an order of magnitude) than the bandwidth renormalisation. 

\begin{figure}[h]
    \centering
    \includegraphics[width=0.6\textwidth]{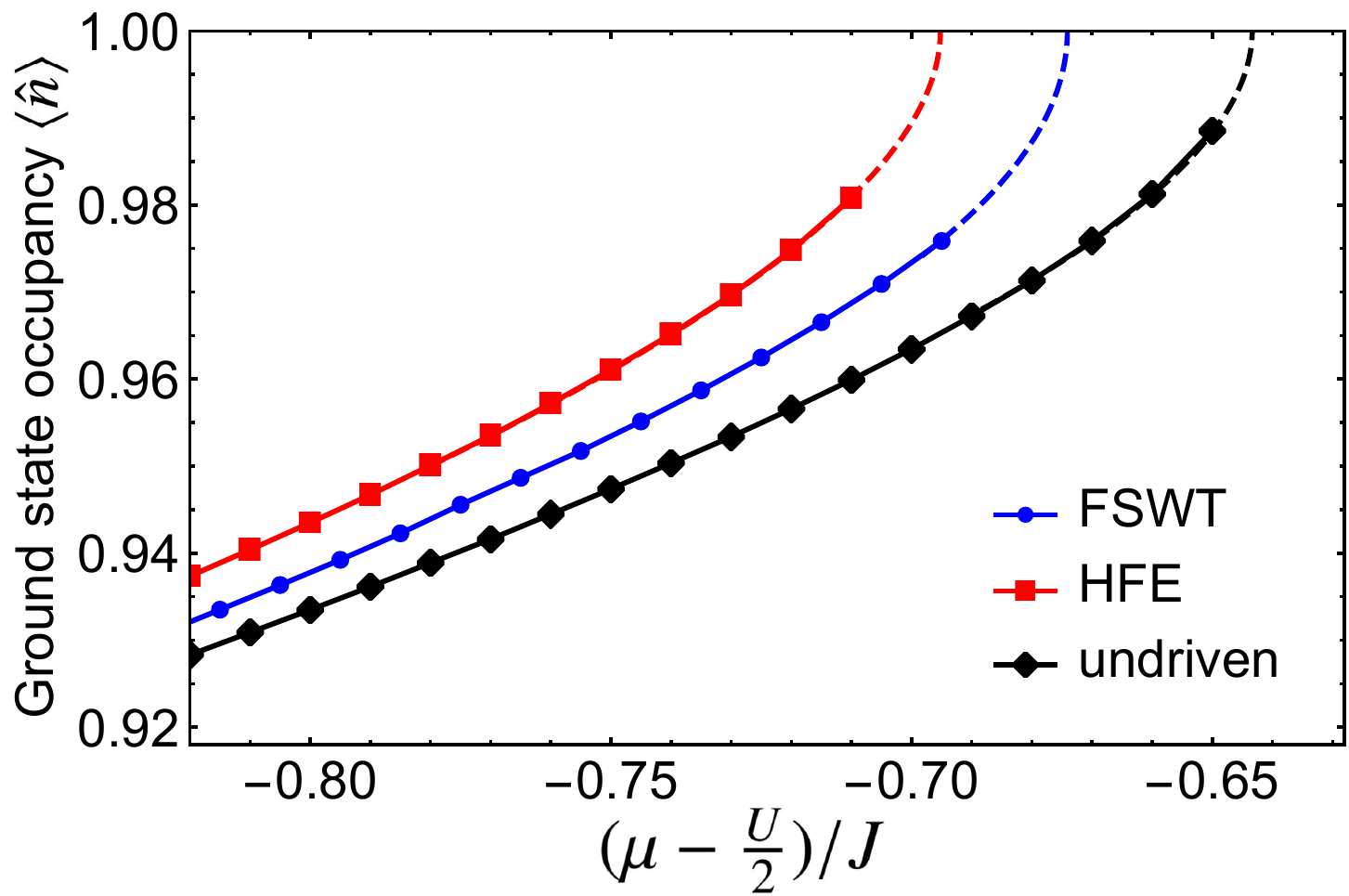}
    \caption{
    %The metal-insulator phase boundary for the ground state of the Floquet Hamiltonian. The phase is indicated by the commensurate/incommensurate filling $n$ when changing the chemical potential $\mu$. 
    The ground state occupancy $\langle \hat{n} \rangle $ is shown vs. the rescaled chemical potential $\mu - U/2$.
    iDMRG simulations with bond dimension 600, which were carried out with TeNPy \cite{tenpy}, are indicated by markers. The extrapolations are obtained from the Lieb-Wu equation. 
    The parameters are chosen as $ U=4J, \omL=12J, g=3J$. 
    %Simulate (for fig2): U=4J, \omL=12J, g= 3J = \omL/4
    }

    %Similutate (not for the paper): U=3, omg=20,g=omg/4 =5
    \label{fig:MIT}
\end{figure}

In Fig.~\ref{fig:MIT} (see Appendix~\ref{appendix:MIT} for details), we carry out infinite DMRG simulations using the TeNPy package~\cite{tenpy} of the ground state occupancy, $\langle \hat{n} \rangle = \lim_{L\to\infty}  \langle \hat{N} \rangle / L$, in a one-dimensional Hubbard chain. 
In these simulations, we assume a non-zero chemical potential $\mu$. This is necessary to observe 
the equilibrium metal-insulator transition in one-dimensional Fermi Hubbard model~\cite{essler2005one}, and it does not alter the solution of any of the Sylvester equations. 
We rescale the horizontal axis to $\mu-U/2$ to match the convention of the Lieb-Wu equation. 
Near criticality, where the charge excitation gap closes and the DMRG algorithm no longer converges, we extrapolate the DMRG result using the Lieb-Wu equation~\cite{essler2005one}. 
%metal-insulator transition in a one-dimensional Hubbard chain~\cite{essler2005one}. 
Our simulations show how the correlated hopping term in our FSWT result Eq.~(\ref{Floquet-H'}), ignored in the lowest order HFE, modifies the occupancy near the Mott transition. 
We see in Fig.~\ref{fig:MIT} that it weakens the Mott insulating phase. It changes the phase boundary as well as the charge susceptibility of the metallic phase (i.e. the slope of the lines) close to the boundary. 
The change is comparable in size to the dynamical localisation effect itself, i.e. the shift from the undriven case to the HFE result. %Apart from the parameters chosen in Fig.~\ref{fig:MIT}, our FSWT result Eq.~(\ref{Floquet-H'}) allows us to simulate this phase diagram for a wide range of interaction strength $U$, including the $U>\omL$ case, where the HFE diverges. 
We expect this modification to be observable in electronic platforms, for instance, in quantum dot Fermi-Hubbard simulators~\cite{hensgens2017quantum} where $\mu$ is controlled by gating.

\subsection{Comparison to exact dynamics}
\begin{figure}[p]
    \centering
        \includegraphics[width=0.6\textwidth]{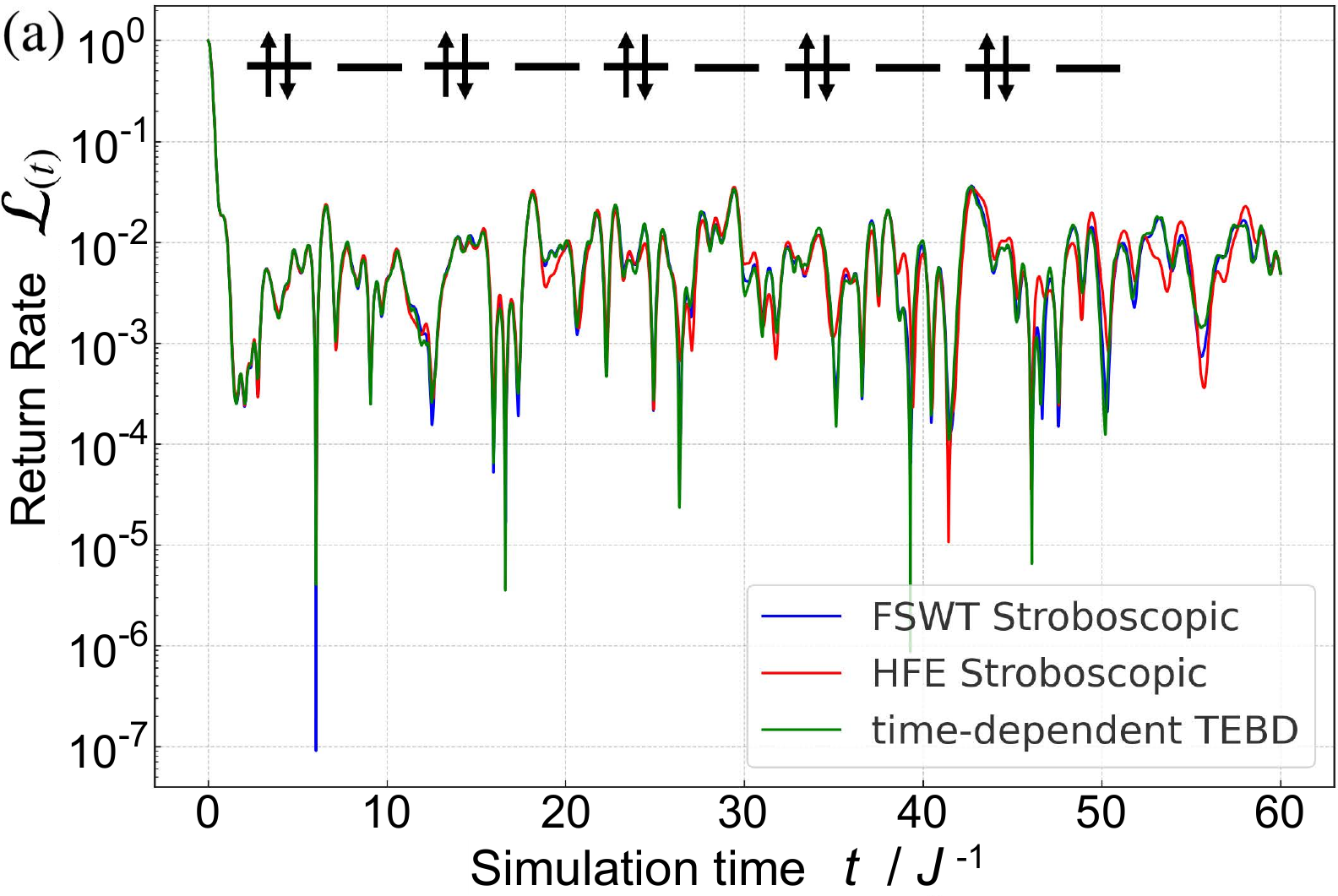}
        \label{fig:subfig1}
\vspace{0.2cm} % Adjust the spacing here
        \includegraphics[width=0.6\textwidth]{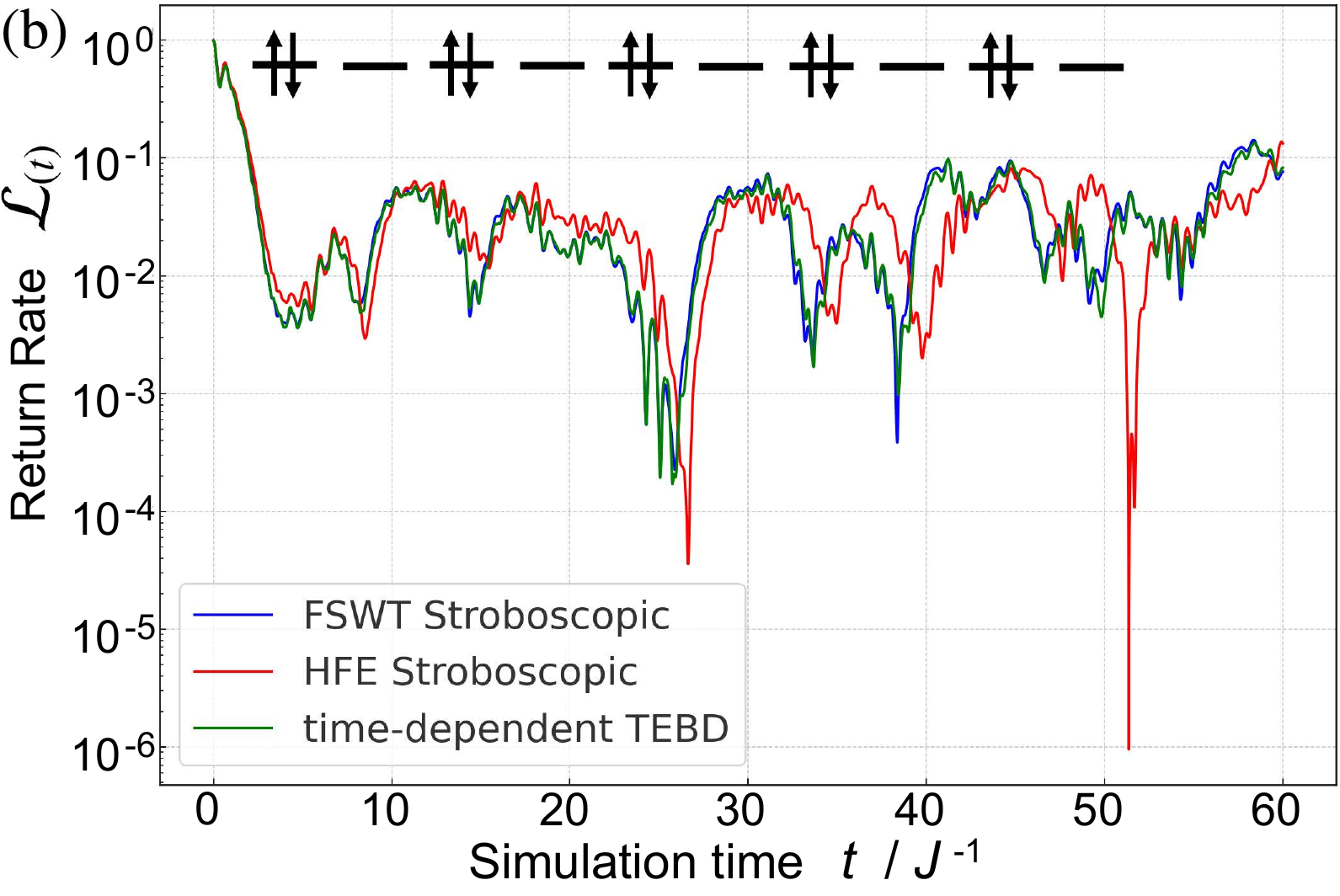}
        \label{fig:subfig2}
\vspace{0.2cm} % Adjust the spacing here
        \includegraphics[width=0.55\textwidth]{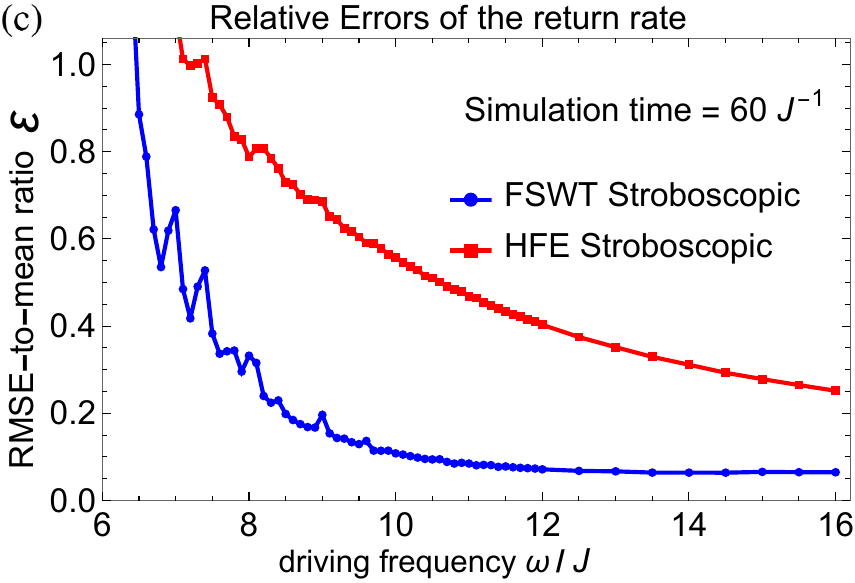}
        \label{fig:subfig3}
    \caption{
    (a) Return rate of CDW state under driving, Eq.~(\ref{eq.Loschmidt}), in an $L=10$ site lattice at driving frequency $\omL = 16 J$ and repulsion $U=3J$, with driving strength fixed at $g = \omL/4$. 
    The Floquet Hamiltonian dynamics are given by TDVP, and the exact time-dependent Hamiltonian dynamics is given by TEBD, with bond-dimension $\chi=300$ and Trotter step size $dt=10^{-3}J^{-1}$. 
    (b) Same as (a) but with increased repulsion $U=8J$. See Appendix \ref{appendix:TrotterCheck} for the convergence check of TEBD over Trotter steps.
    (c) The RMSE of the return rate normalised by its mean value, Eq.~(\ref{eq.NRMSE}), for $U = 3 J$, $g = \omega /4$, and $L=6$ lattice. }
    \label{fig:Evo}
\end{figure}

To gauge the accuracy of our effective Hamiltonian~(\ref{Floquet-H'}), we have carried out extensive simulations comparing it to the exact diagonalisation of small systems and DMRG-based simulations of larger chains. 
We follow Ref.~ \cite{Mendoza-Arenas_2022} and simulate the driven dynamics starting from a product charge density wave (CDW) state of a finite lattice, $\vert \psi_{CDW} \rangle = \Pi_{j=1,3,5,...} \hat{c}_{j,\uparrow}^{\dag} \hat{c}_{j,\downarrow}^{\dag} \vert vac \rangle $, and plot the time-evolution of the return rate~\footnote{The Floquet micro-motion will be ignored in the following, which is a good approximation since the return rate $\mathcal{L}_{(t)}$ we simulate represents an expectation value (of the observable $ \vert \psi_{CDW} \rangle \langle \psi_{CDW} \vert $). If we instead simulated the Floquet fidelity, i.e. the wavefunction overlap between the Floquet evolution and the exact dynamics, then we would have to take the micro-motion into account. },
\begin{align} \label{eq.Loschmidt}
\mathcal{L}_{(t)} &\equiv\big\vert \langle \psi_{CDW} \vert \hat{\mathcal{U}}_{t,t_0} \vert \psi_{CDW} \rangle \big\vert ^2, 
\end{align}
which is closely related to the Loschmidt echo \cite{Sharma_2014,Mendoza-Arenas_2022}. 
The results are shown in Fig.~\ref{fig:Evo}. 
We compare the stroboscopic Floquet Hamiltonian dynamics [where we ignore the micro-motion operators in Eq.~(\ref{eq.propagator})] given by FSWT (\ref{Floquet-H'}) and HFE, with the exact time-dependent Hamiltonian dynamics according to Eq.~(\ref{Ht-example}).

\newpage
In Fig.~\ref{fig:Evo}(a), we chose parameters such that $U\sim J \ll \omL$. Hence, both HFE and FSWT agree very well with the exact dynamics, even though some deviations of the HFE dynamics can be observed after sufficiently long propagation times. 
In Fig.~\ref{fig:Evo}(b), the onsite repulsion is increased, such that $U$ becomes comparable to $\omL$, and the FSWT dynamics still show excellent agreement with the exact result, whereas the HFE expansion deviates substantially. In these simulations, we find that the $\mathcal{O}(J^2)$ part of the FSWT Floquet Hamiltonian (\ref{Floquet-H'}), given by $\frac{1}{2}([\hat{y}_2,\hat{H}^{(1)}_{-1}]+ H.c.)$ in Eq.~(\ref{Floquet-H'2-t2}), is vital for accurately describing the dynamics involving the doublon excitations. 
%On the other hand, when starting from Neel state, the $\mathcal{O}(t^1)$ FSWT Floquet Hamiltonian can already capture the driven physics, regardless of the $U/J$ ratio.
To make this comparison quantitative, we measure the evolution error of FSWT and HFE Hamiltonians, compared to the exact evolution $\mathcal{L}^{exact}_{(t)}$ given by Eq.~(\ref{Ht-example}), using %Floquet fidelity 
\begin{align}\label{eq.NRMSE}
    \mathcal{E} \equiv \bigg( \int_0^{t_f} \frac{\mathrm{d}t}{t_f} ( \mathcal{L}_{(t)} - \mathcal{L}^{exact}_{(t)} )^2 \bigg)^{1/2}  \bigg/ ~  \bigg( \int_0^{t_f} \frac{\mathrm{d}t}{t_f} \mathcal{L}^{exact}_{(t)} \bigg) ,
\end{align}
i.e. the ratio between the averaged root mean squared error (RMSE) of the return rate and the averaged return rate over the simulation time $t_f$. 
The results are shown in Fig.~\ref{fig:Evo}(c), where we fix $U=3J$ and vary the driving frequency $\omL$ under fixed ratio $g/\omL=1/4$ (which keeps the HFE Hamiltonian unchanged). 
Our FSWT result is nearly exact when $\omL\gtrsim 9J$, i.e. $\gamma' J / \omL \lesssim 0.06 $, with an error almost an order of magnitude smaller than that of the HFE result.
For smaller $\omL$, our FSWT result begins to break down, which we attribute to the above-Mott-gap resonances~\cite{okamoto2021floquet} and the $\mathcal{O}(J^3)$ contribution ignored in Eq.~(\ref{Floquet-H'}) [see the discussion in Appendix~\ref{appendix:breakdown} for $\omega=8.5 J$ when the relative error reaches $\mathcal{E} \sim 0.2$]. In the HFE result where the $\mathcal{O}(J)$ and $\mathcal{O}(J^2)$ correlated hopping terms are absent, such a low error can only be reached at a much higher driving frequency $\omL\sim 20 J$.

As the driving frequency decreases further, such that the ratio $ J / \omL$ or $\gamma' J / \omL $ increases to $\mathcal{O} (1)$, the $\hat{y}_3 \sim \mathcal{O}(J^3)$ correction must be included in our FSWT Hamiltonian (\ref{Floquet-H'}). Since hundreds of terms appear in the analytical solution of $\hat{y}_3$, numerical methods may become necessary to solve the Sylvester equation (\ref{FSWT-f^1_1}) in this near-resonant case
\footnote{We can treat $\hat{f}^{(1)}_1$ as a matrix product operator, and solve the Sylvester equation (\ref{FSWT-f^1_1}) using tensor network methods. The corresponding lowest order Floquet Hamiltonian $\hat{H}^{0} + \hat{H}'^{(2)}$ is directly given in an MPO form, which has large bond dimension so that it cannot be explicitly written down analytically, but this MPO can be directly used in the DMRG to predict the steady state under weak drive. This Sylvester-based MPO approach for Floquet Hamiltonian is free from the many-body quasi-energy issue faced by Sambe space DMRG methods like Ref~\cite{sahoo2019periodically}. }, where Floquet heating is expected to appear faster \cite{PhysRevB.93.155132,PhysRevLett.120.197601}, and the pre-thermal states need to be studied in the rotating frame \cite{PhysRevLett.116.125301,Herrmann_2017}. 
Based on the solution of $\hat{f}^{(1)}_1$ in Eq.~(\ref{FSWT-H'-chain-g^2}), we proceed to the second- and third-lowest order FSWT in Appendices~\ref{appendix:higher-order-example}, which results in the $\mathcal{O}(g^4)$ Floquet Hamiltonian $\hat{H}'^{(4)}$. For the parameters considered here, i.e. $g/\omL < 0.5$ and the two-photon process $2\omL$ remaining off-resonant from the doublon energy $U$, we find $\hat{H}'^{(4)}$ is negligible.  

%Using the solution of $\hat{f}^{(1)}_1$ in Eq.~(\ref{FSWT-H'-chain-g^2}), we can proceed to the second-lowest order FSWT, which we present in Appendix \ref{appendix:higher-order-example}. In our driven Hamiltonian (\ref{Ht-example}), we have $\hat{f}^{(2)}_1 =0$ according to Eq.~(\ref{formula-for-f_1^2}), and therefore the $\mathcal{O}(g^3)$ Floquet Hamiltonian vanishes, i.e. $\hat{H}'^{(3)} = 0$. However, Eq.~(\ref{formula-for-f_1^2}) does provide a non-vanishing contribution to the micro-motion, $\hat{f}^{(2)}_2$, which diverges at $\omL=U/2$. It corresponds to the two-photon resonance to the doublon formation (see Appendix \ref{appendix:higher-order-example}). Similarly, we expect the solution to the $l$-th lowest order Sylvester equation to diverge as the driving frequency $\omL$ approaches $ U/l$. See Appendix \ref{appendix:3rd-order-example} for example. 

%\newpage
\subsection{Comparison to Floquet methods in the strongly correlated regime}

We next compare the Floquet Hamiltonian~(\ref{Floquet-H'}) to the one derived in Ref.~\cite{PhysRevLett.116.125301}, where a  Floquet Hamiltonian in driven strongly correlated systems is constructed systematically based on an interaction-dependent frame rotation followed by a high-frequency expansion. 
In this widely used method [categorized as method (b) in Fig.~\ref{fig:illustration}], the transformation to a rotating frame %(accounting for both doublon and driving) 
is carried out, in which the system oscillates at new frequencies $U+l\omL$ with $l$ being an integer. 
This method converges in the strongly correlated limit and yields a Floquet Hamiltonian with coefficients $\frac{J^2}{U+l\omL} \big( \frac{g}{\omL} \big)^{2l}$. 
In this rotating frame, %given by the transform $\hat{U}_{t'}^{\text{rot}} = e ^ { i \sum_{j}  U  \hat{n}_{j,\uparrow} \hat{n}_{j,\downarrow} t' + 2 j g \sin(\omL t') / \omL \sum_s \hat{n}_{j,s} }$, 
the driving terms gain a factor $J$ (see Eq.~(S4) in Ref.~\cite{PhysRevLett.116.125301}), and thus the lowest-order high-frequency expansion in this frame is quadratic in $J$. 
%At first glance, 
In contrast, the Floquet Hamiltonian (\ref{Floquet-H'}) given by our lowest order FSWT gives a contribution which is linear in $J$. %whereas the method of Ref.~\cite{PhysRevLett.116.125301} directly results in a Floquet Hamiltonian which is quadratic in $J$. 
The reason for this discrepancy is that these two Floquet methods are constructed in different frames, and we will show these two Floquet Hamiltonians are equivalent in the limit $U\gg J$, where the two can be compared. 

To show this equivalence in the $\mathcal{O}(g^2)$ Floquet Hamiltonian when $U\gg J$ in a straightforward manner, we consider the half-filling case ($\mu = 0$) and project to the lowest Mott band to obtain effective spin Hamiltonians. 
In our Floquet Hamiltonian (\ref{Floquet-H'}), where we treat the $\mathcal{O}(J)$ part of $(\hat{h}+\hat{H}'^{(2)})$ as the perturbation and $\hat{U}$ as the unperturbed term, %and 
degenerate perturbation theory using the projector $\mathcal{P}$ onto the zero-doublon manifold yields (see Appendix \ref{appendix:project_large_U_limit} for details)
\begin{equation}\label{project_large_U_limit}
\begin{split}
&\mathcal{P} (\hat{h}+\hat{H}'^{(2)}) \frac{1}{-U} (\hat{h}+\hat{H}'^{(2)}) \mathcal{P} \\
&\approx \frac{1}{-U} \mathcal{P} \hat{h}^2 \mathcal{P} + \frac{1}{-U} \mathcal{P} \hat{h} \hat{H}'^{(2)} \mathcal{P} + \frac{1}{-U} \mathcal{P}  \hat{H}'^{(2)} \hat{h} \mathcal{P} + \mathcal{O}(g^4)  \\
&=  \frac{4 J^2}{U} (1-2\frac{g^2}{\omL^2}) \sum\limits_{j=1}^{L-1} {\bf S}_j \cdot {\bf S}_{j+1} \\  
& + 4  \frac{g^2 J^2 }{\omL^2} \left( \frac{1}{U-\omL} + \frac{1}{\omL+U} \right) \sum\limits_{j=1}^{L-1} {\bf S}_j \cdot {\bf S}_{j+1},
\end{split} 
\end{equation}
where $ {\bf S}_j = ( S_j^x , S_j^y , S_j^z  ) $, with $ S_j^x = \frac{1}{2} ( \hat{c}_{j,\uparrow}^{\dag} \hat{c}_{j,\downarrow} + \hat{c}_{j,\downarrow}^{\dag} \hat{c}_{j,\uparrow}  )$, $ S_j^y = \frac{i}{2} ( -\hat{c}_{j,\uparrow}^{\dag} \hat{c}_{j,\downarrow} + \hat{c}_{j,\downarrow}^{\dag} \hat{c}_{j,\uparrow}  )$ and $ S_j^z = \frac{1}{2} ( \hat{n}_{j,\uparrow} - \hat{n}_{j,\downarrow} ) $.
This projected Hamiltonian is equivalent to the expression
%$\mathcal{O}[g^2]$ formula 
for $\hat{H}_{\text{eff}}^{(1)}$ in Ref.~\cite{PhysRevLett.116.125301}. %The details of this derivation are given in 
%, where we use $\hat{n}_{j,\bar{s}} = 1 - \hat{n}_{j,s} $ since we are considering the half-filling (plus $U\gg J$) case.
We thus conclude that in the $U\gg J$ limit, we can derive the previous Hamiltonian (to order $\mathcal{O} (g^2)$) in Ref.~\cite{PhysRevLett.116.125301} from our lowest order FSWT Hamiltonian.
%, although they are of different orders of hopping $J$. 
In other words, the Hamiltonian (\ref{Floquet-H'}) given by our FSWT is the electronic parent Hamiltonian of the spin model derived in Ref.~\cite{PhysRevLett.116.125301}, just like the 
%$\mathcal{O}(t_{\text{hop}})$ 
Hubbard model 
reduces to the 
%is the one behind the $\mathcal{O} (t_{\text{hop}}^2)$ order 
t-J model \cite{essler2005one,spalek2007tj}. 
Moreover, in a two-site lattice, the projected Hamiltonian (\ref{project_large_U_limit}) reduces exactly to the previous Floquet result given by Ref.~\cite{mentink2015ultrafast}. 
Similarly, in the sub-lattice-driven Hubbard model, the enhancement of super-exchange pairing, that was reported in~\cite{PhysRevB.96.085104}, can also be traced back to the correlated hopping terms given by FSWT.
Unlike Refs.~\cite{PhysRevLett.116.125301,mentink2015ultrafast,PhysRevB.96.085104}, our FSWT result can be applied for all ratios $U/J$, as long as $J\ll \omL, \omL/\gamma'$. Our FSWT Hamiltonian (\ref{Floquet-H'}) can also be used in the zero-correlation limit $U\to 0$ and the weakly correlated case $U\sim J$, as shown in Fig.~\ref{fig:Evo}.

%\begin{figure}
%    \centering
%    \includegraphics[width=0.45\textwidth]{applicability_range.png}
%    \caption{Applicability range of the different Floquet methods in the 1d driven Fermi Hubbard chain. Our FSWT result (a) is valid to the right of the vertical dashed line ($\omL \gg t $). For the large-U method (b) in Ref.~\cite{PhysRevLett.116.125301} to work, it requires being not only to the right of the vertical dashed line ($\omL \gg t $) but also to the above of the horizontal dashed line ($U \gg t $). The lowest order HFE result (c) works below the dashed-dotted line ($\omL \gg U$). All methods become resonant close to the inclined solid lines representing the multi-photon resonance. The yellow-shaded region is given by $t \lesssim U \lesssim \omL$ where only our FSWT result remains accurate.}
%    \label{fig:range}
%\end{figure}

\subsection{Applicability of our model}

We illustrate the preceding discussion in Fig.~\ref{fig:illustration}, where we sketch the range of applicability of (a) our FSWT result Eq.~(\ref{Floquet-H'}), 
%given by solving the lowest order Sylvester equation in orders of $J$, 
(b) the lowest order $1 / (U+l\omL)$ large-U method in Ref.~\cite{PhysRevLett.116.125301}, and (c) the lowest order HFE result given by solving the Sylvester equation in (a) in orders of $1/\omL$. 
Method (b) can be used when $U\gg J$, and method (c) can be used when $\omL \gg U$. The FWST result (a) can deal with these regimes as well. 
In addition, we identify a regime, 
%$t \lesssim \omL/4 \lesssim U \lesssim 4 t$, 
$J \lesssim U \lesssim \omL$ with $J\ll \omL, \omL/\gamma'$ [e.g. $U=3J$, $\omL=3 U$ in Fig.~\ref{fig:Evo}(c)], 
where the FSWT method (a) is the only working method among these three methods: 
Here $J\ll \omL, \omL/\gamma'$ guarantees (a) to be applicable, $U \lesssim \omL$ implies that the driving frequency is not sufficiently high for method (c) to be applicable, and $J \lesssim  U$ implies that the undriven system is near the metal-insulator transition where method (b) is not accurate. 
Our method can still be applied in this regime.

\section{Conclusions}\label{conclude}
In Chapter \ref{Chapter5}, we constructed a Floquet Schrieffer Wolff transform to perturbatively generate the many-body Floquet Hamiltonian and micro-motion operators in orders of the driving strength $g$. This transform is obtained from the solution of operator-valued Sylvester equations and does not require knowledge of the eigenbasis of the undriven many-body system.
In this chapter, as an example, we showed how to solve the Sylvester equations perturbatively in driven Hubbard models: By expanding the Sylvester equation in orders of hopping $J$, we derived a Floquet Hamiltonian that may be applied for any interaction strength of $U$. This remains true even when $U\sim J$ or $U>\omL$, provided that $J\ll \omL$ and one does not resonantly drive the Mott gap.
In this FSWT Hamiltonian (\ref{Floquet-H'}), the Floquet-induced interactions play a central role in governing the driven dynamics, whose strength is underestimated by HFE methods.
We used this FSWT Hamiltonian to investigate how Floquet-induced interactions modify the Mott-insulator transition in a one-dimensional chain. We showed that these Floquet-induced interactions in our FSWT Hamiltonian give a large contribution that can be on par with the dominant renormalisation of electronic hopping. 

When the driving strength $g$ becomes larger, a faster convergence can be realised by applying FSWT in the strong driving frame. This strong driving extension of FSWT for Fermi Hubbard models is constructed in Appendix \ref{appendix:StrongDriveFSWT}, where the corresponding Floquet Hamiltonian describes the Floquet-induced interactions under arbitrarily strong laser intensity. The resulting FSWT Hamiltonian is thus essential for understanding the Floquet-induced interactions when the dynamical localisation \cite{PhysRevB.34.3625,PhysRevB.96.085104} is created.

%% file: Chapters/exciton-revisited.tex
%\title{excitonic enhancement revisited:\\ the complete Floquet-induced interactions in the driven semiconductor-cavity systems}

%\maketitle

In Chapter \ref{Chapter6}, we have seen that even in the absence of a cavity, the Floquet-induced interaction can emerge from a driven correlated many-body system. Accurately predicting these cavity-independent Floquet-induced interactions is vital for understanding the dynamics and the pre-thermal states of the driven system. In Chapter \ref{Chapter4}, the cavity-independent Floquet-induced interaction can only be studied within the Hartree-type mean-field approximations in Eq.~(\ref{right-moving}). Under this approximation, the cavity-independent Floquet-induced interaction leads to the excitonically enhanced AC Stark shift, which is a single-particle self-energy effect. Going beyond this mean-field limitation, we will find a Floquet Hamiltonian with complete Floquet-induced two-particle interactions, where the cavity-mediated part coexists with the cavity-independent part. Obtaining such a Floquet Hamiltonian will be necessary for an unbiased prediction of the driving-induced phase transition in the cavity-material setups.

In this final chapter, we apply the FSWT method developed in Chapter \ref{Chapter5} to the generalised driven semiconductor-cavity system in section \ref{sec:general-setup}  with long-range Coulomb interactions and arbitrary laser polarisation. Our FSWT Hamiltonian, at the mean-field level, reduces back to the previous result (given by the Sambe space Gaussian elimination method in Chapter \ref{Chapter4}). Beyond the mean-field treatment, this FSWT method allows us to derive the complete Floquet-induced two-particle interactions in this system, which allows the cavity-mediated interactions to be analysed on equal footing with the cavity-independent Floquet-induced interactions.

In Section \ref{sec:FSWT-completeFMI-intro}, we construct the Sylvester equations in our FSWT for the cavity-material system, and then we decouple these equations in orders of the electron-cavity coupling $g_c$. In Section \ref{suppl-chap6}, we show in detail how to solve these decoupled equations without using the Hartree-type mean-field approximations in Chapter \ref{Chapter4}. In Section \ref{discuss-solution-chap6}, we explain the physical meanings of these solutions and show how can our FSWT result reduce back to the previous Gaussian elimination results in Chapter \ref{Chapter4}. With this solution of the Sylvester equation, we present the FSWT Hamiltonian with complete Floquet-induced interactions in Section \ref{sec:CompleteFMI}, and discuss its relevance in off-resonantly driven TMDC materials in Section \ref{sec.TMDC}. We will set $\hbar=1$.

\section{FSWT for the driven cavity-material system}\label{sec:FSWT-completeFMI-intro}
For the generalised off-resonantly driven cavity-semiconductor system described in Section \ref{sec:general-setup}, the Sylvester equation to the lowest order of driving strength $g$ is given by
\begin{equation}\label{FSWT-f^1_1-chap7}
\hat{H}^{(1)}_{1} + [\hat{f}^{(1)}_1,\hat{H}^{(0)}] - \omL \hat{f}^{(1)}_1 = 0,
\end{equation}
and the corresponding $\mathcal{O}(g^2)$ FSWT Hamiltonian is given by 
\begin{equation}\label{H'-1st-chap7}
\begin{split}
\hat{H}' &= \hat{H}'^{(0)} + \hat{H}'^{(2)} \\
&= \hat{H}^{(0)}  + \frac{1}{2} \big( [\hat{f}^{(1)}_1, \hat{H}^{(1)}_{-1}] + H.c. \big) .
\end{split}
\end{equation}
with the next order correction being $\mathcal{O} (g^4)$. We next solve the Sylvester equation (\ref{FSWT-f^1_1-chap7}) in orders of the electron-cavity coupling $g_c$. We expand the micro-motion operator $\hat{f}^{(1)}_1$ in Eq.~(\ref{FSWT-f^1_1-chap7}) as
\begin{equation}
    \hat{f}^{(1)}_1 = \hat{f}^{(1)}_{1,0} + \hat{f}^{(1)}_{1,1} + \hat{f}^{(1)}_{1,2} + ...
\end{equation}
where $\hat{f}^{(1)}_{1,n} \propto (g_c)^n$. These operators satisfy the following decoupled Sylvester equations
\begin{subequations}\label{decoupled-Sylvester-gc}
\begin{align}
\hat{H}^{(1)}_{1} + [\hat{f}^{(1)}_{1,0},\hat{h} + \hat{U}] - \omL \hat{f}^{(1)}_{1,0} &= 0,  \label{f^1_1,0} \\
[\hat{f}^{(1)}_{1,0},\hat{H}_c] + [\hat{f}^{(1)}_{1,1},\hat{h} + \hat{U}] - \omL \hat{f}^{(1)}_{1,1} &= 0, \label{f^1_1,1} \\
[\hat{f}^{(1)}_{1,1},\hat{H}_c] + [\hat{f}^{(1)}_{1,2},\hat{h} + \hat{U}] - \omL \hat{f}^{(1)}_{1,2} &= 0,  \label{f^1_1,2}
\end{align}
\end{subequations}
which will be solved in the Bloch-electron basis in Section \ref{suppl-chap6} and then discussed in Section \ref{discuss-solution-chap6}. The definitions of operators $\hat{h}$, $\hat{H}_c$, $\hat{U}$ and $\hat{H}^{(1)}_{1}$ are given in Eqs.~(\ref{H1h}), (\ref{H2}),(\ref{U-generalised}) and (\ref{drive}) respectively.

\section{Solving the Sylvester equations in Bloch-electron basis}\label{suppl-chap6}
This technical section shows how to solve the Sylvester equation (\ref{decoupled-Sylvester-gc}) for the cavity-QED setup. Contrary to the example of the driven Hubbard chain in section \ref{sec:general-setup} where we solve the Sylvester equation in orders of electronic hopping $\hat{h}$ (with parameter $J$), for the Coulomb-repulsion system considered in this chapter, we solve the Sylvester equation in orders of electronic interaction $\hat{U}$ (with parameter $V_{\bf q}$). The corresponding derivations are based on an essential property of the Sylvester equation described below.
\subsection{The non-interacting Green function for Sylvester equations}
We can directly check that the solution to 
\begin{equation}\label{green-function}
\hat{c}_{i}^{\dag} \hat{c}_{j}^{\dag}...\hat{c}_{i'}\hat{c}_{j'}... + [\hat{x},\sum_{l} \ep_l \hat{c}_l^{\dag} \hat{c}_l ] -\omega \hat{x} = 0
\end{equation}
where $\hat{c}_l$ is Fermionic operator, is
\begin{equation}\label{decomposed-solution}
\hat{x} = \frac{1}{\omega + \ep_i + \ep_j +... - \ep_{i'}- \ep_{j'}-...} ~ \hat{c}_{i}^{\dag}\hat{c}_{j}^{\dag}...\hat{c}_{i'}\hat{c}_{j'}...
\end{equation}
This can be understood as the non-interacting Green function for Sylvester equations: To solve the Sylvester equation of an arbitrary operator $\hat{T}$ given by
\[
\hat{T} + [\hat{x},\sum_{l} \ep_l \hat{c}_l^{\dag} \hat{c}_l ] -\omega \hat{x} = 0,
\]
we just need to decompose $\hat{T}$ as a summation of terms like $\hat{c}_{i}^{\dag} \hat{c}_{j}^{\dag}...\hat{c}_{i'}\hat{c}_{j'}...$, and write down the solution (\ref{decomposed-solution}) for each term, and then sum them up. Using this property below, we solve the Sylvester equations in Eq.~(\ref{decoupled-Sylvester-gc}).

\subsection{The cavity-independent term}
\begin{figure}[h]
    \centering
    \includegraphics[width=0.8\linewidth]{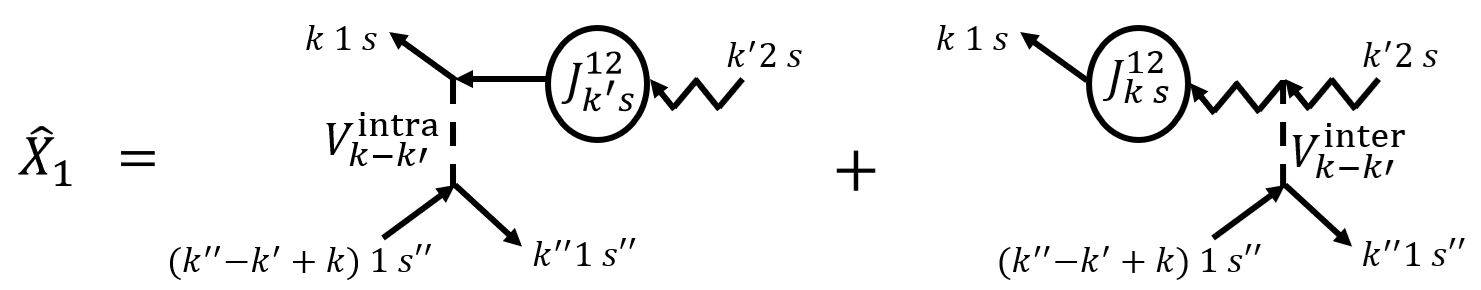}
    \caption{The exact solution of the micro-motion $\hat{X}_1$ in Eq.~(\ref{X1-FSWT-chap5}) contains 2 parts, arising from the intra- and inter-band Coulomb repulsion, respectively.}
    \label{fig:FSWT-X1-chap5}
\end{figure}

To solve the cavity-independent term $\hat{f}^{(1)}_{1,0}$ from Eq.~(\ref{f^1_1,0}), we expand it in orders of interaction $\hat{U}$, 
\begin{equation}\label{f^1_10-expansion-chap5}
    \hat{f}^{(1)}_{1,0} = \hat{X}_0 + \hat{X}_1 + \hat{X}_2 +...
\end{equation}
where $\hat{X}_n \propto U^n$. Inserting Eq.~(\ref{f^1_10-expansion-chap5}) into Eq.~(\ref{f^1_1,0}), we find $\hat{X}_n$ satisfies the following decoupled Sylvester equations
\begin{subequations}\label{X-chap5}
    \begin{align}
& \hat{H}^{(1)}_{1} + [\hat{X}_0,\hat{h}] - \omL \hat{X}_0 = 0  \label{X0} \\
& [\hat{X}_0,\hat{U}] + [\hat{X}_1,\hat{h}] - \omL \hat{X}_1 = 0 \label{X1-subequation-chap5} \\
& [\hat{X}_1,\hat{U}] + [\hat{X}_2,\hat{h}] - \omL \hat{X}_2 = 0,  ...  \label{X2-subequation-chap5}
    \end{align}
\end{subequations} 
Once each $\hat{X}_n$ is obtained, we re-sum all of them to find the final solution to the cavity-independent micro-motion $\hat{f}^{(1)}_{1,0}$. Using the Green function property of Sylvester equations in Eq.~(\ref{green-function}), we can directly solve Eq.~(\ref{X0}), which gives
\begin{equation}
\hat{X}_0=  g  \sum\limits_{{\bf k},s} \sum\limits_{b,b'}  \frac{J^{bb'}_{{\bf k}s}}{\omL+\ep_{{\bf k},bb'}} \hat{c}_{{\bf k}bs}^{\dag} \hat{c}_{{\bf k}b's} 
\end{equation}
where $b$ and $b'$ label the bands in our two-band semiconductor model, and we used the definition $\ep_{{\bf k},bb'} = \ep_{{\bf k},b} - \ep_{{\bf k},b'} $. The driving strength is denoted by $g$, and the electron-laser coupling coefficient $J^{bb'}_{{\bf k}s}$ is a dimensionless factor defined in Section $\ref{sec:general-setup}$.
Based on Eq.~(\ref{X1-subequation-chap5}), we next derive $\hat{X}_1$ in the full particle-hole channel using the following exact relation
\begin{equation}\label{commutator-exact}
[\hat{c}_{{\bf k}b s}^{\dag} \hat{c}_{{\bf k'}b' s'},\hat{U}] =
\sum_{{\bf q}\neq {\bf 0}} \frac{V_{\bf q}}{N} \sum_{ {\bf k''} b'' s''} 
\left(
\hat{c}_{{\bf k} b s}^{\dag} \hat{c}_{{\bf k''} b''s''}^{\dag} \hat{c}_{{\bf k''}-{\bf q}b''s''} \hat{c}_{{\bf k'}+{\bf q} b' s'} 
-
\hat{c}_{{\bf k}-{\bf q} b s}^{\dag} \hat{c}_{{\bf k''} b''s''}^{\dag} \hat{c}_{{\bf k''}-{\bf q}b''s''} \hat{c}_{{\bf k'} b' s'} 
\right)
\end{equation}
this gives
\begin{equation}
[\hat{X}_0,\hat{U}] = g  \sum\limits_{{\bf k},s} \sum\limits_{b,b'}  \frac{J^{bb'}_{{\bf k}s}}{\omL+\ep_{{\bf k},bb'}}
\sum_{{\bf q}\neq {\bf 0}} \frac{V_{\bf q}}{N} \sum_{ {\bf k''} b'' s''} 
\left( -
\hat{c}_{{\bf k} b s}^{\dag} \hat{c}_{{\bf k''}-{\bf q}b''s''} \hat{c}_{{\bf k''} b''s''}^{\dag}  \hat{c}_{{\bf k}+{\bf q} b' s} 
+
\hat{c}_{{\bf k}-{\bf q} b s}^{\dag} \hat{c}_{{\bf k''}-{\bf q}b''s''} \hat{c}_{{\bf k''} b''s''}^{\dag}  \hat{c}_{{\bf k} b' s} 
\right)
\end{equation}
According to the Green function property in Eq.~(\ref{green-function}), we find the solution $\hat{X}_1$ to Eq.~(\ref{X1-subequation-chap5}) has the same form as $[\hat{X}_0,\hat{U}]$ up to some detuning factors as coefficients
(below we show its near-resonant part, i.e., we only consider $b=b''=1,b'=2$)
\begin{equation}
\begin{split}
\hat{X}_1 &= g  \sum\limits_{{\bf k},s}  \frac{J^{12}_{{\bf k}s}}{\omL+\ep_{{\bf k},12}}
\sum_{{\bf q}\neq {\bf 0}} \frac{V_{\bf q}}{N} \sum_{ {\bf k''} s''} 
\bigg( - \frac{1}{\omL+\ep_{{\bf k} 1}-\ep_{{\bf k''}-{\bf q} 1}+\ep_{{\bf k''} 1}-\ep_{{\bf k}+{\bf q} 2}}
\hat{c}_{{\bf k} 1 s}^{\dag} \hat{c}_{{\bf k''}-{\bf q}1s''} \hat{c}_{{\bf k''} 1s''}^{\dag}  \hat{c}_{{\bf k}+{\bf q} 2 s} \\
&~~~~~~~~~~~~~~~~~~~~~~~~~~~~~~~~~~~~~~~   
+ \frac{1}{\omL+\ep_{{\bf k}-{\bf q} 1}-\ep_{{\bf k''}-{\bf q} 1}+\ep_{{\bf k''} 1}-\ep_{{\bf k} 2}}
\hat{c}_{{\bf k}-{\bf q} 1 s}^{\dag} \hat{c}_{{\bf k''}-{\bf q}1s''} \hat{c}_{{\bf k''} 1s''}^{\dag}  \hat{c}_{{\bf k} 2 s} 
\bigg)    
\end{split}
\end{equation}
after changing the dummy index, this becomes
\footnote{
To go back to the previous Hartree-type mean-field result in Chapter \ref{Chapter4}, we can simply approximate the first two operators in Eq.~(\ref{X1-FSWT-chap5}) by their equilibrium expectation value, which means we replace $\hat{c}_{{\bf k} 1 s}^{\dag} \hat{c}_{{\bf k''}-{\bf k'}+{\bf k}1s''}\approx\langle \hat{c}_{{\bf k} 1 s}^{\dag} \hat{c}_{{\bf k''}-{\bf k'}+{\bf k}1s''} \rangle \approx \delta_{{\bf k''},{\bf k'}}\delta_{s,s''}$. However, what we will do next goes beyond this mean-field treatment.
}
\begin{equation}\label{X1-FSWT-chap5}
\begin{split}
\hat{X}_1 &= g  \sum\limits_{{\bf k},s} \sum_{{\bf k'}\neq {\bf k}} \sum_{ {\bf k''} s''}  \left( \frac{J^{12}_{{\bf k'}s}}{\omL+\ep_{{\bf k'},12}}  -  \frac{J^{12}_{{\bf k}s}}{\omL+\ep_{{\bf k},12}} \right)
\frac{V_{{\bf k'}-{\bf k}}}{N}    \frac{1}{\omL + \ep_{{\bf k}1} -  \ep_{{\bf k}+{\bf k''}-{\bf k'}1} + \ep_{{\bf k''}1} - \ep_{{\bf k'}2} } \\
&~~~~~~~~~~~~~~~~~~~~~~~~~~~~~~~~~~~~~~~~~~~~~~~~~~~~~~~~~~~~~~~~~~
\hat{c}_{{\bf k} 1 s}^{\dag} \hat{c}_{{\bf k''}-{\bf k'}+{\bf k}1s''} \hat{c}_{{\bf k''} 1s''}^{\dag}  \hat{c}_{{\bf k'} 2 s}.  \\
\end{split}
\end{equation}
This micro-motion $\hat{X}_1$ includes 2 processes, as depicted in Fig.~\ref{fig:FSWT-X1-chap5}.

\begin{figure}[h]
    \centering
    \includegraphics[width=0.9\linewidth]{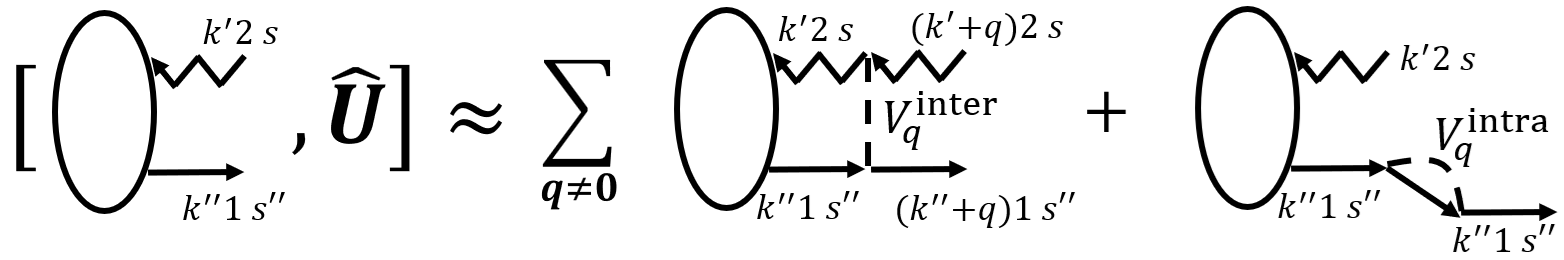}
    \caption{The diagrammatic representation of the approximation Eq.~(\ref{RPA+MF}) made on $[\hat{c}_{{\bf k''} 1s''}^{\dag}  \hat{c}_{{\bf k'} 2 s} ,\hat{U}]$.}
    \label{fig:RPA+MF}
\end{figure}

Next, we calculate $[\hat{X}_1,\hat{U}]$ for Eq.~(\ref{X2-subequation-chap5}) based on the exact solution of $\hat{X}_1$ in Eq.~(\ref{X1-FSWT-chap5}). The particle-hole channel of this commutator includes the following terms
\begin{equation}\label{[x1,U]-chap5-begin}
\begin{split}
[\hat{X}_1,\hat{U}] &\approx g  \sum\limits_{{\bf k},s} \sum_{{\bf k'}\neq {\bf k}} \sum_{ {\bf k''}  s''}  \left( \frac{J^{12}_{{\bf k'}s}}{\omL+\ep_{{\bf k'},12}}  -  \frac{J^{12}_{{\bf k}s}}{\omL+\ep_{{\bf k},12}} \right)
\frac{V_{{\bf k'}-{\bf k}}}{N}    \frac{1}{\omL + \ep_{{\bf k}1} -  \ep_{{\bf k}+{\bf k''}-{\bf k'}1} + \ep_{{\bf k''}1} - \ep_{{\bf k'}2} } \\ 
&~~~~~~~~~~~~~~~~~~~~~~~~~~~~~~~~~~~~~~~~~~~~~~~~~~~~~~~~~~~~~~~ 
\hat{c}_{{\bf k} 1 s}^{\dag} \hat{c}_{{\bf k''}-{\bf k'}+{\bf k}1s''} [\hat{c}_{{\bf k''} 1s''}^{\dag}  \hat{c}_{{\bf k'} 2 s} ,\hat{U}] \\
\end{split}
\end{equation}
where the commutator with the first two operators $[\hat{c}_{{\bf k} 1 s}^{\dag} \hat{c}_{{\bf k''}-{\bf k'}+{\bf k}1s''},\hat{U}]$ is ignored, as it cannot directly provide in-gap resonances. Moreover, in the commutators $[\hat{c}_{{\bf k''} 1s''}^{\dag}  \hat{c}_{{\bf k'} 2 s} ,\hat{U}]$ on the right hand side of Eq.~(\ref{[x1,U]-chap5-begin}), we will only include the following terms 
\begin{equation}\label{RPA-Chap5}
\begin{split}
[\hat{c}_{{\bf k}+{\bf p} 1s'}^{\dag}  \hat{c}_{{\bf k} 2 s} ,\hat{U}] &\approx 
\sum_{{\bf q}\neq {\bf 0}} \frac{V_{\bf q}}{N} 
\left( -
\hat{n}_{{\bf k}+{\bf p} 1 s'} \hat{c}_{{\bf k}+{\bf p}+{\bf q} 1s'}^{\dag}  \hat{c}_{{\bf k}+{\bf q} 2 s} 
+
\hat{n}_{{\bf k}+{\bf p}-{\bf q} 1 s}^{\dag}  \hat{c}_{{\bf k}+{\bf p} 1s'}^{\dag}  \hat{c}_{{\bf k}2 s} 
\right) \\
&\equiv 
\sum\limits_{\bf k'} \hat{\eta}^{1s',2s}_{{\bf p},{\bf k},{\bf k'}} ~ \hat{c}_{{\bf k'}+{\bf p} 1s'}^{\dag}  \hat{c}_{{\bf k'} 2 s}
\end{split}
\end{equation}
where we define
\begin{equation}\label{RPA_new}
\begin{split}
\hat{\eta}^{1s',2s}_{{\bf p},{\bf k},{\bf k'}} &= 
\delta_{{\bf k},{\bf k'}} \sum\limits_{{\bf q}\neq {\bf 0}} \frac{V_{\bf q}}{N} \hat{n}_{{\bf k}+{\bf p}-{\bf q}1s}  
~~ - ~~ (1-\delta_{{\bf k},{\bf k'}}) \frac{V_{{\bf k}-{\bf k'}}}{N}   \hat{n}_{{\bf k}+{\bf p}1s'} 
\end{split}
\end{equation}
Eq.~(\ref{RPA-Chap5}) can be understood as the random phase approximation (RPA) \cite{rowe1968equations,PhysRev.112.1900} on top of the exact commutator Eq.~(\ref{commutator-exact}), where the commutator terms are discarded if they provide vanishing mean-field values in the non-interacting ground state.
Moreover, we assume the driving is off-resonant and the lower-band is almost fully occupied, which allows us to furthermore make the mean-field replacement $\hat{n}_{{\bf k}bs} \to \langle \hat{n}_{bs} \rangle \approx \delta_{b,1}$ in Eq.~(\ref{RPA_new}), such that 
\begin{equation}
\hat{\eta}^{1s',2s}_{{\bf p},{\bf k},{\bf k'}} \to \eta_{{\bf k},{\bf k'}} \equiv
\delta_{{\bf k},{\bf k'}} \sum\limits_{{\bf q}\neq {\bf 0}} \frac{V_{\bf q}}{N} 
~~ - ~~ (1-\delta_{{\bf k},{\bf k'}}) \frac{V_{{\bf k}-{\bf k'}}}{N} 
\end{equation}
The overall approximation (RPA followed by mean-field decoupling) reads
\begin{equation}\label{RPA+MF}
[\hat{c}_{{\bf k}+{\bf p} 1s'}^{\dag}  \hat{c}_{{\bf k} 2 s},\hat{U}] \approx \sum\limits_{\bf k'} \eta_{{\bf k},{\bf k'}} ~ \hat{c}_{{\bf k'}+{\bf p} 1s'}^{\dag}  \hat{c}_{{\bf k'} 2 s}
\end{equation}
which is depicted in Fig.~\ref{fig:RPA+MF}. 

Approximation (\ref{RPA+MF}) makes $[\hat{X}_1,\hat{U}]$ remain in the form of 4-operator interaction, such that
\begin{equation}
\begin{split}
[\hat{X}_1,\hat{U}] &\approx g  \sum\limits_{{\bf k},s} \sum_{{\bf k'}\neq {\bf k}} \sum_{ {\bf k''} s''}  \left( \frac{J^{12}_{{\bf k'}s}}{\omL+\ep_{{\bf k'},12}}  -  \frac{J^{12}_{{\bf k}s}}{\omL+\ep_{{\bf k},12}} \right)
\frac{V_{{\bf k'}-{\bf k}}}{N}    \frac{1}{\omL + \ep_{{\bf k}1} -  \ep_{{\bf k}+{\bf k''}-{\bf k'}1} + \ep_{{\bf k''}1} - \ep_{{\bf k'}2} }
 \\
&~~~~~~~~~~~~~~~~~~~~~~~~~~~~~~~
\hat{c}_{{\bf k} 1 s}^{\dag} \hat{c}_{{\bf k''}-{\bf k'}+{\bf k}1s''} \sum_{\bf k_f} \eta_{{\bf k'},{\bf k_f}} \hat{c}_{{\bf k_f}+{\bf k''}-{\bf k'} 1s''}^{\dag}  \hat{c}_{{\bf k_f} 2 s}
\end{split}
\end{equation}
and thus, according to Eq.~(\ref{green-function}) and (\ref{X2-subequation-chap5}), $\hat{X}_2$ is given by
\begin{equation}
\begin{split}
\hat{X}_2 &\approx g  \sum\limits_{{\bf k},s} \sum_{{\bf k'}\neq {\bf k}} \sum_{ {\bf k''}  s''}  \left( \frac{J^{12}_{{\bf k'}s}}{\omL+\ep_{{\bf k'},12}}  -  \frac{J^{12}_{{\bf k}s}}{\omL+\ep_{{\bf k},12}} \right)
\frac{V_{{\bf k'}-{\bf k}}}{N}    \frac{1}{\omL + \ep_{{\bf k}1} -  \ep_{{\bf k}+{\bf k''}-{\bf k'}1} + \ep_{{\bf k''}1} - \ep_{{\bf k'}2} }
 \\
&~~~~~~~~~ 
\hat{c}_{{\bf k} 1 s}^{\dag} \hat{c}_{{\bf k''}-{\bf k'}+{\bf k}1s''}
\sum_{\bf k_f} \eta_{{\bf k'},{\bf k_f}}
\frac{1}{\omL + \ep_{{\bf k}1} -  \ep_{{\bf k}+{\bf k''}-{\bf k'}1} + \ep_{{\bf k''}-{\bf k'}+{\bf k_f}1} - \ep_{{\bf k_f}2} }
\hat{c}_{{\bf k_f}+{\bf k''}-{\bf k'} 1s''}^{\dag}  \hat{c}_{{\bf k_f} 2 s}
\end{split}
\end{equation}
The same procedure can be conducted to higher orders. Using the following diagonal matrix 
\begin{equation}\label{G-chapter5}
[G^{{\bf k},{\bf q}}]_{{\bf p},{\bf p'}} = \frac{1}{\omL + \ep_{{\bf k}1} -  \ep_{{\bf k}+{\bf q}1} + \ep_{{\bf q}+{\bf p'}1} - \ep_{{\bf p'}2} }
\delta_{{\bf p},{\bf p'}}
\end{equation}
we can simplify the above results to
\begin{equation}
\begin{split}
\hat{X}_1 &= g  \sum\limits_{{\bf k},s} \sum_{{\bf k'}\neq {\bf k}} \sum_{ {\bf k''}  s''}  \left( \frac{J^{12}_{{\bf k'}s}}{\omL+\ep_{{\bf k'},12}}  -  \frac{J^{12}_{{\bf k}s}}{\omL+\ep_{{\bf k},12}} \right)
\frac{V_{{\bf k'}-{\bf k}}}{N}    
\hat{c}_{{\bf k} 1 s}^{\dag} \hat{c}_{{\bf k''}-{\bf k'}+{\bf k}1s''} \\
&~~~~~~~~~~~~~~~~~~~~~~~~~~~~~~~
\sum_{\bf k_f} 
\frac{ \delta_{{\bf k'},{\bf k_f}} }{\omL + \ep_{{\bf k}1} -  \ep_{{\bf k}+{\bf k''}-{\bf k'}1} + \ep_{{\bf k''}-{\bf k'}+{\bf k_f}1} - \ep_{{\bf k_f}2} }
\hat{c}_{{\bf k_f}+{\bf k''}-{\bf k'} 1s''}^{\dag}  \hat{c}_{{\bf k_f} 2 s} \\
&= g  \sum\limits_{{\bf k},s} \sum_{{\bf k'}\neq {\bf k}} \sum_{ {\bf q}  s''}  \left( \frac{J^{12}_{{\bf k'}s}}{\omL+\ep_{{\bf k'},12}}  -  \frac{J^{12}_{{\bf k}s}}{\omL+\ep_{{\bf k},12}} \right)
\frac{V_{{\bf k'}-{\bf k}}}{N}    
\hat{c}_{{\bf k} 1 s}^{\dag} \hat{c}_{{\bf q}+{\bf k}1s''} \\
&~~~~~~~~~~~~~~~~~~~~~~~~~~~~~~~
\sum_{\bf k_f} 
[G^{{\bf k},{\bf q}}]_{{\bf k'},{\bf k_f}}
\hat{c}_{{\bf k_f}+{\bf q} 1s''}^{\dag}  \hat{c}_{{\bf k_f} 2 s}
\end{split}
\end{equation}
and
\begin{equation}
\begin{split}
\hat{X}_2 &\approx g  \sum\limits_{{\bf k},s} \sum_{{\bf k'}\neq {\bf k}} \sum_{ {\bf q} s''}  \left( \frac{J^{12}_{{\bf k'}s}}{\omL+\ep_{{\bf k'},12}}  -  \frac{J^{12}_{{\bf k}s}}{\omL+\ep_{{\bf k},12}} \right)
\frac{V_{{\bf k'}-{\bf k}}}{N}   
\hat{c}_{{\bf k} 1 s}^{\dag} \hat{c}_{{\bf q}+{\bf k}1s''} \\
&~~~~~~~~~~~~~~~~~~~~~~~~~~~~~~~
\sum_{\bf k_f}   [G^{{\bf k},{\bf q}}]_{{\bf k'},{\bf k'}}  \eta_{{\bf k'},{\bf k_f}}
[G^{{\bf k},{\bf q}}]_{{\bf k_f},{\bf k_f}}
\hat{c}_{{\bf k_f}+{\bf q} 1s''}^{\dag}  \hat{c}_{{\bf k_f} 2 s}
\end{split}
\end{equation}
Applying the same approximation on $[\hat{X}_2,\hat{U}]$ again makes $\hat{X}_3$ approximately a 4-operator interaction. We find
\begin{equation}
\begin{split}
\hat{X}_3 &\approx g  \sum\limits_{{\bf k},s} \sum_{{\bf k'}\neq {\bf k}} \sum_{ {\bf q} s''}  \left( \frac{J^{12}_{{\bf k'}s}}{\omL+\ep_{{\bf k'},12}}  -  \frac{J^{12}_{{\bf k}s}}{\omL+\ep_{{\bf k},12}} \right)
\frac{V_{{\bf k'}-{\bf k}}}{N}   
\hat{c}_{{\bf k} 1 s}^{\dag} \hat{c}_{{\bf q}+{\bf k}1s''} \\
&~~~~~~~~~~~~~~~~~~~
\sum_{\bf k_1}   [G^{{\bf k},{\bf q}}]_{{\bf k'},{\bf k'}}  \eta_{{\bf k'},{\bf k_1}}
[G^{{\bf k},{\bf q}}]_{{\bf k_1},{\bf k_1}}
\sum_{\bf k_f} \eta_{{\bf k_1},{\bf k_f}} [G^{{\bf k},{\bf q}}]_{{\bf k_f},{\bf k_f}}
\hat{c}_{{\bf k_f}+{\bf q} 1s''}^{\dag}  \hat{c}_{{\bf k_f} 2 s}
\end{split}
\end{equation}

Re-summing the 4-operator interactions in this particle-hole channel to infinite orders, we find
\begin{equation}\label{X1Screened-final}
\begin{split}
\hat{X}_1^{\text{screened}} 
&=  \hat{X}_1 + \hat{X}_2 + \hat{X}_3 + ... \\
&\approx 
g  \sum\limits_{{\bf k},s} \sum_{{\bf k'}\neq {\bf k}} \sum_{ {\bf q}  s''}  \left( \frac{J^{12}_{{\bf k'}s}}{\omL+\ep_{{\bf k'},12}}  -  \frac{J^{12}_{{\bf k}s}}{\omL+\ep_{{\bf k},12}} \right)
\frac{V_{{\bf k'}-{\bf k}}}{N}   
\hat{c}_{{\bf k} 1 s}^{\dag} \hat{c}_{{\bf q}+{\bf k}1s''} \\
&~~~~~~ \sum_{\bf k_f} 
[G^{{\bf k},{\bf q}} + G^{{\bf k},{\bf q}} \eta G^{{\bf k},{\bf q}} + G^{{\bf k},{\bf q}} \eta G^{{\bf k},{\bf q}} \eta G^{{\bf k},{\bf q}} + ... ]_{{\bf k'},{\bf k_f}}
\hat{c}_{{\bf k_f}+{\bf q} 1s''}^{\dag}  \hat{c}_{{\bf k_f} 2 s} \\
&= 
g  \sum\limits_{{\bf k},s} \sum_{{\bf k'}\neq {\bf k}} \sum_{ {\bf q}  s''}  \left( \frac{J^{12}_{{\bf k'}s}}{\omL+\ep_{{\bf k'},12}}  -  \frac{J^{12}_{{\bf k}s}}{\omL+\ep_{{\bf k},12}} \right)
\frac{V_{{\bf k'}-{\bf k}}}{N}   
\hat{c}_{{\bf k} 1 s}^{\dag} \hat{c}_{{\bf q}+{\bf k}1s''} \\
&~~~~~~ \sum_{\bf k_f} 
\left[ \frac{1}{(G^{{\bf k},{\bf q}})^{-1} - \eta} \right]_{{\bf k'},{\bf k_f}}
\hat{c}_{{\bf k_f}+{\bf q} 1s''}^{\dag}  \hat{c}_{{\bf k_f} 2 s} \\
&\equiv
g  \sum\limits_{{\bf k},s} \sum_{{\bf k'}\neq {\bf k}} \sum_{ {\bf q}  s''}  \left( \frac{J^{12}_{{\bf k'}s}}{\omL+\ep_{{\bf k'},12}}  -  \frac{J^{12}_{{\bf k}s}}{\omL+\ep_{{\bf k},12}} \right)
\frac{V_{{\bf k'}-{\bf k}}}{N}   
\hat{c}_{{\bf k} 1 s}^{\dag} \hat{c}_{{\bf q}+{\bf k}1s''} \\
&~~~~~~ \sum_{\bf k_f} 
\left[ \frac{1}{\Gamma^{12}_{{\bf k},{\bf q}}} \right]_{{\bf k'},{\bf k_f}}
\hat{c}_{{\bf k_f}+{\bf q} 1s''}^{\dag}  \hat{c}_{{\bf k_f} 2 s} \\
\end{split}
\end{equation}
where in the last line, we have defined a matrix $\Gamma^{12}_{{\bf k},{\bf q}}$ with element
\begin{equation}\label{Gamma^12_kq}
\begin{split}
&[\Gamma^{12}_{{\bf k},{\bf q}} ]_{{\bf p},{\bf p'}} = \left[ (G^{{\bf k},{\bf q}})^{-1}\right]_{{\bf p},{\bf p'}} - \left[\eta \right]_{{\bf p},{\bf p'}} \\
&= \bigg(\omL + \ep_{{\bf k}1} -  \ep_{{\bf k}+{\bf q}1} + \ep_{{\bf q}+{\bf p'}1} - \ep_{{\bf p'}2} - \sum\limits_{{\bf q'}\neq {\bf 0}} \frac{V_{\bf q'}}{N}  \bigg) \delta_{{\bf p},{\bf p'}}
~~ + ~~ (1-\delta_{{\bf p},{\bf p'}}) \frac{V_{{\bf p}-{\bf p'}}}{N} 
\end{split}
\end{equation}

The final solution of the Sylvester equation to the 0th order of cavity-electron coupling is 
\begin{equation}\label{FSWT-X0X1MF-detail}
    \hat{f}^{(1)}_{1,0} = \hat{X}_0 + \hat{X}_1^{\text{screened}} 
\end{equation}
Thus, according to Eq.~(\ref{H'-1st-chap7}), $\hat{f}^{(1)}_{1,0}$ will provide two cavity-independent driving-induced effects in the FSWT Hamiltonian: the uncorrelated single-particle term given by $\frac{1}{2} \big( [\hat{X}_0, \hat{H}^{(1)}_{-1}] + H.c. \big)$, and the cavity-independent Floquet-induced interaction given by $\frac{1}{2} \big( [\hat{X}_1^{\text{screened}}, \hat{H}^{(1)}_{-1}] + H.c. \big)$.

\subsection{Mean-field treatments for deriving the cavity-mediated terms}

In the special case where ${\bf q}={\bf 0}$, the above $\Gamma^{12}_{{\bf k},{\bf q}}$ matrix in Eq.~(\ref{Gamma^12_kq}) is greatly simplified, because in this case Eq.~(\ref{G-chapter5}) reads
\begin{equation}
[G^{{\bf k},{\bf q}={\bf 0}}]_{{\bf p},{\bf p'}} =
[G]_{{\bf p},{\bf p'}} 
\equiv \frac{1}{\omL  + \ep_{{\bf p'}12}  }
\delta_{{\bf p},{\bf p'}}
\end{equation}
and thus, we have
\begin{equation}\label{Gamma^12-MF}
\begin{split}
[\Gamma^{12}_{{\bf k},{\bf q}={\bf 0}} ]_{{\bf p},{\bf p'}} = [\Gamma^{12}_{\text{MF}} ]_{{\bf p},{\bf p'}}
&\equiv \bigg(\omL  + \ep_{{\bf p'}12} - \sum\limits_{{\bf q'}\neq {\bf 0}} \frac{V_{\bf q'}}{N}  \bigg) \delta_{{\bf p},{\bf p'}}
~~ + ~~ (1-\delta_{{\bf p},{\bf p'}}) \frac{V_{{\bf p}-{\bf p'}}}{N} 
\end{split}
\end{equation}
These simplified matrices will appear in the following derivation of $\hat{f}^{(1)}_{1,1}$, where an additional Hartree-type mean-field decoupling is conducted on $\hat{f}^{(1)}_{1,0}$ in Eq.~(\ref{FSWT-X0X1MF-detail}).
To proceed to this lowest order cavity-mediated term $\hat{f}^{(1)}_{1,1}$, we will make the following mean-field approximation on the formula of $\hat{X}_1^{\text{screened}} $ in Eq.~(\ref{X1Screened-final}), such that
\begin{equation}\label{MF-back-to-chap4}
\begin{split}
\langle \hat{X}_1^{\text{screened}} \rangle &\approx
g  \sum\limits_{{\bf k},s} \sum_{{\bf k'}\neq {\bf k}} \sum_{ {\bf q}  s''}  \left( \frac{J^{12}_{{\bf k'}s}}{\omL+\ep_{{\bf k'},12}}  -  \frac{J^{12}_{{\bf k}s}}{\omL+\ep_{{\bf k},12}} \right)
\frac{V_{{\bf k'}-{\bf k}}}{N}   \langle
\hat{c}_{{\bf k} 1 s}^{\dag} \hat{c}_{{\bf q}+{\bf k}1s''} 
\rangle \\
&~~~~~~ \sum_{\bf k_f} 
\left[ \frac{1}{\Gamma^{12}_{{\bf k},{\bf q}}} \right]_{{\bf k'},{\bf k_f}}
\hat{c}_{{\bf k_f}+{\bf q} 1s''}^{\dag}  \hat{c}_{{\bf k_f} 2 s} \\
&\approx
g  \sum\limits_{{\bf k},s} \sum_{{\bf k'}\neq {\bf k}}   \left( \frac{J^{12}_{{\bf k'}s}}{\omL+\ep_{{\bf k'},12}}  -  \frac{J^{12}_{{\bf k}s}}{\omL+\ep_{{\bf k},12}} \right)
\frac{V_{{\bf k'}-{\bf k}}}{N}   \langle
\hat{n}_{{\bf k} 1 s}
\rangle \sum_{\bf k_f} 
\left[ \frac{1}{\Gamma^{12}_{{\bf k},{\bf q}={\bf 0}}} \right]_{{\bf k'},{\bf k_f}}
\hat{c}_{{\bf k_f} 1s}^{\dag}  \hat{c}_{{\bf k_f} 2 s} \\
&\approx
g  \sum\limits_{{\bf k},s} \sum_{{\bf k'}\neq {\bf k}}   \left( \frac{J^{12}_{{\bf k'}s}}{\omL+\ep_{{\bf k'},12}}  -  \frac{J^{12}_{{\bf k}s}}{\omL+\ep_{{\bf k},12}} \right)
\frac{V_{{\bf k'}-{\bf k}}}{N}   \sum_{\bf k_f} 
\left[ \frac{1}{\Gamma^{12}_{\text{MF}}} \right]_{{\bf k'},{\bf k_f}}
\hat{c}_{{\bf k_f} 1s}^{\dag}  \hat{c}_{{\bf k_f} 2 s} \\
&=
g  \sum\limits_{{\bf k},s} \sum_{ {\bf k'} }   \frac{J^{12}_{{\bf k}s}}{\omL+\ep_{{\bf k},12}} [\eta]_{{\bf k},{\bf k'}}  \sum_{\bf k_f} 
\left[ \frac{1}{\Gamma^{12}_{\text{MF}}} \right]_{{\bf k'},{\bf k_f}}
\hat{c}_{{\bf k_f} 1s}^{\dag}  \hat{c}_{{\bf k_f} 2 s} \\
&=
g  \sum\limits_{{\bf k},s} J^{12}_{{\bf k}s} \sum_{\bf k_f} 
\left[ G \eta \frac{1}{\Gamma^{12}_{\text{MF}}} \right]_{{\bf k},{\bf k_f}}
\hat{c}_{{\bf k_f} 1s}^{\dag}  \hat{c}_{{\bf k_f} 2 s} \\
\end{split}
\end{equation}
where in the first line, we approximate the operator $\hat{c}_{{\bf k} 1 s}^{\dag} \hat{c}_{{\bf q}+{\bf k}1s''} $ by its expectation value, in the second line, we assume $\langle \hat{c}_{{\bf k} 1 s}^{\dag} \hat{c}_{{\bf q}+{\bf k}1s''} \rangle \approx \langle \hat{n}_{{\bf k} 1 s} \rangle \delta_{{\bf q},{\bf 0}} \delta_{s,s''}$, and in the third line we assume $\langle \hat{n}_{{\bf k} 1 s} \rangle\approx 1$, with matrix $\Gamma^{12}_{\text{MF}}$ defined in Eq.~(\ref{Gamma^12-MF}).
Under this mean-field treatment, we can reduce $\hat{f}^{(1)}_{1,0}$ to 
\begin{equation}\label{f^1_MF_1,0}
\begin{split}
\hat{f}^{(1),MF}_{1,0}  &= \hat{X}_0 + \langle \hat{X}_1^{\text{screened}} \rangle \\
&= g  \sum\limits_{{\bf k},s} J^{12}_{{\bf k}s} \sum_{\bf k_f} 
\left[ G + G \eta \frac{1}{\Gamma^{12}_{\text{MF}}} \right]_{{\bf k},{\bf k_f}}
\hat{c}_{{\bf k_f} 1s}^{\dag}  \hat{c}_{{\bf k_f} 2 s} \\
&= g  \sum\limits_{{\bf k},s} J^{12}_{{\bf k}s} \sum_{\bf k_f} 
\left[ G + G \eta \frac{1}{G^{-1}-\eta } \right]_{{\bf k},{\bf k_f}}
\hat{c}_{{\bf k_f} 1s}^{\dag}  \hat{c}_{{\bf k_f} 2 s} \\
&= g  \sum\limits_{{\bf k},s} J^{12}_{{\bf k}s} \sum_{\bf k_f} 
\left[ \frac{1}{G^{-1}-\eta } \right]_{{\bf k},{\bf k_f}}
\hat{c}_{{\bf k_f} 1s}^{\dag}  \hat{c}_{{\bf k_f} 2 s} \\
&= g  \sum\limits_{{\bf k_f},s} \sum_{\bf k}  J^{12}_{{\bf k}s} 
\left[ \frac{1}{\Gamma^{12}_{\text{MF}}} \right]_{{\bf k},{\bf k_f}}
\hat{c}_{{\bf k_f} 1s}^{\dag}  \hat{c}_{{\bf k_f} 2 s} \\
&\equiv g  \sum\limits_{{\bf k_f},s} \frac{1}{\Delta_{\bf k_f}}
\hat{c}_{{\bf k_f} 1s}^{\dag}  \hat{c}_{{\bf k_f} 2 s} \\
\end{split}
\end{equation}
In the last line, we define a screened denominator $\Delta_{\bf k_f}$, which becomes complex when the electron-light coupling coefficient $J^{12}_{{\bf k}s}$ is complex. In the special case when $J^{12}_{{\bf k}s}=1$ and $V_{\bf q}=U$, this $\Delta_{\bf k_f}$ becomes exactly the previous excitonically screened denominator (\ref{renormalised-denominator}) found in Chapter \ref{Chapter4} using the Gaussian elimination method.

\subsection{The $\mathcal{O}(g_c)$ cavity-mediated term}

The mean-field treatment Eq.~(\ref{f^1_MF_1,0}) will be used to solve $\hat{f}^{(1)}_{1,1}$ from Eq.~(\ref{f^1_1,1}). Besides, we will also ignore the screening effect given by the intrinsic interaction $\hat{U}$. Then, Eq.~(\ref{f^1_1,1}) is simplified to
\begin{equation}
    [\hat{f}^{(1),MF}_{1,0},\hat{H}_c] + [\hat{f}^{(1)}_{1,1},\hat{h} ] - \omL \hat{f}^{(1)}_{1,1} \approx 0
\end{equation}
which gives (we only show the near-resonant term, and $\Delta_c \equiv \omc-\omL$)
\begin{equation}\label{f^1_1-1-chap5}
\hat{f}^{(1)}_{1,1} \approx i \frac{ g_c }{-\sqrt{N} \Delta_c} \hat{a}  \sum\limits_{{\bf k_f},s} \frac{g}{\Delta_{\bf k_f}}
\hat{n}_{{\bf k_f} 1s}   
\end{equation}
This term leads to a driving-induced electron-cavity vertex in the FSWT Hamiltonian (\ref{H'-1st-chap7}), which vanishes as the Floquet Hamiltonian is projected to the low-energy manifold.

\subsection{The $\mathcal{O}(g_c^2)$ cavity-mediated term}
To solve the $\mathcal{O}(g_c^2)$ order micro-motion, $\hat{f}^{(1)}_{1,2}$, we need to solve the Sylvester equation (\ref{f^1_1,2}).
The source term of Eq.~(\ref{f^1_1,2}) is $[\hat{f}^{(1)}_{1,1},\hat{H}_c]$, whose near-resonant part is given by
\begin{equation}
[\hat{f}^{(1)}_{1,1},\hat{H}_c] \approx 
-\frac{ g_c^2 }{N \Delta_c} \sum\limits_{{\bf k'},s'} \hat{c}_{{\bf k'} 1s'}^{\dag} \hat{c}_{{\bf k'} 2s'} \sum\limits_{{\bf k_f},s} \frac{g}{\Delta_{\bf k_f}}
\hat{n}_{{\bf k_f} 1s}   
\end{equation}
Next, assuming $[\sum\limits_{{\bf k_f},s} \frac{1}{\Delta_{\bf k_f}}
\hat{n}_{{\bf k_f} 1s} ,\hat{U}]\approx 0$, we take the same mean-field approximation made in Eq.~(\ref{MF-back-to-chap4}) and find the following solution
\begin{equation}\label{CMI-chap5}
\hat{f}^{(1)}_{1,2} \approx 
-\frac{ g_c^2 }{N \Delta_c} \sum\limits_{{\bf k_f'},s'}  \sum_{\bf k'} 
\left[ \frac{1}{\Gamma^{12}_{\text{MF}}} \right]_{{\bf k'},{\bf k_f'}}
\hat{c}_{{\bf k_f'} 1s'}^{\dag}  \hat{c}_{{\bf k_f'} 2 s'}  \sum\limits_{{\bf k_f},s} \frac{g}{\Delta_{\bf k_f}}
\hat{n}_{{\bf k_f} 1s}   
\end{equation}
which will give rise to the Floquet-induced cavity-mediated interaction in the FSWT Hamiltonian.

\section{Discussion on the solution}\label{discuss-solution-chap6}
In the above technical section, we have shown how to solve the Sylvester equation (\ref{decoupled-Sylvester-gc}) for the generalised driven cavity-semiconductor system described in Section \ref{sec:general-setup}. These solutions represent the Floquet micro-motions in this system, whose physical meanings are discussed below.

\subsubsection{The cavity-independent term}
To solve the cavity-independent micro-motion $\hat{f}^{(1)}_{1,0}$, we expand it in orders of interaction $\hat{U}$ in Eq.~(\ref{f^1_10-expansion-chap5}), such that $\hat{f}^{(1)}_{1,0} = \hat{X}_0 + \hat{X}_1 + \hat{X}_2 +...$, where $\hat{X}_n \propto U^n$.
In its exact solution, $\hat{f}^{(1)}_{1,0}$ already contains infinite-electron terms, including single-electron term $\hat{X}_0$, 2-electron scattering $\hat{X}_1$, 3-electron interaction $\hat{X}_2$, etc. The $(n+1)$-electron term $\hat{X}_n$ can be graphically represented by Feynman-diagram-like scattering diagrams. 

However, our FSWT Hamiltonian method contains several key advantages compared with the commonly used co-rotating frame Matsubara Feynman-diagram method: 
\textit{First}, the Matsubara method works under the rotating wave approximation, which cannot obtain the off-resonant term in $\hat{X}_n$ obtained by FSWT. 
\textit{Second}, in FSWT, each term in $\hat{X}_n$ is generated from the commutator $[\hat{X}_{n-1},\hat{U}]$ when solving the Sylvester equation in orders of $\hat{U}$. This straightforward commutator calculation in FSWT is all we need in order to generate the complete $(n+1)$-electron terms in $\hat{X}_n$. In contrast, to achieve this in the Matsubara method, we have to generate all possible $n$-Coulomb-line diagrams and then pick out the topologically nonequivalent ones to avoid double-counting, which becomes tedious as the number of Coulomb lines goes up. 
\textit{Third}, the Matsubara method only works conveniently in the Bloch-electron basis, while in our FSWT, the Sylvester equation can be solved in other orbitals, for example, in the local Wannier orbital in Chapter \ref{Chapter6}. Thus, FSWT shows stronger flexibility in non-periodic systems or strongly correlated systems where the Bloch basis is no longer the optimal choice. 
\textit{Finally}, the screening calculation in Matsubara formalism requires evaluating the numerically heavy Matsubara frequency-summation. However, in FSWT, by solving the Sylvester equations at the mean-field level, the screened interaction is directly obtained, which no longer requires summing over these internal Matsubara frequencies. 

In this thesis, we only focus on the semiconductor case where the lower-band is almost fully occupied and the upper-band is empty. This filling condition allows us to make a mean-field treatment in the exact solution of $\hat{f}^{(1)}_{1,0}$, which gives the approximation in Eq.~(\ref{FSWT-X0X1MF-detail}), i.e., $\hat{f}^{(1)}_{1,0} \approx \hat{X}_0 + \hat{X}_1^{\text{screened}}$.
Here $\hat{X}_0$ is the 1-electron term independent of interaction $\hat{U}$, and $\hat{X}_1^{\text{screened}}$ is the two-particle term $\hat{X}_1$ screened by the more-than-2-electron terms $\hat{X}_{n\ge 2}$ in the exact solution of $\hat{f}^{(1)}_{1,0}$. 

\subsubsection{The single-particle mean-field effect of the cavity-independent term}
In Chapter \ref{Chapter4}, we found the excitonically enhanced AC Stark shift effect. This previous result can be equivalently derived via FSWT by a further mean-field treatment on $\hat{X}_1^{\text{screened}}$, as scrutinised in Eqs.~(\ref{MF-back-to-chap4}) and (\ref{f^1_MF_1,0}). Under this additional mean-field decoupling, $\hat{f}^{(1)}_{1,0}$ is furthermore reduced to
\begin{equation}\label{FSWT-GE}
    \hat{f}^{(1),MF}_{1,0} =  \hat{X}_0 + \langle \hat{X}_1^{\text{screened}} \rangle = \hat{X}_0^{\text{screened}} 
\end{equation}
According to Eq.~(\ref{f^1_MF_1,0}), we find that our previous Gaussian elimination result in Chapter \ref{Chapter4} is equivalent to the treatment in Eq.~(\ref{FSWT-GE}), where we only took into account the single-particle effect of the two-particle Floquet micro-motion term $\hat{X}_1^{\text{screened}}$. Our FSWT result Eq.~(\ref{FSWT-X0X1MF-detail}) instead goes beyond the previous mean-field picture and provides the cavity-independent Floquet-induced interactions behind this single-particle effect.

\subsubsection{The cavity-mediated term}
From the above solution of cavity-independent $\hat{f}^{(1)}_{1,0}$, we proceed to solve the lowest order cavity-dependent micro-motions $\hat{f}^{(1)}_{1,1} \propto g_c$. From its solution in Eq.~(\ref{f^1_1-1-chap5}), we find $\hat{f}^{(1)}_{1,1}$ only contains photon number changing terms, thus according to Eq.~(\ref{H'-1st-chap7}) it gives rise to a driving-induced electron-cavity coupling term in the Floquet Hamiltonian. However, after projecting to the empty-cavity manifold, we find this driving-induced coupling has no contribution to the low-energy part of the Floquet Hamiltonian. 

In the next order solution $\hat{f}^{(1)}_{1,2} \propto (g_c)^2$ in Eq.~(\ref{CMI-chap5}), we find a term which still exists in the manifold that contains 0 cavity photon. According to Eq.~(\ref{H'-1st-chap7}), this 2-particle term in $\hat{f}^{(1)}_{1,2}$ results in the Floquet-induced cavity-mediated interaction in the FSWT Hamiltonian. In the exact solution of this cavity-mediated micro-motion $\hat{f}^{(1)}_{1,2}$, there will also be cavity-mediated more-than-two electron interactions. According to our mean-field approximation (\ref{f^1_MF_1,0}) made in deriving $\hat{f}^{(1)}_{1,1}$, these multi-particle interactions are effectively replaced by a screening effect (i.e. the excitonic enhancement) on the cavity-mediated two-electron interactions. Under this mean-field treatment, the Floquet-induced cavity-mediated interaction in $\frac{1}{2} \big( [\hat{f}^{(1)}_{1,2}, \hat{H}^{(1)}_{-1}] + H.c. \big)$ reduces to a long-ranged two-particle term. 

\subsubsection{Treating cavity-independent terms at the same level of cavity-mediated term}
Although both treatments (on the cavity-independent term), (\ref{FSWT-X0X1MF-detail}) and (\ref{FSWT-GE}), can reveal the excitonic enhancement effect, it is preferable to use (\ref{FSWT-X0X1MF-detail}) but not (\ref{FSWT-GE}) when predicting the phase transition triggered by the Floquet-induced cavity-mediated interactions. The reason is explained below. The cavity-mediated interaction is a 2-particle term in the Floquet Hamiltonian (renormalised by more-than-2-particle terms). Likewise, the cavity-independent $\hat{X}_1^{\text{screened}}$ provides a same-level 2-particle interaction in the Floquet Hamiltonian, i.e., $\frac{1}{2} \big( [\hat{X}_1^{\text{screened}}, \hat{H}^{(1)}_{-1}] + H.c. \big)$. Before cavity-mediated interaction triggers an instability, this cavity-independent Floquet-induced interaction may already push the electrons out of the Fermi liquid phase. Thus, treatment (\ref{FSWT-X0X1MF-detail}) is necessary to capture the complete Floquet-induced interactions that will be used in the future study of driving-induced phase transitions.

\section{The complete Floquet-induced interactions}\label{sec:CompleteFMI}

Inserting the approximated solution of the micro-motion obtained above
\begin{equation}
    \hat{f}^{(1)}_1 \approx    \hat{X}_0 + \hat{X}_1^{\text{screened}}    + \hat{f}^{(1)}_{1,2}
\end{equation}
into Eq.~(\ref{H'-1st-chap7}), we obtained the following FSWT Hamiltonian
\begin{equation}\label{H'-Chap5-result}
\begin{split}
\hat{H}' 
&= \hat{H}^{(0)}  + \frac{1}{2} \big( [\hat{f}^{(1)}_1, \hat{H}^{(1)}_{-1}] + H.c. \big) \\
&= \hat{h} + \hat{H}_c + \hat{U} + \hat{H}_{\text{U=0}} + \hat{H}_{\text{cav-indep FII}} + \hat{H}_{\text{cav-med-FII}}.
\end{split}
\end{equation}
Apart from the undriven Hamiltonian, we obtain 3 driving-induced terms. The main purpose of this chapter is to derive these 3 terms by FSWT beyond the Hartree-type mean-field treatment in Chapter \ref{Chapter4}. We will briefly explain each of the terms in the following.

\subsubsection{The uncorrelated single-particle energy shift}
The first driving-induced term comes from $\hat{X}_0$ in $\hat{f}^{(1)}_{1,0}$, and it represents the uncorrelated single-particle energy shift, 
\begin{equation}\label{H_U=0}
\begin{split}
\hat{H}_{\text{U=0}} &= \frac{1}{2} \big( 
g  \sum\limits_{{\bf k},s} \sum\limits_{b,b'}  \frac{J^{bb'}_{{\bf k}s}}{\omL+\ep_{{\bf k},bb'}} [ \hat{c}_{{\bf k}bs}^{\dag} \hat{c}_{{\bf k}b's} 
, \hat{H}^{(1)}_{-1}] + H.c. \big) \\
&= g^2 \sum\limits_{{\bf k},s} \left( \frac{\big\vert J^{12}_{{\bf k}s} \big\vert^2}{\omL+\ep_{{\bf k},12}} - \frac{\big\vert J^{21}_{{\bf k}s} \big\vert^2}{\omL+\ep_{{\bf k},21}} \right) \hat{n}_{{\bf k} 1 s} 
~~ + ~~ \text{upper-band processes}
\end{split}
\end{equation}
where the driving term $\hat{H}^{(1)}_{-1}$ is defined in Eq.~(\ref{drive}). In the second line, we project to the lower-band and find the uncorrelated AC Stark shift and the uncorrelated Bloch-Siegert shift.

\subsubsection{Cavity-independent Floquet-induced interaction}
The second driving-induced term in Eq.~(\ref{H'-Chap5-result}) comes from $\hat{X}_1^{\text{screened}}$ in $\hat{f}^{(1)}_{1,0}$, and it represents the cavity-independent Floquet-induced interaction \footnote{These two terms $\hat{H}_{\text{U=0}}$ and $\hat{H}_{\text{cav-indep FII}}$, at the mean-field level, together provide the excitonic enhancement of the AC Stark shift found in Chapter \ref{Chapter4}.}.
\begin{equation}\label{cav-indep-FII-chap5}
\begin{split}
\hat{H}_{\text{cav-indep FII}}&=
\frac{1}{2} \big( [\hat{X}_1^{\text{screened}}, \hat{H}^{(1)}_{-1}] + H.c. \big) \\
&= 
\frac{1}{2} \sum_{{\bf k},{\bf k_1},{\bf q},s,s'}
\left(
V^{ss'}_{ {\bf k},{\bf k_1},{\bf q} } ~ (J^{12}_{{\bf k_1} s})^*
~ + ~ 
J^{12}_{{\bf k} s} ~ (V^{ss'}_{ {\bf k_1},{\bf k},{\bf q} })^* 
\right)
\hat{c}_{{\bf k} 1 s}^{\dag} \hat{c}_{{\bf k}+{\bf q} 1 s'} \hat{c}_{{\bf k_1}+{\bf q} 1 s'}^{\dag} \hat{c}_{{\bf k_1} 1 s} 
\end{split}
\end{equation}
Here, we only show the near-resonant part in the lower-band, where we have defined a scattering strength
\begin{equation}\label{cav-indep-scatter-chap5}
V^{ss'}_{ {\bf k},{\bf k_1},{\bf q} } \equiv
g^2 \sum_{\bf k'} 
 \left( \frac{J^{12}_{{\bf k'}s}}{\omL+\ep_{{\bf k'},12}}  -  \frac{J^{12}_{{\bf k}s}}{\omL+\ep_{{\bf k},12}} \right)
\frac{V_{{\bf k'}-{\bf k}}}{N}   
\left[ \frac{1}{\Gamma^{12}_{{\bf k},{\bf q}}} \right]_{{\bf k'},{\bf k_1}}
\end{equation}
which contains the inversion of a $\Gamma^{12}$ matrix representing the excitonic enhancement effects. As defined in Eq.~(\ref{Gamma^12_kq}), the matrix element of this $\Gamma^{12}$ matrix at row-index $\bf p$ and column-index $\bf p'$ is given by
\begin{equation}\label{Gamma12_kq}
\begin{split}
[\Gamma^{12}_{{\bf k},{\bf q}} ]_{{\bf p},{\bf p'}} 
\equiv
\bigg(\omL + \ep_{{\bf k}1} -  \ep_{{\bf k}+{\bf q}1} + \ep_{{\bf q}+{\bf p'}1} - \ep_{{\bf p'}2} - \sum\limits_{{\bf q'}\neq {\bf 0}} \frac{V_{\bf q'}}{N}  \bigg) \delta_{{\bf p},{\bf p'}}
~~ + ~~ (1-\delta_{{\bf p},{\bf p'}}) \frac{V_{{\bf p}-{\bf p'}}}{N} 
\end{split}
\end{equation}
where $V_{\bf q}$ represents the screened Coulomb interaction introduced in Section \ref{sec:general-setup}.

Below we analyse the resonances of the scattering strength $V^{ss'}_{ {\bf k},{\bf k_1},{\bf q} }$ in Eq.~(\ref{cav-indep-scatter-chap5}) as a function of driving frequency $\omL$. As the driving frequency approaches an exciton resonance, the matrix-inversion in Eq.~(\ref{cav-indep-scatter-chap5}) tends to diverge, which represents the excitonic enhancement on the cavity-independent Floquet-induced interaction. We note that the exciton involved in the Floquet-induced interaction (\ref{cav-indep-FII-chap5}) can contain non-zero center-of-mass momentum $\bf q$. As the driving frequency exceeds such an exciton resonance, the element of the matrix-inversion $ 1/ \Gamma^{12}_{{\bf k},{\bf q}}$ in Eq.~(\ref{cav-indep-scatter-chap5}) in general changes sign. To see this, note that the matrix defined in Eq.~(\ref{Gamma12_kq}) is real and symmetric, i.e., $[\Gamma^{12}_{{\bf k},{\bf q}} ]_{{\bf p},{\bf p'}} = [\Gamma^{12}_{{\bf k},{\bf q}} ]_{{\bf p'},{\bf p}} $, thus it can be diagonalised as
\begin{equation}
[\Gamma^{12}_{{\bf k},{\bf q}} ]_{{\bf p},{\bf p'}} = \sum_i (\omL-E_{{\bf k},{\bf q}}^{(i)}) \phi_i({\bf p}) \phi_i({\bf p'})
\end{equation}
where $(\omL-E_{{\bf k},{\bf q}}^{(i)})$ represents the $i$-th eigenvalue, and the eigenvectors satisfy $\sum_{\bf p} \phi_i({\bf p}) \phi_j({\bf p}) = \delta_{i,j} $. Thus the matrix-inversion in Eq.~(\ref{cav-indep-scatter-chap5}) reads
\begin{equation}\label{inverseGamma-eigendecompose}
\left[\frac{1}{\Gamma^{12}_{{\bf k},{\bf q}} }\right]_{{\bf p},{\bf p'}} = \sum_i \frac{1}{\omL-E_{{\bf k},{\bf q}}^{(i)}} \phi_i({\bf p}) \phi_i({\bf p'})
\approx \frac{1}{\omL-E_{{\bf k},{\bf q}}^{(j)}} \phi_j({\bf p}) \phi_j({\bf p'})
\end{equation}
where in the approximation symbol, we assume that the driving frequency $\omL$ approaches the $j$-th resonance $E_{{\bf k},{\bf q}}^{(j)}$, which allows us to ignore the contribution from other eigenvectors. From Eq.~(\ref{inverseGamma-eigendecompose}) we see that the matrix-inversion in the expression of $V^{ss'}_{ {\bf k},{\bf k_1},{\bf q} }$ in Eq.~(\ref{cav-indep-scatter-chap5}) changes sign as $\omL$ exceeds an excitonic resonance $E_{{\bf k},{\bf q}}^{(j)}$. This indicates that a sign change will also occur in the scattering strength $V^{ss'}_{ {\bf k},{\bf k_1},{\bf q} }$ in Eq.~(\ref{cav-indep-scatter-chap5}), and thus in the entire interaction $\hat{H}_{\text{cav-indep FII}}$ in Eq.~(\ref{cav-indep-FII-chap5}). 

In the minimal model considered in Chapter \ref{Chapter4}, a similar sign change has been found in the excitonically renormalised denominator. However, for the generalised model considered in this chapter, we note that the excitonic enhancement effect on the scattering strength $V^{ss'}_{ {\bf k},{\bf k_1},{\bf q} }$ depends crucially on the electron-light coupling coefficient $J^{12}_{{\bf k} s}$. According to Eq.~(\ref{cav-indep-scatter-chap5}), the phase of this coefficient $J^{12}_{{\bf k} s}$ will in general cause non-trivial interference effects in the scattering strength $V^{ss'}_{ {\bf k},{\bf k_1},{\bf q} }$ of the Floquet-induced interaction.

\subsubsection{Cavity-mediated Floquet-induced interaction}
The third driving-induced term in Eq.~(\ref{H'-Chap5-result}) comes from  $\hat{f}^{(1)}_{1,2}$, and it represents the excitonically enhanced cavity-mediated Floquet-induced interaction.
\begin{equation}
\begin{split}
\hat{H}_{\text{cav-med-FII}} &= \frac{1}{2} \big( [ \hat{f}^{(1)}_{1,2} , \hat{H}^{(1)}_{-1}] + H.c. \big) 
= \sum\limits_{{\bf k_f},{\bf k_f'},s,s'} 
U^{ss'}_{{\bf k_f},{\bf k_f'}} \hat{n}_{{\bf k_f} 1s} \hat{n}_{{\bf k_f'} 1s'}     
\end{split}
\end{equation}
This is a global-range interaction with strength
\begin{equation}
U^{ss'}_{{\bf k_f},{\bf k_f'}} \equiv
-\frac{ g^2 g_c^2 }{N \Delta_c}   \sum_{{\bf k},{\bf k'}} 
\left[ \frac{1}{\Gamma^{12}_{\text{MF}}} \right]_{{\bf k},{\bf k_f}} \left[ \frac{1}{\Gamma^{12}_{\text{MF}}} \right]_{{\bf k'},{\bf k_f'}}
\Re \left[ J^{12}_{{\bf k}s} ~ ( J^{12}_{{\bf k_f'}s'} )^* \right]
\end{equation}
where the mean-field matrix $\Gamma_{\text{MF}}$, defined in Eq.~(\ref{Gamma^12-MF}), has the following matrix element
\begin{equation}
[\Gamma^{12}_{\text{MF}} ]_{{\bf p},{\bf p'}}
\equiv \bigg(\omL  + \ep_{{\bf p'}12} - \sum\limits_{{\bf q'}\neq {\bf 0}} \frac{V_{\bf q'}}{N}  \bigg) \delta_{{\bf p},{\bf p'}}
~~ + ~~ (1-\delta_{{\bf p},{\bf p'}}) \frac{V_{{\bf p}-{\bf p'}}}{N} 
\end{equation}
This global-range interaction takes the form of a Landau interaction \cite{chubukov2018fermi,engelbrecht1992landau}, and thus, it favours a Pomeranchuk instability at low temperatures \cite{lamas2008fermi,rodriguez2013fermi}, which breaks the symmetry of the Fermi surface.
%(see Appendix~\ref{appendix-chap5} for an estimation of critical temperature). 
However, a detailed comparison between other possible instabilities triggered by the complete Floquet-induced interactions needs to be conducted in future works. In particular, using the FSWT Hamiltonian Eq.~(\ref{H'-Chap5-result}), we can predict whether the cavity-independent Floquet-induced interaction enhances the Pomaranchuk instability, or it already leads to another instability at a higher temperature.

\section{Application in driven TMDC materials}\label{sec.TMDC}
Below, we focus on a specific setup where a circularly polarised laser virtually excites Wannier excitons in TMDC materials. In this system, we discuss a physical consequence given by the cavity-independent Floquet-induced interaction in Eq.~(\ref{cav-indep-FII-chap5}). This effect cannot be captured by the results of the Gaussian elimination method in Chapter \ref{Chapter4}, as it requires going beyond the Hartree-type mean-field approximation.

In what follows, within the general model in Section \ref{sec:general-setup}, we will adopt the bandstructure $\ep_{{\bf k},b}$ for 2d Transition Metal Dichalcogenide (TMDC) materials \cite{chaves2020bandgap}. In TMDC, the bandgap reaches a minimum ($E_{gap}\sim 2.4$eV for WS$_2$ \cite{chernikov2014exciton}) at two inequivalent quasi-momenta, $\bf K$ and $\bf K'$. These valley points are often denoted by the K and K' points.
In TMDC, the Wannier excitons are formed around the valley points, with a typical 1s exciton resonance $E^{(1)}_{ex}\sim2.1$eV and 2s exciton resonance $E^{(2)}_{ex}\sim2.26$eV (for WS$_2$ \cite{chernikov2014exciton}). For the driving to be off-resonant, we require the driving strength $g$ to be much smaller than the laser's detuning to the resonances in the 2d material. For example, the driving frequency $\omL$ can either be red-detuned relative to the first exciton resonance or lie between the first and second exciton resonances, provided that $g\ll \vert \hbar \omL - E^{(1)}_{ex} \vert $ and $g\ll \vert E^{(2)}_{ex} - \hbar \omL \vert $.

\subsection{Coulomb mixing effect in the Floquet-induced interactions}
\begin{figure}
    \centering
    \includegraphics[width=0.7\linewidth]{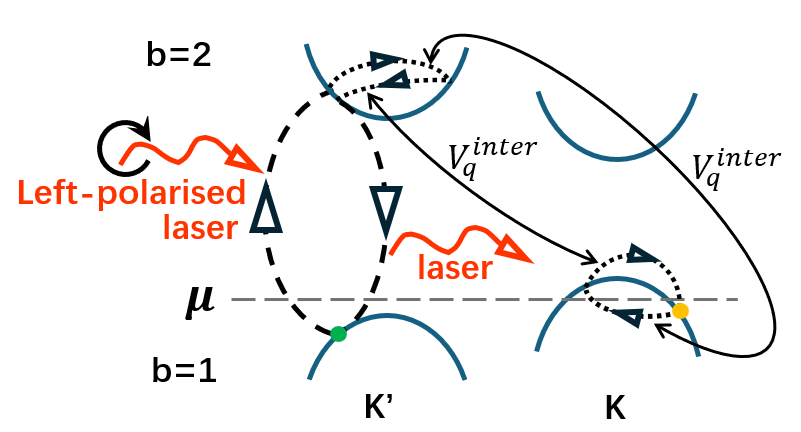}
    \caption{Coulomb mixing effect of the Floquet-induced interaction.}
    \label{fig:Coulomb-Mixing}
\end{figure}

In TMDC materials driven by circularly polarised lasers, the electron-light coupling coefficient $J^{bb'}_{{\bf k}s}$ shows a valley-selective property. For example, a left-handed circularly polarised light cannot resonantly couple the inter-band transition at the K-point \cite{zeng2012valley,cao2012valley}, because $J^{12}_{{\bf k}s}\to0$ as ${\bf k}\to{\bf K}$
\footnote{The calculation of the Rabi frequency $g J^{bb'}_{{\bf k}s}$ for TMDC models can be found in, e.g., Refs.~\cite{PhysRevB.100.235423,PhysRevLett.108.196802,PhysRevB.88.045416}.}.
This means the uncorrelated AC Stark shift in Eq.~(\ref{H_U=0}) cannot affect the electron's self-energy at the K-valley. Thus, when electron-electron interactions are negligible, the driving-induced energy shift at the K-valley consists only of a weak Bloch-Siegert shift. 
However, the Coulomb interaction will couple the electrons at different momenta, making the driving-induced energy shift at the K-valley no longer weak. This is identified as the Coulomb mixing effect \cite{doi:10.1021/nl500595u}. It cannot be observed from the excitonically enhanced AC Stark shift derived in Chapter \ref{Chapter4}. However, in the Floquet-induced interaction Eq.~(\ref{cav-indep-FII-chap5}) given by our FSWT which goes beyond the mean-field used in Chapter \ref{Chapter4}, this Coulomb mixing effect can be directly observed: For example, by taking ${\bf k_1}={\bf k}$ in Eq.~(\ref{cav-indep-FII-chap5}), we find electrons at arbitrary momentum $\bf k$ can contribute to the self-energy at K-valley, with strength $\text{Re} [ V^{ss'}_{ {\bf k},{\bf k},{\bf K}-{\bf k} } ~ (J^{12}_{{\bf k} s})^* ]$. Next, we explain this effect using Fig.~(\ref{fig:Coulomb-Mixing}).

In Fig.~(\ref{fig:Coulomb-Mixing}), we depict the processes involved in the Floquet-induced interaction (\ref{cav-indep-FII-chap5}) that lead to the Coulomb mixing effect. We assume the laser is left-polarised, such that the uncorrelated AC Stark shift in Eq.~(\ref{H_U=0}) only appears around the K'-point. We assume the AC Stark shift near K'-point is strong enough, such that it pushes the entire K' valley below the chemical potential. On the other hand, the electrons near the K valley only experience the weak Bloch-Siegert shift in Eq.~(\ref{H_U=0}), thus the lower-band $\ep_{{\bf k},1}$ still crosses the chemical potential $\mu$ near the K-point. In this scenario, an electron near K'-point (represented by the green dot) can still interact with another electron near K-point (represented by the yellow dot), according to the following process in the Floquet induced interaction: 
The off-resonant laser first virtually excite the K'-point electron to the upper band, and then this virtual excitation in the upper-band scatters with the K-point electron via the inter-band Coulomb repulsion $V_{\bf q}$. After scattering for more than 1 time, these two electrons can both be scattered back to their initial momenta 
\footnote{For example, in Fig.~(\ref{fig:Coulomb-Mixing}), we see that the two electrons can be scattered back to their initial momenta via scattering twice. The requirement of going back to the initial momenta means taking ${\bf k_1}={\bf k}$ in the Floquet-induced interaction (\ref{cav-indep-FII-chap5}).}.
When this happens, the virtual upper-band electron becomes able to return to the hole it left in the lower band and transfer its energy back to the off-resonant laser. The overall process generates an effective (momentum-space) density-density inter-valley coupling. This effective interaction can be found in the ${\bf k_1}={\bf k}$ channel of Eq.~(\ref{cav-indep-FII-chap5}), which reads
\begin{equation}\label{cav-indep-FII-CoulMix}
\hat{H}_{\text{cav-indep FII}}^{\text{Coulomb mix}}=
\sum_{{\bf k},{\bf q},s,s'}
\text{Re} \left[ V^{ss'}_{ {\bf k},{\bf k},{\bf q} } ~ (J^{12}_{{\bf k} s})^* \right]
\hat{c}_{{\bf k} 1 s}^{\dag} \hat{c}_{{\bf k}+{\bf q} 1 s'} \hat{c}_{{\bf k}+{\bf q} 1 s'}^{\dag} \hat{c}_{{\bf k} 1 s} .
\end{equation}
Taking ${\bf q}={\bf K}-{\bf k}$ in Eq.~(\ref{cav-indep-FII-CoulMix}), we find this interaction can modify the self-energy of electrons near the K-point, and thus it contributes to the Coulomb mixing effect. 
Moreover, since Eq.~(\ref{cav-indep-FII-CoulMix}) exhibits excitonic enhancement (due to the presence of $V^{ss'}_{ {\bf k},{\bf k},{\bf q} }$), we expect the Coulomb mixing effect to dominate the Bloch-Siegert shift as the driving frequency approaches an exciton resonance.

%Additionally, excitonic enhancement also appears in this Coulomb mixing term: As the driving frequency $\omL$ exceeds the lowest exciton resonance $E_{ex}^{(1)}$, the whole term in Eq.~(\ref{cav-indep-FII-CoulMix}) diverges and changes sign.

This Coulomb mixing effect provides an explanation for the significant enhancement of the Bloch-Siegert shift found in the recent experiment \cite{Conway2023} on monolayer WS$_2$: At the K-valley, where the AC Stark shift is absent in Eq.~(\ref{H_U=0}), the Bloch-Siegert shift becomes the only term in $\hat{H}_{\text{U=0}}$ that can influence the electron self-energy. However, as discussed above, the Floquet-induced interaction $\hat{H}_{\text{cav-indep FII}}$ can also shift the electron self-energy at this K-valley. When this additional $\mathcal{O}(g^2)$ self-energy contribution from $\hat{H}_{\text{cav-indep FII}}$ is attributed to the Bloch-Siegert shift, the latter can effectively be enhanced, compared with the prediction of the two-level atom model in Ref.~\cite{Conway2023}. This demonstrates that the Floquet-induced interaction in Eq.~(\ref{cav-indep-FII-chap5}) is relevant for future research on analysing the light-induced shifts in TMDC materials.

\subsection{Destructive interference in the Floquet-induced interactions}

To estimate the strength of the Coulomb mixing effect given by Eq.~(\ref{cav-indep-FII-chap5}), the phase property of $J^{12}_{{\bf k}s}$ becomes crucial. In TMDC driven by circularly polarised laser, near the K'-point, the phase of the coefficient $J^{12}_{{\bf k}s}$ could become non-analytic, such that $J^{12}_{{\bf k}s} \to   e^{i \varphi_{{\bf k}-{\bf K'}}}$ as ${\bf k}\to{\bf K'}$, where $\varphi_{{\bf k}-{\bf K'}} = \arg \big( ({\bf k}-{\bf K'})_x + i({\bf k}-{\bf K'})_y  \big) $ represents the direction from which $\bf k$ approaches the K'-point \cite{PhysRevLett.108.196802}. 
In Eq.~(\ref{cav-indep-scatter-chap5}), as we sum over $\bf k'$, this non-analytic behavior of the coefficient $J^{12}_{{\bf k} s}$ around the K'-point leads to a destructive interference in $V^{ss'}_{ {\bf k},{\bf k_1},{\bf q} }$. Accurate DFT simulations \cite{marzari2012maximally} on the phase properties of $J^{12}_{{\bf k} s}$ are needed for future studies to understand this non-trivial interference in Floquet-induced interactions, which is crucial for estimating the strength of Coulomb mixing effect and developing proposals on driving-induced pairing in TMDC materials.

%Wrong Explaination since in the non-interacing case we can always rotate the overall phase of an eigenstate to eliminate this strange phase
%\footnote{This non-analytic behaviour can be easily understood using a 2-level system: The two-level Hamiltonian $\hat{H}_{TL}= \Delta \hat{\sigma}_z + q_x \hat{\sigma}_x + q_y \hat{\sigma}_y $ has two eigenstates, labelled by $\vert + \rangle$ and $\vert - \rangle$. Here we have used the Pauli matrices, and we define $\hat{\sigma}_+ = (\hat{\sigma}_x + i \hat{\sigma}_y)/2$. The matrix element $ \langle + \vert \hat{\sigma}_+ \vert - \rangle $ always contains a phase factor $e^{i \varphi}$ where $\varphi=\arg(q_x + i q_y)$. Consequently, when $q_x , q_y$ goes to 0, the norm of $ \langle + \vert \hat{\sigma}_+ \vert - \rangle $ converges to 1 while its phase $\varphi$ cannot converge. This is precisely the situation for the phase factor in $J^{bb'}_{{\bf k}s}$ under circularly polarised laser drive.} 

\section{Conclusion}
In this chapter, using the FSWT based on Sylvester equations, we once again obtained the excitonic enhancement effect on the cavity-mediated Floquet-induced interaction. Our FSWT Hamiltonian also shows that, apart from the cavity-mediated interaction, the laser drive simultaneously creates another type of Floquet-induced interaction, which always exists even in the absence of the cavity. This cavity-independent Floquet-induced interaction, at the mean-field level, gives rise to the excitonic enhancement of AC Stark shift found by the previous Gaussian elimination method in Chapter \ref{Chapter4}. Going beyond the mean-field treatment, this Floquet-induced interaction becomes able to reveal the Coulomb mixing effects in the off-resonantly driven TMDC materials. This effect is relevant for understanding the anomalously large Bloch-Siegert shift observed in previous experiments.

Our FSWT Hamiltonian treats the two types of Floquet-induced interactions, the cavity-mediated term and the cavity-independent term, on equal footing. It provides a foundational tool for future predictions of driving-induced Fermi surface instabilities triggered by the competition among the intrinsic Coulomb interactions and the two types of Floquet-induced interactions. 

In the FSWT Hamiltonian, increasing the driving frequency above an exciton resonance can still lead to a sign change of Floquet-induced interactions, consistent with the previous findings in the minimal model in Chapter \ref{Chapter4}. However, we find the phase property of the electron-light coupling coefficient can result in non-trivial interference effects in the Floquet-induced interactions. Our FSWT Hamiltonian provides a practical starting point for future analyses of this interference effect, where numerical simulations of the phase properties of $J^{12}_{{\bf k} s}$ become relevant.

%% file: Chapters/Conclusions.tex
In the research area of driven quantum many-body systems, using the example of excitonic enhancement in Chapter \ref{Chapter4}, this thesis shows the necessity of studying the Floquet-induced interactions in the many-body picture. Then, an advanced method of finding the Floquet-induced interactions in the many-body picture, i.e., FSWT, is constructed in Chapter \ref{Chapter5}. Using this FSWT method, we have obtained the Floquet-induced interactions in various experimentally-relevant models in Chapters \ref{Chapter6} and \ref{Chapter7}. The main achievements of the thesis are summarised as follows.

\subsubsection{Summary of contributions}
In Chapter \ref{Chapter4}, we investigated the effects of excitonic enhancement on Floquet-induced interactions. In a cavity-material setup with two-band on-site interactions, we find that the Floquet-induced interactions provide low-energy Floquet engineering effects, such as the renormalised Stark shift, Bloch-Siegert shift, and the cavity-mediated interaction. 
We showed how the exciton formation due to electronic interactions enhances these low-energy effects by carrying out an inter-band screening calculation in combination with a mean-field decoupling of the interactions.  
Altogether, we find that the screened Floquet Hamiltonian looks superficially similar to one of a noninteracting system (which we obtain readily by simply setting $\hat{U} = 0$). However, the Floquet-induced band structure change is enhanced by the interaction across much of reciprocal space.
In tetracene-type materials, both the electronic dispersion and the cavity-mediated interactions are strongly changed in both amplitude and range: 
In the direct vicinity of the $\Gamma$-point, the dynamical localisation due to Stark and Bloch-Siegert shifts is reduced, and their relative strength is shown to depend on the electronic interaction strengths. 
Additionally, the Floquet-induced cavity-mediated interactions in the conduction band can be enhanced by up to one order of magnitude and broadened in reciprocal space, and the enhancement is even greater in other parts of the Brillouin zone. 
This excitonic enhancement strengthens the cavity-mediated interaction at an anomalously broad range of incoming electron momenta, and thus efficiently couples the Fermi surface with electrons far from it. According to Refs.~\cite{lamas2008fermi,rodriguez2013fermi}, the forward-scattering part of this broadened interaction is likely to significantly enhance the Pomeranchuk instability in the driven material, resulting in a Fermi surface deformation. Our screened Floquet Hamiltonian thus serves as a foundation for future proposals on driving-induced phase transitions facilitated by the excitonic enhancement mechanism.
Our observations in Chapter \ref{Chapter4} are readily explained by the mixing of momenta associated with the exciton. This introduces a dispersionless resonance in the Hamiltonian, which can be targeted by an optimised choice of driving parameters from any point in reciprocal space and thus strongly enhances the interactions. 
They are reminiscent of the well-established Coulomb enhancement of light-matter coupling in 2D semiconductors~\cite{Wang2018,RevModPhys.82.1489}. 

Our results in Chapter \ref{Chapter4} show how it is possible to incorporate static screening effects into effective low-energy Floquet Hamiltonians without using the rotating wave approximation. The current Hartree-type screening calculation needs to be refined to account for additional many-body orders in the bare ground state of a correlated material. Going beyond the low-energy projection we used in Chapter \ref{Chapter4}, it will be interesting to analyse these screening effects in stronger-driving or ultrastrong cavity-coupling regimes, where the present approximations seize to be valid and a self-consistent evaluation of the effective Hamiltonian will be essential. 
Furthermore, we have focused on Frenkel excitons in Chapter \ref{Chapter4}, which emerge from local interactions. It will be interesting to extend our study to Wannier excitons and investigate how the excitonic enhancement is related to the screening of the Coulomb interaction in materials. This will also be necessary to allow for quantitative comparisons with recent experiments in dichalcogenides~\cite{Conway2023} as well as for future optical control applications which exploit the coupling to cavities, such as the proposed cavity quantum spin liquids~ \cite{chiocchetta2021cavity}. 

To understand the Floquet-mediated interactions in these generalised models, the projector-based Sambe space Gaussian elimination Floquet theory developed in Chapter \ref{Chapter4} is no longer sufficient. Moreover, we note that the excitonically enhanced AC Stark shift found in Chapter \ref{Chapter4} is actually a mean-field remnant of the cavity-independent Floquet-induced interaction. To predict the driving-induced phase transitions triggered by the cavity-mediated interaction, we should simultaneously take this cavity-independent Floquet-induced interaction into account, such that these two Floquet-induced interactions are treated on an equal footing. This cannot be realised straightforwardly by the method developed in Chapter \ref{Chapter4}, which is limited by mean-field approximations.  

The above limitations of the Gaussian elimination method motivate us to develop a systematic Floquet block-diagonalisation method to understand the Floquet-induced interactions in many-body systems. This method is constructed in Chapter \ref{Chapter5}, where we presented a Floquet Schrieffer Wolff transform (FSWT) to obtain effective Floquet Hamiltonians and micro-motion operators of periodically driven many-body systems for any non-resonant driving frequency. The FSWT perturbatively eliminates the oscillatory components in the driven Hamiltonian by solving operator-valued Sylvester equations with well-controlled approximations. FSWT is an advanced method for obtaining the Floquet-induced interactions in driven many-body systems since it no longer suffers from the mean-field limitations. Meanwhile, by solving the Sylvester equation, FSWT can provide the self-consistent Floquet-induced interactions which remain unchanged at all quasi-energies $\Ea$. 

FSWT goes beyond various high-frequency expansion (HFE) methods commonly used in Floquet theory, as we demonstrate with the example of the driven Fermi-Hubbard model in Chapter \ref{Chapter6}. In this driven system, we solve the many-body Sylvester equation in perturbative orders of electron hopping. This allows us to derive the FSWT Hamiltonian, from which the Floquet-induced interactions are identified as correlated hoppings. In the limit of high driving frequencies, the FSWT Hamiltonian can reduce to the widely used HFE result, yet it provides a more accurate prediction of the driven dynamics, particularly in the regime of non-resonant driving parameters. We thus anticipate the FSWT method to be of practical use in designing Rydberg multi-qubit gates, controlling correlated hopping in quantum simulations in optical lattices and describing multi-orbital and long-range interacting systems driven in-gap. 

We envision several extensions of our FSWT formalism that can treat important scenarios of driven many-body physics: 
For near-resonant driving, it will be convenient to construct the FSWT in the co-rotating frame, where the solution to the Sylvester equation has no divergence. The corresponding Floquet Hamiltonian then describes how off-resonant driving terms, which give rise to, e.g., the Bloch-Siegert shift \cite{Sie2018} in TMDC, influence the resonant dynamics. 
For deep strong driving (relevant for rf-driven superconducting qubits \cite{doi:10.1126/science.1119678}), where $g\gg \omL$, one can apply our FSWT in a nonuniformly rotating frame~\cite{doi:10.1126/science.1119678,PhysRevLett.125.195301,10.21468/SciPostPhys.5.2.017} (as achieved in Appendix \ref{appendix:StrongDriveFSWT}), where the time dependence is transferred from $g$ to smaller parameters in the undriven system, e.g. the hopping $J$ (we assume the driving term commutes with the interaction): Constructing the FSWT expansion in powers of $J$, one can describe how the electron hopping perturbatively modifies the strong driving physics. 

The FSWT expansion method presented in Chapter \ref{Chapter6} could be directly adapted to other driven many-body systems with more interaction terms (such as longer-ranged density-density repulsions or multi-orbital Kanamori interactions \cite{PhysRevB.100.220403}). This method is also suitable for systems driven by time-dependent interactions~\cite{PhysRevLett.109.203005,meinert2016floquet}, where the Floquet Hamiltonian can be obtained by solving Eq.~(\ref{y2-part}). The crucial advantage of our method is to provide a single formula which is not restricted to the driving frequency being the largest energy scale involved. This advantage becomes more prominent when more and more interacting terms are taken into account. Finally, we note that the FSWT provides a convenient and accurate method to analyse driven correlated materials near the Mott transition. This regime is particularly relevant for several unconventional superconductors, where light-induced superconductivity was observed in recent experiments in fullerides~\cite{mitrano2016possible, Budden2021, Rowe2023} and charge transfer salts~\cite{Buzzi2020, Buzzi2021}.

In Chapter \ref{Chapter7}, we apply our FSWT to the generalised driven cavity-semiconductor setup described in Chapter \ref{Chapter3}. With the help of FSWT, we can study systems with long-range Coulomb interactions driven by lasers with arbitrary polarisation. The corresponding Floquet Hamiltonian treats the cavity-independent Floquet-induced interaction on equal foot with the cavity-mediated Floquet-induced interaction. This FSWT Hamiltonian offers a systematic way to predict the driving-induced phase transitions in the cavity-QED setup. 

We utilise our results in Chapter \ref{Chapter7} to analyse the cavity-independent Floquet-induced interactions in TMDC materials driven by a left circularly polarised laser. In this context, the Floquet-induced interaction contains a Coulomb mixing term, which facilitates a notable self-energy shift at the otherwise forbidden K valley, with the prohibition arising from the valley selection rule in the non-interacting system. This Coulomb mixing term describes the driving effect that transcends the mean-field approximation, thus it cannot be found using the Sambe space Gaussian elimination method in Chapter \ref{Chapter4}. This Coulomb mixing term also exhibits excitonic enhancement: as the driving frequency approaches the exciton resonance, the Coulomb mixing can exert a more significant influence on the self-energy near the K valley than the Bloch-Siegert shift. This indicates that the FSWT Hamiltonian offers a potential explanation for the anomalously enhanced Bloch-Siegert shift observed in TMDC materials, suggesting that this anomalous enhancement arises from the Coulomb mixing term in the Floquet-induced interaction. Therefore, we anticipate that our FSWT method will be instrumental in future studies of light-induced shifts in TMDC materials.
Moreover, the formula for the Floquet-induced interaction reveals a non-trivial interference effect stemming from the momentum dependence of the electron-laser coupling coefficient. Our findings highlight the need for future numerical calculations of this coupling coefficient to quantitatively assess the strength of this Floquet-induced interaction, which is crucial for understanding the Coulomb mixing effect and proposing driving-induced phase transitions.

\newpage
\subsubsection{Outlooks on many-body Sylvester equations}
\begin{figure}[h]
    \centering
    \includegraphics[width=0.8\linewidth]{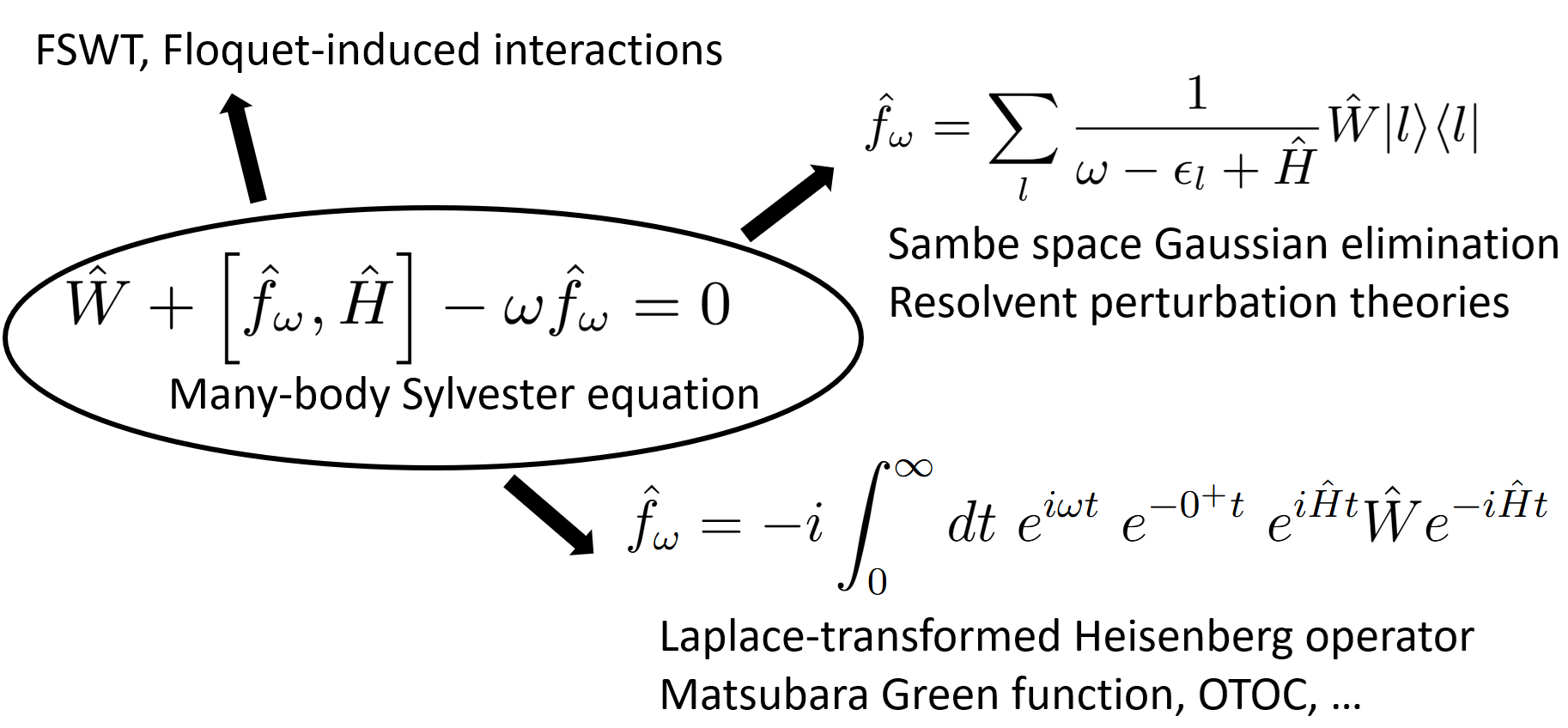}
    \caption{Sylvester equation plays a central role in quantum many-body engineering.}
    \label{fig:Conclusion}
\end{figure}

Throughout the above works, we have revealed the central role played by the solution of the Sylvester equation, i.e., the micro-motion operator $\hat{f}$, in quantum many-body engineering, as depicted in Fig.~\ref{fig:Conclusion}. Our FSWT constructed in Chapter \ref{Chapter5} shows that the stroboscopic Floquet Hamiltonian (and the corresponding Floquet-induced interactions) can be directly obtained from the micro-motion operator $\hat{f}$. This micro-motion $\hat{f}$ can be written in the Green operator form, whose low-energy limit reduces back to the Sambe space Gaussian elimination Floquet formalism studied in Chapter \ref{Chapter4}. Most importantly, this micro-motion $\hat{f}$ turns out to be the Laplace transform of the Heisenberg operator of the driving operator. This connection indicates that the calculations in Matsubara equilibrium Green function formalism could be replaced by solving the Sylvester equations, which becomes more flexible in treating strongly correlated systems or in the absence of spatial translational invariance. Furthermore, this connection also offers a new perspective to understand higher-order correlators relevant to quantum information scrambling: For example, the spectrum of the out-of-time-order correlator (OTOC) \cite{swingle2018unscrambling,yuan2022quantum} can be understood as the self-convolution of the micro-motion operator $\hat{f}$. Consequently, solving the Sylvester equations in various quantum information processing platforms, in particular the Rydberg arrays and ion traps, will be essential to understanding the controlled dynamics of these engineered many-body systems.

\subsubsection{Parallel evaluation of Heisenberg operators using Sylvester equations}
We end by proposing a promising application of numerically solving the many-body Sylvester equation, e.g., by a tensor network ansatz. To numerically evolve a Heisenberg operator $e^{i \hat{H} t}\hat{W}e^{-i \hat{H} t}$, we can instead numerically solve the following Sylvester equation in the complex frequency domain
\begin{equation}\label{Sylvester-complex-frequency}
\hat{W} + [\hat{f}_{\omL+i\gamma},\hat{H}] - (\omL+i\gamma) \hat{f}_{\omL+i\gamma} = 0
\end{equation}
for $\omega\in (-\infty,\infty)$. Here, $\gamma >0$ is a fixed finite positive number representing the imaginary part of the frequency. According to Section \ref{sec:FSWT=vanVleck}, we know that $\gamma >0$ makes $\hat{f}_{\omL+i\gamma}$ analytic for arbitrary $\omL$, which should guarantee an efficient convergence when  numerically solving Eq.~(\ref{Sylvester-complex-frequency}). Moreover, the formal solution of the Sylvester equation in Appendix ~\ref{linear-response} remains valid for $\gamma >0$, which gives 
\begin{equation}\label{parallel-Heisenberg}
    \forall t>0, ~~~~~~ 
    e^{i \hat{H} t}\hat{W}e^{-i \hat{H} t} = \frac{i}{2\pi}  e^{\gamma  t}  \int_{-\infty}^{\infty} d \omL ~ e^{-i \omL t} \hat{f}_{\omL+i\gamma}
\end{equation}
This equation shows that, to obtain the Heisenberg operator $e^{i \hat{H} t}\hat{W}e^{-i \hat{H} t}$, instead of evolving the matrix exponential $\hat{e}^{\pm i \hat{H} t}$, we can solve the Sylvester equation (\ref{Sylvester-complex-frequency}) \textit{in parallel} for different $\omL$ with fixed $\gamma$. As the simulation time $t$ increases, the conventional real-time approach (such as TEBD) requires more Trotter-gate layers. In the parallel approach using the Sylvester equations, we instead need to simulate $\hat{f}_{\omL+i\gamma}$ on a finer grid of $\omL$ by using more computing nodes in parallel. 

The errors of the parallel approach are increased by the exponential factor $e^{\gamma t}$ in Eq.~(\ref{parallel-Heisenberg}). For this reason, we may also need to decrease $\gamma$, which can be achieved using the following trick: We first numerically solve Eq.~(\ref{Sylvester-complex-frequency}) with a large $\gamma$. Once the solution for all $\omL$ has been obtained in parallel for this $\gamma$, the derivative $\frac{\partial}{\partial \omL} \hat{f}_{\omL+i\gamma}$ can be calculated. Then, without the need to solve Eq.~(\ref{Sylvester-complex-frequency}), we can use the relation (known as the complex form of Cauchy–Riemann equations)
\begin{equation}
\frac{\partial}{\partial \gamma} \hat{f}_{\omL+i\gamma} = i \frac{\partial}{\partial \omL} \hat{f}_{\omL+i\gamma}   
\end{equation}
to get the solution of $\hat{f}_{\omL+i(\gamma-d \gamma)}$ with a smaller imaginary frequency $(\gamma-d \gamma)$.
With this trick, we anticipate that Eq.~(\ref{parallel-Heisenberg}) could provide a new way to decouple the strongly correlated Heisenberg operator $e^{i \hat{H} t}\hat{W}e^{-i \hat{H} t}$ into a continuous sum of less complicated operators.

%% file: Chapters/appendix_A.tex
%\appendix

%\section{Appendix for the driven cavity-material Hamiltonian}\label{appendix:dipolar}

%\section{Appendix for Sambe Space Gaussian elimination}

\section{Useful relations in the detuned, low-energy limit} \label{appendix:useful-relations}
Before going ahead with the derivation of the effective Hamiltonian, we list simplifications that will be used repeatedly in Chapter \ref{Chapter4}. 

In the in-gap driving regime (see the red region in Fig.~\ref{fig:Frenkel-resonance}), the cavity-material system can only borrow energy virtually from the laser (shown by the dashed lines in Fig.~\ref{fig:band}), 
%creating virtual processes as shown in fig(\ref{fig:band}), 
instead of actually absorbing it. 
%Therefore, there will be longer time for our driven system to condense into a low-energy space (spanned by the low-energy eigenstates of the effective Floquet Hamiltonian in Eq\ref{projected-Heff}), before the laser eventually heat our system away from this space. 
This means we can focus on the low-energy sector of the static Hamiltonian $\hat{H}_{0}$ (\ref{H0}).
The low-energy Floquet states contain neither cavity nor band excitation. Hence, we can require the projector $\PEa$ introduced in Section~\ref{appendix-filter} to additionally satisfy the low-energy conditions
\begin{equation}\label{low-energy-approx}
    \begin{split}
        \hat{a}                      \PEa &= 0   ~~~~ \text{and} ~~~~
        \hat{c}_{ {\bf k} , 2, s }   \PEa = 0    ~~~  \forall {\bf k}, s
    \end{split}. 
\end{equation}
When the electron-cavity coupling goes to ultra-strong regime $g_c \sim (\omc + \ep_{21})$, so that the ground state $\vert G \rangle$ of the static Hamiltonian $\hat{H}_{0}$ already contains photons and upper-band electrons, the condition (\ref{low-energy-approx}) will no longer hold. 
%In this regime, one-band models won't be sufficient anymore to describe the low-energy Floquet Hamiltonian. 
In this regime, polaritonic transformations\cite{ashida2022nonperturbative,sheremet2023waveguide} (or polaritonic projectors) might be necessary to derive the low-energy Floquet Hamiltonian. 
In this work we will not consider such strong electron-cavity coupling.

We further list three mathematical relations that will be repeatedly used to obtain the effective Floquet Hamiltonian. 
The first relation is the Dyson expansion of the Green operators
\[
    \hat{G}^{0}_{(E)} = \hat{G}^{b}_{(E)} + \hat{G}^{b}_{(E)} \hat{H}_c \hat{G}^{0}_{(E)} = \hat{G}^{b}_{(E)} + \hat{G}^{0}_{(E)} \hat{H}_c \hat{G}^{b}_{(E)},
\]
\begin{equation}\label{Dyson}
    \begin{split}
        \hat{G}^{b}_{(E)} &= \hat{g}_{(E)}    + \hat{g}_{(E)} \hat{U} \hat{G}^{b}_{(E)}       = \hat{g}_{(E)} + \hat{G}^{b}_{(E)} \hat{U} \hat{g}_{(E)},
    \end{split}
\end{equation}
which will be applied recursively to simplify the expression of $\hat{H}^{\text{eff}}_{(\Ea)}$ in Eq.~(\ref{Heff_weak_drive}). Here, we have defined
\begin{equation}\label{definition-G1-g}
    \begin{split}
        \hat{G}^{b}_{(E)} &= \frac{1}{E-\hat{H}_b} ~~~~ \text{and} ~~~~
        \hat{g}_{(E)}     = \frac{1}{E-\hat{h}}
    \end{split}
\end{equation}
as the Green operator of $\hat{H}_b$ (\ref{H1}) and $\hat{h}$ (\ref{H1h}) at energy $E$, respectively. Our definition of $\hat{g}$, $\hat{G}^{b}$, and $\hat{G}^{0}$ refers to the Green operators in exciton literature, see e.g.  Ref.~\cite{PhysRevB.29.4401}.

Second, two exact commutation relations of the non-interacting Green operator $\hat{g}$ (see Appendix \ref{appendix:useful-relations}.1 for the proof) will be used 
\begin{equation}\label{g-commutator}
    \begin{split}
        \hat{g}_{(E)} \hat{c}_{{\bf k} b s}^{\dag} \hat{c}_{{\bf k} b' s} &= \hat{c}_{{\bf k} b s}^{\dag} \hat{c}_{{\bf k} b' s} \hat{g}_{(E - \ep_{{\bf k}, b} + \ep_{{\bf k}, b'} )}, \\
        \hat{g}_{(E)} \hat{a}^{\dag} &= \hat{a}^{\dag} \hat{g}_{(E - \omc )}.
    \end{split}
\end{equation}
%They are exact . 
These relations allow to move $\hat{G}^{b}$ to the right of the driving operator $\hat{H}_{-1}$ in structures like $\PEa \hat{H}_{1}\hat{G}^{b}\hat{H}_{-1} \PEa$ which we will encounter below.

Third, in the low-energy limit, the laser detuning results in the third relation (see Appendix \ref{appendix:useful-relations}.2 for the proof)
\begin{equation}\label{G1-reduced-to-number}
\begin{split}
    \hat{G}^{b}_{( \Ea - \Delta_{\bf k}^0 )}  \PEa
    &= \big( 1+O[\frac{g_c}{ 2 \omL }] \big) \hat{G}^{0}_{( \Ea - \Delta_{\bf k}^0 )} \PEa  \\
    &~~~~~~~~~~~~~~~~~~~   = - ( \Delta_{\bf k}^0 )^{-1} \PEa 
\end{split}
\end{equation}
which plays the same role as taking the static limit in the Green function of the lower band: 
It reduces the Green operator $\hat{G}^{b}$ to a commuting c-number (independent of $\Ea$), %the denominator related to detuning, 
when $\hat{G}^{b}$ appears adjacent to the projector $\PEa$.
% the first '=' comes from a Dyson series expansion of G1, by G0 and H2. 
%We explain these relations in Appendix \ref{appendix:useful-relations}.
Here $\Delta^{0}_{\bf k}= \ep_{{\bf k}, 2} - \ep_{{\bf k}, 1}  - \omL$ denotes the laser-bandgap detuning.

\subsection{The commutation relations of $\hat{g}$ in Eq.(\ref{g-commutator})}
\paragraph{Proving $\hat{g}_{(E)} \hat{c}_{{\bf k} b s}^{\dag} \hat{c}_{{\bf k} b' s} = \hat{c}_{{\bf k} b s}^{\dag} \hat{c}_{{\bf k} b' s} \hat{g}_{(E - \ep_{{\bf k}, b} + \ep_{{\bf k}, b'} )} $}
We first evaluate $\hat{g}_{(\omega)} \hat{c}_{{\bf q} b s}^{\dag} \hat{c}_{{\bf q} b' s}$ for an arbitrary electron quasi-momentum $\bf q$, band-index $b,b'$ and spin $s$. The non-interacting Green operator $\hat{g}_{(\omega)}$ is defined in Eq.~(\ref{definition-G1-g}). The free Hamiltonian $\hat{h}$ commutes with the kinetic Hamiltonian $\hat{H}_{{\bf q} ,s}$, $ [ \hat{h} , \hat{H}_{{\bf q} ,s} ] =0  $, where we define 
\begin{equation}\label{def-kinetic-Hamiltonian}
    \hat{H}_{{\bf q},s} \equiv \sum_b (\ep_{\mathbf{q}, b} -\mu ) \hat{c}_{\mathbf{q} b s}^{\dag} \hat{c}_{\mathbf{q} b s}.
\end{equation}
This allows us to write down the following simplified Dyson expansion
\begin{equation}
    \begin{split}
        \hat{g}_{(\omega)} &= \frac{1}{\omega - (\hat{h} - \hat{H}_{{\bf q}, s} ) - \hat{H}_{{\bf q}, s} } \\
        &= \sum\limits_{n=0}^{\infty} (\frac{1}{\omega - (\hat{h} - \hat{H}_{{\bf q}, s} )})^{n+1} (\hat{H}_{{\bf q}, s})^{n},
    \end{split}
\end{equation}
and therefore 
\begin{equation}
    \begin{split}
        \hat{g}_{(x)} \hat{c}_{\mathbf{q} b s}^{\dag} \hat{c}_{\mathbf{q} b' s}  &= \sum\limits_{n=0}^{\infty} (\frac{1}{\omega - (\hat{h} - \hat{H}_{{\bf q}, s} )})^{n+1}   ( \hat{H}_{{\bf q}, s} )^{n} \hat{c}_{\mathbf{q} b s}^{\dag} \hat{c}_{\mathbf{q} b' s} \\
        &= \sum\limits_{n=0}^{\infty} (\frac{1}{\omega - (\hat{h} - \hat{H}_{{\bf q}, s} )})^{n+1} \hat{c}_{\mathbf{q} b s}^{\dag} \hat{c}_{\mathbf{q} b' s}  ( \hat{H}_{{\bf q}, s} + \epsilon_{\mathbf{q}, bb'})^{n}  \\
        &= \hat{c}_{\mathbf{q} b s}^{\dag} \hat{c}_{\mathbf{q} b' s} \sum\limits_{n=0}^{\infty} (\frac{1}{\omega - (\hat{h} - \hat{H}_{{\bf q}, s} )})^{n+1}   ( \hat{H}_{{\bf q}, s} + \epsilon_{\mathbf{q}, bb'})^{n}.
    \end{split}
\end{equation}
In the second line we define $ \epsilon_{\mathbf{q}, bb'} \equiv \epsilon_{\mathbf{q}, b} - \epsilon_{\mathbf{q}, b'} $, and then we use the commuting relation $[\hat{H}_{{\bf q}, s} , \hat{c}_{\mathbf{q} b s}^{\dag} \hat{c}_{\mathbf{q} b' s}]= \epsilon_{\mathbf{q}, bb'} \hat{c}_{\mathbf{q} b s}^{\dag} \hat{c}_{\mathbf{q} b' s}$ to repeatedly move $\hat{c}_{\mathbf{q} b s}^{\dag} \hat{c}_{\mathbf{q} b' s}$ leftward. In the third line, we use the fact that $\hat{H}_{{\bf q} ,s}$ is the only part in $\hat{h}$ that acts non-trivially on $\hat{c}_{\mathbf{q} b s}^{\dag} \hat{c}_{\mathbf{q} b' s}$, i.e. we use $ [ \hat{h} - \hat{H}_{{\bf q} ,s} , ~ \hat{c}_{\mathbf{q} b s}^{\dag} \hat{c}_{\mathbf{q} b' s} ] =0  $ to move $\hat{c}_{\mathbf{q} b s}^{\dag} \hat{c}_{\mathbf{q} b' s}$ to the very left. 

However, since $ [ \hat{h} , \hat{H}_{{\bf q} ,s} ] =0  $, we can reverse the following Dyson expansion
\begin{equation}
    \begin{split}
        \sum\limits_{n=0}^{\infty} (\frac{1}{\omega - (\hat{h} - \hat{H}_{{\bf q},s} )})^{n+1} (\hat{H}_{{\bf q}, s} + \epsilon_{\mathbf{q}, bb'} )^{n}  &=  
        \frac{1}{\omega - (\hat{h} - \hat{H}_{{\bf q},s} ) - (\hat{H}_{{\bf q},s}  +  \epsilon_{\mathbf{q}, bb'}   )}  \\
        &= \frac{1}{(\omega -\epsilon_{\mathbf{q}, bb'}  ) - \hat{h}      } \\
        &= \hat{g}_{(\omega -\epsilon_{\mathbf{q}, bb'}  )}.
    \end{split}
\end{equation}
Thus we derive the exact equation in (\ref{g-commutator}) which reads
\begin{equation}
    \hat{g}_{(\omega)} \hat{c}_{\mathbf{q} b s}^{\dag} \hat{c}_{\mathbf{q} b' s} = \hat{c}_{\mathbf{q} b s}^{\dag} \hat{c}_{\mathbf{q} b' s} \hat{g}_{(\omega -\epsilon_{\mathbf{q}, b} +  \epsilon_{\mathbf{q}, b'} )}. \label{eq.B6}
\end{equation}
We note that the chemical-potential dependence cancels out.

\newpage
\paragraph{Proving $\hat{g}_{(E)} \hat{a}^{\dag} = \hat{a}^{\dag} \hat{g}_{(E - \omc )}$}
Similarly, we next evaluate $\hat{g}_{(\omega)} \hat{a}^{\dag} $, where this time $\hat{a}^{\dag}$ is a bosonic operator creating a cavity photon. Since $ [ \hat{h} , \omega_{c} \hat{a}^{\dag}\hat{a}  ] =0  $, we have the following simplified Dyson expansion
\begin{equation}
    \begin{split}
        \hat{g}_{(\omega)} &= \frac{1}{\omega - (\hat{h} - \omega_{c} \hat{a}^{\dag}\hat{a} ) - \omega_{c} \hat{a}^{\dag}\hat{a} } \\
        &= \sum\limits_{n=0}^{\infty} (\frac{1}{\omega - (\hat{h} - \omega_{c} \hat{a}^{\dag}\hat{a} )})^{n+1} (\omega_{c} \hat{a}^{\dag}\hat{a})^{n},
    \end{split}
\end{equation}
and thus
\begin{equation}
    \begin{split}
        \hat{g}_{(\omega)} \hat{a}^{\dag}
        &= \sum\limits_{n=0}^{\infty} (\frac{1}{\omega - (\hat{h} - \omega_{c} \hat{a}^{\dag}\hat{a} )})^{n+1} (\omega_{c})^{n} (\hat{a}^{\dag}\hat{a})^{n} \hat{a}^{\dag} \\
        &= \sum\limits_{n=0}^{\infty} (\frac{1}{\omega - (\hat{h} - \omega_{c} \hat{a}^{\dag}\hat{a} )})^{n+1} (\omega_{c})^{n} \hat{a}^{\dag}(\hat{a} \hat{a}^{\dag})^{n}\\
        &= \sum\limits_{n=0}^{\infty} (\frac{1}{\omega - (\hat{h} - \omega_{c} \hat{a}^{\dag}\hat{a} )})^{n+1} (\omega_{c})^{n} \hat{a}^{\dag}(\hat{a}^{\dag} \hat{a} + 1)^{n}\\
        &= \hat{a}^{\dag} \sum\limits_{n=0}^{\infty} (\frac{1}{\omega - (\hat{h} - \omega_{c} \hat{a}^{\dag}\hat{a} )})^{n+1}  (\omega_{c} \hat{a}^{\dag} \hat{a} + \omega_{c})^{n} \\
        &= \hat{a}^{\dag} \frac{1}{\omega - (\hat{h} - \omega_{c} \hat{a}^{\dag}\hat{a} ) - (\omega_{c} \hat{a}^{\dag} \hat{a} + \omega_{c})}.
    \end{split}
\end{equation}
In the second line we again shift the position of $n$ parentheses without moving any operator. In the third line we used the bosonic commutation relation $[\hat{a},\hat{a}^{\dag}]=1$, in the fourth line we used the commuting relation $[\hat{h} - \omega_{c} \hat{a}^{\dag}\hat{a}, ~ \hat{a}^{\dag}]=0$ to move $\hat{a}^{\dag}$ to the very left. In the fifth line we reverse the Dyson expansion, whose validity is guaranteed by the commutation relation $ [ \hat{h} , \omega_{c} \hat{a}^{\dag}\hat{a}  ] =0  $. Thus,we derive the following equation
\begin{equation}
    \hat{g}_{(\omega)} \hat{a}^{\dag} = \hat{a}^{\dag} \hat{g}_{(\omega - \omc)}. \label{eq.C3}
\end{equation}
Taking the Hermitian conjugate of both sides followed by a shift of argument $\omega \to \omega + \omc$, we find
\begin{equation}
    \hat{g}_{(\omega)} \hat{a} = \hat{a} \hat{g}_{(\omega + \omc)}. \label{eq.C4}
\end{equation}
These two equations (\ref{eq.C3}) and (\ref{eq.C4}) for the cavity photon operators, combined with the corresponding previous equation (\ref{eq.B6}) for electron operators, allow us to move $\hat{g}_{(x)}$ to the very left/right of each term of the effective Hamiltonian (\ref{Heff_weak_drive}) expanded by (\ref{Dyson}).

\subsection{Approximations in Eq.~(\ref{G1-reduced-to-number})}
Eq.~(\ref{G1-reduced-to-number}) allows us to replace the Green operator $\hat{G}^{b}$ by a number when it lies adjacent to the energy projector $\PEa$. To prove this, we make the following Dyson expansion
\begin{equation}\label{G1-reduce-explain}
\begin{split}
\hat{G}^{b}_{( \Ea - \Delta_{\bf k}^0 )}  \PEa &= 
\frac{1}{\Ea - \Delta_{\bf k}^0 - \hat{H}_0 + \hat{H}_{c} } \PEa  \\
&= \hat{G}^{0}_{( \Ea - \Delta_{\bf k}^0 )}  \PEa - \hat{G}^{b}_{( \Ea - \Delta_{\bf k}^0 )} \hat{H}_{c} \hat{G}^{0}_{( \Ea - \Delta_{\bf k}^0 )} \PEa \\
&\approx \frac{-1}{\Delta_{\bf k}^0} \PEa - \hat{G}^{b}_{( \Ea - \Delta_{\bf k}^0 )} \hat{\mathcal{P}}_{1,1} \hat{H}_{c} \frac{-1}{\Delta_{\bf k}^0} \PEa
\end{split}
\end{equation}
where in the first line we use $\hat{H}_{b} = \hat{H}_0 - \hat{H}_{c}$, in the second line is we make a Dyson expansion. In the third line we use the definition of the projector $\PEa \equiv \delta(\Ea-\hat{H}_0)$ in Section~\ref{appendix-filter} so that $\hat{G}^{0}_{( \Ea - \Delta_{\bf k}^0 )}  \PEa \approx \frac{-1}{\Delta_{\bf k}^0} \PEa $. Meanwhile, according to the definition of the electron-cavity interaction $\hat{H}_{c}$ in (\ref{H0}), we know that when $\Ea$ belongs to the low-energy limit (\ref{low-energy-approx}), $\hat{H}_{c} \PEa$ lives in the ``1 band-excitation, 1 cavity-photon" Hilbert space, denoted by the projector $\hat{\mathcal{P}}_{1,1}$. The space $\hat{\mathcal{P}}_{1,1}$ is off-resonant when acting on the Green operator $\hat{G}^{b}_{( \Ea - \Delta_{\bf k}^0 )}$, meaning that $\hat{G}^{b}_{( \Ea - \Delta_{\bf k}^0 )} \hat{\mathcal{P}}_{1,1} \hat{H}_{c} \PEa \approx \frac{1}{-\Delta_{\bf k}^0 - \omc - \ep_{21}} \hat{H}_{c} \PEa \propto \frac{g_c}{2\omL} $, thus we have
\begin{equation*}
    \hat{G}^{b}_{( \Ea - \Delta_{\bf k}^0 )}  \PEa \approx \frac{-1}{\Delta_{\bf k}^0} \big( 1 + O[\frac{g_c}{2\omL}] \big) \PEa \approx \frac{-1}{\Delta_{\bf k}^0} \PEa
\end{equation*}
which is exactly (\ref{G1-reduced-to-number}). Besides, there is another way to argue why we can ignore the term $\hat{G}^{b}_{( \Ea - \Delta_{\bf k}^0 )} \hat{H}_{c} \hat{G}^{0}_{( \Ea - \Delta_{\bf k}^0 )} \PEa$ in (\ref{G1-reduce-explain}), note that if we apply a rotating wave approximation to $\hat{H}_{c}$, then $\hat{H}_{c}$ can no longer simultaneously create a cavity photon and a band-excitation, thus the term $\hat{H}_{c} \PEa$ directly vanishes in the low-energy limit.

\newpage

\section{The non-interacting model: deriving the Floquet Hamiltonian}
\label{sec:zero-correlation-model}

In this section, we analyze the low-energy sector of the Floquet Hamiltonian in the absence of electronic interactions, i.e. $\hat{U} = 0$. 
In this non-interacting case, inserting the Dyson expansion~(\ref{Dyson}) into the expression of the effective Hamiltonian~$\hat{H}^{\text{eff}}_{(\Ea)}$ in (\ref{Heff_weak_drive}), we obtain various contributions, which are further sandwiched by the projector $\PEa$ in the low-energy limit,
\begin{equation}\label{Heff-expansion-terms}
\begin{split}
    \hat{H}^{\text{eff}}_{(\Ea)} 
    &\approx \PEa \hat{H}^{\text{eff}}_{(\Ea)} \PEa + ... \\
    &\approx \PEa \hat{H}_{0} \PEa + \PEa \hat{H}_{1}\hat{g}_{(\Ea +\omL)}\hat{H}_{-1} \PEa \\
    & ~~ + \PEa \hat{H}_{1}\hat{g}_{(\Ea +\omL)} \hat{H}_{c} \hat{g}_{(\Ea +\omL)} \hat{H}_{-1} \PEa \\
    & ~~ + \PEa \hat{H}_{1}\hat{g}_{(\Ea +\omL)} \hat{H}_{c} \hat{g}_{(\Ea +\omL)} \hat{H}_{c} \hat{g}_{(\Ea +\omL)} \hat{H}_{-1} \PEa \\
    & ~~ + \PEa \hat{H}_{-1} \hat{g}_{(\Ea -\omL)}\hat{H}_{1} \PEa + ...
\end{split}
\end{equation}
Here, $\PEa \hat{H}_{0} \PEa$ is the direct projection of the static Hamiltonian into the low-energy limit. 
We have further used that $\hat{G}^{b} = \hat{g}$ when $\hat{U}=0$.
Each of the remaining terms in Eq.~(\ref{Heff-expansion-terms}) %, after the following simplifications, 
can be ascribed a straightforward physical meaning, as we will show in the following.  
%For example, $\PEa \hat{H}_{1}\hat{g}\hat{H}_{-1} \PEa$ represents the optical Stark shift, $\PEa \hat{H}_{-1} \hat{g}\hat{H}_{1} \PEa$ represents the Bloch-Siegert shift, $\hat{H}_{1}\hat{g}_{(\Ea +\omL)} \hat{H}_{c} \hat{g}_{(\Ea +\omL)} \hat{H}_{-1}$ is an off-resonant term contributing to the Floquet heating, $\PEa \hat{H}_{1}\hat{g} \hat{H}_{c} \hat{g} \hat{H}_{c} \hat{g} \hat{H}_{-1} \PEa$ represents the cavity-mediated interaction under RWA.
%In (\ref{Heff-expansion-terms}), 
The expansion terms from $\hat{H}_{1}\hat{G}^{0}_{(\Ea +\omL)}\hat{H}_{-1}$ are shown only up to the second order of cavity-electron coupling $\hat{H}_{c}$, and the expansion terms from $\hat{H}_{-1}\hat{G}^{0}_{(\Ea -\omL)}\hat{H}_{1}$ are shown only up to the 0-th order of $\hat{H}_{c}$. 
As will be shown the following, other expansion terms can be neglected as they are much weaker than the terms we retained in Eq.~(\ref{Heff-expansion-terms}).

\subsection{The Optical Stark Shift}\label{U=0_ACS}
We first analyze the term $\PEa \hat{H}_{1}\hat{g}\hat{H}_{-1} \PEa$ in Eq.~(\ref{Heff-expansion-terms}). It stems from inserting the Dyson expansion (\ref{Dyson}) of $\hat{G}^{0}$ in (\ref{Heff_weak_drive}), and then truncating to the 0-th order of $\hat{H}_{c}$ in this expansion. Explicitly, we find
\begin{equation}\label{Stark}
    \begin{split}
         & \PEa \hat{H}_{1}\hat{g}_{(\Ea +\omL)}\hat{H}_{-1} \PEa \\ 
         &= 
          \vert g \vert^2
          \PEa 
          \sum_{{\bf k},{\bf k}', s, s'}
          \hat{c}_{{\bf k}' 1 s'}^{\dag} \hat{c}_{{\bf k}' 2 s'}
          \hat{g}_{(\Ea +\omL)}
          \hat{c}_{{\bf k} 2 s}^{\dag} \hat{c}_{{\bf k} 1 s}   \PEa \\
          &=
         \vert g \vert^2
          \PEa 
          \sum_{{\bf k},{\bf k}', s, s'}
          \hat{c}_{{\bf k}' 1 s'}^{\dag} \hat{c}_{{\bf k}' 2 s'} \hat{c}_{{\bf k} 2 s}^{\dag} \hat{c}_{{\bf k} 1 s} 
          \hat{g}_{(\Ea - \Delta_{\bf k}^0 ) }    \PEa \\
         &=
         - \vert g \vert^2
          \PEa 
          \sum_{{\bf k}, s}
          \hat{c}_{{\bf k} 1 s}^{\dag} \hat{c}_{{\bf k} 1 s}
          \frac{1}{ \Delta_{\bf k}^0 }
          \PEa.
    \end{split}
\end{equation}
%To derive (\ref{Stark}), 
Here we insert the definition of $\hat{H}_{-1}$ (\ref{D}) in the first line, and the low-energy limit (\ref{low-energy-approx}) allows us to discard the de-excitation term in $\hat{H}_{-1}$. In the second line we use (\ref{g-commutator}) to move $g$ adjacent to $\PEa$, and we have used the definition of the laser-bandgap detuning $\Delta^{0}_{\bf k}= \ep_{{\bf k}, 2} - \ep_{{\bf k}, 1}  - \omL$. In the third line we use (\ref{G1-reduced-to-number}) to reduce $\hat{g}$ by the detuning denominator, and we also use the relation $ \PEa \hat{c}_{{\bf k}' 2 s'} \hat{c}_{{\bf k} 2 s}^{\dag} = \delta_{{\bf k},{\bf k'}} \delta_{s,s'} \PEa $ which directly follows from the low-energy limit (\ref{low-energy-approx}).

We identify this term as the lower-band's optical Stark shift, i.e., laser-induced renormalization of the electronic band energy. In semiconductor quantum wells, this effect was first reported in Ref.~\cite{mysyrowicz1986dressed}.
%when driven by a laser, the energy of an electron in the lower band with momentum $\bf k$ decreases by $\frac{\vert g \vert ^2 }{\Delta{\bf k}}$.

\subsection{The Bloch-Siegert shift}\label{BSS_U=0}
There is another term $\PEa \hat{H}_{-1} \hat{g}\hat{H}_{1} \PEa$ in (\ref{Heff-expansion-terms}), which also arises form the Dyson expansion of Eq.~(\ref{Heff_weak_drive}),
\begin{equation}\label{Bloch-Siegert}
    \begin{split}
         & \PEa \hat{H}_{-1}\hat{g}_{(\Ea -\omL)}\hat{H}_{1} \PEa \\ 
         &= 
          \vert g \vert^2
          \PEa 
          \sum_{{\bf k},{\bf k}', s, s'}
          \hat{c}_{{\bf k}' 1 s'}^{\dag} \hat{c}_{{\bf k}' 2 s'}
          \hat{g}_{(\Ea -\omL)}
          \hat{c}_{{\bf k} 2 s}^{\dag} \hat{c}_{{\bf k} 1 s}   \PEa \\
          &=
         \vert g \vert^2
          \PEa 
          \sum_{{\bf k},{\bf k}', s, s'}
          \hat{c}_{{\bf k}' 1 s'}^{\dag} \hat{c}_{{\bf k}' 2 s'} \hat{c}_{{\bf k} 2 s}^{\dag} \hat{c}_{{\bf k} 1 s} 
          \hat{g}_{(\Ea - \Delta_{\bf k}^0 -2\omL ) }    \PEa \\
         &\approx
         - \vert g \vert^2
          \PEa 
          \sum_{{\bf k}, s}
          \hat{c}_{{\bf k} 1 s}^{\dag} \hat{c}_{{\bf k} 1 s}
          \frac{1}{ \Delta_{\bf k}^0 +2\omL }
          \PEa.
    \end{split}
\end{equation}
This is the Bloch-Siegert shift in materials~\cite{PhysRev.57.522, Sie2018} which further decreases the energy of electrons in the lower band. It stems from the non-RWA terms in the laser-electron coupling. The Bloch-Siegert shift in 2-level systems has been previously derived by various Floquet methods~\cite{shirley1965solution,PhysRevA.81.022117,PhysRevA.79.032301,PhysRevLett.105.257003}, our Floquet Hamiltonian method generalizes this derivation to a many-body system.

%If we apply the rotating wave approximation (RWA) to the driving term $\hat{H}_{-1} \to -i \sum_{{\bf{k}}, s} g \hat{c}_{{\bf{k}} 2 s}^{\dag} \hat{c}_{{\bf{k}} 1 s}$, this Bloch-Siegert shift will vanish, because after RWA there will only be de-excitation term in $\hat{H}_{1}$, and then the low-energy limit (\ref{low-energy-approx}) means $\hat{H}_{1} \PEa \to 0$, and thus $\PEa \hat{H}_{-1} \hat{g}\hat{H}_{1} \PEa \to 0$. 
%Although it is much weaker than the optical Stark shift in the red-detuned condition, this shift becomes stronger in the blue-detuned case where this formula sill holds.

%Similarly, any term expanded from $\hat{H}_{-1}\hat{G}^{0}_{(\Ea -\omL)}\hat{H}_{1}$ in (\ref{Heff_weak_drive}) becomes much weaker than its counterpart in $\hat{H}_{1}\hat{G}^{0}_{(\Ea +\omL)}\hat{H}_{-1}$. Within RWA, the term $\hat{H}_{-1}\hat{G}^{0}_{(\Ea -\omL)}\hat{H}_{1}$ completely disappears in the low-energy limit.

\subsection{Cavity-electron vertex and cavity-mediated interactions}\label{appendix:subsec:cav-med-int_U=0}
We finally consider the term $\PEa \hat{H}_{1} \hat{g} \hat{H}_{c} \hat{g} \hat{H}_{c} \hat{g}\hat{H}_{-1} \PEa$ in (\ref{Heff-expansion-terms}), which appears as we expand the Floquet Hamiltonian~(\ref{Heff_weak_drive}) to the second order in $\hat{H}_{c}$ with the Dyson series~(\ref{Dyson}). 

For convenience, we first analyze a part of this term, $\hat{H}_{c} \hat{g} \hat{H}_{-1} \PEa$, which we identify as an effective cavity-electron vertex,
\begin{equation}\label{lower-band-vertex}
    \begin{split}
         \hat{H}_{c} \hat{g}_{(\Ea +\omL)}\hat{H}_{-1} \PEa 
        &=   \frac{  g }{\sqrt{N}} \hat{a}^{\dag}  \sum_{{\bf k},s,{\bf k'},{\bf q'}, s'} 
        \big( g_{c,{\bf q'}}^{*}  \hat{c}_{{\bf k'}-{\bf q'} 1 s'}^{\dag} \hat{c}_{{\bf k'} 2 s'}  + h.c.   \big)  \hat{g}_{(\Ea +\omL)}
        \hat{c}_{{\bf k} 2 s}^{\dag} \hat{c}_{{\bf k} 1 s} \PEa \\
        &=  \frac{ g }{\sqrt{N}} \hat{a}^{\dag}  \sum_{{\bf k},s, {\bf k'},{\bf q'}, s'}   
        \big( g_{c,{\bf q'}}^{*}  \hat{c}_{{\bf k'}-{\bf q'} 1 s'}^{\dag} \hat{c}_{{\bf k'} 2 s'}  + h.c.   \big)  \hat{c}_{{\bf k} 2 s}^{\dag} \hat{c}_{{\bf k} 1 s} \hat{g}_{(\Ea -\Delta_{\bf k}^0)}  \PEa \\
        & \approx \frac{ g }{\sqrt{N}} \hat{a}^{\dag}  \sum_{{\bf k},{\bf q'}, s} g_{c,{\bf q'}}^{*} \hat{c}_{{\bf k}-{\bf q'} 1 s}^{\dag}  \hat{c}_{{\bf k} 1 s} 
        \frac{1}{ - \Delta_{\bf k}^0 } \PEa  
        ~~~~~~~  + \hat{\mathcal{P}}_{2,1} \hat{H}_{c} \hat{g}_{(\Ea +\omL)}\hat{H}_{-1} \PEa  \\
        & \approx -  \hat{a}^{\dag}  \sum_{{\bf k},{\bf q'}, s}  \frac{g g_{c,{\bf q'}}^{*} }{\sqrt{N} \Delta_{\bf k}^0 }  \hat{c}_{{\bf k}-{\bf q'} 1 s}^{\dag}  \hat{c}_{{\bf k} 1 s} 
         \PEa .
    \end{split}
\end{equation}
The resulting formula describes the scattering vertex of a laser photon into the cavity, which is mediated by a lower-band electron. Here, the summation over the transferred momenta ${\bf q'}$ doesn't come from the electron's coupling to different bosonic modes, but from the spatial inhomogeneity of the single cavity mode, whose mode-wavefunction has a finite shape as shown in Fig.~(\ref{fig:setup}).

To derive (\ref{lower-band-vertex}), in the first line, we insert the definition of $\hat{H}_{c}$, and then the low-energy limit (\ref{low-energy-approx}) allows us to discard the annihilator $\hat{a}$ in $\hat{H}_{c}$. In the second line, we use (\ref{g-commutator}) to move $\hat{g}$ rightward, so that $\hat{g}$ can then be reduced by (\ref{G1-reduced-to-number}). In the third line, the band de-excitation part of $\hat{H}_{c}$ is simplified using $\hat{c}_{{\bf k'},2,s'} \hat{c}_{{\bf k},2,s}^{\dag} \PEa = \delta_{{\bf k},{\bf k'}} \delta_{s,s'} \PEa$, while the ``band excitation part" (denoted as h.c. in above lines) is separated out by another projector $\hat{\mathcal{P}}_{2,1}$ projecting to the off-resonant subspace containing two band-excitations and one cavity-photon. In the last line, we discard this off-resonant part associated with $\hat{\mathcal{P}}_{2,1}$, since its contribution in the term $\PEa \hat{H}_{1} \hat{g} \hat{H}_{c} \hat{g}\hat{H}_{-1} \PEa$ is
\begin{equation*}
\begin{split}
    & \PEa \hat{H}_{1} \hat{g} \hat{H}_{c} \hat{\mathcal{P}}_{2,1} \hat{g}_{(\Ea +\omL)} \hat{\mathcal{P}}_{2,1} \hat{H}_{c} \hat{g} \hat{H}_{-1} \PEa  \\
    \approx &  \PEa \hat{H}_{1} \hat{g} \hat{H}_{c} \hat{\mathcal{P}}_{2,1}  \frac{1}{\omL-2\ep_{21}-\omc}  \hat{\mathcal{P}}_{2,1} \hat{H}_{c} \hat{g} \hat{H}_{-1} \PEa \\
    \approx & - \frac{1}{2~\omL} \PEa \hat{H}_{1} \hat{g} \hat{H}_{c} \hat{\mathcal{P}}_{2,1} \hat{H}_{c} \hat{g} \hat{H}_{-1} \PEa
\end{split}
\end{equation*}
where $\ep_{21}$ denotes the scale of the band gap. We will see from the following calculation that the $\frac{1}{\omL}$ contribution is much weaker than the contribution given by the cavity-electron vertex part (with no band excitations and one cavity-photon), which we retained in the last line of Eq.~(\ref{lower-band-vertex}).

Having derived the cavity-electron vertex, we come back to the term $\PEa \hat{H}_{1} \hat{g} \hat{H}_{c} \hat{g} \hat{H}_{c} \hat{g} \hat{H}_{-1} \PEa$ in Eq.~(\ref{Heff-expansion-terms}),
\begin{equation}\label{cavity-meidated-interaction-U=0}
    \begin{split}
        & \PEa \hat{H}_{1} \hat{g}_{(\Ea + \omL)}  \hat{H}_{c} \hat{g}_{(\Ea + \omL)}  \hat{H}_{c} \hat{g}_{(\Ea + \omL)}\hat{H}_{-1} \PEa \\
        &\approx   
        \PEa \frac{1}{\sqrt{N}} \sum_{{\bf k'},{\bf q'}, s'}   \bigg( \frac{g_{c,{\bf q'}} g^{*}}{\Delta_{\bf k'}^0} \bigg) \hat{c}_{{\bf k'} 1 s'} ^{\dag}  \hat{c}_{{\bf k'}-{\bf q'} 1 s'} 
        \times \hat{a} \hat{g}_{(\Ea + \omL)} \hat{a}^{\dag} 
        \times \frac{1}{\sqrt{N}} \sum_{{\bf k},{\bf q}, s}   \bigg( \frac{g_{c,{\bf q}}^{*} g}{-\Delta_{\bf k}^0} \bigg) \hat{c}_{{\bf k}-{\bf q} 1 s}^{\dag}  \hat{c}_{{\bf k} 1 s}  \PEa \\
        &= \PEa  \frac{1}{N}  \sum_{ \substack{ {\bf k},{\bf q}, s \\ {\bf k'},{\bf q'}, s'} } \frac{\vert g \vert^2 g_{c,{\bf q'}} g_{c,{\bf q}}^{*} }{\Delta_{\bf k'}^0 \Delta_{\bf k}^0}
        \hat{c}_{{\bf k'} 1 s'} ^{\dag}  \hat{c}_{{\bf k'}-{\bf q'} 1 s'}
        \hat{c}_{{\bf k}-{\bf q} 1 s}^{\dag}  \hat{c}_{{\bf k} 1 s} 
        \times \hat{a} \hat{a}^{\dag} \hat{g}_{(\Ea + \omL - \omc - \ep_{{\bf k}-{\bf q},1} + \ep_{{\bf k},1} )}  \PEa \\
        &\approx \PEa  \frac{1}{N}  \sum_{ \substack{ {\bf k},{\bf q}, s \\ {\bf k'},{\bf q'}, s'} } \frac{\vert g \vert^2 g_{c,{\bf q'}} g_{c,{\bf q}}^{*} }{\Delta_{\bf k'}^0 \Delta_{\bf k}^0}
        \hat{c}_{{\bf k'} 1 s'} ^{\dag}  \hat{c}_{{\bf k'}-{\bf q'} 1 s'}
        \hat{c}_{{\bf k}-{\bf q} 1 s}^{\dag}  \hat{c}_{{\bf k} 1 s} 
        \times \hat{a} \hat{a}^{\dag} \frac{-1}{ \Delta_c + \ep_{{\bf k}-{\bf q},1} - \ep_{{\bf k},1}  } \PEa \\
        &\approx -\frac{1}{N}  \sum_{ \substack{ {\bf k},{\bf q}, s \\ {\bf k'},{\bf q'}, s'} } \frac{\vert g \vert^2 g_{c,{\bf q'}} g_{c,{\bf q}}^{*} }{\Delta_c \Delta_{\bf k'}^0 \Delta_{\bf k}^0}
        \hat{c}_{{\bf k'} 1 s'} ^{\dag}  \hat{c}_{{\bf k'}-{\bf q'} 1 s'}
        \hat{c}_{{\bf k}-{\bf q} 1 s}^{\dag}  \hat{c}_{{\bf k} 1 s} \PEa.
    \end{split}
\end{equation}
In deriving (\ref{cavity-meidated-interaction-U=0}), in the first line we insert into the formula of the vertex (\ref{lower-band-vertex}), in the second line we move $\hat{g}$ to the far-right using (\ref{g-commutator}), which is then reduced to a denominator by (\ref{G1-reduced-to-number}) in the third line. In the last line, we use $\hat{a}\hat{a}^{\dag} \PEa = \PEa$ which follows from (\ref{low-energy-approx}). Here we also assume
\begin{equation*}
    \frac{ g_{c,{\bf q}}^{*} }{ \Delta_c + \ep_{{\bf k}-{\bf q},1} - \ep_{{\bf k},1} } \approx \frac{ g_{c,{\bf q}}^{*} }{ \Delta_c }
\end{equation*}
which is very accurate, because the cavity mode wavefunction changes slowly at the distance of lattice constant, so that $g_{c,{\bf q}}$ always vanish whenever $\bf q$ becomes large enough to make $( \ep_{{\bf k}-{\bf q},1} - \ep_{{\bf k},1} )$ comparable to $\Delta_c$.

In (\ref{cavity-meidated-interaction-U=0}), we obtain a laser-assisted cavity-mediated interaction between lower-band electrons, which is structurally similar to the cavity-mediated interactions in cold atoms~\cite{LandigRenate2016Qpfc}, 1-band electron jellium~\cite{GaoHongmin2020Pepi}, and multi-band spin systems~\cite{chiocchetta2021cavity}. To the best of our knowledge, the laser-assisted cavity-mediated interaction has not been described by a Floquet method before. The virtual processes mediating this interaction are shown in Fig.~(\ref{fig:band}): A lower-band electron is first virtually excited to the upper band by a laser photon, and then it decays back to the lower band by emitting a cavity photon. These virtual processes together form the electron-cavity vertex described by Eq.~(\ref{lower-band-vertex}). The time-reverse of these virtual processes is described by the Hermitian conjugate of (\ref{lower-band-vertex}). %, thus we denote them as 'conjugated vertex'. 
When the photon emitted by a vertex is reabsorbed by another conjugated vertex, a cavity-mediated electron interaction is established in the lower band. In this cavity-mediated interaction (\ref{cavity-meidated-interaction-U=0}), the total quasi-momentum of the two lower-band electrons is not exactly conserved, but it changes by $({\bf q'}-{\bf q})$, because the cavity mode in our model breaks translational invariance. 

For an estimation of the strength of the cavity-mediated interaction (\ref{cavity-meidated-interaction-U=0}), we choose the similar driving strength and detuning as Ref.~\cite{chiocchetta2021cavity} considers, so that an interaction strength $\sim 100$K is reached when $\Delta_{\bf q}^0\sim 1$THz and $\Delta_c\sim 0.1$THz. According to Ref.~\cite{yamase2005mean}, such a strong cavity-mediated interaction with tiny electron-momentum-transfer may lead to the spontaneous Fermi surface symmetry breaking, known as the Pomeranchuk instability.

\subsection{Non-RWA corrections to the cavity-mediated interaction}
The non-RWA part of the cavity-mediated interaction comes from multiple sources in the expansion (\ref{Heff-expansion-terms}). First, in $\PEa \hat{H}_{1} \hat{g} \hat{H}_{c} \hat{g} \hat{H}_{c} \hat{g}\hat{H}_{-1} \PEa$, the non-RWA part is already mentioned above
\begin{equation}\label{non-RWA-cavity-med-interaction-U=0}
\begin{split}
    & \PEa \hat{H}_{1} \hat{g} \hat{H}_{c} \hat{\mathcal{P}}_{2,1} \hat{g}_{(\Ea +\omL)} \hat{\mathcal{P}}_{2,1} \hat{H}_{c} \hat{g} \hat{H}_{-1} \PEa  \\
    & \approx  \frac{1}{N}  \sum_{ \substack{ {\bf k},{\bf q}, s \\ {\bf k}_1,{\bf q'}, s_1  }} \frac{-\vert g \vert^2 g_{c,{\bf q'}}^{*} g_{c,{\bf q}} }{(2 \omL+\Delta_{{\bf k}_1}^0 + \Delta_{\bf k}^0 + \Delta_c) (\Delta_{{\bf k}_1}^0)^2  } 
    \times    \hat{c}_{{\bf k}_1 1 s_1} ^{\dag}  \hat{c}_{{\bf k}+{\bf q}-{\bf q'} 1 s}^{\dag}
        \hat{c}_{{\bf k} 1 s}  \hat{c}_{{\bf k}_1 1 s_1} \\
    &~~ +\frac{1}{N}  \sum_{ \substack{ {\bf k},{\bf q}, s \\ {\bf k}_1,{\bf q'}, s_1  }} \frac{-\vert g \vert^2 g_{c,{\bf q'}}^{*} g_{c,{\bf q}} }{(2 \omL+\Delta_{{\bf k}_1}^0 + \Delta_{\bf k}^0 + \Delta_c) \Delta_{{\bf k}_1}^0 \Delta_{{\bf k}+{\bf q}}^0  } 
    \times    \hat{c}_{{\bf k}+{\bf q} 1 s} ^{\dag}  \hat{c}_{{\bf k}_1 -{\bf q'} 1 s_1 }^{\dag}
        \hat{c}_{{\bf k} 1 s}  \hat{c}_{{\bf k}_1 1 s_1} \\
    &=\frac{1}{N}  \sum_{ \substack{ {\bf k},{\bf q}, s \\ {\bf k}_1,{\bf q'}, s_1  }} \frac{-\vert g \vert^2 g_{c,{\bf q'}}^{*} g_{c,{\bf q}} }{(2 \omL+\Delta_{{\bf k}_1}^0 + \Delta_{\bf k}^0 + \Delta_c) \Delta_{{\bf k}_1}^0  }
    ( \frac{1}{ \Delta_{{\bf k}_1}^0 } \hat{c}_{{\bf k}_1 1 s_1} ^{\dag}  \hat{c}_{{\bf k}+{\bf q}-{\bf q'} 1 s}^{\dag} 
    + \frac{1}{ \Delta_{{\bf k}+{\bf q}}^0 } \hat{c}_{{\bf k}+{\bf q} 1 s} ^{\dag}  \hat{c}_{{\bf k}_1 -{\bf q'} 1 s_1 }^{\dag} )
    \hat{c}_{{\bf k} 1 s}  \hat{c}_{{\bf k}_1 1 s_1} \\
\end{split}
\end{equation}
where the summations exclude the case when the two spin-orbitals $({\bf k}_1 , s_1) = ({\bf k}+{\bf q} , s)$ are identical. This exclusion follows from the Pauli exclusion principle, meaning that we cannot simultaneously excite two upper-band electrons with same spin and momentum.

Second, the full term $\PEa \hat{H}_{-1} \hat{g} \hat{H}_{c} \hat{g} \hat{H}_{c} \hat{g}\hat{H}_{1} \PEa$ also contributes to the non-RWA part of cavity-mediated interaction. However, a direct estimation of this term gives
\begin{equation*}
    \PEa \hat{H}_{-1} \hat{g}_{(\Ea - \omL)}  \hat{H}_{c} \hat{g}_{(\Ea - \omL)}  \hat{H}_{c} \hat{g}_{(\Ea - \omL)}\hat{H}_{1} \PEa
    \propto \frac{-1}{(2\omL)^3}
\end{equation*}
which is much smaller than the first non-RWA part of the cavity-mediated interaction in (\ref{non-RWA-cavity-med-interaction-U=0}), and thus can be directly ignored.

As we will see below, the non-RWA cavity-mediated interaction in (\ref{non-RWA-cavity-med-interaction-U=0}) is in general much weaker than the interaction induced by the leading off-resonant terms in Eq.~(\ref{off-resonant-leading-order-induced-int}), 
unless the electron-cavity coupling constant $g_c$ becomes comparable to the electron-laser coupling strength $g $.
Moreover, when the correction Eq.~(\ref{non-RWA-cavity-med-interaction-U=0}) becomes strong, as discussed below (\ref{low-energy-approx}), other electron-cavity non-RWA effect will arise, which cannot be captured by the low-energy condition (\ref{low-energy-approx}). This means that not all ultra-strong electron-cavity coupling effects to the same order of (\ref{non-RWA-cavity-med-interaction-U=0}) are captured by our treatment. We left these missing effects, together with the dipolar self-energy in $\hat{H}_{c}$, for future studies.

\subsection{Inter-manifold terms in the Floquet Hamiltonian}

\paragraph{(1) Inter-manifold terms from $\hat{H}_{1}\hat{g}_{(\Ea +\omL)}\hat{H}_{-1}$}\label{off-resonant-leading-order}
In the absence of rotating wave approximation, there exists a term in $\hat{H}_{1}\hat{g}_{(\Ea +\omL)}\hat{H}_{-1}$ which scatters two electrons into the upper electronic band. 
%the system from the low-energy limit $\PEa$ to the off-resonant space containing 2 upper-band electrons, denoted by the projector $\hat{\mathcal{P}}_{2,0}$. 
This term reads
\begin{equation}\label{inter-manifold-band-excitation}
    \begin{split}
         \hat{\mathcal{P}}_{2,0} \hat{H}_{1}\hat{g}_{(E_\alpha +\omL)}\hat{H}_{-1} \PEa
         &= 
          g^2
          \hat{\mathcal{P}}_{2,0}
          \sum_{{\bf q},{\bf q}', s, s'}
          \hat{c}_{\mathbf{q}' 2 s'}^{\dagger} \hat{c}_{\mathbf{q}' 1 s'}
          \hat{g}
          \hat{c}_{\mathbf{q} 2 s}^{\dagger} \hat{c}_{\mathbf{q} 1 s}  \PEa \\
          &= 
          g^2
          \sum_{{\bf q},{\bf q}', s, s'}
          \hat{c}_{\mathbf{q}' 2 s'}^{\dagger} \hat{c}_{\mathbf{q}' 1 s'}\hat{c}_{\mathbf{q} 2 s}^{\dagger} \hat{c}_{\mathbf{q} 1 s}  \hat{g}_{( \Ea -\Delta_{\bf q} ) }
           \PEa \\
          &\approx
          \sum_{{\bf q},{\bf q}', s, s'}
          \frac{- g^2}{\Delta_{\bf q}^0}
          \hat{c}_{\mathbf{q}' 2 s'}^{\dagger} \hat{c}_{\mathbf{q} 2 s}^{\dagger} \hat{c}_{\mathbf{q} 1 s} \hat{c}_{\mathbf{q}' 1 s'}  \PEa
    \end{split}
\end{equation}
where $\hat{\mathcal{P}}_{2,0}$ is the projector onto the corresponding excited subspace. 
In the first line, we take into account the excitation-creating part of both $\hat{H}_{-1}$ and $\hat{H}_{1}$ [see Eq.~(\ref{D})]. Naturally, the Hermitian conjugate term 
\begin{equation}\label{inter-manifold-band-de-excitation}
    \begin{split}
        \PEa \hat{H}_{1}\hat{g}_{(E_\alpha +\omL)}\hat{H}_{-1} \hat{\mathcal{P}}_{2,0} 
        &\approx
          \sum_{{\bf q},{\bf q}', s, s'}
          \frac{-( g^* )^2}{\Delta_{\bf q}^0}
          \hat{\mathcal{P}}_{00}
          \hat{c}_{\mathbf{q}' 1 s'}^{\dagger} \hat{c}_{\mathbf{q} 1 s}^{\dagger} \hat{c}_{\mathbf{q} 2 s} \hat{c}_{\mathbf{q}' 2 s'}  
    \end{split}
\end{equation}
also exists in $\hat{H}_{1}\hat{g}_{(\Ea +\omL)}\hat{H}_{-1}$. Eqs.~(\ref{inter-manifold-band-excitation}) and (\ref{inter-manifold-band-de-excitation}) are the two leading off-resonant terms in the Floquet Hamiltonian which are independent of the electron-cavity coupling. They create/annihilate two inter-band excitations, respectively, thereby heating the driven material. They are suppressed by the detuning $\sim g ^2 / \Delta_{\bf q}^0$.

By virtually creating 2 band-excitations and annihilating them, the off resonant terms (\ref{inter-manifold-band-excitation}) and (\ref{inter-manifold-band-de-excitation}) cooperatively induce an effective interaction in the low-energy limit, which scales as
\begin{equation} \label{off-resonant-leading-order-induced-int}
    \frac{-2\vert g \vert^4}{(\Delta_{\bf q}^0)^2 (\Delta_{\bf q}^0 + \Delta_{\bf q'}^0 + 2\omL)} \hat{c}_{\mathbf{q}' 1 s'}^{\dagger} \hat{c}_{\mathbf{q} 1 s}^{\dagger} \hat{c}_{\mathbf{q} 1 s} \hat{c}_{\mathbf{q}' 1 s'}
\end{equation}
This induced interaction is strongly suppressed by the factor $\frac{1}{\omL}$, and it cannot transfer momentum between electrons.

\paragraph{(2) Inter-manifold terms from $\hat{H}_{1} \hat{g} \hat{H}_{c} \hat{g}\hat{H}_{-1}$}
We next consider the term $\hat{H}_{1} \hat{g} \hat{H}_{c} \hat{g}\hat{H}_{-1}$ in $\hat{H}^{\text{eff}}_{(\Ea)}$, given by expanding $\hat{G}^{0}$ in Eq.~(\ref{Heff_weak_drive}) to the first order in the cavity-electron coupling $\hat{H}_{c}$. This term is completely off-resonating, i.e. it vanishes in the low-energy limit,
\begin{equation*}
    \begin{split}
        & \PEa \hat{H}_{1} \hat{g}_{(\Ea +\omL)} \hat{H}_{c} \hat{g}_{(\Ea +\omL)}\hat{H}_{-1} \PEa = 0,
    \end{split}
\end{equation*}
because $\hat{H}_{c}$ creates a cavity photon that cannot be annihilated by any other operators
\footnote{More generally, all expansion terms containing an odd number of $\hat{H}_{c}$ vanish in the low-energy limit.}
in $\hat{H}_{1} \hat{g} \hat{H}_{c} \hat{g}\hat{H}_{-1}$.

Similar to Section~\ref{off-resonant-leading-order}, we introduce the projector $\hat{\mathcal{P}}_{m,n}$ onto the excited subspace with $m$ upper-band electrons and $n$ cavity photons. Then this term reads
\begin{equation}
\begin{split}
& \hat{H}_{1} \hat{g}_{(\Ea +\omL)} \hat{H}_{c} \hat{g}_{(\Ea +\omL)}\hat{H}_{-1} \PEa \\
&= \hat{\mathcal{P}}_{1,1} \hat{H}_{1} \hat{g}_{(\Ea +\omL)} \hat{\mathcal{P}}_{0,1} \hat{H}_{c} \hat{g}_{(\Ea +\omL)} \hat{\mathcal{P}}_{1,0} \hat{H}_{-1} \PEa \\
& ~ + \hat{\mathcal{P}}_{1,1} \hat{H}_{1} \hat{g}_{(\Ea +\omL)} \hat{\mathcal{P}}_{2,1} \hat{H}_{c} \hat{g}_{(\Ea +\omL)} \hat{\mathcal{P}}_{1,0} \hat{H}_{-1} \PEa \\
& ~ + \hat{\mathcal{P}}_{3,1} \hat{H}_{1} \hat{g}_{(\Ea +\omL)} \hat{\mathcal{P}}_{2,1} \hat{H}_{c} \hat{g}_{(\Ea +\omL)} \hat{\mathcal{P}}_{1,0} \hat{H}_{-1} \PEa \\
&\approx \hat{\mathcal{P}}_{1,1} \hat{H}_{1} \hat{g}_{(\Ea +\omL)} \hat{\mathcal{P}}_{0,1} \hat{H}_{c} \hat{g}_{(\Ea +\omL)} \hat{\mathcal{P}}_{1,0} \hat{H}_{-1} \PEa \\
& \approx -  \hat{\mathcal{P}}_{1,1} \hat{H}_{1} \hat{g}_{(\Ea +\omL)} \hat{a}^{\dag}  \sum_{{\bf k},{\bf q'}, s}  \frac{g g_{c,{\bf q'}}^{*} }{\sqrt{N} \Delta_{\bf k}^0 }  \hat{c}_{{\bf k}-{\bf q'} 1 s}^{\dag}  \hat{c}_{{\bf k} 1 s} \PEa \\
& \approx  \hat{a}^{\dag} \sum_{{\bf{k'}}, s'}  \hat{c}_{{\bf{k'}} 2 s'}^{\dag} \hat{c}_{{\bf{k'}} 1 s'}  \sum_{{\bf k},{\bf q'}, s}  \frac{(g)^2 g_{c,{\bf q'}}^{*} }{\sqrt{N} \Delta_{\bf k}^0 \Delta_c }  \hat{c}_{{\bf k}-{\bf q'} 1 s}^{\dag}  \hat{c}_{{\bf k} 1 s} \PEa \\
\end{split}
\end{equation}
Here in the first line, we use the projectors to distinguish different processes in $\hat{H}_{1} \hat{g} \hat{H}_{c} \hat{g}\hat{H}_{-1} \PEa$. All three processes disappear if we make RWA. Only the first process is kept into the second line, as the other two processes are further suppressed by a factor $\sim 1/\omL$. The only kept term projects the system from the low-energy subspace $\PEa$ to $\hat{\mathcal{P}}_{1,1}$. Note from the final line that, according to the relation (\ref{cavity-detuning-requirement}), this off-resonating term $\hat{H}_{1} \hat{g} \hat{H}_{c} \hat{g}\hat{H}_{-1} \PEa$ is much weaker (by a factor of $g_c/\Delta_c$) than the previous off-resonating term $\hat{\mathcal{P}}_{2,0} \hat{H}_{1}\hat{g}\hat{H}_{-1} \PEa$ in (\ref{inter-manifold-band-excitation}).
%This term vanishes, even without projector, in the rotating wave approximation of $\hat{H}_{c}$.

\subsection{The Floquet Low-energy Hamiltonian}
Collecting these results, we find the leading-order Floquet low-energy Hamiltonian in the non-interacting case, % that is correct to the lowest order $\vert g \vert^2$ of driving strength,
\begin{equation}\label{final-result-U=0}
    \begin{split}
       &\hat{H}^{\text{eff}}_{\text{non-int}} = \PEa \hat{H}^{\text{eff}}_{(\Ea)} \PEa \\
        &\approx
        \sum_{{\bf k},  s} \big( \ep_{{\bf k},1} - \frac{\vert g \vert^2}{\Delta_{\bf k}^0 } -\mu \big) \hat{c}_{{\bf k} 1 s}^{\dag} \hat{c}_{{\bf k} 1 s}   ~ + ~ \frac{1}{N}  \sum_{\substack{{\bf k},{\bf q}, s \\ {\bf k'},{\bf q'}, s'}} \frac{\vert g \vert^2 g_{c,{\bf q'}} g_{c,{\bf q}}^{*} }{ -\Delta_c  \Delta_{\bf k'}^0   \Delta_{\bf k}^0  } 
        \hat{c}_{{\bf k'} 1 s'} ^{\dag}  \hat{c}_{{\bf k'}-{\bf q'} 1 s'}
        \hat{c}_{{\bf k}-{\bf q} 1 s}^{\dag}  \hat{c}_{{\bf k} 1 s} 
    \end{split}
\end{equation}
where we find that the optical Stark shift $\frac{\vert g \vert^2}{\Delta_{\bf q}^0 }$ comes from the term $\hat{H}_{1}\hat{G}^{b}_{(\Ea +\omL)}\hat{H}_{-1}$, and the cavity-mediated interaction (final line) comes from the term $\hat{H}_{1}\hat{G}^{b}\hat{H}_{c}\hat{G}^{b}\hat{H}_{c}\hat{G}^{b}\hat{H}_{-1}$. All other driving induced terms are strongly suppressed by a factor $\frac{1}{\omL}$ and thus ignored under the detuning condition (\ref{band-detuning-requirement}) and (\ref{cavity-detuning-requirement}).

\section{The interacting model: deriving the screened Floquet Hamiltonian}\label{appendix:correlated-model-k-GRPA}

Next we study how electronic interactions (i.e. a finite $\hat{U}$ in $\hat{H}_{b}$) influence the effective Floquet Hamiltonian~(\ref{final-result-U=0}). To obtain the new effective Hamiltonian $\hat{H}^{\text{eff}}_{\text{int}} = \PEa \hat{H}^{\text{eff}}_{(\Ea)} \PEa$, the Dyson expansion (\ref{Dyson}) is again inserted into (\ref{Heff_weak_drive}), resulting in
\begin{equation} \label{appendix:eq.H_eff_int}
\begin{split}
    &\hat{H}^{\text{eff}}_{\text{int}} = \PEa \hat{H}^{\text{eff}}_{(\Ea)} \PEa \\
    &\approx \PEa \hat{H}_{0} \PEa + \PEa \hat{H}_{1}\hat{G}^{b}_{(\Ea +\omL)}\hat{H}_{-1} \PEa 
     ~~ + \PEa \hat{H}_{1}\hat{G}^{b}_{(\Ea +\omL)} \hat{H}_{c} \hat{G}^{b}_{(\Ea +\omL)} \hat{H}_{c} \hat{G}^{b}_{(\Ea +\omL)} \hat{H}_{-1} \PEa 
     ~~ + ...
\end{split}
\end{equation}
In contrast to the non-interacting case in Section \ref{sec:zero-correlation-model}, here we have to deal with the many-body Green operator $\hat{G}^{b}$, where we encounter additional expansion terms. These additional terms contain non-zero order of $\hat{U}$.

Similar to how we proceeded before, in order to move $\hat{G}^{b}$ adjacent to $\PEa$, we need to move the driving operator $\hat{H}_{-1}$ from the right to the left. In this process, we will encounter the commutator $[\hat{U},\hat{H}_{-1}]$. In this commutator, we make the following approximation which moves $\hat{b}^{\dag}$ leftward
\begin{equation}\label{U-b-commutator-reduced}
    \begin{split}
        &\hat{U} \hat{b}_{{\bf q},s}^{\dag} \approx \sum\limits_{\bf k} \hat{b}_{{\bf k},s}^{\dag} \hat{f}_{{\bf k},{\bf q}}^{s}
    \end{split}
\end{equation}
where $\hat{b}_{{\bf k},s}^{\dag} \equiv \hat{c}_{{\bf k},2,s}^{\dag} \hat{c}_{{\bf k},1,s}$ denotes the inter-band excitation operator at momentum $\bf k$ and spin $s$. In (\ref{U-b-commutator-reduced}) we define
\begin{equation}\label{fkqs}
    \begin{split}
        &\hat{f}_{{\bf k},{\bf q}}^{s} \equiv  
        \delta_{{\bf k},{\bf q}}  
        \big( \hat{U} 
        - U_{11} \hat{\nu}_{\Bar{s}}
        + U_{12}\sum\limits_{s'} \hat{\nu}_{s'}
        \big) - \frac{U_{12}}{N} \hat{n}_{{\bf q},s}
    \end{split}
\end{equation}
where $\hat{n}_{{\bf q},s}\equiv \hat{c}_{{\bf q} 1 s}^{\dag} \hat{c}_{{\bf q} 1 s}$ is the electron number operator in the lower band with momentum $\bf q$ and spin $s$. Here $\Bar{s}$ represents the opposite spin of $s$, and we define the spin-resolved filling operator in the lower band
\begin{equation}\label{filling-operator}
    \hat{\nu}_{s} \equiv \frac{1}{N}\sum\limits_{{\bf k'}}\hat{n}_{{\bf k'},s}.
\end{equation}
The approximation~(\ref{U-b-commutator-reduced}) is scrutinized in Appendix \ref{appendix:U-b-commutator}. It is conceptually very similar to the linearization process made by Anderson to perform RPA in his seminal BCS paper~\cite{PhysRev.112.1900}. 

Meanwhile, (\ref{U-b-commutator-reduced}) omits the on-site repulsion $\hat{U}_{22}$ between two upper-band electrons on the same site with opposite spin, which is accurate because this same-site double-excitation cannot be created by the single $\hat{H}_{-1}$ in the $\vert g \vert^2$ order 
\footnote{However, for sufficiently strong driving $g  \sim \Delta_{\bf k}^0$, we need to consider the $\vert g \vert^4$ order effects in (\ref{Heff}), and then $\hat{U}_{22}$ can no longer be ignored in (\ref{U-b-commutator-reduced}), which will show bi-exciton effects. 
Meanwhile, in the ultra-strong electron-cavity coupling regime $g_c\sim\omc$, the $\hat{U}_{22}$ will lead to bi-exciton effects even in the $\vert g \vert^2$ order, because the non-RWA part of $\hat{H}_{c}$ will create the second upper-band electron in the term $\hat{H}_{1}\hat{G}^{b}\hat{H}_{c}\hat{G}^{b}\hat{H}_{c}\hat{G}^{b}\hat{H}_{-1}$ of the lowest-order Floquet Hamiltonian (\ref{Heff_weak_drive}). These bi-exciton effects will be analyzed in future works.} 
Floquet Hamiltonian (\ref{Heff_weak_drive}). 

Below we evaluate the various terms in Eq.~(\ref{appendix:eq.H_eff_int}) in detail, following the same analysis as in the previous noninteracting model, and obtain a screened Floquet Hamiltonian.  
During this derivation, since $\hat{U}\neq0$, $\hat{g}\PEa$ can no longer be reduced by Eq.~(\ref{G1-reduced-to-number}) because $\hat{G}^{b}\neq\hat{g}$. Therefore, we need to apply a re-summation over infinite orders of $\hat{U}$, followed by a mean-field decoupling, so as to turn $\hat{g}\PEa$ back to the form of $\hat{G}^{b}\PEa$, which can then be reduced by (\ref{G1-reduced-to-number}). 
This mean-field decoupling contains two steps: Firstly, in (\ref{reduce-gHs-filling}) the above-mentioned filling operator will be replaced by its expectation value, 
\begin{equation}\label{mean-field-global}
    \hat{\nu}_s \rightarrow \nu_s = \langle \hat{\nu}_s \rangle,
\end{equation}
and secondly, inside the momentum-summation in (\ref{reduce-gHs-single-occupation}), the microscopic electronic occupation will be replaced by its expectation value,
\begin{equation}\label{mean-field-local}
    \hat{n}_{{\bf q},s} \rightarrow  \langle  \hat{n}_{{\bf q},s}  \rangle.
\end{equation}

Finally, we will compare our approach with the multi-band generalized random phase approximation (GRPA), a Feynman diagram summation procedure previously used to study excitons in semiconductors\cite{PhysRevB.40.3802} and itinerant antiferromagnets\cite{PhysRevB.80.174401}. It can be used to calculate the RWA contributions to the effective Hamiltonian in the adiabatic limit. 
%The re-summation based on the approximation (\ref{U-b-commutator-reduced}), is structurally similar to the multi-band generalized random phase approximation (GRPA), a Feynman diagram summation procedure previously used to study semiconductor excitons\cite{PhysRevB.40.3802}. 
We compare our Floquet method with this GRPA Feynman diagram method, and find that these two methods give consistent results for the RWA terms. Additionally, we calculate the screening of the non-RWA terms, which can not be accomplished with the equilibrium GRPA approach.

\subsection{Screened optical Stark shift}\label{appendixsec:Screened_ACS}
We first consider the term $\PEa \hat{H}_{1} \hat{G}^{b} \hat{H}_{-1} \PEa$, which appears when we expand (\ref{Heff_weak_drive}) by (\ref{Dyson}) and then truncate to the 0-th order of $\hat{H}_{c}$. It contains infinite orders of the interaction $\hat{U}$, and evaluates to
    \begin{equation}\label{k-space-GRPA-Stark-begin}
    \begin{split}
        &\PEa \hat{H}_{1}\hat{G}^{b}_{(E_\alpha +\omL)}\hat{H}_{-1} \PEa \\
          &=\vert g \vert^2
          \PEa 
          \sum_{{\bf q},{\bf q}', s, s'}
          \hat{b}_{{\bf q'},s'}
          \sum\limits_{n=0}^{\infty} (\hat{g}_{(\Ea + \omL)} \hat{U})^n \hat{g}_{(\Ea + \omL)}
          \hat{b}_{{\bf q},s}^{\dag}
          \PEa \\
          &=\vert g \vert^2
          \PEa 
          \sum_{{\bf q},{\bf q}', s, s'}
          \hat{b}_{{\bf q'},s'}
          \sum\limits_{n=0}^{\infty} (\hat{g}_{(\Ea + \omL)} \hat{U})^n \hat{b}_{{\bf q},s}^{\dag}
          \hat{g}_{(\Ea - \Delta_{\bf q})}
          \PEa \\
          &\approx \vert g \vert^2
          \PEa 
          \sum_{{\bf q},{\bf q}', s, s'}
          \hat{b}_{{\bf q'},s'} 
          \sum\limits_{n=0}^{\infty}   ~~
          \sum\limits_{ {\bf k}_1 , {\bf k}_2 , ... , {\bf k}_n } \hat{b}_{{\bf k}_n,s}^{\dag} \\
          & ~~~~ \times 
          ( \hat{g}_{(\Ea - \Delta_{{\bf k}_n}) }   \hat{f}_{{\bf k}_n,{\bf k}_{n-1}}^{s}   )
          ( \hat{g}_{(\Ea - \Delta_{{\bf k}_{n-1}})}   \hat{f}_{{\bf k}_{n-1},{\bf k}_{n-2}}^{s}   ) 
          ~ ... ~
          ( \hat{g}_{(\Ea - \Delta_{{\bf k}_2})}   \hat{f}_{{\bf k}_2,{\bf k}_1}^{s}   )
          ( \hat{g}_{(\Ea - \Delta_{{\bf k}_1})}   \hat{f}_{{\bf k}_1,{\bf q}}^{s}   )  
          ~
          \hat{g}_{(\Ea - \Delta_{\bf q})}
          \PEa \\
          &= \vert g \vert^2
          \PEa 
          \sum_{{\bf q},{\bf q}', s, s'}
          \hat{b}_{{\bf q'},s'} ~
          \sum\limits_{n=0}^{\infty}   ~~
          \sum\limits_{ {\bf k}_n } ~~ \hat{b}_{{\bf k}_n,s}^{\dag} ~
          [({\rm g}*{\rm f}^{s})^{n} * {\rm g}]_{ {\bf k}_n , {\bf q} }
          \PEa\\
          &= \vert g \vert^2
          \sum_{{\bf q},{\bf q}', s,s'}
          \sum\limits_{n=0}^{\infty} \sum\limits_{ {\bf k}_n }
          \delta_{{\bf q'},{\bf k}_n} \delta_{s,s'} \PEa \hat{n}_{{\bf q'},s} ~
          [({\rm g} * {\rm f}^{s})^{n} * {\rm g}]_{ {\bf q'} , {\bf q} }
          \PEa\\
          &= \vert g \vert^2 \PEa
          \sum_{{\bf q}', s} \hat{n}_{{\bf q'},s} ~ 
          \bigg( \sum_{\bf q}
          [  ({\rm g}^{-1} - {\rm f}^{s})^{-1} ]_{ {\bf q}' , {\bf q} }
          \bigg)   \PEa  
    \end{split}
\end{equation}
In the first line of Eq.~(\ref{k-space-GRPA-Stark-begin}), we insert the definition~(\ref{D}) of the driving operator $\hat{H}_{-1}$, and then discard the de-excitation part of $\hat{H}_{-1}$ because $\hat{b}_{{\bf q},s}\PEa=0$, which follows from the low-energy limit~(\ref{low-energy-approx}). 
We also expand $\hat{G}^{b}$ into a Born series containing infinite orders of $\hat{U}$, according to the Dyson expansion (\ref{Dyson}). In the second line, we use (\ref{g-commutator}) to move the far-right $\hat{b}^{\dag}$ leftward.
In the third line, we keep moving this $\hat{b}^{\dag}$ further leftward until it is adjacent to the $\hat{b}$ at the far-left. We use Eq.~(\ref{g-commutator}) whenever $\hat{b}^{\dag}$ crosses $\hat{g}$, and the approximation~(\ref{U-b-commutator-reduced})  whenever $\hat{b}^{\dag}$ crosses $\hat{U}$. %After $\hat{b}^{\dag}$ completely crosses a full term (this term is the $n$-th order expansion of $\hat{G}^{b}$ over $\hat{U}$), 
For the $n$-th order expansion of $\hat{G}^{b}$, 
the repeated application of (\ref{U-b-commutator-reduced}) creates summations over $n$ internal momenta (${\bf k_1},{\bf k_2}, ... ,{\bf k_n}$), and the momentum carried by $\hat{b}^{\dag}$ is changed from the initial $\bf q$ to the final $\bf k_n$.
In the fourth line, we introduce the matrices ${\rm f}^s$ and ${\rm g}$, with operator-valued matrix elements
\footnote{Here, we define the multiplication between two matrices (with operator-valued matrix elements) $\rm A$ and $\rm B$ as
$$ [{\rm A} * {\rm B}]_{{\bf k'},{\bf k}} \equiv \sum\nolimits_{\bf k''} 
    ~ [{\rm A} ]_{{\bf k'},{\bf k''}}  ~ [{\rm B}]_{{\bf k''},{\bf k}} $$
where we sum over $\bf k''$ in the first Brillioun zone.}
defined as
\begin{equation}\label{operator-valued-matrix-element}
    \begin{split}
        [{\rm g}]_{ {\bf k} , {\bf q} } &\equiv \delta_{ {\bf k} , {\bf q} } \hat{g}_{(\Ea - \Delta_{\bf q}^0)}, \\
        [{\rm f}^{s}]_{ {\bf k} , {\bf q} } &\equiv \hat{f}_{{\bf k},{\bf q}}^{s},
    \end{split}
\end{equation}
%The matrix $\rm g$ is diagonal in its momentum indices.
where $\hat{f}_{{\bf k},{\bf q}}^{s}$ is defined in (\ref{fkqs}). Hence, the inverse matrix of $\rm g$ is simply
\begin{equation*}
    [ {\rm g}^{-1} ]_{{\bf k},{\bf q}} = \delta_{{\bf k},{\bf q}} \hat{g}_{(\Ea - \Delta_{\bf q}^0)}^{-1},
\end{equation*}
we can directly check that this ${\rm g}^{-1}$ satisfies the definition of inverse, i.e., $[{\rm g} * {\rm g}^{-1}]_{{\bf k'},{\bf k}}=[{\rm g}^{-1} * {\rm g}]_{{\bf k'},{\bf k}}=\delta_{{\bf k'},{\bf k}}$. 
%Moreover, we require this multiplication to satisfy associativity, ${\rm A}*({\rm B}+{\rm C})={\rm A}*{\rm B}+{\rm A}*{\rm C}$, which is trivial because we require the addition of our matrix to satisfy
%\begin{equation*}
%    [({\rm A}+{\rm B})]_{{\bf k'},{\bf k}} = 
%    [{\rm A}]_{{\bf k'},{\bf k}} + [{\rm B}]_{{\bf k'},{\bf k}}.
%\end{equation*}
In the fifth line of (\ref{k-space-GRPA-Stark-begin}), we use $\PEa \hat{b}_{{\bf q'},s'} \hat{b}_{{\bf k}_n,s}^{\dag} = \delta_{{\bf q'},{\bf k}_n} \delta_{s,s'} \PEa \hat{n}_{{\bf q'},s}$ which follows directly from the low-energy limit (\ref{low-energy-approx}). Then $\delta_{{\bf q'},{\bf k}_n}$ allows us to switch the first index of the matrix from ${\bf k}_n$ to ${\bf q'}$.
In the sixth (final) line of (\ref{k-space-GRPA-Stark-begin}), we use the formula
\begin{equation*}
    \sum\limits_{n=0}^{\infty}  ({\rm g} * {\rm f}^{s})^{n} * {\rm g} =    ({\rm g}^{-1} - {\rm f}^{s})^{-1} 
\end{equation*}
which is a matrix Taylor expansion (i.e. a matrix Born series). It is a result of the identity
%as a result of the following equation
\begin{equation*}
\begin{split}
    & \big( \sum\limits_{n=0}^{\infty}  ({\rm g} * {\rm f}^{s})^{n} * {\rm g} \big)
    * ({\rm g}^{-1} - {\rm f}^{s}) 
     = \sum\limits_{n=0}^{\infty}  ({\rm g} * {\rm f}^{s})^{n} -
    \sum\limits_{n=1}^{\infty}  ({\rm g} * {\rm f}^{s})^{n} = 1.
\end{split}
\end{equation*}
%where we used the associativity of the multiplication mentioned above.

To further simplify the structure in the final line of (\ref{k-space-GRPA-Stark-begin}), we recombine the matrix ${\rm f}^s$ and ${\rm g}$ by two new matrices ${\rm f}_{F}^s$ and ${\rm g}_{H}^{s}$, respectively defined as
\begin{equation}\label{split-gdelta-fF}
\begin{split}
    [{\rm f}_{F}^s]_{ {\bf k} , {\bf q} } 
    &= - \frac{U_{12}}{N} \hat{n}_{{\bf q},s} \equiv [{\rm f}_{F}^s]_{ * , {\bf q} } ~~ \text{independent of} ~ {\bf k} \\
    [{\rm g}_{H}^{s}]_{ {\bf k} , {\bf q} } 
    &=  \delta_{{\bf k},{\bf q}} 
    \bigg(   \big(\hat{G}^{b}_{(\Ea - \Delta_{\bf q}^0)} \big)^{-1}
    + U_{11}\hat{\nu}_{\Bar{s}}
    - U_{12}\sum\limits_{s'}\hat{\nu}_{s'}
    \bigg)^{-1}
\end{split}
\end{equation}
Here the character $F$ and $H$ stands for ``Fock" and ``Hartree". The reason for these names will become apparent in Appendix \ref{appendix:Naming-of-Hartree/Fock}. Note that the new matrix element $[{\rm f}_{F}^s]_{ {\bf k} , {\bf q} }$ depends only on its second momentum index $\bf q$, thus it can be represented by a simpler symbol, denoted by $[{\rm f}_{F}^s]_{ * , {\bf q} }$ in (\ref{split-gdelta-fF}).
%this behaviour indicates the selection of terms with random phase for summation.

According to (\ref{split-gdelta-fF}), the expression $\big( {\rm g}^{-1} - {\rm f}^{s} \big)$ in (\ref{k-space-GRPA-Stark-begin}) can be recombined as $\big( ({\rm g}_{H}^{s})^{-1} - {\rm f}_{F}^s \big)$, because
\begin{equation*}
    \begin{split}
        [{\rm g}^{-1}]_{ {\bf k} , {\bf q} } - [({\rm g}_{H}^{s})^{-1}]_{ {\bf k} , {\bf q} } &= 
        \delta_{ {\bf k} , {\bf q} } \big( \hat{U} - U_{11}\hat{\nu}_{\Bar{s}} + U_{12}\sum\limits_{s'}\hat{\nu}_{s'} \big) \\
        &= [{\rm f}^{s}]_{ {\bf k} , {\bf q} }  -  [{\rm f}_{F}^s]_{ {\bf k} , {\bf q} } 
    \end{split}
\end{equation*}
where in the first line we use $(\hat{G}^{b})^{-1} = \hat{g}^{-1} - \hat{U}$ stemming from the definition (\ref{definition-G1-g}), and the second line directly follows from the definition (\ref{fkqs}). 

This recombination (\ref{split-gdelta-fF}) allows the following evaluation of the parenthesis in the last line of (\ref{k-space-GRPA-Stark-begin}),
\begin{equation}\label{k-space-GRPA-collection}
    \begin{split}
        &\sum_{\bf q} [  ({\rm g}^{-1} - {\rm f}^{s})^{-1} ]_{ {\bf q}' , {\bf q} } 
        = \sum_{\bf q} [ \bigg( ({\rm g}_{H}^{s})^{-1} - {\rm f}_{F}^s \bigg)^{-1} ]_{ {\bf q}' , {\bf q} } \\
        &= \sum\limits_{\bf q} [{\rm g}_{H}^{s}]_{ {\bf q}' , {\bf q} } + \sum\limits_{\bf q} [{\rm g}_{H}^{s} * {\rm f}_{F}^s * {\rm g}_{H}^{s} ]_{ {\bf q}' , {\bf q} }
        + \sum\limits_{\bf q} [{\rm g}_{H}^{s} * {\rm f}_{F}^s * {\rm g}_{H}^{s}* {\rm f}_{F}^s * {\rm g}_{H}^{s}]_{ {\bf q}' , {\bf q} } + ~ ... \\
        & = ~~~   [{\rm g}_{H}^{s}]_{ {\bf q'} , {\bf q'} }  
        ~ + ~ [{\rm g}_{H}^{s}]_{ {\bf q'} , {\bf q'} } \sum\limits_{\bf q} [{\rm f}_{F}^s]_{ {\bf q'} , {\bf q} } [{\rm g}_{H}^{s}]_{ {\bf q} , {\bf q} } \\
        & ~~~   
        + [{\rm g}_{H}^{s}]_{ {\bf q'} , {\bf q'} }
        \sum\limits_{{\bf q}_1}
        [{\rm f}_{F}^s]_{ {\bf q'} , {\bf q}_1 }  [{\rm g}_{H}^{s}]_{ {\bf q}_1 , {\bf q}_1 }
        \sum\limits_{\bf q}
        [{\rm f}_{F}^s]_{ {\bf q}_1 , {\bf q} } [{\rm g}_{H}^{s}]_{ {\bf q} , {\bf q} } \\
        & ~~~   
        + [{\rm g}_{H}^{s}]_{ {\bf q'} , {\bf q'} }
        \sum\limits_{{\bf q}_1 }
        [{\rm f}_{F}^s]_{ {\bf q'} , {\bf q}_1 }  [{\rm g}_{H}^{s}]_{ {\bf q}_1 , {\bf q}_1 }
        \sum\limits_{ {\bf q}_2 }
        [{\rm f}_{F}^s]_{ {\bf q}_1 , {\bf q}_2 } [{\rm g}_{H}^{s}]_{ {\bf q}_2 , {\bf q}_2 }
        \sum\limits_{\bf q}
        [{\rm f}_{F}^s]_{ {\bf q}_2 , {\bf q} } [{\rm g}_{H}^{s}]_{ {\bf q} , {\bf q} }
        ~~~ + ...\\
        &=  [{\rm g}_{H}^{s}]_{ {\bf q'} , {\bf q'} } 
        ~~ + ~~
        [{\rm g}_{H}^{s}]_{ {\bf q'} , {\bf q'} } \bigg( \sum\limits_{\bf q} [{\rm f}_{F}^s]_{ * , {\bf q} } [{\rm g}_{H}^{s}]_{ {\bf q} , {\bf q} } \bigg) \\
        &~~~ +
        [{\rm g}_{H}^{s}]_{ {\bf q'} , {\bf q'} }
        \bigg( \sum\limits_{{\bf q}_1}
        [{\rm f}_{F}^s]_{ * , {\bf q}_1 }  [{\rm g}_{H}^{s}]_{ {\bf q}_1 , {\bf q}_1 } \bigg)
        \bigg( \sum\limits_{\bf q}
        [{\rm f}_{F}^s]_{ * , {\bf q} } [{\rm g}_{H}^{s}]_{ {\bf q} , {\bf q} } \bigg) \\
        &~~~ + 
        [{\rm g}_{H}^{s}]_{ {\bf q'} , {\bf q'} }
        \bigg( \sum\limits_{{\bf q}_1}
        [{\rm f}_{F}^s]_{ * , {\bf q}_1 }  [{\rm g}_{H}^{s}]_{ {\bf q}_1 , {\bf q}_1 }
        \bigg)
        \bigg( \sum\limits_{ {\bf q}_2 }
        [{\rm f}_{F}^s]_{ * , {\bf q}_2 } [{\rm g}_{H}^{s}]_{ {\bf q}_2 , {\bf q}_2 }\bigg) 
        \bigg( \sum\limits_{\bf q} [{\rm f}_{F}^s]_{ * , {\bf q} } [{\rm g}_{H}^{s}]_{ {\bf q} , {\bf q} } \bigg) 
         + ...\\
        & =[{\rm g}_{H}^{s}]_{ {\bf q'} , {\bf q'} } \sum\limits_{n=0}^{\infty} \bigg( \sum\limits_{ {\bf q}'' }  [{\rm f}_{F}^s]_{ * , {\bf q}'' }  [{\rm g}_{H}^{s}]_{ {\bf q}'' , {\bf q}'' }  \bigg)^n \\
        & = [{\rm g}_{H}^{s}]_{ {\bf q'} , {\bf q'} }  \bigg( 1 - \sum\limits_{ {\bf q}'' }  [{\rm f}_{F}^s]_{ * , {\bf q}'' }  [{\rm g}_{H}^{s}]_{ {\bf q}'' , {\bf q}'' }  \bigg)^{-1}. \\
        %&= \bigg( [{\rm g}_{H}^{s}]^{-1}_{ {\bf q'} , {\bf q'} }  - \sum\limits_{ {\bf q}'' }  [{\rm f}_{F}^s]_{ * , {\bf q}'' }  [{\rm g}_{H}^{s}]_{ {\bf q}'' , {\bf q}'' } [{\rm g}_{H}^{s}]^{-1}_{ {\bf q'} , {\bf q'} }  \bigg)^{-1} \\
    \end{split}
\end{equation}
In the first line we replace ${\rm g}^{-1} - {\rm f}^{s} $ by $ ({\rm g}_{H}^{s})^{-1} - {\rm f}_{F}^s $, in the second line we use the matrix Taylor expansion. In the third line we expand the matrix multiplication, and then we use the diagonal property of matrix ${\rm g}_{H}^{s}$ to reduce the number of momentum indices to be summed. In the fourth line we use $[{\rm f}_{F}^s]_{ {\bf k} , {\bf q} } = [{\rm f}_{F}^s]_{ * , {\bf q} }$, then the summation over momentum indices decouple from one another, and can be evaluated separately (as we did in this line). All these parenthesised terms are equivalent, and thus in the fifth line, we collect them order by order. In the sixth (final) line, we use the Taylor expansion formula to replace this infinite summation by the inverse of a single operator.

Inserting (\ref{k-space-GRPA-collection}) back to (\ref{k-space-GRPA-Stark-begin}), we obtain
\begin{equation}\label{k-space-GRPA-Stark-middle}
    \begin{split}
        &\PEa \hat{H}_{1}\hat{G}^{b}_{(E_\alpha +\omL)}\hat{H}_{-1} \PEa 
        = \vert g \vert^2 \PEa
          \sum_{{\bf q}', s} \hat{n}_{{\bf q'},s} ~
          [{\rm g}_{H}^{s}]_{ {\bf q'} , {\bf q'} } 
          \bigg( 1 - \sum\limits_{ {\bf q}'' }  [{\rm f}_{F}^s]_{ * , {\bf q}'' }  [{\rm g}_{H}^{s}]_{ {\bf q}'' , {\bf q}'' }  \bigg)^{-1}
            \PEa.  
    \end{split}
\end{equation}
See Eq.~(\ref{definition-G1-g}) for the definition of $\hat{G}^{b}$ and Eq.~(\ref{split-gdelta-fF}) for the definition of ${\rm g}_{H}^{s}$ and ${\rm f}_{F}^s$. With Eq.~(\ref{k-space-GRPA-Stark-middle}), we have obtained a resummation of the interaction terms in the Stark Hamiltonian. 

Up to now we haven't made any approximation apart from Eq.~(\ref{U-b-commutator-reduced}). To further simplify Eq.~(\ref{k-space-GRPA-Stark-middle}), we next 
%eliminate the upper electronic band 
eliminate the operator $\hat{G}^{b}$ using Eq.~(\ref{G1-reduced-to-number}) and introduce the mean-field decoupling~(\ref{mean-field-global}). We find
%we introduce the mean-field decoupling in the final line of the following evaluation
\begin{equation}\label{reduce-gHs-filling}
\begin{split}
    [{\rm g}_{H}^{s}]_{ {\bf q} , {\bf q} }  \PEa  
    &=  \bigg(   \big(\hat{G}^{b}_{(\Ea - \Delta_{\bf q}^0)} \big)^{-1}
    + U_{11}\hat{\nu}_{\Bar{s}}
    - U_{12}\sum\limits_{s'}\hat{\nu}_{s'}
    \bigg)^{-1} \PEa \\
    &= \sum_{n=0}^{\infty} \big(-U_{11}\hat{\nu}_{\Bar{s}}+U_{12}\sum\limits_{s'}\hat{\nu}_{s'} \big)^{n} \big( \hat{G}^{b}_{(\Ea - \Delta_{\bf q}^0)} \big)^{n+1}  \PEa  \\
    &\approx \sum_{n=0}^{\infty} \big(-U_{11}\hat{\nu}_{\Bar{s}}+U_{12}\sum\limits_{s'}\hat{\nu}_{s'} \big)^{n} (\frac{-1}{\Delta_{\bf q}^0})^{n+1}  \PEa  \\
    &=\big(   -\Delta_{\bf q}^0
    + U_{11}\hat{\nu}_{\Bar{s}}
    - U_{12}\sum\limits_{s'}\hat{\nu}_{s'}
    \big)^{-1} \PEa \\
    &\approx \big(   -\Delta_{\bf q}^0
    + U_{11}\nu_{\Bar{s}}
    - U_{12}\sum\limits_{s'}\nu_{s'}
    \big)^{-1} \PEa \\
\end{split}
\end{equation}
where in the first line we insert into the definition of ${\rm g}_{H}^{s}$ in (\ref{split-gdelta-fF}). In the second line we apply a Dyson expansion of the operator, in this expansion, we can put all $\hat{G}^{b}$ to the far-right because $\hat{G}^{b}$ commutes with the electronic occupation in the lower band, $[\hat{G}^{b},\hat{\nu}_{s}]=0$. In the third line we use (\ref{G1-reduced-to-number}) to reduce $\hat{G}^{b}$ into the denominator, which eliminates the degree of freedom of the upper band and the cavity. In the fourth line we turn the infinite Dyson expansion back to the inverse of a single operator. In the last line, we replace each filling operator $\hat{\nu}_s$ by its expectation value, which is the spin-resolved filling factor $\nu_s$. This approximation (\ref{mean-field-global}) treats filling factors as mean fields, which reduces $[{\rm g}_{H}^{s}]_{ {\bf q} , {\bf q} }$ to a screened denominator when it lies adjacent to $\PEa$.

Based on the approximation in Eq,~(\ref{reduce-gHs-filling}), we further simplify
\begin{equation}\label{reduce-gHs-single-occupation}
    \begin{split}
        \sum\limits_{ {\bf q} }  [{\rm f}_{F}^s]_{ * , {\bf q} }  [{\rm g}_{H}^{s}]_{ {\bf q} , {\bf q} } \PEa  
        &\approx - \frac{U_{12}}{N} \sum_{ {\bf q} } \hat{n}_{{\bf q},s}  \frac{ 1 } 
        {   -\Delta_{\bf q}^0 + U_{11}\nu_{\Bar{s}} - U_{12}\sum\limits_{s'}\nu_{s'}
        } \PEa  \\
        &\approx - \frac{U_{12}}{N} \sum_{ {\bf q} } \langle \hat{n}_{{\bf q},s} \rangle \frac{ 1 } {   -\Delta_{\bf q}^0 + U_{11}\nu_{\Bar{s}} - U_{12}\sum\limits_{s'}\nu_{s'}
        } \PEa  \\
    \end{split}
\end{equation}
where in the first line we insert the approximation (\ref{reduce-gHs-filling}). In the second line we further replace the number operator of the lower-band electron $\hat{n}_{{\bf q},s}$ by its expectation value. This is the second step of the Hartree-type mean-field decoupling, as previously mentioned in (\ref{mean-field-local}). Based on the approximation (\ref{reduce-gHs-single-occupation}) and (\ref{reduce-gHs-filling}), the expansion term~(\ref{k-space-GRPA-Stark-middle}) in the effective Hamiltonian is finally reduced to
\begin{equation}\label{k-space-GRPA-Stark-final}
\begin{split}
        \PEa \hat{H}_{1}\hat{G}^{b}_{(E_\alpha +\omL)}\hat{H}_{-1} \PEa 
        &\approx \vert g \vert^2 \PEa
        \frac{
        \sum_{{\bf q}', s} \hat{n}_{{\bf q'},s} ~
          \big(   -\Delta_{\bf q'}^0
    + U_{11}\nu_{\Bar{s}}
    - U_{12}\sum\limits_{s'}\nu_{s'}
    \big)^{-1}
        }{ 1 + \frac{U_{12}}{N} \sum_{ {\bf q''} } \langle \hat{n}_{{\bf q''},s} \rangle (   -\Delta_{\bf q''}^0 + U_{11}\nu_{\Bar{s}} - U_{12}\sum\limits_{s'}\nu_{s'}
        )^{-1} }
            \PEa  \\
        &\equiv - \vert g \vert^2 \PEa
        \sum_{{\bf q}, s}
        \frac{1}{\Delta_{{\bf q},s} }
        \hat{n}_{{\bf q},s} \PEa
\end{split}
\end{equation}

where in the last line, the screened denominator $\Delta_{{\bf q},s} $ is defined as
    \begin{equation}\label{appendix:renormalised-denominator}
        \begin{split}
            \Delta_{{\bf q},s} \equiv \Delta_{\bf q}^0 - U_{11} \nu_{\Bar{s}} + U_{12} \sum\limits_{s'}\nu_{s'}
            - \frac{U_{12}}{N} \sum\limits_{ {\bf q}'' }  \langle \hat{n}_{{\bf q}'',s} \rangle
            \frac
            {  \Delta_{\bf q}^0 - U_{11} \nu_{\Bar{s}}
            + U_{12} \sum\limits_{s'}\nu_{s'} }
            { \Delta_{\bf q''}^0 - U_{11} \nu_{\Bar{s}}
            + U_{12} \sum\limits_{s'}\nu_{s'}}
        \end{split}
    \end{equation}
Comparing (\ref{k-space-GRPA-Stark-final}) with (\ref{Stark}), we see that after including the electron repulsion $\hat{U}$, under the approximation (\ref{U-b-commutator-reduced}) and the mean-field decoupling (\ref{mean-field-global}) and (\ref{mean-field-local}), the expansion term $\PEa \hat{H}_{1} \hat{G}^{b} \hat{H}_{-1} \PEa$ in the Floquet Hamiltonian is still reduced to an optical Stark shift effect, albeit with a screened detuning in the denominator. 
The screened denominator Eq.~(\ref{appendix:renormalised-denominator}) is the central result of this work. 
%\end{figure}

\subsection{Screened Bloch-Siegert shift}\label{appendix:screened-Bloch-Siegert}
Following the same calculation as above, the screened Bloch-Siegert shift, $\PEa \hat{H}_{-1} \hat{G}^{b}\hat{H}_{1} \PEa$, has the same form as the screened Stark shift (\ref{k-space-GRPA-Stark-final}), except for a substitution $\Delta_{\bf q}^0 \to \Delta_{\bf q}^0 + 2\omL$ in the expression of $\Delta_{{\bf q},s}$ therein. The resulting denominator $\Delta_{{\bf q},s}^{BS}$ for the screened Bloch-Siegert shift reads
\begin{equation*}
    \begin{split}
        \Delta_{{\bf q},s}^{BS}&=
        \Delta_{\bf q}^0 +2\omL - U_{11} \nu_{\Bar{s}}
        + U_{12} \sum\limits_{s'}\nu_{s'} 
        ~~  - ~~ \frac{U_{12}}{N} \sum\limits_{ {\bf q}'' }  \langle \hat{n}_{{\bf q}'',s} \rangle
        \frac
        { - \Delta_{\bf q}^0 -2\omL + U_{11} \nu_{\Bar{s}}
        - U_{12} \sum\limits_{s'}\nu_{s'} }
        {- \Delta_{\bf q''}^0 -2\omL + U_{11} \nu_{\Bar{s}}
        - U_{12} \sum\limits_{s'}\nu_{s'}} \\
        &\approx  \Delta_{\bf q}^0 +2\omL - (U_{11} - U_{12}) \nu_{\Bar{s}}
    \end{split}
\end{equation*}
This derivation is equivalent to Eq.~(\ref{k-space-GRPA-Stark-begin}) $\sim$ (\ref{k-space-GRPA-Stark-final}), and we only need to make the substitution in the Green operators therein.

\subsection{Screened cavity-electron vertex and screened cavity-mediated interaction}
We next study the term $\PEa \hat{H}_{1} \hat{G}^{b} \hat{H}_{c} \hat{G}^{b} \hat{H}_{c} \hat{G}^{b} \hat{H}_{-1} \PEa$, which appears from the expansion of Eq.~(\ref{Heff_weak_drive}) with the Dyson series~(\ref{Dyson}). Again, we first consider a part of this term, $\hat{H}_{c} \hat{G}^{b} \hat{H}_{-1} \PEa$, which reads,
\begin{equation}\label{renormalised-lower-band-vertex}
    \begin{split}
        \hat{H}_{c} \hat{G}^{b}_{(E_\alpha +\omL)}\hat{H}_{-1} \PEa 
        &\approx    \hat{a}^{\dag}  \sum_{{\bf k}, s, {\bf k'},{\bf q'},s'} \frac{ g_{c,{\bf q'}}^{*} g }{\sqrt{N}}  \hat{c}_{{\bf k'}-{\bf q'} 1 s'}^{\dag} \hat{c}_{{\bf k'} 2 s'} 
        \sum\limits_{n=0}^{\infty} (\hat{g} \hat{U})^n \hat{g}
        \hat{b}_{{\bf k},s}^{\dag}
        \PEa \\
        &=  \hat{a}^{\dag}  
        \sum_{{\bf k}, s, {\bf k'},{\bf q'},s'} \frac{ g_{c,{\bf q'}}^{*} g }{\sqrt{N}}  \hat{c}_{{\bf k'}-{\bf q'} 1 s'}^{\dag} \hat{c}_{{\bf k'} 2 s'}  
        \sum\limits_{{\bf q}}
        \hat{b}_{{\bf q},s}^{\dag}
        [ \sum\limits_{n=0}^{\infty} ({\rm g} * {\rm f}^{s})^{n} * {\rm g}]_{ {\bf q} , {\bf k} }
        \PEa\\
        &=  \hat{a}^{\dag}  
        \sum_{{\bf q'},{\bf q}, s} \frac{ g_{c,{\bf q'}}^{*} g }{\sqrt{N}}  \hat{c}_{{\bf q}-{\bf q'} 1 s}^{\dag} \hat{c}_{{\bf q} 1 s}       
        \sum\limits_{\bf k}
        [ ({\rm g}^{-1} - {\rm f}^{s})^{-1}  ]_{ {\bf q} , {\bf k} }
        \PEa\\
        &\approx  \hat{a}^{\dag}  
        \sum_{{\bf q'},{\bf q}, s} \frac{ g_{c,{\bf q'}}^{*} g }{\sqrt{N}}  \hat{c}_{{\bf q}-{\bf q'} 1 s}^{\dag} \hat{c}_{{\bf q} 1 s}  
        ( - \Delta_{{\bf q},s} )^{-1}
        \PEa\\
        & = - \hat{a}^{\dag} \sum_{{\bf k},{\bf q'}, s}   \bigg( \frac{ g g_{c,{\bf q'}}^{*} }{ \sqrt{N} \Delta_{{\bf k},s}   } \bigg) \hat{c}_{{\bf k}-{\bf q'} 1 s}^{\dag}  \hat{c}_{{\bf k} 1 s}  \PEa
    \end{split}
\end{equation}
where in the first line we insert the definition of $\hat{H}_{-1}$ and $\hat{H}_{c}$, expand $\hat{G}^{b}$ over infinite orders of $\hat{U}$, and reduce the terms using $\hat{b}_{{\bf k},s} \PEa =0$. Here we also discard the off-resonant part in $\hat{H}_{c}$, as we did in the last line of (\ref{lower-band-vertex}), so we restrict ourselves to the single-excitation subspace. 
%so that the remaining part lives in the ``0 band-exciton, 1 photon" Hilbert space. 
In the second line, we move $\hat{b}^{\dag}$ to the left, exactly as what we did in (\ref{k-space-GRPA-Stark-begin}). In the third line, we use $\hat{c}_{{\bf k'} 2 s'} \hat{b}_{{\bf q},s}^{\dag} \PEa = \delta_{{\bf k'},{\bf q}}\delta_{s',s} \hat{c}_{{\bf q} 1 s} \PEa$ which follows directly from (\ref{low-energy-approx}). We use the matrix Taylor expansion to replace the infinite summation by the inverse of a single matrix. In the fourth line, we repeat the evaluation (\ref{k-space-GRPA-collection}), and again make the mean-field decoupling (\ref{mean-field-global}) and (\ref{mean-field-local}), which results in the same renormalised denominator $\Delta_{{\bf q},s}$ in (\ref{appendix:renormalised-denominator}).

Comparing (\ref{renormalised-lower-band-vertex}) with (\ref{lower-band-vertex}), we see that after including $\hat{U}$, 
%under the approximation (\ref{U-b-commutator-reduced}) based on the mean-field approximation (\ref{reduce-gHs-filling}) and (\ref{reduce-gHs-single-occupation}), 
the term $\hat{H}_{c} \hat{G}^{b} \hat{H}_{-1} \PEa$ is still reduced to a scattering vertex between cavity photon and a lower-band electron, albeit the laser-bandgap detuning in its denominator is screened to Eq.~(\ref{appendix:renormalised-denominator}).
Having derived the screened cavity-electron vertex, we come back to the term $\PEa \hat{H}_{1} \hat{G}^{b} \hat{H}_{c} \hat{G}^{b} \hat{H}_{c} \hat{G}^{b} \hat{H}_{-1} \PEa$,
\begin{equation}\label{cavity-meidated-interaction-k-GRPA}
    \begin{split}
        & \PEa \hat{H}_{1} \hat{G}^{b}_{(\Ea + \omL)}  \hat{H}_{c} \hat{G}^{b}_{(\Ea + \omL)}  \hat{H}_{c} \hat{G}^{b}_{(\Ea + \omL)}\hat{H}_{-1} \PEa \\
        &\approx   
        \PEa \frac{1}{\sqrt{N}} \sum_{{\bf k'},{\bf q'}, s'}   \bigg( \frac{g_{c,{\bf q'}} g^{*}}{\Delta_{{\bf k'},s'}} \bigg) \hat{c}_{{\bf k'} 1 s'} ^{\dag}  \hat{c}_{{\bf k'}-{\bf q'} 1 s'} \\
        &~~~~~~~~~~~~~~~~~~~~~~~ \times \hat{a} \sum\limits_{n=0}^{\infty} (\hat{g}_{(\Ea + \omL)} \hat{U})^n \hat{g}_{(\Ea + \omL)} \hat{a}^{\dag} 
        ~~ \times \frac{1}{\sqrt{N}} \sum_{{\bf k},{\bf q}, s}   \bigg( \frac{g_{c,{\bf q}}^{*} g}{\Delta_{{\bf k},s}} \bigg) \hat{c}_{{\bf k}-{\bf q} 1 s}^{\dag}  \hat{c}_{{\bf k} 1 s}  \PEa \\
        &= \PEa \frac{1}{N}  \sum_{ \substack{ {\bf k},{\bf q}, s \\ {\bf k'},{\bf q'}, s'} } \frac{\vert g \vert^2 g_{c,{\bf q'}} g_{c,{\bf q}}^{*} }{\Delta_{{\bf k'},s'} \Delta_{{\bf k},s}} \hat{c}_{{\bf k'} 1 s'} ^{\dag}  \hat{c}_{{\bf k'}-{\bf q'} 1 s'} 
        \sum\limits_{n=0}^{\infty} (\hat{g}_{(\Ea -\Delta_c)}  \hat{U})^n \hat{g}_{(\Ea -\Delta_c)} 
        \hat{c}_{{\bf k}-{\bf q} 1 s}^{\dag}  \hat{c}_{{\bf k} 1 s} \PEa \\
        &\approx \PEa \frac{1}{N}  \sum_{ \substack{ {\bf k},{\bf q}, s \\ {\bf k'},{\bf q'}, s'} } \frac{\vert g \vert^2 g_{c,{\bf q'}} g_{c,{\bf q}}^{*} }{\Delta_{{\bf k'},s'} \Delta_{{\bf k},s}} \hat{c}_{{\bf k'} 1 s'} ^{\dag}  \hat{c}_{{\bf k'}-{\bf q'} 1 s'} 
        \hat{c}_{{\bf k}-{\bf q} 1 s}^{\dag}  \hat{c}_{{\bf k} 1 s}  \hat{G}^{b}_{(\Ea -\Delta_c)}  \PEa \\
        &\approx - \frac{1}{N}  \sum_{ \substack{ {\bf k},{\bf q}, s \\ {\bf k'},{\bf q'}, s'} } \frac{\vert g \vert^2 g_{c,{\bf q'}} g_{c,{\bf q}}^{*} }{\Delta_c \Delta_{{\bf k'},s'} \Delta_{{\bf k},s}}
        \hat{c}_{{\bf k'} 1 s'} ^{\dag}  \hat{c}_{{\bf k'}-{\bf q'} 1 s'}
        \hat{c}_{{\bf k}-{\bf q} 1 s}^{\dag}  \hat{c}_{{\bf k} 1 s} \PEa.
    \end{split}
\end{equation}
where in the first line we use the expression (\ref{renormalised-lower-band-vertex}) for the screened vertex. In the second line we move $\hat{a}^{\dag}$ to the far-left using (\ref{g-commutator}), and then we reduce this operator using $\PEa \hat{a} \hat{a}^{\dag} = \PEa$ which follows from the low-energy limit (\ref{low-energy-approx}). 
In the third line, we use $[\hat{G}^{b}_{(\Ea -\Delta_c)},\hat{c}_{{\bf k}-{\bf q} 1 s}^{\dag}  \hat{c}_{{\bf k} 1 s}]\approx0$ to switch these two operators. As explained in Appendix \ref{appendix:GRPA-cavity-screening}, this neglects the screening of the cavity-electron vertex, which is appropriate as long as $g_{c,{\bf q}}$ remains non-zero only for extremely small $\bf q$.
In the fourth line, we use Eq.~(\ref{G1-reduced-to-number}) to replace $\hat{G}^{b}$ by a denominator related to the laser-cavity detuning $\Delta_c$. Again, comparing Eq.~(\ref{cavity-meidated-interaction-k-GRPA}) with the unscreened interaction~(\ref{cavity-meidated-interaction-U=0}), the expansion term \\ $\PEa \hat{H}_{1} \hat{G}^{b} \hat{H}_{c} \hat{G}^{b} \hat{H}_{c} \hat{G}^{b} \hat{H}_{-1} \PEa$ still results in the cavity-mediated interaction, albeit with renormalized interaction strength. 

%This screening result could be straightforwardly applied to analyse the multi-band screening effect on the cavity-mediated interaction in Ref.~\cite{chiocchetta2021cavity}. In that work, the lower-band is correlated. Thus, it is also necessary to consider inter-band repulsion, which is ignored therein.

\subsection{Screened non-RWA corrections to the cavity-mediated interaction}
The leading screening effect on the non-RWA corrections to the cavity-mediated interaction (\ref{non-RWA-cavity-med-interaction-U=0}) stems from the term $\hat{H}_{1}\hat{G}^{0}_{(\Ea +\omL)}\hat{H}_{-1}$ in (\ref{Heff_weak_drive}), instead of the term $\hat{H}_{-1}\hat{G}^{0}_{(\Ea -\omL)}\hat{H}_{1}$ which gives the Bloch-Siegert shift. 
The expression of this screened interaction
\begin{equation*}
    \PEa \hat{H}_{1} \hat{G}^{b} \hat{H}_{c} \hat{\mathcal{P}}_{2,1} \hat{G}^{b}_{(\Ea +\omL)} \hat{\mathcal{P}}_{2,1} \hat{H}_{c} \hat{G}^{b} \hat{H}_{-1} \PEa
\end{equation*}
is obtained by a substitution $\Delta_{\bf q}^0 \to \Delta_{{\bf q},s}$ in (\ref{non-RWA-cavity-med-interaction-U=0}). Here the screened denominator $\Delta_{{\bf q},s}$ is the same as in (\ref{appendix:renormalised-denominator}).

\subsection{Screened inter-manifold terms in the Floquet Hamiltonian}\label{screened-off-resonant-leading-order}
In the presence of on-site repulsion, in the Floquet Hamiltonian, the leading inter-manifold term $\hat{H}_{1}\hat{G}^{b}_{(\Ea +\omL)}\hat{H}_{-1}$ becomes screened. This term reads
\begin{equation}\label{screened-inter-manifold-band-excitation}
    \begin{split}
         \hat{\mathcal{P}}_{2,0} \hat{H}_{1}\hat{G}^{b}_{(E_\alpha +\omL)}\hat{H}_{-1} \PEa 
         &= 
          g^2
          \hat{\mathcal{P}}_{2,0}
          \sum_{{\bf q},{\bf q}', s, s'}
          \hat{c}_{\mathbf{q}' 2 s'}^{\dagger} \hat{c}_{\mathbf{q}' 1 s'}
          \hat{G}^{b}
          \hat{c}_{\mathbf{q} 2 s}^{\dagger} \hat{c}_{\mathbf{q} 1 s}  \PEa \\
          &\approx
         - \sum_{{\bf q},{\bf q}', s, s'}
          \frac{g^2}{\Delta_{{\bf q},s}}
          \hat{c}_{\mathbf{q}' 2 s'}^{\dagger} \hat{c}_{\mathbf{q} 2 s}^{\dagger} \hat{c}_{\mathbf{q} 1 s} \hat{c}_{\mathbf{q}' 1 s'}  \PEa
    \end{split}
\end{equation}
Comparing (\ref{screened-inter-manifold-band-excitation}) with the unscreened result (\ref{inter-manifold-band-excitation}), we see that the off-resonant effects are screened according to the renormalisation on the denominator (\ref{appendix:renormalised-denominator}).

\section{Approximations on the commutator $[\hat{U},\hat{b}^\dag]$}\label{appendix:U-b-commutator}

\subsection{Commutator terms}
The commutation relation~(\ref{U-b-commutator-reduced}) of the main text is our central approximation in Chapter \ref{Chapter4}. It discards all non-commuting terms other than the Hartree-/Fock-type terms. 
We justify it in detail here: Using the definition of $\hat{U}=\hat{U}_{11}+\hat{U}_{12}+\hat{U}_{22}$ in momentum space (\ref{U11-term}) and (\ref{U12-term}), we have
\begin{equation}\label{U-b-commutator-appendix}
    \begin{split}
        &\hat{U} \hat{c}_{\mathbf{q} 2 s}^{\dag} \hat{c}_{\mathbf{q} 1 s}
        = \hat{c}_{\mathbf{q} 2 s}^{\dag} \hat{c}_{\mathbf{q} 1 s} \hat{U} + \big[ \hat{U}_{11}, \hat{c}_{\mathbf{q} 2 s}^{\dag} \hat{c}_{\mathbf{q} 1 s} \big] + \big[ \hat{U}_{12}, \hat{c}_{\mathbf{q} 2 s}^{\dag} \hat{c}_{\mathbf{q} 1 s} \big] + \big[ \hat{U}_{22}, \hat{c}_{\mathbf{q} 2 s}^{\dag} \hat{c}_{\mathbf{q} 1 s} \big] \\
        &= \hat{c}_{\mathbf{q} 2 s}^{\dag} \hat{c}_{\mathbf{q} 1 s} \hat{U} - 
           2\frac{U_{11}}{2N} \hat{c}_{\mathbf{q} 2 s}^{\dag} \sum\limits_{{\bf k'},{\bf q'}} \hat{c}_{{\bf k'}-{\bf q'}, 1, \Bar{s}}^{\dag} \hat{c}_{{\bf k'}, 1, \Bar{s}} \hat{c}_{{\bf q}-{\bf q'}, 1, s}  \\
        &~~~~~~~~ -\frac{U_{12}}{N} \sum\limits_{\bf q'} \hat{c}_{{\bf q}+{\bf q'}, 2, s}^{\dag} \hat{c}_{{\bf q}+{\bf q'}, 1, s} + \frac{U_{12}}{N} \sum\limits_{{\bf k'},{\bf q'},s'} \hat{c}_{{\bf q}+{\bf q'}, 2, s }^{\dag} \hat{c}_{{\bf q}, 1, s} \hat{c}_{{\bf k'}-{\bf q'}, 1, s'}^{\dag} \hat{c}_{{\bf k'}, 1, s'} 
        - \hat{c}_{{\bf q}, 2, s}^{\dag} \hat{c}_{{\bf k'}+{\bf q'}, 2, s'}^{\dag} \hat{c}_{{\bf k'}, 2, s'} \hat{c}_{{\bf q}+{\bf q'}, 1, s} \\
        &~~~~~~~~ - 2\frac{U_{22}}{2N}  \sum\limits_{{\bf k'},{\bf q'} } \hat{c}_{{\bf k'}, 2, \Bar{s}}^{\dag} \hat{c}_{{\bf q}-{\bf q'}, 2, s}^{\dag}  \hat{c}_{{\bf k'}-{\bf q'}, 2, \Bar{s}} \hat{c}_{\mathbf{q} 1 s} \\
        &= \hat{c}_{\mathbf{q} 2 s}^{\dag} \hat{c}_{\mathbf{q} 1 s} 
        \big( \hat{U} 
        - \frac{U_{11}}{N}\sum\limits_{\bf k}\hat{n}_{{\bf k},1,\Bar{s}} 
        + \frac{U_{12}}{N}\sum\limits_{{\bf k},s'}\hat{n}_{{\bf k},1,s'}
        \big) 
        -  \frac{ U_{12} }{N} 
        \big(
        \sum\limits_{\bf k} \hat{c}_{{\bf k} 2 s}^{\dag} \hat{c}_{{\bf k} 1 s}
        \big)
        \hat{n}_{{\bf q},1,s}
        ~~ + ~~ ...
    \end{split}
\end{equation}
where in the last line we keep only three terms: 

1) In $\big[ \hat{U}_{11}, \hat{c}_{\mathbf{q} 2 s}^{\dag} \hat{c}_{\mathbf{q} 1 s} \big]$, we only keep ${\bf q'}=0$ term
\begin{equation*}
    -\frac{U_{11}}{N} \hat{c}_{\mathbf{q} 2 s}^{\dag} \sum\limits_{{\bf k'},{\bf q'}=0} \hat{c}_{{\bf k'}-{\bf q'}, 1, \Bar{s}}^{\dag} \hat{c}_{{\bf k'}, 1, \Bar{s}} \hat{c}_{{\bf q}-{\bf q'}, 1, s} ~ = ~
    -\frac{U_{11}}{N} \hat{c}_{\mathbf{q} 2 s}^{\dag} \hat{c}_{{\bf q}, 1, s} \sum\limits_{{\bf k'}} \hat{n}_{{\bf k'}, 1, \Bar{s}} 
\end{equation*}
corresponding to the intra-band Hartree term. The semi-classical justification for this treatment is as follows: When ${\bf q'}\neq0$, the expectation value $\sum_{\bf k'} \langle \hat{c}_{{\bf k'}-{\bf q'}, 1, \Bar{s}}^{\dag} \hat{c}_{{\bf k'}, 1, \Bar{s}} \rangle$ represents the charge density wave of the lower-band electron with spin $\Bar{s}$ at wave-vector $\bf q'$. However, this expectation value becomes much stronger at ${\bf q'}=0$, because in this case the expectation value $\sum_{\bf k'} \langle \hat{c}_{{\bf k'}, 1, \Bar{s}}^{\dag} \hat{c}_{{\bf k'}, 1, \Bar{s}} \rangle$ represents the total electron number in the lower-band with spin $\Bar{s}$. Thus when the electron density is high, we can ignore all terms in $\big[ \hat{U}_{11}, \hat{c}_{\mathbf{q} 2 s}^{\dag} \hat{c}_{\mathbf{q} 1 s} \big]$ except for the ${\bf q'}=0$ contribution.

2) In $\big[ \hat{U}_{12}, \hat{c}_{\mathbf{q} 2 s}^{\dag} \hat{c}_{\mathbf{q} 1 s} \big]$, we only keep two terms: the first term is the inter-band Hartree term where (as justified above) we take ${\bf q'}=0 $ in the following summation,
\begin{equation*}
    \frac{U_{12}}{N} \sum\limits_{{\bf k'},{\bf q'}=0,s'} \hat{c}_{{\bf q}+{\bf q'}, 2, s }^{\dag} \hat{c}_{{\bf q}, 1, s} \hat{c}_{{\bf k'}-{\bf q'}, 1, s'}^{\dag} \hat{c}_{{\bf k'}, 1, s'} ~ = ~
     \frac{U_{12}}{N} \hat{c}_{\mathbf{q} 2 s}^{\dag} \hat{c}_{{\bf q}, 1, s} \sum\limits_{{\bf k'},s'} \hat{n}_{{\bf k'}, 1, s'}.
\end{equation*}
The second term is the inter-band Fock term, where we take ${\bf q'}={\bf k'}-{\bf q} $ in the following part of $\big[ \hat{U}_{12}, \hat{c}_{\mathbf{q} 2 s}^{\dag} \hat{c}_{\mathbf{q} 1 s} \big]$
\begin{equation*}
\begin{split}
    & -\frac{U_{12}}{N} \sum\limits_{\bf q'} \hat{c}_{{\bf q}+{\bf q'}, 2, s}^{\dag} \hat{c}_{{\bf q}+{\bf q'}, 1, s} + \frac{U_{12}}{N} \sum\limits_{{\bf q'}={\bf k'}-{\bf q} \atop {\bf k'},s'=s} \hat{c}_{{\bf q}+{\bf q'}, 2, s }^{\dag} \hat{c}_{{\bf q}, 1, s} \hat{c}_{{\bf k'}-{\bf q'}, 1, s'}^{\dag} \hat{c}_{{\bf k'}, 1, s'} \\
    &= -\frac{U_{12}}{N} \sum\limits_{\bf q'} \hat{c}_{{\bf q}+{\bf q'}, 2, s}^{\dag} \hat{c}_{{\bf q}+{\bf q'}, 1, s} + \frac{U_{12}}{N} \sum\limits_{{\bf k'}} \hat{c}_{{\bf k'}, 2, s }^{\dag} (1-\hat{n}_{{\bf q}, 1, s}) \hat{c}_{{\bf k'}, 1, s} \\
    &= -\frac{U_{12}}{N}  \sum\limits_{{\bf k'}} \hat{c}_{{\bf k'}, 2, s }^{\dag} \hat{n}_{{\bf q}, 1, s}  \hat{c}_{{\bf k'}, 1, s}  \\
    &=  -\frac{U_{12}}{N} \big( \sum\limits_{{\bf k}} \hat{c}_{{\bf k}, 2, s }^{\dag} \hat{c}_{{\bf k}, 1, s} \big) \hat{n}_{{\bf q}, 1, s} ~ + ~ \frac{U_{12}}{N} \hat{c}_{{\bf q}, 2, s }^{\dag} \hat{c}_{{\bf q}, 1, s} \\
    &\approx  -\frac{U_{12}}{N} \big( \sum\limits_{{\bf k}} \hat{c}_{{\bf k}, 2, s }^{\dag} \hat{c}_{{\bf k}, 1, s} \big) \hat{n}_{{\bf q}, 1, s}  ~~ \text{when the electron number} \gg 1
\end{split}
\end{equation*}
In the last line we discard the term $\frac{U_{12}}{N} \hat{c}_{{\bf q}, 2, s }^{\dag} \hat{c}_{{\bf q}, 1, s}$, as it will be much smaller than the inter-band Hartree term $-\frac{U_{11}}{N} \hat{c}_{\mathbf{q} 2 s}^{\dag} \hat{c}_{{\bf q}, 1, s} \sum_{{\bf k'}} \hat{n}_{{\bf k'}, 1, \Bar{s}}  $ (which has the same form), when the total electron number in the lower-band with spin $\bar{s}$ is large. The semi-classical justification for only keeping the ${\bf q'}={\bf k'}-{\bf q} \And s'=s$ part in the above summation is that, when ignoring the electron correlation in the lower-band, the expectation value will vanish $\langle \hat{c}_{{\bf q}, 1, s} \hat{c}_{{\bf k'}-{\bf q'}, 1, s'}^{\dag} \rangle = 0$ unless ${\bf q'}={\bf k'}-{\bf q} \And s'=s$. Thus, when the electron correlation effect in the lower-band is not strong, we can discard all terms other than ${\bf q'}={\bf k'}-{\bf q} \And s'=s$ in the summation.

Besides, in $\big[ \hat{U}_{12}, \hat{c}_{\mathbf{q} 2 s}^{\dag} \hat{c}_{\mathbf{q} 1 s} \big]$, we completely discard the last term 
\begin{equation*}
    \hat{c}_{{\bf q}, 2, s}^{\dag} \hat{c}_{{\bf k'}+{\bf q'}, 2, s'}^{\dag} \hat{c}_{{\bf k'}, 2, s'} \hat{c}_{{\bf q}+{\bf q'}, 1, s}
\end{equation*}
because this term vanishes in the low-energy limit where $\hat{c}_{{\bf k'}, 2, s'} \PEa =0$, which is the only case we will encounter in the following evaluations. 

3) For the same reason, the commutator $\big[ \hat{U}_{22}, \hat{c}_{\mathbf{q} 2 s}^{\dag} \hat{c}_{\mathbf{q} 1 s} \big]$ can be ignored in the low-energy limit. 

Altogether, in the last line of (\ref{U-b-commutator-appendix}), we keep the intra-band Hartree term, inter-band Hartree term and the inter-band Fock term of the commutator $\big[ \hat{U}, \hat{c}_{\mathbf{q} 2 s}^{\dag} \hat{c}_{\mathbf{q} 1 s} \big]$. The approximation made in (\ref{U-b-commutator-appendix}) is analogous to the random phase approximation in the equation-of-motion method, as elucidated in \cite{rowe1968equations}, where the commutator terms are discarded if their expectation value (under the non-interacting ground state) is 0.

\subsection{Naming of the Hartree \& Fock commutator terms}\label{appendix:Naming-of-Hartree/Fock}
\begin{figure}[h!]
    \centering
    \includegraphics[width=0.3\textwidth]{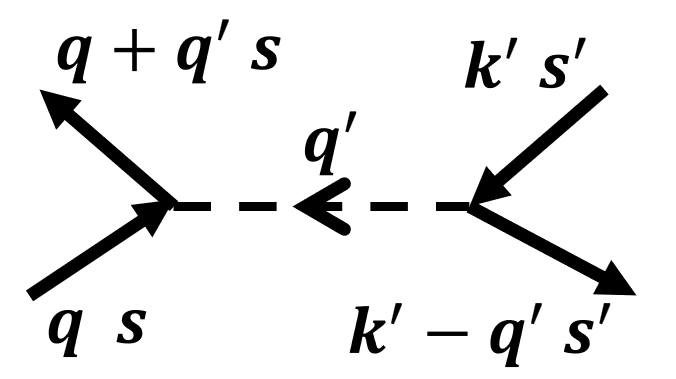}
    \caption{diagrammatic illustration of the commutator term (\ref{U12-bqs-commutator-example}) in $\big[ \hat{U}_{12}, \hat{c}_{\mathbf{q} 2 s}^{\dag} \hat{c}_{\mathbf{q} 1 s} \big]$, where the different band-indices are not distinguished out.}
    \label{fig:U12_illu}
\end{figure}

We next explain why we call the terms in Eq.~(\ref{U-b-commutator-appendix}) ``Hartree"/``Fock" terms, this naming convention resembles the ``Direct"/``Exchange" commuting terms in Ref~\cite{PhysRev.112.1900}: For example, in the commutator $\big[ \hat{U}_{12}, \hat{c}_{\mathbf{q} 2 s}^{\dag} \hat{c}_{\mathbf{q} 1 s} \big]$, the term
\begin{equation}\label{U12-bqs-commutator-example}
    \frac{U_{12}}{N} \sum\limits_{{\bf k'},{\bf q'},s'} \hat{c}_{{\bf q}+{\bf q'}, 2, s }^{\dag} \hat{c}_{{\bf q}, 1, s} \hat{c}_{{\bf k'}-{\bf q'}, 1, s'}^{\dag} \hat{c}_{{\bf k'}, 1, s'} 
\end{equation}
can be diagrammatically represented by Fig.~(\ref{fig:U12_illu}), where the band-index is not distinguished, and the incoming (outgoing) arrow represents the annihilation (creation) operator in (\ref{U12-bqs-commutator-example}).

\begin{figure}[h!]
    \centering
    \includegraphics[width=0.8\textwidth]{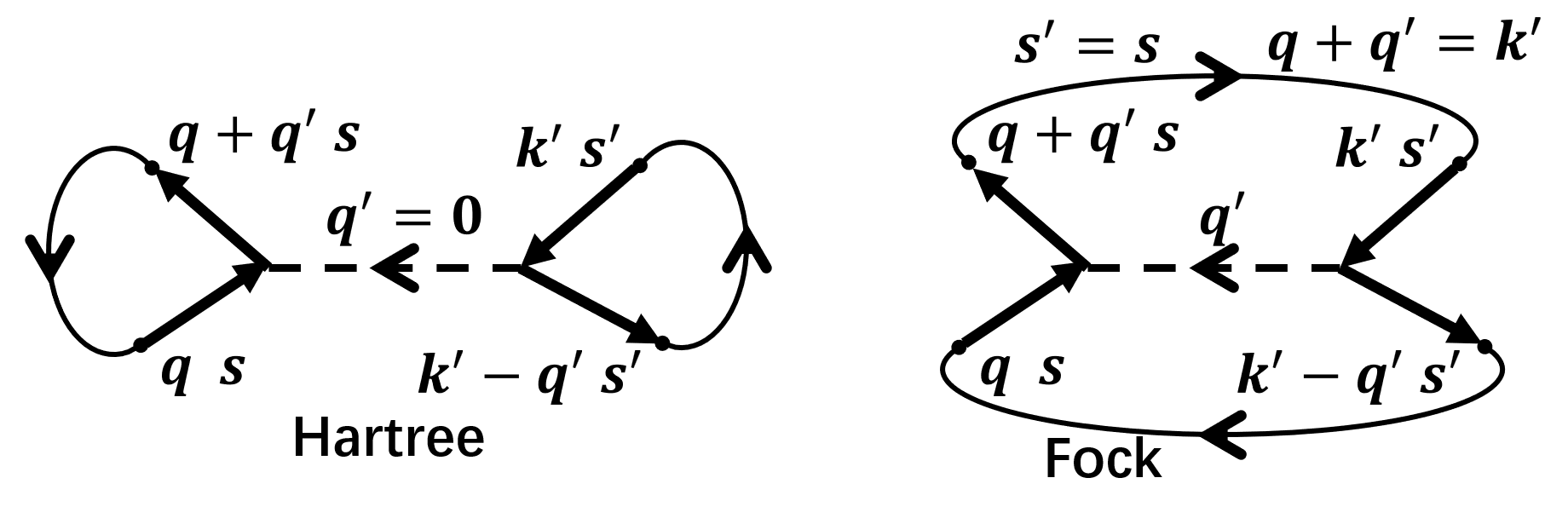}
    \caption{The ``Hartree" and ``Fock" terms in the commutator $\big[ \hat{U}_{12}, \hat{c}_{\mathbf{q} 2 s}^{\dag} \hat{c}_{\mathbf{q} 1 s} \big]$, these terms respectively corresponds to two possible ways to contract the momentum and spin, making the diagram closed.}
    \label{fig:H-F_illu}
\end{figure}

In our approximation , we only keep two graphs where the incoming and outgoing momentum lines are contracted, as shown in fig(\ref{fig:H-F_illu}).
Cutting one fermion line in these close diagrams respectively results in the Hartree/Fock-like self energy diagrams. In fig(\ref{fig:H-F_illu}), we find the left (``Hartree") diagram is the ${\bf q'}=0$ part of fig(\ref{fig:U12_illu}), and the right (``Fock") diagram is the ${\bf k'}={\bf q}+{\bf q'} ,s'=s$ part of fig(\ref{fig:U12_illu}). This is equivalent to the approximation we make in the last line of (\ref{U-b-commutator-appendix}), and thus we denote the terms we kept there as ``Hartree"/``Fock" terms.

In Eq.~(\ref{split-gdelta-fF}), ${\rm f}_{F}^s$ only contains the contribution from the inter-band Fock term, while ${\rm g}_{H}^s$ contains the contribution from the intra- and inter-band Hartree terms. Consequently, these two quantities are respectively represented by ``F" and ``H".

\section{Screening of the cavity mode in Eq.~(\ref{cavity-meidated-interaction-k-GRPA})}
\label{appendix:GRPA-cavity-screening}

\subsection{Screening of the cavity-electron interaction vertex}
In the third line of Eq.~(\ref{cavity-meidated-interaction-k-GRPA}), we switch $\hat{G}^{b}_{(\Ea-\Delta_c)}$ and $\hat{c}_{{\bf k}-{\bf q} 1 s}^{\dag}  \hat{c}_{{\bf k} 1 s}$, which ignores the screening effect on the cavity mode given by the electron on-site repulsion. To understand why this approximation is valid in the parameter regimes considered in this paper, we study the following commuting relation of the intra-band electron-hole creation operator
\begin{equation}\label{Hartree-Fock-commuting-relation}
\begin{split}
     \hat{U}_{11} \hat{c}_{{\bf k}-{\bf q} 1 s}^{\dag}  \hat{c}_{{\bf k} 1 s} 
    &= \hat{c}_{{\bf k}-{\bf q} 1 s}^{\dag}  \hat{c}_{{\bf k} 1 s} \hat{U}_{11} 
    + \frac{U_{11}}{N} \sum\limits_{{\bf k'},{\bf q'}} \hat{c}_{{\bf k'}-{\bf q'} 1 \Bar{s}}^{\dag}  \hat{c}_{{\bf k'} 1 \Bar{s}}
    ( \hat{c}_{{\bf k}-{\bf q}+{\bf q'} 1 s}^{\dag} \hat{c}_{{\bf k} 1 s}
    -
    \hat{c}_{{\bf k}-{\bf q} 1 s}^{\dag} \hat{c}_{{\bf k}-{\bf q'} 1 s}) \\
    &\approx \hat{c}_{{\bf k}-{\bf q} 1 s}^{\dag}  \hat{c}_{{\bf k} 1 s} \hat{U}_{11}  
    + \frac{U_{11}}{N} \sum\limits_{ {\bf k'} } \hat{c}_{{\bf k'}-{\bf q} 1 \Bar{s}}^{\dag}  \hat{c}_{{\bf k'} 1 \Bar{s}} ( \hat{n}_{ {\bf k}1s  }  -   \hat{n}_{ {\bf k}-{\bf q}  1s  }  ) \\
    &= \sum\limits_{ {\bf k'} , s' }
    \hat{c}_{\mathbf{k'}-{\bf q} 1 s}^{\dag}  \hat{c}_{{\bf k'} 1 s}
    \bigg(
     \delta_{  {\bf k'}  , {\bf k} } \delta_{s', s}   \hat{U}_{11}
    + \delta_{s', \Bar{s}} \frac{U_{11}}{N}  ( \hat{n}_{ {\bf k}1s  }  -   \hat{n}_{ {\bf k}-{\bf q}  1s  }  )
    \bigg)
\end{split}
\end{equation}
where in the second line, to be consistent with the approximation (\ref{U-b-commutator-reduced}), we only keep the contribution from ${\bf q'}={\bf q}$ (note that the contribution from ${\bf q'}={\bf 0}$ automatically vanishes). After this approximation, we see from the third line that, this commutation relation has a similar momentum-summation structure as Eq.~(\ref{U-b-commutator-reduced}), allowing the evaluation of
\begin{equation*}
    \sum\limits_{n=0}^{\infty} (\hat{g}_{(\Ea-\Delta_c)} \hat{U}_{11})^n \hat{g}_{(\Ea-\Delta_c)} 
    \sum\limits_{\bf k, \bf q} 
    \frac{ g_{c,{\bf q}}^{*} }{  \Delta_{{\bf k},s}  }
    \hat{c}_{{\bf k}-{\bf q} 1 s}^{\dag}  \hat{c}_{{\bf k} 1 s} 
    \PEa
\end{equation*}
in the second line of (\ref{cavity-meidated-interaction-k-GRPA}), by introducing another two matrices whose elements are operators, similar to (\ref{operator-valued-matrix-element}). However, we can avoid this complex evaluation by noting that the spatial scale of the cavity mode is much larger than the lattice constant, so that $g_{c,{\bf q}}$ remains non-zero only for extremely small $\bf q$, and thus
\begin{equation}\label{narrow-k-space-distribution-gc}
\begin{split}
    \sum\limits_{{\bf k},{\bf q}} \frac{ g_{c,{\bf q}}^{*} }{  \Delta_{{\bf k},s}  } \bigg(
    \frac{U_{11}}{N} \sum\limits_{ \bf k' } \hat{c}_{{\bf k'}-{\bf q} 1 \Bar{s}}^{\dag}  \hat{c}_{{\bf k'} 1 \Bar{s}} 
    ( \hat{n}_{ {\bf k}1s  }  -   \hat{n}_{ {\bf k}-{\bf q}  1s  }  )  \bigg)
    &\approx 
    \frac{U_{11}}{N} \sum\limits_{ {\bf k'} , \bf q } \hat{c}_{{\bf k'}-{\bf q} 1 \Bar{s}}^{\dag}  \hat{c}_{{\bf k'} 1 \Bar{s}} 
    \sum\limits_{\bf k} 
    ( \frac{ g_{c,{\bf q}}^{*} }{  \Delta_{{\bf k},s}  }
    \hat{n}_{ {\bf k}1s  } 
    - \frac{ g_{c,{\bf q}}^{*} }{  \Delta_{{\bf k}-{\bf q},s}  }
    \hat{n}_{ {\bf k}-{\bf q}  1s  }  ) \\
    &= \frac{U_{11}}{N} \sum\limits_{ {\bf k'} , \bf q } \hat{c}_{{\bf k'}-{\bf q} 1 \Bar{s}}^{\dag}  \hat{c}_{{\bf k'} 1 \Bar{s}} 
    \times 0 ~~ = ~~ 0
\end{split}
\end{equation}
where in the first line we assume $\Delta_{{\bf k},s}\approx\Delta_{{\bf k}-{\bf q},s}$ since $\bf q$ can only be a tiny number when $g_{c,{\bf q}} \neq 0$. Combining the Hartree-Fock treatment (\ref{Hartree-Fock-commuting-relation}) with (\ref{narrow-k-space-distribution-gc}), we directly have the following commuting relation
\begin{equation}\label{U11-cdag1c1-commuting}
    \begin{split}
    & \hat{U}_{11} 
    \sum\limits_{\bf k, \bf q} 
    \frac{ g_{c,{\bf q}}^{*} }{  \Delta_{{\bf k},s}  }
    \hat{c}_{{\bf k}-{\bf q} 1 s}^{\dag}  \hat{c}_{{\bf k} 1 s} \approx
    \sum\limits_{\bf k, \bf q} 
    \frac{ g_{c,{\bf q}}^{*} }{  \Delta_{{\bf k},s}  }
    \hat{c}_{{\bf k}-{\bf q} 1 s}^{\dag}  \hat{c}_{{\bf k} 1 s} \hat{U}_{11}
    \end{split}
\end{equation}
this commuting relation means that
\begin{equation}
\begin{split}
    \hat{G}^{b}_{(\Ea-\Delta_c)}
    \sum\limits_{\bf k, \bf q} 
    \frac{ g_{c,{\bf q}}^{*} }{  \Delta_{{\bf k},s}  }
    \hat{c}_{\mathbf{k}-{\bf q} 1 s}^{\dagger}  \hat{c}_{{\bf k} 1 s} 
    & \approx
    \sum\limits_{\bf k, \bf q} 
    \frac{ g_{c,{\bf q}}^{*} }{  \Delta_{{\bf k},s}  }
    \hat{c}_{\mathbf{k}-{\bf q} 1 s}^{\dagger}  \hat{c}_{{\bf k} 1 s} 
    \sum\limits_{n=0}^{\infty} 
    (\hat{g}_{(E_\alpha - \Delta_c - \epsilon_{{\bf k}-{\bf q}1} + \epsilon_{{\bf k}1})} \hat{U})^n 
    \hat{g}_{(E_\alpha - \Delta_c - \epsilon_{{\bf k}-{\bf q}1} + \epsilon_{{\bf k}1})} 
      \\
    &\approx 
    \sum\limits_{\bf k, \bf q} 
    \frac{ g_{c,{\bf q}}^{*} }{  \Delta_{{\bf k},s}  }
    \hat{c}_{\mathbf{k}-{\bf q} 1 s}^{\dagger}  \hat{c}_{{\bf k} 1 s} 
    \sum\limits_{n=0}^{\infty} 
    (\hat{g}_{(E_\alpha - \Delta_c )} \hat{U})^n 
    \hat{g}_{(E_\alpha - \Delta_c)} 
      \\
    &=\sum\limits_{\bf k, \bf q} 
    \frac{ g_{c,{\bf q}}^{*} }{  \Delta_{{\bf k},s}  }
    \hat{c}_{\mathbf{k}-{\bf q} 1 s}^{\dagger}  \hat{c}_{{\bf k} 1 s} 
    \hat{G}^{b}_{(E_\alpha - \Delta_c )}
      \\
\end{split}
\end{equation}
where in the first line, we apply the Dyson expansion to $\hat{G}^{b}$, and then, Eq.(\ref{g-commutator}) and (\ref{U11-cdag1c1-commuting}) together allow us to move $\hat{G}^{b}$ rightward. In the second line we use $\Delta_c + \epsilon_{{\bf k}-{\bf q}1} - \epsilon_{{\bf k}1} \approx \Delta_c $, which is again very accurate since $\bf q$ can only be a tiny number when $g_{c,{\bf q}} \neq 0$, so that $\Delta_c \gg \vert \epsilon_{{\bf k}-{\bf q}1} - \epsilon_{{\bf k}1} \vert$. 
This is the reason why we can switch $\hat{G}^{b}_{(\Ea-\Delta_c)}$ and $\hat{c}_{{\bf k}-{\bf q} 1 s}^{\dag}  \hat{c}_{{\bf k} 1 s}$ in the third line of Eq.~(\ref{cavity-meidated-interaction-k-GRPA}).

\subsection{Screening by inter-band polarization bubbles}
On top of the above screening of the cavity-electron vertex, throughout this work, we also ignore another type of expansion terms containing higher orders of $\hat{H}_{c}$, for example,
\begin{equation*}
\PEa \hat{H}_{1} \hat{G}^{b}_{(\Ea + \omL)}  \hat{H}_{c} \hat{G}^{b}_{(\Ea + \omL)}  \hat{H}_{c} \hat{G}^{b}_{(\Ea + \omL)}\hat{H}_{c} \hat{G}^{b}_{(\Ea + \omL)}  \hat{H}_{c} \hat{G}^{b}_{(\Ea + \omL)}\hat{H}_{-1} \PEa
\end{equation*}
which also screens the cavity-mediated interaction (\ref{cavity-meidated-interaction-k-GRPA}). These screening terms can be understood as the inter-band polarization bubble screening, because the iterative application of the operator $\hat{H}_{c} \hat{G}^{b}$ corresponds to the consecutive creation and annihilation of virtual inter-band excitons.

The cavity-mediated interaction cannot be strongly screened by these inter-band polarization bubbles, because this bubble comprises two electron-cavity vertices connected by an inter-band electron-hole propagator, giving rise to a factor $\sim \vert g_c \vert^2 / (\Delta_c \Delta_{{\bf q },s}^{U})$. This coefficient is small under the off-resonating driving condition considered in this work, as shown in fig(\ref{fig:Frenkel-resonance}).

\section{Low-temperature absorbance}\label{sec.absorbance}
The optical absorbance spectrum $\alpha(\omega)$ for our model considered in Chapter \ref{Chapter4} is defined as the imaginary part of the dipole-dipole correlation function in frequency domain, which, in the low-temperature limit, reads~\cite{matsueda2005excitonic,PhysRevB.29.4401}
\begin{equation}\label{eq.absorbance}
\begin{split}
\alpha(\omega) &= -\frac{1}{\pi} \text{Im} \langle G \vert \frac{1}{g ^*} \hat{H}_{1} \frac{1}{\omega+E_G-\hat{H}_0 + i \gamma } \frac{1}{g } \hat{H}_{-1} \vert G \rangle \\
&= - \frac{1}{\vert g  \vert^{2} \pi} \text{Im} \langle G \vert \hat{\mathcal{P}}_{E_G} \hat{H}_{1} \hat{G}_{(\omega+E_G + i \gamma)}^0 \hat{H}_{-1} \hat{\mathcal{P}}_{E_G} \vert G \rangle. 
\end{split}
\end{equation}
Here $\gamma$ is a tiny positive number broadening the absorbance spectrum, $\vert G \rangle$ is the ground state of the static Hamiltonian $\hat{H}$ with eigenenergy $E_G$, and in the second line we use $\hat{\mathcal{P}}_{E_G} \vert G \rangle=\vert G \rangle$ which follows from the definition of the projector $\hat{\mathcal{P}}_{E_G} = \delta( E_G - \hat{H}_0 ) $. Compared with Eq.~(\ref{Heff_weak_drive}), the same Floquet low-energy Hamiltonian structure $\hat{\mathcal{P}}\hat{H}_{1}\hat{G}\hat{H}_{-1}\hat{\mathcal{P}}$ appears in $\alpha(\omega)$ in Eq.~(\ref{eq.absorbance}). This means that, to determine the absorbance at the driving frequency $\alpha(\omL)$, we just need to calculate the ground-state expectation value of (the RWA part of) our Floquet Hamiltonian.

When the driving frequency $\omL$ becomes large enough so that $\Delta_{{\bf q},s}$ approaches 0 (at an arbitrary momentum $\bf q$ with finite lower-band population) in Eq.~(\ref{renormalised-denominator}), the screened optical Stark shift in $\hat{\mathcal{P}}\hat{H}_{1}\hat{G}\hat{H}_{-1}\hat{\mathcal{P}}$ will diverge (because its strength is inversely proportional to $\Delta_{{\bf q},s}$). This divergence of our Floquet Hamiltonian propagates into its ground-state expectation value, $\langle G \vert \hat{\mathcal{P}}\hat{H}_{1}\hat{G}\hat{H}_{-1}\hat{\mathcal{P}} \vert G \rangle $. 
Thus, the absorbance spectrum $\alpha(\omL)$ in Eq.~(\ref{eq.absorbance}) will also peak (or diverge if we take $\gamma \to 0$). An optical absorption peak at in-gap frequency indicates the exciton resonance.
Consequently, the excitonic resonance frequency $\omega_{\text{ex}}$ is the smallest $\omL$ for which the screened denominator $\Delta_{{\bf q},s}$ equals to 0.

\section{Diagrammatic GRPA calculation}
\label{sec.GRPA}

Below we compare our screened Floquet Hamiltonian~(\ref{eq.main-result}) with an alternative method using the Matsubara formalism, where we construct Feynman diagrams in the laser-rotating frame. We move to the laser-rotating frame by applying the following unitary transformation to the original driven Hamiltonian (\ref{H-dip-full-terms}), $H(t) \to U_{t} H(t) U_{t}^{\dag} + i\hbar (\partial_{t} U_{t})U^{\dag}_{t}$, and $\vert\psi\rangle_{t} \to U_{t}\vert\psi\rangle_{t}$, where
\begin{equation} \label{eq.U_t}
    U_{t}=e^{i \omega_{L} t (\hat{a}^{\dag}\hat{a} + \sum\limits_{\mathbf{q}, s} \hat{c}_{\mathbf{q} 2 s}^{\dag} \hat{c}_{\mathbf{q} 2 s} )}.
\end{equation}
Then, after discarding the counter-rotating terms (i.e applying the RWA in the laser-matter interaction), the dipolar Hamiltonian becomes static in the rotating frame, 
\begin{equation}\label{H_Dip_r}
\begin{split}
\hat{H}_{\text {dip}}^{\text{rot}} &= \sum_{\mathbf{q}, s} \big( \epsilon_{\mathbf{q}, 1} -\mu \big) \hat{n}_{\mathbf{q} 1 s} + \big( \epsilon_{\mathbf{q}, 2} -\mu - \omL \big) \hat{n}_{\mathbf{q} 2 s} + (\omega_c-\omL) ~ \hat{a}^{\dag}\hat{a} \\
& +  \sum_{\mathbf{q}, s}   ~ (g +  g_{c} \hat{a} ) \hat{c}_{\mathbf{q} 2 s}^{\dag} \hat{c}_{\mathbf{q} 1 s}+\text { h.c. }   \\
& + \hat{U}_{11} + \hat{U}_{22} + \hat{U}_{12}.
\end{split}
\end{equation}
    \begin{figure}[t]
    \centering
        \includegraphics[width=0.4\textwidth]{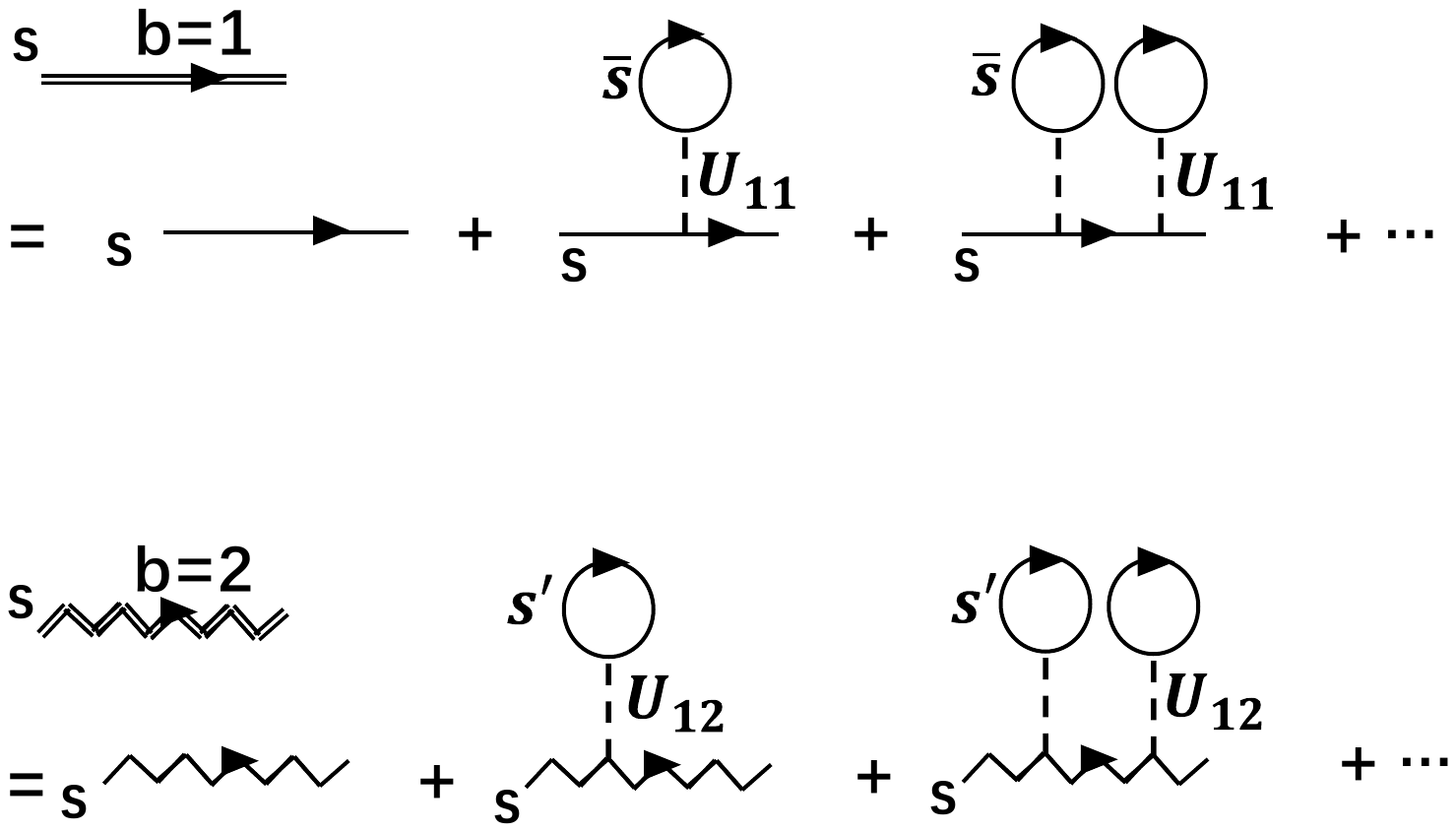}
        \captionsetup{justification=raggedright,singlelinecheck=false}
        \caption{The dressed single-particle propagators of two bands in the rotating frame. Note that the b=1 line will be used as the \textit{hole} propagator in the lower-band. The Hartree contribution of $U_{22}$ term is ignored in the b=2 line, for the reason discussed below Eq.~(\ref{U-b-commutator-reduced}).}
        \label{fig:single-particle-propagator}
    \end{figure}
   \begin{figure}[t]
   \centering
        \includegraphics[width=0.4\textwidth]{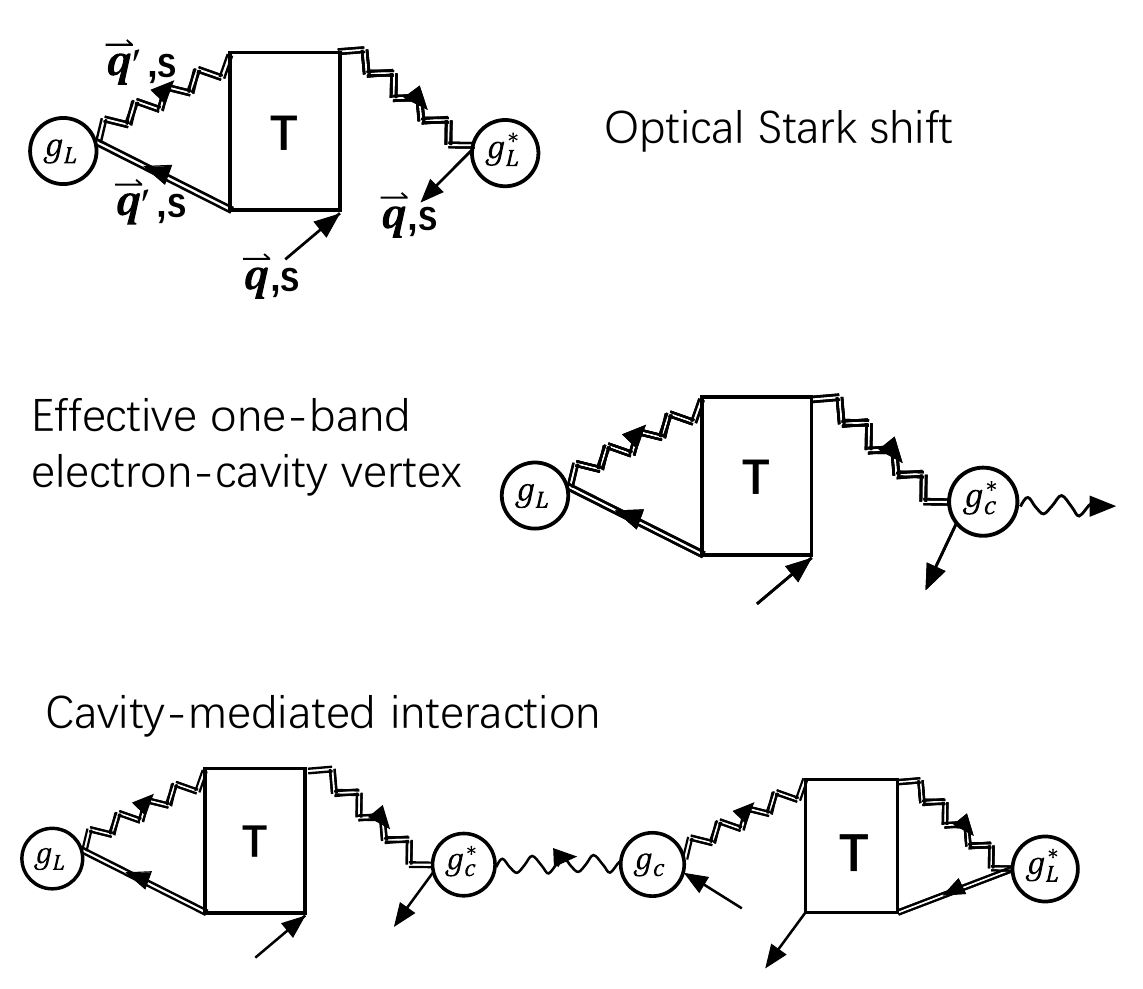}
        \captionsetup{justification=raggedright,singlelinecheck=false}
        \caption{The renormalised optical Stark shift, effective one-band electron-cavity vertex, and cavity-mediated interaction in the rotating frame. The wavy arrow denotes the cavity photon. The incoming and outgoing straight arrows denote the lower-band electron.
        %, while the upper-band hole is not shown here for simplicity.
        }
        \label{fig:renormalisation-by-T}
    \end{figure}
We next apply the usual Feynman diagrammatic approach to this Hamiltonian to study its effective low-energy response. 
We will sum over a series of Feynman diagrams in the irreducible two-particle vertex, which in the static limit gives the equivalent result to Eq.~(\ref{renormalised-denominator}). As shown in Fig.~\ref{fig:single-particle-propagator}, the electron and hole propagators in the Hamiltonian~(\ref{H_Dip_r}) are first dressed by the Hartree terms. The dressed propagators for the lower- and upper-band respectively read
\begin{equation}
    \begin{split}
        \mathcal{G}_{1,{\bf q},s,i q_n} &= \frac{1}{i q_n - (\epsilon_{{\bf q},1} -\mu + U_{11} \nu_{1,\Bar{s}}) } \\
        \mathcal{G}_{2,{\bf q},s,i q_n} &= \frac{1}{i q_n - (\epsilon_{{\bf q},2} - \omL -\mu + U_{12} \sum\limits_{s'}\nu_{1,s'}) }
    \end{split}
\end{equation}
where the fermionic Matsubara Frequency is defined as $q_n = \frac{(2n+1)\pi}{\beta}$ for all integer $n \in (-\infty,\infty)$. At low temperature we assume $\beta\to\infty$. The asymmetry between $\mathcal{G}_{1}$ and $\mathcal{G}_{2}$ arises from the fact that the upper-band is empty, so that $\nu_{2,s}=0$. The Fock self-energy disappears in these single particle propagators, because we only consider on-site electron-electron interactions.

\begin{figure}
    \centering
        \includegraphics[width=0.4\textwidth]{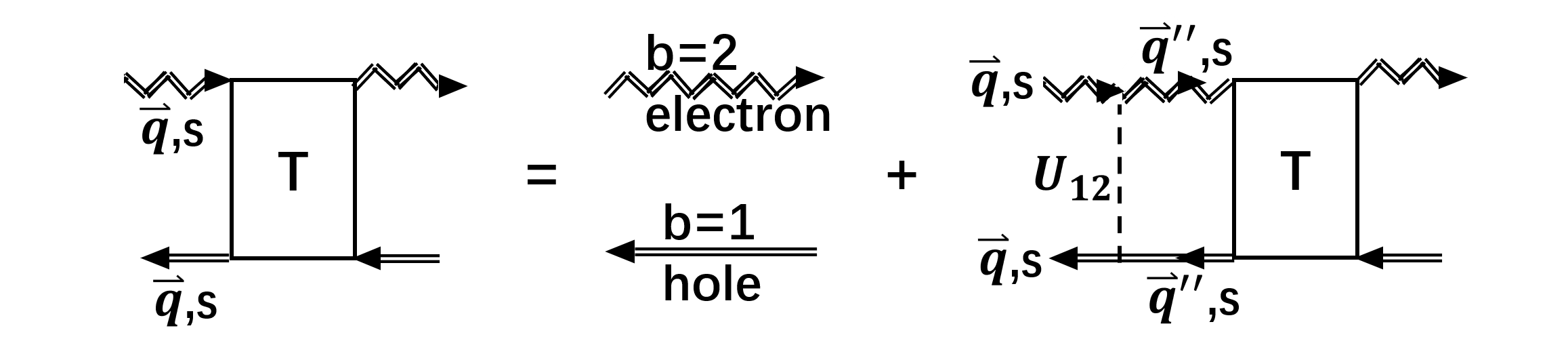}
        \captionsetup{justification=raggedright,singlelinecheck=false}
        \caption{The electron-hole t-matrix, which describes the attraction between the upper-band electron and the lower-band hole. The Matsubara frequency is not shown for simplicity.}
        \label{fig:particle-hole-transfer-matrix}
    \end{figure}

The effective attraction between the screened electrons and holes, represented by the so-called GRPA polarisation diagrams \cite{PhysRevB.40.3802}, are shown in Fig.~\ref{fig:renormalisation-by-T}. All of these diagrams contain an opened polarization bubble~\cite{PhysRevB.80.174401} with an inter-band electron-hole t-matrix~\cite{zimmermann1985mass}, as shown in Fig.~\ref{fig:particle-hole-transfer-matrix}. It reads
 
\begin{equation}
    \begin{split}
        T &= \sum\limits_{n=0}^{\infty} \bigg( \frac{U_{12}}{\beta N} \sum\limits_{{\bf q''},i q_n} \frac{1}{i q_n - (\epsilon_{{\bf q''},1} -\mu + U_{11} \nu_{1,\Bar{s}}) } \frac{1}{i q_n - (\epsilon_{{\bf q''},2} - \omL -\mu + U_{12} \sum\limits_{s'}\nu_{1,s'}) } \bigg)^n \\
        &= \sum\limits_{n=0}^{\infty} \bigg( \frac{U_{12}}{ N} \sum\limits_{{\bf q''} } \frac{- n_F( \epsilon_{{\bf q''},1} -\mu + U_{11} \nu_{1,\Bar{s}} ) + n_F(\epsilon_{{\bf q''},2} - \omL -\mu) }{ (\epsilon_{{\bf q''},1} -\mu + U_{11} \nu_{1,\Bar{s}}) - (\epsilon_{{\bf q''},2} - \omL -\mu + U_{12} \sum\limits_{s'}\nu_{1,s'} )  } \bigg)^n \\
        &= \sum\limits_{n=0}^{\infty} \bigg( -\frac{U_{12}}{ N} \sum\limits_{{\bf q''} } \frac{ \langle \hat{n}_{{\bf q}'',1,s} \rangle }{ - \Delta_{\bf q''}^0 + U_{11} \nu_{1,\Bar{s}}
        - U_{12} \sum\limits_{s'}\nu_{1,s'}  } \bigg)^n \\
        &= \bigg(1 + \frac{U_{12}}{ N} \sum\limits_{{\bf q''} } \frac{ \langle \hat{n}_{{\bf q}'',1,s} \rangle }{ - \Delta_{\bf q''}^0 + U_{11} \nu_{1,\Bar{s}}
        - U_{12} \sum\limits_{s'}\nu_{1,s'}  } \bigg)^{-1} \\
    \end{split}
\end{equation}
where the Fermi distribution function is defined as $n_F(x) \equiv 1/(1+\exp(\beta x))$. In the third line, a geometric series similar to the infinite sum in Eq.~(\ref{k-space-GRPA-collection}) appears. Note that in the non-interacting model where $U_{12}\to0$, this electron-hole t-matrix reduces to unity $T\to1$, which is understood as a two-particle delta function (for both the upper-band electron line and the lower-band hole line). Meanwhile, note that when the exciton resonance requirement (\ref{exciton-resonance}) is fulfilled, it diverges, $T\to\infty$.

The $t$-matrix allows us to calculate the renormalised optical Stark shift, the effective one-band electron-cavity vertex, and the cavity-mediated interaction (but not the Bloch-Siegert shift). The expressions are represented by the Feynman diagrams in Fig.~\ref{fig:renormalisation-by-T}.
For example, the closed GRPA bubble for the optical Stark shift in Fig.~\ref{fig:renormalisation-by-T}, contributes to a self-energy term for the lower-band electron, whose value at the 4-momentum ${\bf q},i q_n$ evaluates to 
\begin{equation}
    \begin{split}
        \Sigma_{{\bf q}, iq_n} ~ = ~ \vert g \vert^2  \frac{1}{i q_n - (\epsilon_{{\bf q},2} - \omL -\mu + U_{12} \sum\limits_{s'}\nu_{1,s'}) } T ,
    \end{split}
\end{equation}
and thus the laser-dressed propagator of the lower-band electron reads
\begin{equation}
    \begin{split}
        & \mathcal{G}_{1,{\bf q},s,i q_n}^{\text{Fl}} = \frac{1}{\mathcal{G}_{1,{\bf q},s,i q_n}^{-1} - \Sigma_{{\bf q}, iq_n} }  \\
        &=    \frac{1}{i q_n - (\epsilon_{{\bf q},1} -\mu + U_{11} \nu_{1,\Bar{s}}  +   \frac{ \vert g \vert^2 T}{i q_n - (\epsilon_{{\bf q},2} - \omL -\mu + U_{12} \sum\limits_{s'}\nu_{1,s'}) }   ) }   
    \end{split}
\end{equation}
which has a pole at
 
\begin{equation}
    \begin{split}
        i q_n & \to \epsilon_{{\bf q},1} -\mu + U_{11} \nu_{1,\Bar{s}}  +   \frac{ \vert g \vert^2 T}{i q_n - (\epsilon_{{\bf q},2} - \omL -\mu + U_{12} \sum\limits_{s'}\nu_{1,s'}) }     \\
        &\approx  \epsilon_{{\bf q},1} -\mu + U_{11} \nu_{1,\Bar{s}}  +   \frac{ \vert g \vert^2 T}{(\epsilon_{{\bf q},1} -\mu + U_{11} \nu_{1,\Bar{s}}) - (\epsilon_{{\bf q},2} - \omL -\mu + U_{12} \sum\limits_{s'}\nu_{1,s'}) } \\
        &=  \epsilon_{{\bf q},1} -\mu + U_{11} \nu_{1,\Bar{s}}  
        - \frac{ \vert g \vert^2 }
        { \big(\Delta_{\bf q}^0  - U_{11} \nu_{1,\Bar{s}}  + U_{12} \sum\limits_{s'}\nu_{1,s'}\big)
        \bigg(1 + \frac{U_{12}}{ N} \sum\limits_{{\bf q''} } \frac{ \langle \hat{n}_{{\bf q}'',1,s} \rangle }{ - \Delta_{\bf q''}^0 + U_{11} \nu_{1,\Bar{s}}
        - U_{12} \sum\limits_{s'}\nu_{1,s'}  } \bigg) }  \\
        &=  \epsilon_{{\bf q},1} -\mu + U_{11} \nu_{1,\Bar{s}}  - \frac{ \vert g \vert^2 }{ \Delta_{{\bf q},s}  }
    \end{split}
\end{equation}
which exactly reveals the lower-band's energy shift given by the screened optical Stark effect at spin-orbital $({\bf q},s)$ in Eq.~(\ref{eq.h_eff}). Compared with the optical Stark shift in the non-interacting case, an excitonic enhancement with factor $T$ is observed.
This shows that the renormalised denominator $\Delta_{{\bf q },s}$, which appears in the renormalized optical Stark shift and the cavity-mediated interaction in the effective Floquet Hamiltonian~(\ref{eq.main-result}), corresponds to the GRPA graphs in the retarded interaction formalism. 

    \begin{figure}[h]
    \centering
        \includegraphics[width=0.45\textwidth]{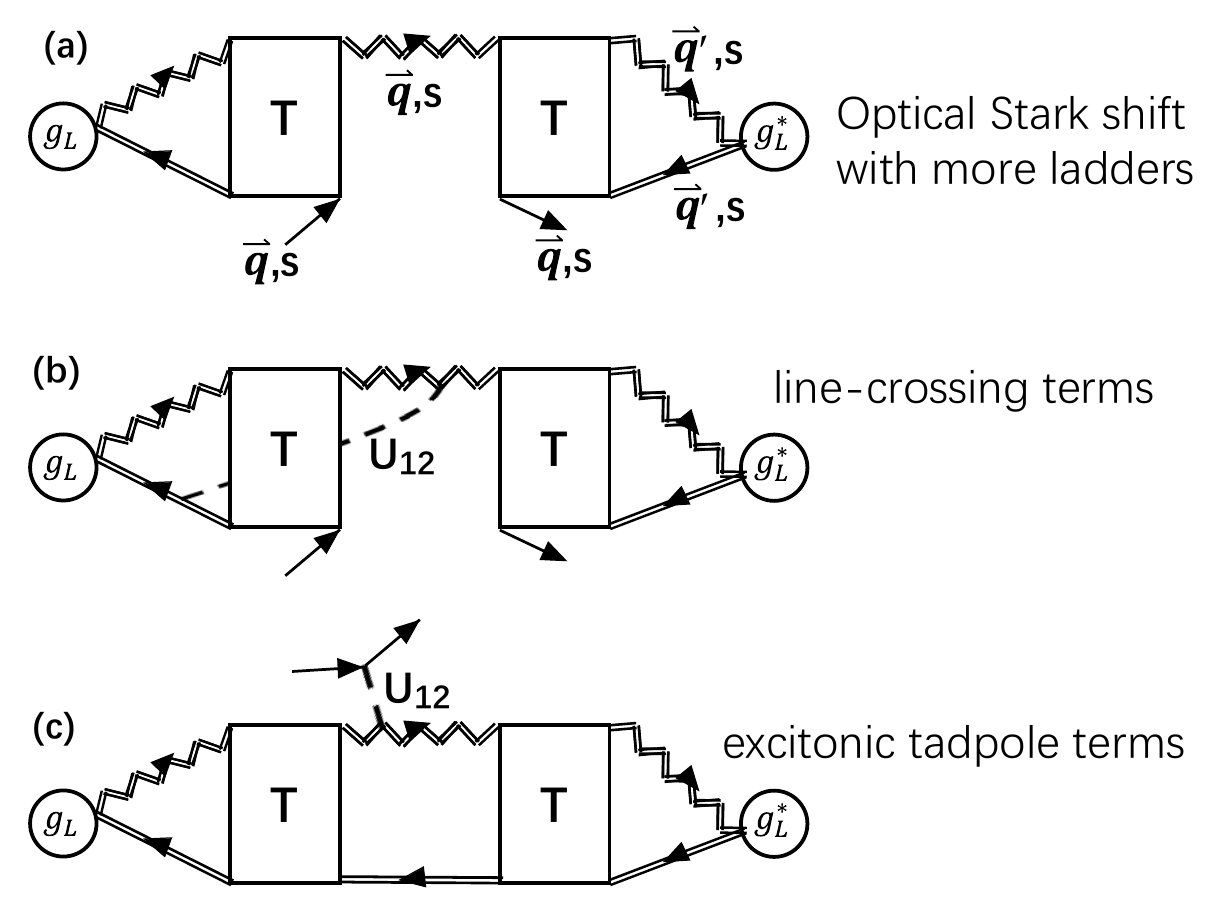}
        \captionsetup{justification=raggedright,singlelinecheck=false}
        \caption{The additional Ladder diagrams for the self-energy in the lower-band, which are not included in our result. Merely including the analytically tractable diagrams in (a) will give non-physical result, $(\Delta_{{\bf q},s})^{-1} \propto T^2$.}
        \label{fig:more-Ladder-graphs}
    \end{figure}

One can include more ladder diagrams in the self-energy calculation, e.g. as shown in Fig.\ref{fig:more-Ladder-graphs}(a), however, this would end up in a $T^2$ enhancement on the optical Stark shift $\Delta_{{\bf q},s}$ showing several nonphysical features: this enhancement is not consistent with the exact result in the flat-bandgap limit (see the right 3 panels in Fig.~\ref{fig.denominator}), and also, it restricts the sign of the detuning $\Delta_{{\bf q},s}$ to be non-negative even when $\omL > \omega_{ex}$. As discussed in Ref.~\cite{PhysRevB.40.3788}, in order to include all possible ladder diagrams, one must simultaneously include other diagrams, e.g. the ``line-crossing" graphs in Fig.\ref{fig:more-Ladder-graphs}(b) and the ``excitonic tadpole" graphs in Fig.\ref{fig:more-Ladder-graphs}(c), which could cancel the nonphysical effects, but are analytically uneasy to tract. In our Green operator approach, the $T^2$ enhancement can be reproduced by a consecutive mean-field decoupling on Eq.~(\ref{k-space-GRPA-Stark-middle}), but before including these additional terms, we must first retain more terms in the commutator relation Eq.~(\ref{U-b-commutator-appendix}), so as to keep the resulting Floquet Hamiltonian valid in the flat-bandgap limit. This improvement is left for future study.

\section{The self-consistent derivation of $\vert g \vert^4$ order Floquet Hamiltonian}
\label{appendix:self-consistent-gL4-Hamiltonian}

\subsection{
The spurious Floquet-induced interaction in the in $\vert g \vert^4$ order Floquet Hamiltonian}\label{subsec:leading-gL4}
All of the driving-induced effects studied in Chapter \ref{Chapter4} are to the second order of driving strength $\vert g \vert^2$ within the approximation (\ref{Heff_weak_drive}). Next we go beyond (\ref{Heff_weak_drive}) to analyse the leading effect in the fourth order $\vert g \vert^4$, as these will be of similar size as the cavity-mediated interaction described above. According to Eq.~(\ref{Heff}), this leading term comes from the expansion $\hat{H}_{1} \hat{G}^{0} \hat{H}_{1} \hat{G}^{0} \hat{H}_{-1} \hat{G}^{0} \hat{H}_{-1}$, which reads (we assume the absence of electron interaction $\hat{U}=0$)
\begin{equation}\label{leading-gL4}
    \begin{split}
        &\PEa \hat{H}_{1} \hat{g}_{(\Ea+\omL)} \hat{H}_{1} \hat{g}_{(\Ea+2\omL)} \hat{H}_{-1} \hat{g}_{(\Ea+\omL)} \hat{H}_{-1} \PEa \\
        &\approx \sum_{{\bf k}, s, {\bf k'}, s'} \frac{-\vert g \vert^4 }{ (\Delta_{\bf k}^0)^2 \Delta_{\bf k'}^0  } 
        \hat{c}_{{\bf k'} 1 s'} ^{\dag}  \hat{c}_{{\bf k'} 1 s'}
        \hat{c}_{{\bf k} 1 s}^{\dag}  \hat{c}_{{\bf k} 1 s} 
    \end{split}
\end{equation}
This expression holds only after excluding the case when $({\bf k} , s) = ({\bf k'}, s')$ in the summation, for the same reason explained below Eq.~(\ref{non-RWA-cavity-med-interaction-U=0}). In the derivation of Eq.~(\ref{leading-gL4}), we use (\ref{g-commutator}) to move all three $\hat{g}$ to the right, and then use (\ref{G1-reduced-to-number}) to reduce them into denominators. In the virtual process described in Eq.~(\ref{leading-gL4}), two upper-band electrons are sequentially excited by two $\hat{H}_{-1}$, and then annihilated by two $\hat{H}_{1}$. The resulting Floquet-induced interaction (\ref{leading-gL4}) has the same form as the interaction~(\ref{off-resonant-leading-order-induced-int}), but a much larger coefficient. Below we explain why Eq.~(\ref{leading-gL4}) is a spurious Floquet-induced interaction, which hinders the application of the Gaussian elimination method in understanding the higher order driving effects in Chapter \ref{Chapter4}.
%The coefficient in (\ref{leading-gL4}) is comparable to the coefficient of the cavity-mediated interaction in (\ref{cavity-meidated-interaction-U=0}), but these two driving-induced effects cannot completely cancel with each other, as the electron cannot transfer momentum in (\ref{leading-gL4}).

First, note that the interaction term (\ref{leading-gL4}) is not a unique result of our Floquet method: In the alternative rotating-frame RWA method in appendix \ref{sec.GRPA}, this leading $\vert g \vert^4$ term also appears, as long as a projector-based perturbation technique is used to derive the low-energy effective Hamiltonian. Next, we explain the appearance of (\ref{leading-gL4}) in our Floquet formalism.

%At mean-field level, this term (\ref{leading-gL4}) renormalises the lower-band's dispersion energy. Its influence is much weaker than the optical stark shift (\ref{Stark}) when $ \mathcal{N} \vert g \vert^2 \ll (\Delta_{\bf k})^2 $, where $\mathcal{N}$ is the total electron number in the lower-band. Note that for fixed filling factor, $\mathcal{N}$ is proportional to the area of the material driven by laser, thus $\mathcal{N} \vert g \vert^2$ is proportional to the power of the driving laser. When the laser power becomes large enough so that $ \mathcal{N} \vert g \vert^2 \sim (\Delta_{\bf k})^2 $, the $\vert g \vert^4$ order effect (\ref{leading-gL4}) becomes comparable to the $\vert g \vert^2$ effects (for example the optical Stark shift), then the expression of the Floquet Hamiltonian in (\ref{Heff}) diverges.\footnote{In the alternative rotating-frame RWA method in appendix \ref{appendix:FeynmanDiagramGRPACalc}, the same divergence problem appears when $ \mathcal{N} \vert g \vert^2 \sim (\Delta_{\bf k})^2 $, thus this divergence is not a unique problem of our Floquet method. In fact, this strong-driving divergence doesn't correspond to any physical phase transition, it is just an nonphysical divergence of the many-body perturbation theory.} 

Terms like (\ref{leading-gL4}) and (\ref{off-resonant-leading-order-induced-int}) are interactions mediated by Floquet ``photons". These interactions still appear in the Floquet Hamiltonian, even though the undriven many-body system is non-interacting. This strange behaviour is thoroughly explained in Ref.~\cite{PhysRevB.93.144307}: %, as briefly reviewed below: 
The interaction (\ref{leading-gL4}) generates electron correlations in the eigenstate $\vert \alpha,0 \rangle$ of our Floquet Hamiltonian $\hat{H}^{\text{eff}}_{(\Ea)}$. However, this correlation in $\vert \alpha,0 \rangle$ arises merely because we map the many-body dynamics onto the 0-th Floquet sector. If we re-construct all $\vert \alpha,n \rangle$ from $\vert \alpha,0 \rangle$ according to %their relation 
Eq.~(\ref{Sambe}), we will find that the electrons in the $\alpha$-th Floquet state $\vert \alpha \rangle_{(t)}$ described in (\ref{eigen-oscillation}) are again not correlated. In other words, in a driven many-body non-interacting system, it is the interaction (\ref{leading-gL4}) in the 0-th Floquet sector that prevents our Floquet method from predicting nonphysical electron-correlations in the real-time evolution.

If we treat the Sambe space eigenvalue problem (\ref{Sambe}) as a gauge theory with gauge freedom (\ref{gauge-freedom}), whose gauge fixing process is described by (\ref{gauge_requirement}), then the generated interaction (\ref{leading-gL4}) plays a similar role as the gauge ghost~\cite{aguilar2008gluon}: a gauge ghost cannot be physically observed, but it cancels the nonphysical effects arising from the gauge fixing process. 

In addition to (\ref{leading-gL4}), there is another leading term in $\vert g \vert^4$ order, which appears as we take into account the self-consistent change of $\Ea$ in Eq.~(\ref{Heff_weak_drive}), as shown below. Thus in total there are two leading terms in $\vert g \vert^4$ order, and their divergent behaviour cancel with each other. 

% This divergence also exists in the alternative many-body method (i.e. perturbing the static many-body RWA Hamiltonian in the rotating frame)

\subsection{Curing the issue within the Sambe space Gaussian elimination formalism remains challenging}
Beyond the high-frequency limit, the shift of the eigenvalue $\Ea$ must also be self-consistently tracked \& recorded when deriving the Floquet Hamiltonian $\hat{H}^{\text{eff}}_{(\Ea)}$. We start from the self-consistent eigenvalue problem (\ref{0-HarmonicEigenProblem}) where $\hat{H}^{\text{eff}}_{(\Ea)}$ takes the form (\ref{Heff}). In a self-consistent order-expansion of the eigenvalue problem (\ref{0-HarmonicEigenProblem}) over $\vert g \vert$, we have
\begin{equation}\label{self-consistent-expansion-gL4}
\begin{split}
\hat{H}^{\text{eff}}_{(\Ea)} &= \hat{H}_{0} + \vert g \vert^2 \hat{H}^{(2)} + \vert g \vert^4 \hat{H}^{(4)} + O[\vert g \vert^6] \\
\Ea &= E_{\alpha}^{(0)} + \vert g \vert^2 E_{\alpha}^{(2)} + \vert g \vert^4 E_{\alpha}^{(4)} + O[\vert g \vert^6] \\
\vert \alpha,0 \rangle &= \vert \alpha,0 \rangle^{(0)} + \vert g \vert^2 \vert \alpha,0 \rangle^{(2)} + \vert g \vert^4 \vert \alpha,0 \rangle^{(4)} + O[\vert g \vert^6]
\end{split}
\end{equation}
where $\Ea$, $E_{\alpha}^{(0)}$ and $\vert g \vert$ have the dimension of energy, but $E_{\alpha}^{(2)}$ has the dimension of one-over-energy. The reason why $\vert g \vert^{1}$ order disappear in this expansion (\ref{self-consistent-expansion-gL4}) is that, in the expression of $\hat{H}^{\text{eff}}_{(\Ea)}$ in (\ref{Heff}), the driving operator $\hat{H}_{-1}\sim \vert g \vert$ will always come in pair with its Hermitian conjugate $\hat{H}_{1}\sim \vert g \vert$, thus $\hat{H}^{\text{eff}}_{(\Ea)}$ only contains perturbations to the even orders of $\vert g \vert$. In future work, if we work in Coulomb gauge and take into account the diamagnetic coupling, the $e^{\pm i 2\omL t}$ driving frequency will appear in (\ref{H_Dip_newform}), which will make the structure of (\ref{self-consistent-expansion-gL4}) more complex.

Expanding $\hat{H}^{\text{eff}}_{(\Ea)}$ to the $\vert g \vert^4$ order, we have
\begin{equation}\label{self-consistent-expansion-Heff}
\begin{split}
\hat{H}^{\text{eff}}_{(\Ea)} &= \hat{H}_0 ~ + ~ \hat{H}_{1} \frac{1}{\Ea-\hat{H}_0 +\omL-\hat{H}_{1} \frac{1}{\Ea-\hat{H}_0 +2 \omL-\ldots} \hat{H}_{-1}} \hat{H}_{-1} \\
& ~~~~~~~~~ + ~ 
\hat{H}_{-1} \frac{1}{\Ea-\hat{H}_0 -\omL-\hat{H}_{-1} \frac{1}{\Ea-\hat{H}_0 -2 \omL-\ldots} \hat{H}_{1}} \hat{H}_{1} \\
&\approx  \hat{H}_0 + \hat{H}_{1} \frac{1}{E_{\alpha}^{(0)} + \vert g \vert^2 E_{\alpha}^{(2)} +\omL -\hat{H}_0  }  \hat{H}_{-1}   + \hat{H}_{-1} \frac{1}{E_{\alpha}^{(0)} + \vert g \vert^2 E_{\alpha}^{(2)}  -\omL -\hat{H}_0 }  \hat{H}_{1}  \\
&~~~~~~~~+ \hat{H}_{1} \frac{1}{E_{\alpha}^{(0)} + \omL -\hat{H}_0  }  \hat{H}_{1} \frac{1}{E_{\alpha}^{(0)} + 2\omL -\hat{H}_0  }  \hat{H}_{-1} \frac{1}{E_{\alpha}^{(0)} + \omL -\hat{H}_0  }  \hat{H}_{-1} \\
&~~~~~~~~+ \hat{H}_{-1} \frac{1}{E_{\alpha}^{(0)} - \omL -\hat{H}_0  }  \hat{H}_{-1} \frac{1}{E_{\alpha}^{(0)} - 2\omL -\hat{H}_0  }  \hat{H}_{1} \frac{1}{E_{\alpha}^{(0)} - \omL -\hat{H}_0  }  \hat{H}_{1} \\
&\approx \hat{H}_0 + \hat{H}_{1} \frac{1}{E_{\alpha}^{(0)} +\omL -\hat{H}_0  }  \hat{H}_{-1} + \hat{H}_{1} \frac{-\vert g \vert^2 E_{\alpha}^{(2)}}{\big(E_{\alpha}^{(0)} +\omL -\hat{H}_0\big)^2  }  \hat{H}_{-1}  \\
& ~~~~~~~~
+ \hat{H}_{1} \frac{1}{E_{\alpha}^{(0)} + \omL -\hat{H}_0  }  \hat{H}_{1} \frac{1}{E_{\alpha}^{(0)} + 2\omL -\hat{H}_0  }  \hat{H}_{-1} \frac{1}{E_{\alpha}^{(0)} + \omL -\hat{H}_0  }  \hat{H}_{-1} \\
& ~~~~~~~~ + \hat{H}_{-1} \frac{1}{E_{\alpha}^{(0)}  -\omL -\hat{H}_0 }  \hat{H}_{1} + \hat{H}_{-1} \frac{-\vert g \vert^2 E_{\alpha}^{(2)}}{\big(E_{\alpha}^{(0)} -\omL -\hat{H}_0\big)^2  }  \hat{H}_{1} \\
& ~~~~~~~~
+ \hat{H}_{-1} \frac{1}{E_{\alpha}^{(0)} - \omL -\hat{H}_0  }  \hat{H}_{-1} \frac{1}{E_{\alpha}^{(0)} - 2\omL -\hat{H}_0  }  \hat{H}_{1} \frac{1}{E_{\alpha}^{(0)} - \omL -\hat{H}_0  }  \hat{H}_{1} \\
\end{split}
\end{equation}
where in the first line we insert the formula (\ref{Heff}) for $\hat{H}^{\text{eff}}_{(\Ea)}$. In the second line we expand to the $\vert g \vert^4$ order, and then we insert the expansion of $\Ea$ in (\ref{self-consistent-expansion-gL4}). In the third line we use Taylor expansion to move $E_{\alpha}^{(2)}$ outside the denominator. Comparing (\ref{self-consistent-expansion-Heff}) with (\ref{self-consistent-expansion-gL4}), we identify $\hat{H}^{(2)}$ and $\hat{H}^{(4)}$ as
\begin{equation}\label{self-consistent-expansion-H2}
\begin{split}
\vert g \vert^2 \hat{H}^{(2)} &= \hat{H}_{1} \frac{1}{E_{\alpha}^{(0)} + \omL -\hat{H}_0  } \hat{H}_{-1} + \hat{H}_{-1} \frac{1}{E_{\alpha}^{(0)} - \omL -\hat{H}_0  } \hat{H}_{1}
\end{split}
\end{equation}
\begin{equation}\label{self-consistent-expansion-H4}
\begin{split}
\vert g \vert^4 \hat{H}^{(4)} &= \hat{H}_{1} \frac{- \vert g \vert^2 E_{\alpha}^{(2)}}{\big( E_{\alpha}^{(0)} + \omL -\hat{H}_0 \big)^{2}  } \hat{H}_{-1} + \hat{H}_{1} \frac{1}{E_{\alpha}^{(0)} + \omL -\hat{H}_0  } \hat{H}_{1}  \frac{1}{E_{\alpha}^{(0)} + 2\omL -\hat{H}_0  } \hat{H}_{-1} \frac{1}{E_{\alpha}^{(0)} + \omL -\hat{H}_0  } \hat{H}_{-1} \\
&+ \hat{H}_{-1} \frac{-\vert g \vert^2 E_{\alpha}^{(2)}}{\big( E_{\alpha}^{(0)} - \omL -\hat{H}_0 \big)^{2}  } \hat{H}_{1} + \hat{H}_{-1} \frac{1}{E_{\alpha}^{(0)} - \omL -\hat{H}_0  } \hat{H}_{-1}  \frac{1}{E_{\alpha}^{(0)} - 2\omL -\hat{H}_0  } \hat{H}_{1} \frac{1}{E_{\alpha}^{(0)} - \omL -\hat{H}_0  } \hat{H}_{1} \\ 
\end{split}
\end{equation}
We see that, $\hat{H}^{(2)}$ is nothing but our lowest-order Floquet Hamiltonian in (\ref{Heff_weak_drive}). In $\hat{H}^{(4)}$, the term $\hat{H}_{1} \hat{G}^{0}\hat{H}_{1} \hat{G}^{0}\hat{H}_{-1} \hat{G}^{0}\hat{H}_{-1}$ gives rise to the spurious Floquet-induced interaction in (\ref{leading-gL4}). 

However, in $\hat{H}^{(4)}$, the term $-E_{\alpha}^{(2)}\hat{H}_{1} \big(\hat{G}^{0}\big)^2 \hat{H}_{-1}$ cannot be derived by our projector-based technique developed in the main text. This is because the energy filter $\PEa$ introduced in Section \ref{appendix-filter}  cannot capture the self-consistent change of $\Ea$ in the eigenvalue problem (\ref{0-HarmonicEigenProblem}). Consequently, the current method based on $\PEa$ cannot capture any $\vert g \vert^4$ term which contains the self-consistent energy shift $E_{\alpha}^{(2)}$. In future works, this drawback could possibly be fixed by choosing a more sophisticated many-body projector than our energy-filter $\PEa$: For example in Ref.~\cite{klein1974degenerate}, the many-body projectors used therein also undergo a self-consistent shift as the interaction strength is turned on.

The procedure of deriving $\hat{H}^{(4)}$ is: First, use the 0-th order of the self-consistent eigenvalue problem (\ref{0-HarmonicEigenProblem})
$$ \hat{H}_{0} \vert \alpha,0 \rangle^{(0)} = E_{\alpha}^{(0)} \vert \alpha,0 \rangle^{(0)}$$
to decide the 0-th order eigenvalue $E_{\alpha}^{(0)}$. Then, use $E_{\alpha}^{(0)}$ to construct $\hat{H}^{(2)}$ via (\ref{self-consistent-expansion-H2}). Then, use the 2-nd order of the eigenvalue problem (\ref{0-HarmonicEigenProblem}) 
$$ \big(\hat{H}_{0} + \vert g \vert^2 \hat{H}^{(2)} \big) \big( \vert \alpha,0 \rangle^{(0)} + \vert g \vert^2 \vert \alpha,0 \rangle^{(2)} \big) 
= \big( E_{\alpha}^{(0)} + \vert g \vert^2 E_{\alpha}^{(2)} \big)  \big( \vert \alpha,0 \rangle^{(0)} + \vert g \vert^2 \vert \alpha,0 \rangle^{(2)} \big)  $$
to decide the self-consistent energy shift $E_{\alpha}^{(2)}$. At last, construct $\hat{H}^{(4)}$ by inserting $E_{\alpha}^{(2)}$ into (\ref{self-consistent-expansion-H4}). When $\hat{H}_{0}$ describes a many-body system, the exact diagonalization of the above two eigenvalue problems becomes impossible. Thus we need to introduce a many-body projector, which not only circumvents the need of exact diagonalization (this circumvention is achieved by $\PEa$ in the main text), but also undergoes self-consistent shift. We leave this projector-based self-consistent derivation of $\hat{H}^{(4)}$ for future works, and stick to the $\vert g \vert^2$ effects in which no ``gauge ghost effect" will be encountered.

At last we note that, for a many-body system with $\mathcal{N}$ electrons in the lower-band, the ``Floquet-induced interaction" term (\ref{leading-gL4}) in $\hat{H}^{(4)}$, at mean-field level, is $\frac{\mathcal{N} \vert g \vert^2 }{(\Delta_2)^2}$-times stronger than the optical Stark shift (\ref{Stark}) in $\hat{H}^{(2)}$. In other words, the term $\hat{H}_{1} \hat{G}^{0}\hat{H}_{1} \hat{G}^{0}\hat{H}_{-1} \hat{G}^{0}\hat{H}_{-1}$ in $\hat{H}^{(4)}$ shows a divergence (proportional to the particle-number $\mathcal{N}$). Meanwhile, the other term $-E_{\alpha}^{(2)}\hat{H}_{1} \big(\hat{G}^{0}\big)^2 \hat{H}_{-1}$ in $\hat{H}^{(4)}$ also shows a divergence to the same order of particle-number, because $E_{\alpha}^{(2)}\propto \mathcal{N}$ is the shift of the many-body eigen-energy. Future works should look into the cancellation between these two divergences, in analogy to the similar cancellations in equilibrium quantum many-body systems \cite{shavitt_bartlett_2009,TAKAYANAGI2020168119}. 
%Cancellation-of-infinities has been well-understood in equilibrium quantum many-body systems \cite{shavitt_bartlett_2009,TAKAYANAGI2020168119}, where the disconnected diagrams (becoming divergent as $\mathcal{N}\to\infty$) in the numerator of the expectation value $Tr[\hat{\rho}\hat{O}]/\mathcal{Z}$ are exactly cancelled out by the partition function $\mathcal{Z}=Tr[\hat{\rho}]$ in the denominator. We expect similar behaviour of infinity-cancellation in our $\vert g \vert^4$ order Floquet Hamiltonian.

It remains challenging to cure these self-consistency-related issues within the Gaussian elimination Floquet method developed in Chapter \ref{Chapter4}. However, these issues are completely circumvented by the upgraded Floquet theory developed in Chapter \ref{Chapter5}.

\newpage

%% file: Chapters/appendix_B.tex
\section{The formal solution to the Sylvester equation} \label{linear-response}

Here, we formally solve the Sylvester equations in our FSWT, based on the following mathematical relation (see Theorem 9.2 in Ref.~\cite{bhatia1997and}): For 3 arbitrary operators $\hat{a}$, $\hat{b}$ and $\hat{c}$, where each eigenvalue of $\hat{a}$ has positive real part, and each eigenvalue of $\hat{b}$ is purely imaginary, the solution to the Sylvester equation $\hat{a} \hat{f} - \hat{f} \hat{b} = \hat{c} $ is given by
\begin{equation}\label{math-solution}
\begin{split}
\hat{f} &= - e^{-t \hat{a}} \hat{f} e^{t \hat{b}} \big\vert_{t=0}^{\infty} = - \int_{0}^{\infty} dt ~ \partial_t \big( e^{-t \hat{a}} \hat{f} e^{t \hat{b}} \big) \\
&= - \int_{0}^{\infty} dt ~ \partial_t \big( e^{-t \hat{a}} \big) \hat{f} e^{t \hat{b}} + e^{-t \hat{a}} \hat{f} \partial_t \big( e^{t \hat{b}} \big) \\
&=  \int_{0}^{\infty} dt ~   e^{-t \hat{a}}~ \hat{a} \hat{f} ~ e^{t \hat{b}} - e^{-t \hat{a}} ~ \hat{f} \hat{b} ~ e^{t \hat{b}}  \\
&= \int_{0}^{\infty} dt ~   e^{-t \hat{a}}~ \hat{c} ~ e^{t \hat{b}} 
\end{split}
\end{equation}
If we take $\hat{a} = -i ( \hat{H}^{(0)} + j \omL ) + 0^+ $, $\hat{b} = -i  \hat{H}^{(0)} $ and $\hat{c} = -i  \hat{H}^{(1)}_j $, we can identify the lowest-order Sylvester equation (\ref{formula-for-f_1^1}) with this form, and thus we can directly write down its solution $\hat{f}^{(1)}_j$ using Eq.~(\ref{math-solution}): $\forall j \neq 0$,  
\begin{equation}\label{formal-solu-f^1_j}
    \hat{f}_j^{(1)} = -i \int_0^{\infty} dt ~ e^{i j \omL t} ~ e^{- 0^+ t} ~ e^{i \hat{H}^{(0)} t} \hat{H}_j^{(1)}  e^{-i \hat{H}^{(0)} t}
\end{equation}

Eq.~(\ref{math-solution}) further provides the formal solution to the Sylvester equation at higher orders. For example, to solve the second-order equation (\ref{formula-for-f_1^2}), we take $\hat{a} = -i ( \hat{H}^{(0)} + j \omL ) + 0^+ $, $\hat{b} = -i  \hat{H}^{(0)} $ and $\hat{c} = -i  \hat{S}^{(2)}_j $ where 
\begin{equation*}
\hat{S}^{(2)}_j = \hat{H}^{(2)}_j + \frac{1}{2} \sum\limits_{j' \neq 0} [\hat{f}_{j'}^{(1)},\hat{H}_{j-j'}^{(1)}] + \frac{1}{2} [\hat{f}_j^{(1)},\hat{H}_0^{(1)}]
\end{equation*}
denotes the source term in the Sylvester equation (\ref{formula-for-f_1^2}). Eq.~(\ref{math-solution}) then gives, $\forall j \neq 0$,
\begin{equation}
\hat{f}^{(2)}_j = -i \int_0^{\infty} dt ~ e^{i j \omL t} ~ e^{- 0^+ t} ~ e^{i \hat{H}^{(0)} t} \hat{S}_j^{(2)}  e^{-i \hat{H}^{(0)} t}
\end{equation}
In general, once the $n$th order Sylvester equation is constructed from which source term $\hat{S}_j^{(n)}$ is identified, the formal solution is directly given by,  $\forall j \neq 0$,
\begin{equation}
\hat{f}^{(n)}_j = -i \int_0^{\infty} dt ~ e^{i j \omL t} ~ e^{- 0^+ t} ~ e^{i \hat{H}^{(0)} t} \hat{S}_j^{(n)}  e^{-i \hat{H}^{(0)} t}
\end{equation}
which always takes the form of a frequency-domain (i.e. Laplace-transformed) Heisenberg operator under the undriven Hamiltonian $\hat{H}^{(0)}$.

\subsection{Link to the retarded Green function}
The above formal solution links $\hat{f}^{(n)}_j$ to the retarded Green function of the undriven system.
To show this, we consider the correlator $\langle [\hat{f}^{(1)}_1,\hat{A}] \rangle$ where $\hat{A}$ is an observable and the expectation value is taken over the thermal/ground state of the undriven system $\hat{H}^{(0)}$. Then according to Eq.~(\ref{formal-solu-f^1_j}) we have
\begin{equation}
\langle [\hat{f}^{(1)}_1,\hat{A}] \rangle = -i \int_{-\infty}^{\infty} dt~ e^{i (\omL + i 0^+) t} ~\theta_t  \langle [\big( \hat{H}_1^{(1)} \big)^{H}_t ,\hat{A}] \rangle
\end{equation}
where $\theta_t$ is the step function at $t=0$, $ \big( \hat{H}_1^{(1)} \big)^{H}_t = e^{i \hat{H}^{(0)} t} \hat{H}_1^{(1)}  e^{-i \hat{H}^{(0)} t}$ represents the Heisenberg operator of $\hat{H}_1^{(1)}$. 
Hence, $\langle [\hat{f}^{(1)}_1,\hat{A}] \rangle$ is a frequency-domain retarded Green function of the undriven system $\hat{H}^{(0)}$. 
%$\langle [\hat{f}^{(1)}_1,\hat{A}] \rangle$ represents the linear response of the observable $\hat{A}$ in the system $\hat{H}^{(0)}$ monochromatically driven by $\hat{H}_1^{(1)}$. 
This direct link between $\hat{f}^{(n)}_j$ and the retarded Green function suggests that we can obtain the linear and non-linear responses of the system 
\footnote{For example, in a monochromatically driven system described by $\hat{H}_t = \hat{H}^{(0)} + \hat{H}^{(1)}_{-1}e^{-i\omega t} + \hat{H}^{(1)}_1e^{i\omega t}$, by expanding the expectation value $\langle  \hat{\mathcal{U}}_{t,t_0}^{\dag} \hat{A} \hat{\mathcal{U}}_{t,t_0}  \rangle$ in orders of $g$ and picking out terms oscillating at $e^{ij\omL t}$ where $j\in \mathbb{Z}$, we find that the $\mathcal{O}(g)$ linear response of an observable $\hat{A}$ is given by $\delta^{(1)} \langle \hat{A} \rangle = \langle  [\hat{F}^{(1)}_t,\hat{A}] \rangle$, and the $\mathcal{O}(g^2)$ response is given by $\delta^{(2)} \langle \hat{A} \rangle = \langle  [\hat{F}^{(2)}_t,\hat{A}]  \rangle  + \frac{1}{2} \langle [\hat{F}^{(1)}_t,[\hat{F}^{(1)}_t,\hat{A}]]  \rangle$. This means we can directly compute these responses when $\hat{f}^{(n)}_j$ is found.}
from the solutions of the Sylvester equations.

\subsection{Obtaining the spectral function by solving the Sylvester equation}

In equilibrium-state many-body systems, quantities such as the spectral function can also be obtained by solving the Sylvester equations. According to Eq.~(\ref{math-solution}), the Laplace-transformed Heisenberg operator 
\begin{equation}
    \hat{f}_\omL = -i \int_0^{\infty} dt ~ e^{i \omL t} ~ e^{- 0^+ t} ~ e^{i \hat{H} t} \hat{W}  e^{-i \hat{H} t}
\end{equation}
is the formal solution to the following Sylvester equation 
\begin{equation}\label{Sylv-general}
    \hat{W} + [\hat{f}_\omL,\hat{H}] - (\omL + i 0^+) \hat{f}_\omL =0.
\end{equation}
The frequency-domain retarded Green function $G^R$ can be directly obtained from $\hat{f}_\omL$, such that
\footnote{More explicitly, $ G^R = \langle [\hat{f}_\omL, \hat{W}^{\dag}]_{\pm} \rangle $ where $\hat{W}=\hat{\psi}_{{\bf k},\sigma}$ is the annihilation operator, and $\pm$ represents the anti-commutator for Fermions and the commutator for Bosons, respectively.}
$ G^R \sim \langle \hat{f}_\omL \hat{W}^{\dag} \rangle $, thus the spectral function is given by $A\equiv-\frac{1}{\pi} \text{Im}[G^R]\sim \text{Im}[ \langle \hat{f}_\omL \hat{W}^{\dag} \rangle ]$. We note that the imaginary part $i0^+$ enters in Eq.~(\ref{Sylv-general}) as we take $\hat{a} = -i ( \hat{H} + \omL ) + 0^+ $ in Eq.~(\ref{math-solution}). Keeping track of this $i0^+$ in Eq.~(\ref{Sylv-general}) is necessary to obtain the spectral function $A$, but it causes no difficulty in solving the Sylvester equation: All the methods developed for solving the Sylvester equation in this thesis allows $\omL$ to be complex.

\section{The Sylvester Equation of the driven Hubbard chain}

\subsection{The $\mathcal{O}(g)$ lowest order} \label{appendix:first-order-example}
Here we solve the lowest-order Sylvester equation (\ref{FSWT-f^1_1}) for the driven $L$-site Hubbard chain described by Eqs.~(\ref{H^0-example}) and (\ref{H_1-example}), based on the approach used for the driven Hubbard dimer in Section~\ref{subsec:driven-dimer}. We expand the solution $\hat{f}^{(1)}_{1}$ in orders of hopping $J$, i.e. $\hat{f}^{(1)}_1 = \sum_{n=0}^{\infty} \hat{y}_n$ where $\hat{y}_n \sim J^n$, and then Eq.~(\ref{FSWT-f^1_1}) is decoupled into the set of Eqs.~(\ref{y}). 

Due to the commutation relation $[\hat{H}^{(1)}_{1},\hat{U}]=0$, the solution to Eq.~(\ref{y0}) is exactly given by
\begin{equation}\label{result-y0}
\hat{y}_0 = \frac{\hat{H}^{(1)}_{1}}{\omL} = \frac{g}{\omL} \sum\limits_{s} \sum\limits_{j=1}^{L} j ~ \hat{n}_{j,s}
\end{equation}
Then, to solve Eq.~(\ref{y1}) for $\hat{y}_1$, we need to calculate the commutator
\begin{equation}\label{[y0,h]}
[\hat{y}_0,\hat{h}] = \frac{J g}{\omL} \sum\limits_{s} \sum\limits_{i,j =1}^{L} \left( \delta_{i-j,1} - \delta_{j-i,1} \right) \hat{c}_{j,s}^{\dag} \hat{c}_{i,s}
\end{equation}
which acts as the \textit{source} of the $J^1$-order in Eq.~(\ref{y1}). Since the Sylvester equation is linear, 
%this source means that, to solve Eq.~(\ref{y1}), 
we only need to solve the following part (which sums to the final result for $\hat{y}_1$),
\begin{equation}\label{y1-part}
    \hat{c}_{j,s}^{\dag} \hat{c}_{i,s} + [ \hat{x}_1 , \hat{U} ] - \omL \hat{x}_1 = 0.
\end{equation}
Since $\hat{U}$ only contains local operators, %we know that 
the solution $\hat{x}_1$ can only contain the degree of freedom of sites $i$ and $j$, and thus we need to solve (for $i\neq j$)
\begin{equation}\label{x1-equation}
    \hat{c}_{j,s}^{\dag} \hat{c}_{i,s} + [ \hat{x}_1 , U \hat{n}_{i,\uparrow} \hat{n}_{i,\downarrow} + U \hat{n}_{j,\uparrow} \hat{n}_{j,\downarrow}  ] - \omL \hat{x}_1 = 0
\end{equation}
which is no longer a many-body problem. %and has been solved in the previous two-site driven system in \ref{subsec:driven-dimer}. 
According to the symmetry argument in Section~\ref{subsec:driven-dimer}, the solution $\hat{x}_1$ can only take the following form
\begin{equation}\label{x-form}
\hat{x}_1 = \hat{c}_{j,s}^{\dag} \hat{c}_{i,s} \frac{1}{\omL} ( 1 + \beta' \hat{n}_{j,\bar{s}} + \gamma' \hat{n}_{i,\bar{s}} + \delta' \hat{n}_{j,\bar{s}} \hat{n}_{i,\bar{s}} )
\end{equation}
whose parameters can be determined by inserting Eq.~(\ref{x-form}) into Eq.~(\ref{x1-equation}) and then requiring the prefactors of each operator in Eq.~(\ref{x1-equation}) to vanish. The result is given in Eq.~(\ref{parameters'}).
From this result Eq.~(\ref{x-form}), we know that the Sylvester equation Eq.~(\ref{y1}) with source term Eq.~(\ref{[y0,h]}) has the solution
\begin{equation}\label{y1-solu}
\begin{split}
\hat{y}_1 &= \frac{J g}{\omL^2} \sum\limits_{s} \sum\limits_{i,j =1}^{L} \left( \delta_{i-j,1} - \delta_{j-i,1} \right) \hat{c}_{j,s}^{\dag} \hat{c}_{i,s} \\
&~~~~~~~~~~~ \times ( 1 + \beta' \hat{n}_{j,\bar{s}} + \gamma' \hat{n}_{i,\bar{s}} + \delta' \hat{n}_{j,\bar{s}} \hat{n}_{i,\bar{s}} )
\end{split}
\end{equation}
Collecting $\hat{y}_0$ and $\hat{y}_1$ gives the solution Eq.~(\ref{FSWT-H'-chain-g^2}) shown in the main part.
%In this appendix 
Next we further outline how to solve Eq.~(\ref{y2}) for $\hat{y}_2 \sim J^2$. %Again, i
Its source term reads
 
\begin{equation}\label{[y1,h]}
\begin{split}
[\hat{y}_1,\hat{h}] 
% two-site terms
&= \frac{2 J^2 g}{\omL^2}  \sum\limits_{j =1}^{L-1}
\bigg\{ ~ (\beta'-\gamma') (\hat{c}_{j \uparrow}^{\dag} \hat{c}_{j \downarrow}^{\dag}\hat{c}_{j+1 \uparrow} \hat{c}_{j+1 \downarrow} - \hat{c}_{j+1 \uparrow}^{\dag} \hat{c}_{j+1 \downarrow}^{\dag}\hat{c}_{j \uparrow} \hat{c}_{j \downarrow} ) ~~ - \hat{n}_j + \hat{n}_{j+1}  \\
&~~~~~~~~~~~~~~~~~~~~~~~~ 
+ (\beta'+\gamma') \big( ~ \hat{n}_{j+1\uparrow} \hat{n}_{j+1\downarrow} (1-\hat{n}_j) ~ - ~ \hat{n}_{j\uparrow} \hat{n}_{j\downarrow} (1-\hat{n}_{j+1})  ~ \big) \bigg\}
\\
% three-site terms
&+ \frac{J^2 g}{\omL^2}  \sum\limits_{j =2}^{L-1} \sum\limits_{s} \bigg\{ ~
\hat{c}_{j-1 s}^{\dag} \hat{c}_{j+1 s} \big( (\beta' - \gamma') \hat{n}_{j,\bar{s}} - \beta' \hat{n}_{j-1,\bar{s}} + \gamma' \hat{n}_{j+1,\bar{s}} + (\beta' + \gamma') \hat{n}_{j,\bar{s}} ( \hat{n}_{j-1,\bar{s}} - \hat{n}_{j+1,\bar{s}} ) \big) \\
&~~~~~~~~~~~~~~~~~~~~ 
+ \hat{c}_{j+1 s}^{\dag} \hat{c}_{j-1 s} \big( (\gamma' - \beta') \hat{n}_{j,\bar{s}} - \gamma' \hat{n}_{j-1,\bar{s}} + \beta' \hat{n}_{j+1,\bar{s}} + (\beta' + \gamma') \hat{n}_{j,\bar{s}} ( \hat{n}_{j-1,\bar{s}} - \hat{n}_{j+1,\bar{s}} ) \big) \\
&~~~~~~~~~~~~~~~~~~~~
+ \hat{c}_{j s}^{\dag}  \hat{c}_{j-1 \bar{s}}^{\dag} \hat{c}_{j \bar{s}} \hat{c}_{j+1 s} ~ \big( (\beta' - \gamma') + (\beta' + \gamma') (\hat{n}_{j-1 s} - \hat{n}_{j+1 \bar{s}} )  \big) \\
&~~~~~~~~~~~~~~~~~~~~
+ \hat{c}_{j+1 s}^{\dag}  \hat{c}_{j \bar{s}}^{\dag} \hat{c}_{j-1 \bar{s}} \hat{c}_{j s} ~ \big( ( \gamma' - \beta' ) + (\beta' + \gamma') (\hat{n}_{j-1 s} - \hat{n}_{j+1 \bar{s}} )  \big) \\
&~~~~~~~~~~~~~~~~~~~~ 
+ ( \hat{c}_{j s}^{\dag}  \hat{c}_{j \bar{s}}^{\dag} \hat{c}_{j-1 \bar{s}} \hat{c}_{j+1 s} 
+ \hat{c}_{j+1 s}^{\dag}  \hat{c}_{j-1 \bar{s}}^{\dag} \hat{c}_{j \bar{s}} \hat{c}_{j s} ) 
~ (\beta' + \gamma') (\hat{n}_{j+1 \bar{s}} - \hat{n}_{j-1 s} )   ~ \bigg\} 
\end{split}
\end{equation}
%where in the first line the operator $\hat{T}_{ji,s}$ is defined from $  \hat{T}_{ji,s} \equiv t^{-1}  [\hat{c}_{j,s}^{\dag} \hat{c}_{i,s},\hat{h}]  $, which reads
%\begin{equation}
%\hat{T}_{jis} = 
%(1-\delta_{j,N}) \hat{c}_{j+1,s}^{\dag} \hat{c}_{i,s}  +  (1-\delta_{j,1}) \hat{c}_{j-1,s}^{\dag} \hat{c}_{i,s}  - (1-\delta_{i,N}) \hat{c}_{j,s}^{\dag} \hat{c}_{i+1,s} - (1-\delta_{i,1}) \hat{c}_{j,s}^{\dag} \hat{c}_{i-1,s} ,
%\end{equation}

where we use $\delta'=-\beta'-\gamma'$ and $\hat{n}_j = \sum_s \hat{n}_{j,s}$ to simplify the result.
In the non-interacting limit $U\to0$ (where $\beta'=\gamma'=\delta'=0$), we find the commutator $[\hat{y}_1,\hat{h}]$ reduces completely to boundary effects, which only contain the operators at the boundary sites $j=1,L$. 
However, for $U\neq0$, $[\hat{y}_1,\hat{h}]$ will become non-vanishing in the bulk. Since the Sylvester equation (\ref{y2}) is linear, we can find the solution for each individual term in this $[\hat{y}_1,\hat{h}]$, and then sum them up to get $\hat{y}_2$, just as what we did above to find $\hat{y}_1$. This means that to find $\hat{y}_2$, we need to solve 3-site Sylvester equations, such as 
\begin{equation}\label{y2-part}
\begin{split}
\hat{c}_{j,s}^{\dag} \hat{c}_{k,\bar{s}}^{\dag} \hat{c}_{j,\bar{s}} \hat{c}_{i,s}  + [ \hat{x}_2 , \hat{U} ] - \omL \hat{x}_2 &= 0, \\
\hat{c}_{j,s}^{\dag} \hat{c}_{j,\bar{s}}^{\dag} \hat{c}_{k,\bar{s}} \hat{c}_{i,s}  + [ \hat{x}_3 , \hat{U} ] - \omL \hat{x}_3 &= 0,
\end{split}
\end{equation}
which is similar to Eq.~(\ref{y1-part}). According to the same symmetry argument, the solution to Eq.~(\ref{y2-part}) takes the form 
\begin{equation}
\begin{split}
\hat{x}_2 &= \hat{c}_{j,s}^{\dag} \hat{c}_{k,\bar{s}}^{\dag} \hat{c}_{j,\bar{s}} \hat{c}_{i,s} \frac{1}{\omL}  (1 + \beta' \hat{n}_{k,s} + \gamma' \hat{n}_{i,\bar{s}} + \delta' \hat{n}_{k,s} \hat{n}_{i,\bar{s}}  ) \\
\hat{x}_3 &= \hat{c}_{j,s}^{\dag} \hat{c}_{j,\bar{s}}^{\dag} \hat{c}_{k,\bar{s}} \hat{c}_{i,s} \frac{1}{\omL}
\big( \frac{\omL}{\omL+U} -\beta' (\hat{n}_{k,s} + \hat{n}_{i,\bar{s}}) -\delta' \hat{n}_{k,s} \hat{n}_{i,\bar{s}}  \big) 
\end{split}
\end{equation}
where the coefficients are decided by Eq.~(\ref{y2-part}). The solution of $\hat{y}_2$ is then constructed from $\hat{x}_2$, according to the source term $[\hat{y}_1,\hat{h}]$. In the final solution of $\hat{y}_2$, each term will contain a common factor $J^2 g / \omL^3$, together with a term-specific factor that makes $\hat{y}_2$ accurate at arbitrary $U/J$ ratio. In our driven Hubbard chain, $\hat{y}_2$ is found to be
 
\begin{equation}\label{y2-full-result}
\begin{split}
\hat{y}_2 &= \frac{2 J^2 g}{\omL^3} \sum\limits_{j =1}^{L-1}
\bigg\{ ~ (\beta'-\gamma') (\hat{c}_{j \uparrow}^{\dag} \hat{c}_{j \downarrow}^{\dag}\hat{c}_{j+1 \uparrow} \hat{c}_{j+1 \downarrow} - \hat{c}_{j+1 \uparrow}^{\dag} \hat{c}_{j+1 \downarrow}^{\dag}\hat{c}_{j \uparrow} \hat{c}_{j \downarrow} ) ~~ - \hat{n}_j + \hat{n}_{j+1}  \\
&~~~~~~~~~~~~~~~~~~~~~~~~ 
+ (\beta'+\gamma') \big( ~ \hat{n}_{j+1\uparrow} \hat{n}_{j+1\downarrow} (1-\hat{n}_j) ~ - ~ \hat{n}_{j\uparrow} \hat{n}_{j\downarrow} (1-\hat{n}_{j+1})  ~ \big) \bigg\} \\
%3-site terms
&+ \frac{J^2 g}{\omL^3}  \sum\limits_{j =2}^{L-1} \sum\limits_{s} \\
&\bigg\{ ~
\hat{c}_{j-1 s}^{\dag} \hat{c}_{j+1 s} \big( (\beta' - \gamma') \hat{n}_{j,\bar{s}} - \beta' \hat{n}_{j-1,\bar{s}} + \gamma' \hat{n}_{j+1,\bar{s}} - \delta' \hat{n}_{j,\bar{s}} ( \hat{n}_{j-1,\bar{s}} - \hat{n}_{j+1,\bar{s}} ) \big) \\
& ~~~~~~~~~~~~~~~~~~~~~~~~~~~~~~~~~~~~~~~~~~~~~
\times( 1 + \beta' \hat{n}_{ j-1 \bar{s}} + \gamma' \hat{n}_{j+1 \bar{s}} + \delta' \hat{n}_{ j-1 \bar{s}} \hat{n}_{j+1 \bar{s}} )
\\
& 
+ \hat{c}_{j+1 s}^{\dag} \hat{c}_{j-1 s} \big( (\gamma' - \beta') \hat{n}_{j,\bar{s}} - \gamma' \hat{n}_{j-1,\bar{s}} + \beta' \hat{n}_{j+1,\bar{s}} - \delta' \hat{n}_{j,\bar{s}} ( \hat{n}_{j-1,\bar{s}} - \hat{n}_{j+1,\bar{s}} ) \big) \\
& ~~~~~~~~~~~~~~~~~~~~~~~~~~~~~~~~~~~~~~~~~~~~~
\times( 1 + \beta' \hat{n}_{ j+1 \bar{s}} + \gamma' \hat{n}_{j-1 \bar{s}} + \delta' \hat{n}_{ j+1 \bar{s}} \hat{n}_{j-1 \bar{s}} )
\\
& 
+ \hat{c}_{j s}^{\dag}  \hat{c}_{j-1 \bar{s}}^{\dag} \hat{c}_{j \bar{s}} \hat{c}_{j+1 s} ~ \big( (\beta' - \gamma') -\delta' (\hat{n}_{j-1 s} - \hat{n}_{j+1 \bar{s}} )  \big)
( 1 + \beta' \hat{n}_{ j-1 s} + \gamma' \hat{n}_{j+1 \bar{s}} + \delta' \hat{n}_{ j-1 s} \hat{n}_{j+1 \bar{s}} )
\\
& 
+ \hat{c}_{j+1 s}^{\dag}  \hat{c}_{j \bar{s}}^{\dag} \hat{c}_{j-1 \bar{s}} \hat{c}_{j s} ~ \big( ( \gamma' - \beta' ) -\delta' (\hat{n}_{j-1 s} - \hat{n}_{j+1 \bar{s}} )  \big) 
( 1 + \beta' \hat{n}_{j+1 \bar{s}}  + \gamma' \hat{n}_{ j-1 s}  + \delta' \hat{n}_{j+1 \bar{s}} \hat{n}_{ j-1 s}  )
\\
& 
+  \hat{c}_{j s}^{\dag}  \hat{c}_{j \bar{s}}^{\dag} \hat{c}_{j-1 \bar{s}} \hat{c}_{j+1 s} 
~ \delta' ( \hat{n}_{j-1 s} - \hat{n}_{j+1 \bar{s}}  )
\big( \frac{\omL}{\omL+U} - \beta' (\hat{n}_{j-1 s} + \hat{n}_{j+1 \bar{s}} ) - \delta' \hat{n}_{j-1 s} \hat{n}_{j+1 \bar{s}}  \big)
\\
& 
+ \hat{c}_{j+1 s}^{\dag}  \hat{c}_{j-1 \bar{s}}^{\dag} \hat{c}_{j \bar{s}} \hat{c}_{j s} 
~ \delta' ( \hat{n}_{j-1 s} - \hat{n}_{j+1 \bar{s}}  )
\big( \frac{\omL}{\omL-U} - \gamma' (\hat{n}_{j-1 s} + \hat{n}_{j+1 \bar{s}} ) - \delta' \hat{n}_{j-1 s} \hat{n}_{j+1 \bar{s}}  \big)
~ \bigg\} 
\end{split}
\end{equation}   % I have checked (using Mathematica) that this y2 indeed satisfies the Sylvester equation for y2
In this $\mathcal{O}(J^2)$ order micro-motion $\hat{y}_2$, we find driving-induced two-site correlated processes, including doublon-holon exchange and doublon-holon density-density interactions, as well as three-site correlated processes, including next-nearest neighbour hopping, two-electron hopping, doublon formation and dissociation.
This $\hat{y}_2$ contributes to the following $\mathcal{O}(J^2)$ term in the lowest order Floquet Hamiltonian $\hat{H}'^{(2)}$ in Eq.~(\ref{Floquet-H'}), which reads
\begin{equation}\label{Floquet-H'2-t2}
\begin{split}
& \frac{1}{2}([\hat{y}_2,\hat{H}^{(1)}_{-1}]+ H.c.) \\
% simplify the 3-site term
&= \frac{4 J^2 g^2}{\omL^3} (\beta'-\gamma') \sum\limits_{j =1}^{L-1} \big(
\hat{c}_{j \uparrow}^{\dag} \hat{c}_{j \downarrow}^{\dag}\hat{c}_{j+1 \uparrow} \hat{c}_{j+1 \downarrow} + \hat{c}_{j+1 \uparrow}^{\dag} \hat{c}_{j+1 \downarrow}^{\dag}\hat{c}_{j \uparrow} \hat{c}_{j \downarrow} \big) \\
%3-site terms
&+ \frac{J^2 g^2}{\omL^3} (\beta' - \gamma')  \sum\limits_{j =2}^{L-1} \sum\limits_{s} \\
&\bigg\{ ~
(\hat{c}_{j-1 s}^{\dag} \hat{c}_{j+1 s} + \hat{c}_{j+1 s}^{\dag} \hat{c}_{j-1 s} )  \big( 2 \hat{n}_{j,\bar{s}} - \hat{n}_{j-1,\bar{s}} - \hat{n}_{j+1,\bar{s}} + \delta' (1- 2\hat{n}_{j,\bar{s}}) ( \hat{n}_{j-1,\bar{s}} + \hat{n}_{j+1,\bar{s}} - 2 \hat{n}_{ j-1 \bar{s}} \hat{n}_{j+1 \bar{s}} ) \big)
\\
&~~ 
+ 2  (\hat{c}_{j s}^{\dag}  \hat{c}_{j-1 \bar{s}}^{\dag} \hat{c}_{j \bar{s}} \hat{c}_{j+1 s} + \hat{c}_{j+1 s}^{\dag}  \hat{c}_{j \bar{s}}^{\dag} \hat{c}_{j-1 \bar{s}} \hat{c}_{j s})
( 1 - \delta' \hat{n}_{ j-1 s} - \delta' \hat{n}_{j+1 \bar{s}} + 2 \delta' \hat{n}_{ j-1 s} \hat{n}_{j+1 \bar{s}} )
~ \bigg\} 
\end{split}
\end{equation}
where the coefficients are given in Eq.~(\ref{parameters'-main}). We find this $\mathcal{O}(J^2)$ Floquet Hamiltonian contains fewer terms than the micro-motion $\hat{y}_2$: According to our driving term $\hat{H}^{(1)}_{-1}$, if a term in $\hat{y}_2$ conserves the center-of-mass position, it will commute with $\hat{H}^{(1)}_{-1}$, and thus it will have no impact on the Floquet Hamiltonian (\ref{Floquet-H'2-t2}).

\subsection{The $\mathcal{O}(g^2)$ second-lowest order}\label{appendix:higher-order-example}

For the second-lowest order ($n=2$) driving effects, we need to solve two Sylvester equations according to Eq.~(\ref{formula-for-f_1^2}) for the system described by Eq.~(\ref{Ht-example}),
\begin{subequations}\label{FSWT-f^2}
\begin{align}
&  [\hat{f}^{(2)}_1,\hat{H}^{(0)}] - \omL \hat{f}^{(2)}_1 = 0  \label{FSWT-f^2_1} \\
&  \frac{1}{2} [\hat{f}^{(1)}_1,\hat{H}^{(1)}_1] + [\hat{f}^{(2)}_2,\hat{H}^{(0)}] - 2 \omL \hat{f}^{(2)}_2 = 0.  \label{FSWT-f^2_2}
\end{align}
\end{subequations}
Form these, we find $\hat{f}^{(2)}_{\pm1} =0$ in this model, and thus according to Eq.~(\ref{H'3}), there is no $\mathcal{O}(g^3)$-correction to the Floquet Hamiltonian, $\hat{H}'^{(3)}=0$. More generally, only the even-order $\hat{H}'^{(2n)}$ Floquet Hamiltonians do not vanish in the driven system described by the Hamiltonian Eq.~(\ref{Ht-example}). The vanishing of odd orders
%This absence of $\hat{H}'^{(2n+1)}$ 
can be verified using the Floquet method \cite{PhysRevB.101.024303} based on Gaussian elimination.

Here we solve Eq.~(\ref{FSWT-f^2_2}) in the $J\ll \omL$ limit, where we expand the solution $\hat{f}^{(2)}_2$ in orders of hopping $J$, i.e. $\hat{f}^{(2)}_2 = \sum_{n=0}^{\infty} \hat{z}_n$ where $\hat{z}_n \sim J^n$. This decomposes the Sylvester equation  (\ref{FSWT-f^2_2}) in orders of $J$, e.g. 
\begin{subequations}\label{z}
    \begin{align}
& \frac{1}{2}[\hat{y}_0,\hat{H}^{(1)}_{1}] + [\hat{z}_0,\hat{U}] - 2\omL \hat{z}_0 = 0  \label{z0} \\
& \frac{1}{2}[\hat{y}_1,\hat{H}^{(1)}_{1}] +  [\hat{z}_0,\hat{h}] + [\hat{z}_1,\hat{U}] - 2\omL \hat{z}_1 = 0 \label{z1} \\
& \frac{1}{2}[\hat{y}_2,\hat{H}^{(1)}_{1}] +  [\hat{z}_1,\hat{h}] + [\hat{z}_2,\hat{U}] - 2\omL \hat{z}_2 = 0 
    \end{align}
\end{subequations}
and so on. Here we also used the expansion $\hat{f}^{(1)}_1 = \sum_{n=0}^{\infty} \hat{y}_n$ where $\hat{y}_0$ and $\hat{y}_1$ is solved in Appendix \ref{appendix:first-order-example}.
According to the result Eq.~(\ref{result-y0}), we find $[\hat{y}_0,\hat{H}^{(1)}_{1}]=0$, and thus Eq.~(\ref{z0}) gives $\hat{z}_0=0$. The source term in Eq.~(\ref{z1}) is then given by
 
\begin{equation}
\begin{split}
& \frac{1}{2}[\hat{y}_1,\hat{H}^{(1)}_{1}]  = \frac{J g^2}{2 \omL^2} \sum\limits_{s} \sum\limits_{i,j =1}^{L} \left( \delta_{i-j,1} + \delta_{j-i,1} \right) 
\hat{c}_{j,s}^{\dag} \hat{c}_{i,s}  ( 1 + \beta' \hat{n}_{j,\bar{s}} + \gamma' \hat{n}_{i,\bar{s}} + \delta' \hat{n}_{j,\bar{s}} \hat{n}_{i,\bar{s}} ).
\end{split}
\end{equation}
Similar to what we did in Appendix \ref{appendix:first-order-example}, we solve the Sylvester equation for each term in $\frac{1}{2}[\hat{y}_1,\hat{H}^{(1)}_{1}]$ and then sum them up, which gives the solution to Eq.~(\ref{z1}). It reads explicitly
\begin{equation}\label{z1-solu}
\begin{split}
\hat{z}_1 &= \frac{J g^2}{4 \omL^3} \sum\limits_{s} \sum\limits_{i,j =1}^{L} \left( \delta_{i-j,1} + \delta_{j-i,1} \right) 
\hat{c}_{j,s}^{\dag} \hat{c}_{i,s}  ( 1 + \beta' \hat{n}_{j,\bar{s}} + \gamma' \hat{n}_{i,\bar{s}} + \delta' \hat{n}_{j,\bar{s}} \hat{n}_{i,\bar{s}} ) 
( 1 + \beta'' \hat{n}_{j,\bar{s}} + \gamma'' \hat{n}_{i,\bar{s}} + \delta'' \hat{n}_{j,\bar{s}} \hat{n}_{i,\bar{s}} ) \\
&\equiv \frac{J g^2}{4 \omL^3} \sum\limits_{s} \sum\limits_{i,j =1}^{L} \left( \delta_{i-j,1} + \delta_{j-i,1} \right) \hat{c}_{j,s}^{\dag} \hat{c}_{i,s} 
( 1 + \beta_2 \hat{n}_{j,\bar{s}} + \gamma_2 \hat{n}_{i,\bar{s}} + \delta_2 \hat{n}_{j,\bar{s}} \hat{n}_{i,\bar{s}} ) 
\end{split}
\end{equation}    

where in the first line $\beta'',\gamma'', \delta''$ is are simply the coefficients $\beta',\gamma', \delta'$ with $\omL$ replaced by $2\omL$. Since $\hat{f}^{(2)}_2 = \hat{z}_1 + \mathcal{O}(J^2)$, we thus see from Eq.~(\ref{z1-solu}) that $\hat{f}^{(2)}_2$ diverges not only when $\omL=U$ (when $\gamma'$ diverges), but also when $\omL=U/2$ (when $\gamma''$ diverges). The divergence at $\omL=U/2$ is understood as the two-photon resonance to excite a doublon. In the second line of Eq.~(\ref{z1-solu}), we have evaluated the multiplication in the first line, from which the parameters $\beta_2,\gamma_2,\delta_2$ are defined for future use, i.e. $\beta_2\equiv \beta'+\beta''+\beta'\beta''$, $\gamma_2\equiv \gamma'+\gamma''+\gamma'\gamma''$ and $\delta_2 \equiv \delta' + \delta'' + \gamma'\beta'' + \gamma''\beta' + \gamma'\delta'' + \gamma''\delta' + \beta'\delta'' + \beta''\delta' + \delta'\delta'' $.

\subsection{The $\mathcal{O}(g^3)$ third-lowest order}\label{appendix:3rd-order-example}
To find the third lowest order ($n=3$) driving effect in the driving systems described by Eq.~(\ref{Ht-example}), we need to solve 3 Sylvester equations

\begin{subequations}\label{FSWT-f^3}
\begin{align}
 \frac{1}{2} [\hat{f}^{(2)}_2,\hat{H}^{(1)}_{-1}] 
+ \frac{1}{12} [\hat{f}^{(1)}_{-1},[\hat{f}^{(1)}_{1},\hat{H}^{(1)}_{1}]] 
+ \frac{2}{3} [\hat{f}^{(1)}_1,\hat{H}'^{(2)}] +  [\hat{f}^{(3)}_1,\hat{H}^{(0)}] - \omL \hat{f}^{(3)}_1 
&= 0  \label{FSWT-f^3_1} \\
[\hat{f}^{(3)}_2,\hat{H}^{(0)}] - 2 \omL \hat{f}^{(3)}_2 &= 0  \label{FSWT-f^3_2} \\
\frac{1}{2} [\hat{f}^{(2)}_2,\hat{H}^{(1)}_1] + \frac{1}{12} [\hat{f}^{(1)}_{1},[\hat{f}^{(1)}_{1},\hat{H}^{(1)}_{1}]] 
+ [\hat{f}^{(3)}_3,\hat{H}^{(0)}] - 3 \omL \hat{f}^{(3)}_3 &= 0  \label{FSWT-f^3_3} 
\end{align}
\end{subequations}
whose solutions determine the $\mathcal{O}(g^4)$ order Floquet Hamiltonian correction, %which is
\begin{equation}\label{H'4}
\hat{H}'^{(4)} = 
\left( \frac{1}{2} [\hat{f}^{(3)}_1, \hat{H}^{(1)}_{-1}] 
+ \frac{1}{12} [\hat{f}^{(2)}_2,[\hat{f}^{(1)}_{-1}, \hat{H}^{(1)}_{-1}]] 
+ \frac{1}{12} [\hat{f}^{(1)}_{-1},[\hat{f}^{(2)}_{2}, \hat{H}^{(1)}_{-1}]] 
- \frac{1}{12} [\hat{f}^{(1)}_{1},[\hat{f}^{(1)}_{-1}, \hat{H}'^{(2)}]]  \right)
+ H.c.
\end{equation}    
These higher-order solutions can be derived in the same way. We find $\hat{f}^{(3)}_1$ diverges at $\omL=U, U/2$, $\hat{f}^{(3)}_2 = 0$, and $\hat{f}^{(3)}_3$ diverges at $\omL=U, U/2, U/3$. This means $\hat{H}'^{(4)}$ diverges at $\omL=U, U/2$.
Explicitly, by solving Eq.~(\ref{FSWT-f^3_1}), we find
\begin{equation}\label{f^3_1-solu}
\begin{split}
\hat{f}^{(3)}_1 &= \frac{J g^3}{\omL^4} \sum\limits_{s} \sum\limits_{i,j =1}^{L} \left( \delta_{i-j,1} - \delta_{j-i,1} \right) \hat{c}_{j,s}^{\dag} \hat{c}_{i,s} 
( -\frac{11}{24} + \beta_3 \hat{n}_{j,\bar{s}} + \gamma_3 \hat{n}_{i,\bar{s}} + \delta_3 \hat{n}_{j,\bar{s}} \hat{n}_{i,\bar{s}} ) ~~ + \mathcal{O}(J^2)
\end{split}
\end{equation}
where the parameters $\beta_3,\gamma_3,\delta_3$ are determined from
\begin{equation}
\begin{split}
&( -\frac{11}{24} + \beta_3 \hat{n}_{j,\bar{s}} + \gamma_3 \hat{n}_{i,\bar{s}} + \delta_3 \hat{n}_{j,\bar{s}} \hat{n}_{i,\bar{s}} ) \\
&~~~~= \frac{1}{8} ( 1 + \beta' \hat{n}_{j,\bar{s}} + \gamma' \hat{n}_{i,\bar{s}} + \delta' \hat{n}_{j,\bar{s}} \hat{n}_{i,\bar{s}} )^2  ~ 
( -1 + \beta'' \hat{n}_{j,\bar{s}} + \gamma'' \hat{n}_{i,\bar{s}} + \delta'' \hat{n}_{j,\bar{s}} \hat{n}_{i,\bar{s}} ) \\
&~~~~~~ - \frac{1}{3} ( 1 + \beta' \hat{n}_{j,\bar{s}} + \gamma' \hat{n}_{i,\bar{s}} + \delta' \hat{n}_{j,\bar{s}} \hat{n}_{i,\bar{s}} ) ~ 
( 1 + \beta' \hat{n}_{j,\bar{s}} + \gamma' \hat{n}_{i,\bar{s}} + \delta' \hat{n}_{j,\bar{s}} \hat{n}_{i,\bar{s}} ).
\end{split}
\end{equation}    
Inserting the solution of $\hat{f}^{(2)}_2$ and $\hat{f}^{(3)}_1$, i.e. Eqs.~(\ref{z1-solu}) and (\ref{f^3_1-solu}), into Eq.~(\ref{H'4}), we find
\begin{equation}\label{H'4-result}
\begin{split}
\hat{H}'^{(4)} &= \frac{g^4 J}{2\omL^4} \sum\limits_{s} \sum\limits_{i,j =1}^{L} 
\left( \delta_{i-j,1} + \delta_{j-i,1} \right) 
(\hat{c}_{j,s}^{\dag} \hat{c}_{i,s} + \hat{c}_{i,s}^{\dag} \hat{c}_{j,s} )
\big(
-\frac{1}{4} + \beta_4 \hat{n}_{j,\bar{s}} + \gamma_4 \hat{n}_{i,\bar{s}} + \delta_4 \hat{n}_{j,\bar{s}} \hat{n}_{i,\bar{s}}
\big)
\\
&= \frac{g^4 J}{\omL^4} \sum\limits_{s} \sum\limits_{j=1}^{L-1}  ( \hat{c}_{j,s}^{\dag} \hat{c}_{j+1,s} + \hat{c}_{j+1,s}^{\dag} \hat{c}_{j,s}   )  
\left( -\frac{1}{4} + \frac{\beta_4 + \gamma_4}{2} (\hat{n}_{j,\bar{s}} +  \hat{n}_{j+1,\bar{s}} ) +  \delta_4 \hat{n}_{j,\bar{s}} \hat{n}_{j+1,\bar{s}} \right) ~~ + \mathcal{O}(J^2)
\end{split}
\end{equation}
where we have defined $\beta_4 \equiv \beta_3 + \beta_2 / 24 + \beta' / 6 $, $\gamma_4 \equiv \gamma_3 + \gamma_2 / 24 + \gamma' / 6 $ and $\delta_4 \equiv \delta_3 + \delta_2 / 24 + \delta' / 6 $. In the non-interacting case $U=0$, these 3 parameters vanish, and then $\hat{H}'^{(4)}$ reduces exactly to the non-interacting dynamical localisation effect predicted before in Ref.~\cite{PhysRevB.34.3625}: The bandwidth renormalisation factor $\frac{1}{4} g^4/\omL^4$ in Eq.~(\ref{H'4-result}) matches the Taylor expansion of the Bessel function $J_0 (2g/\omL) = 1 - g^2/\omL^2 + \frac{1}{4} g^4/\omL^4 + \mathcal{O}(g^6)$.

\section{Simulations on the driven Hubbard model}
\subsection{DMRG simulation of the metal-insulator phase transition}
\label{appendix:MIT}
Here, we present details of the iDMRG simulation in Fig.~\ref{fig:MIT}. 
Unlike the Bose-Hubbard model \cite{jaksch1998cold}, the one-dimensional Fermi-Hubbard model shows a metal-insulator transition at $U/J=0$ at half-filling \cite{RevModPhys.70.1039,essler2005one}.
This makes it difficult to witness the phase transition by modifying the ratio $U/J$ and calculating the single-particle gap. Therefore, we instead include a non-zero chemical potential $\mu$ in Eq.~(\ref{H^0-example}),  
%\begin{equation}
%\hat{U} \to U \sum\limits_{j=1}^{L} (\hat{n}_{j,\uparrow} -\frac{1}{2}) ( \hat{n}_{j,\downarrow} - \frac{1}{2}) - \mu \sum\limits_{j=1}^{L} \sum\limits_{s} \hat{n}_{j,s}
%\end{equation}
and witness the (filling-control) phase transition by decreasing the chemical potential $\mu$, during which the occupancy changes from $\langle \hat{n}_j \rangle = 1$ to incommensurate values \cite{essler2005one}. 
%All the previous FSWT results, like $\hat{f}^{(1)}_1$, 
%The solution of the Sylvester equations $\hat{f}^{(1)}_1$ (and thus the Floquet Hamiltonian) remains unchanged after adding this chemical potential term to $\hat{H}^{(0)}$ since it conserves the total particle number. 
We fix the parameters to $ U=4J, \omL=12J, g=3J$, which is within the applicability range of our FSWT result Eq.~(\ref{Floquet-H'}).
We use infinite-DMRG to simulate the Floquet Hamiltonian's ground state, truncating the bond dimension to $\chi=600$. The exploration around the transition shows a shift in non-commensurate occupation values and critical behaviour due to the additional correlated hopping terms compared to the HFE. To obtain the change in the critical point of the transition, we interpolated each point (in the FSWT curve) near the transition with a function $1-A\sqrt{\mu_c-\mu}$, typical of the commensurate-incommensurate phase transition of the Hubbard model \cite{essler2005one}. The interpolation of HFE and the undriven curve is based on the exact Lieb-Wu equation \cite{essler2005one}.

\begin{figure}[t]
    \centering
    \includegraphics[width=0.5\textwidth]{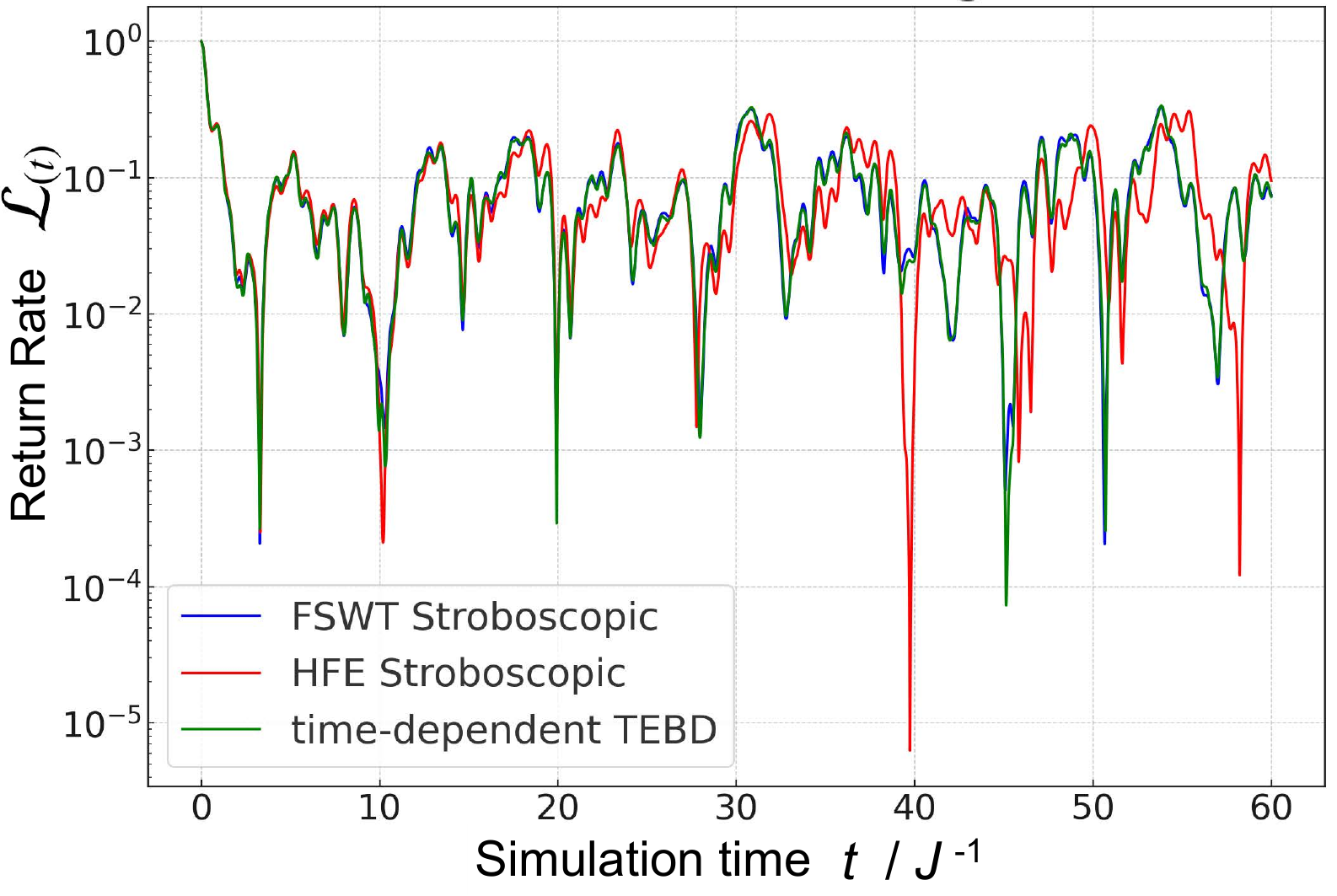}
    \caption{The return rate, Eq.~(\ref{eq.Loschmidt}) of a $L=6$ lattice for the parameters in Fig.~\ref{fig:MIT}, i.e., $\omL=12 J$, $U=4 J$ and $g=3 J$.}
    \label{fig:assure}
\end{figure}

To confirm the applicability of our FSWT Hamiltonian (\ref{Floquet-H'}) for Fig.~\ref{fig:MIT}, we show in Fig.~\ref{fig:assure} the return rate defined in Eq.~(\ref{eq.Loschmidt}) on a finite $L=6$ lattice, using the same parameters chosen in Fig.~\ref{fig:MIT}. We see that the FSWT result Eq.~(\ref{Floquet-H'}), which includes the $\mathcal{O}(J)$ and $\mathcal{O}(J^2)$ correlated hopping terms, matches very well with the exact dynamics throughout the simulation time, while the HFE result obviously deviates from the exact dynamics at early simulation time. 
We find that the $\mathcal{O}(J^2)$ term, Eq.~(\ref{Floquet-H'2-t2}), is vital for the FSWT Hamiltonian (\ref{Floquet-H'}) to describe the ground state and evolution accurately.  

\subsection{The convergence check of TEBD over Trotter steps}\label{appendix:TrotterCheck}
\begin{figure}[t]
    \centering
    \includegraphics[width=0.45\textwidth]{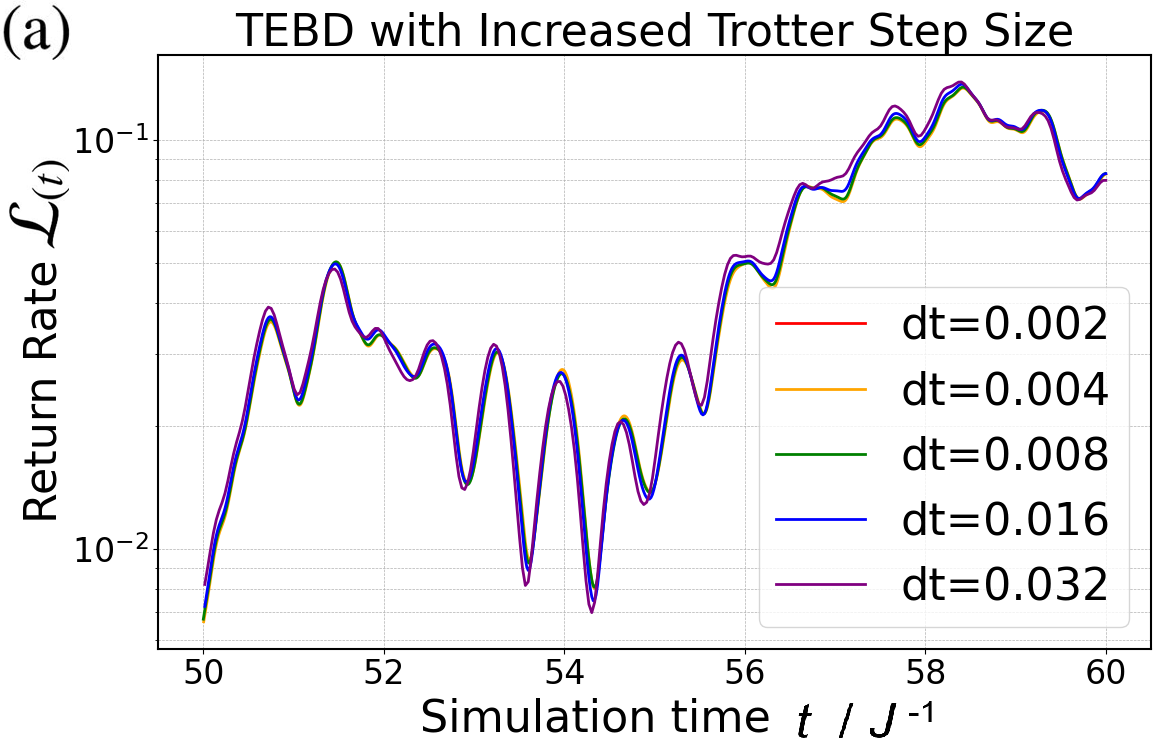}
    \hspace{0.3cm}
    \includegraphics[width=0.45\textwidth]{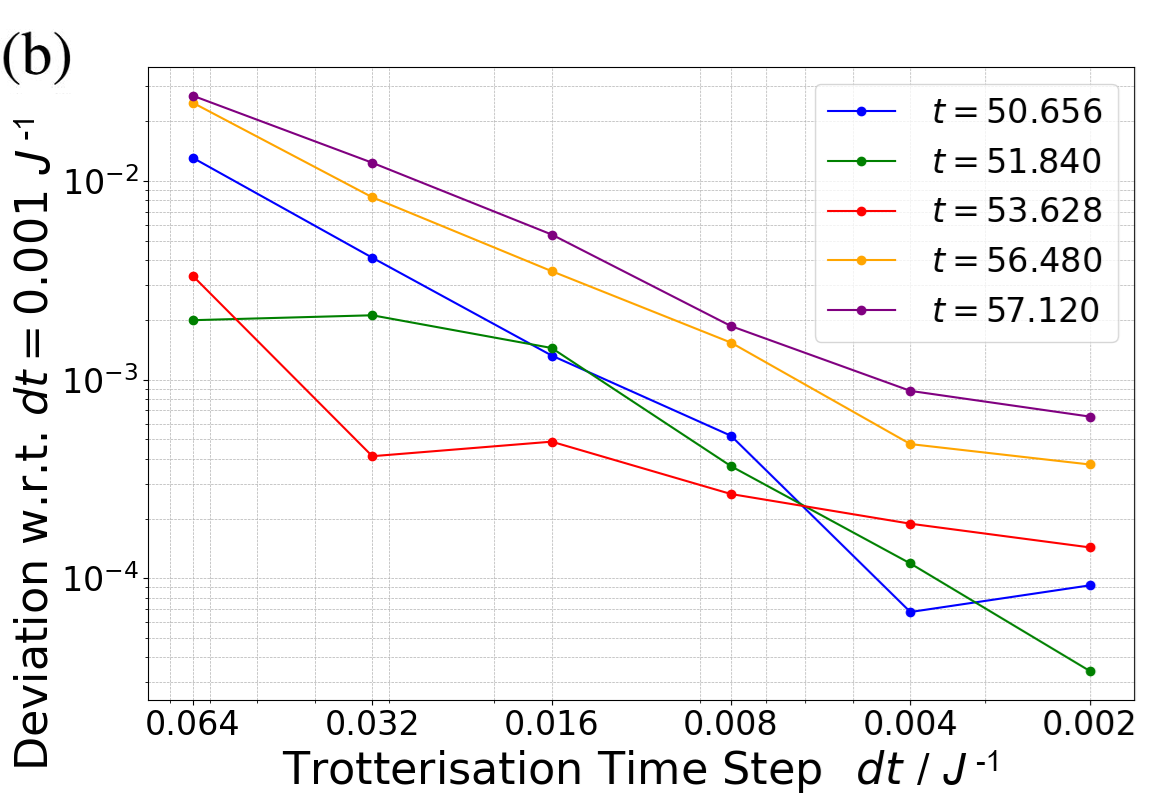}
    \caption{(a) The TEBD simulation of return rate, Eq.~(\ref{eq.Loschmidt}) for $L=10$, $\omL=16J$, $U=8J$ and $g=\omL/4$. Noticeable deviations appear only when the Trotterisation step size $dt\gtrsim0.004 J^{-1}$. 
    (b) The deviation of (a) compared to the $dt = 10^{-3} J^{-1}$ TEBD simulation in Fig.~\ref{fig:Evo}(b) at several simulation time points.
    }
    \label{fig:TEBDConvCheck}
\end{figure}

In the simulation of return rates in Fig.~\ref{fig:Evo}, the exact time-dependent Hamiltonian evolution is simulated using order-4 TEBD in TeNPy with a Trotterisation step of $dt=10^{-3} J^{-1}$. The numerical convergence of TEBD with respect to $dt$ is verified in Fig.~\ref{fig:TEBDConvCheck}(a), where $dt$ is gradually increased, and the same TEBD simulation as in Fig.~\ref{fig:Evo}(b) is plotted. 
%We see that the simulation still achieves convergence when $dt$ increases to $0.008 J^{-1}$. 
In Fig.~\ref{fig:TEBDConvCheck}(b), we focus on several time points in Fig.~\ref{fig:TEBDConvCheck}(a) where the deviation compared to the $dt=10^{-3} J^{-1}$ simulation becomes obvious. We see that these deviations in general reduce to $10^{-3} \sim 10^{-4}$ at $dt=10^{-3} J^{-1}$, indicating that the TEBD Trotterisation error can be ignored in Fig.~\ref{fig:Evo} (where this Trotterisation error is by orders of magnitude smaller than the deviation between TEBD simulation and the HFE stroboscopic Floquet Hamiltonian simulations).

\subsection{The accuracy of FSWT Hamiltonian at low frequencies}\label{appendix:breakdown}
\begin{figure}[h]
    \centering
    \includegraphics[width=0.5\textwidth]{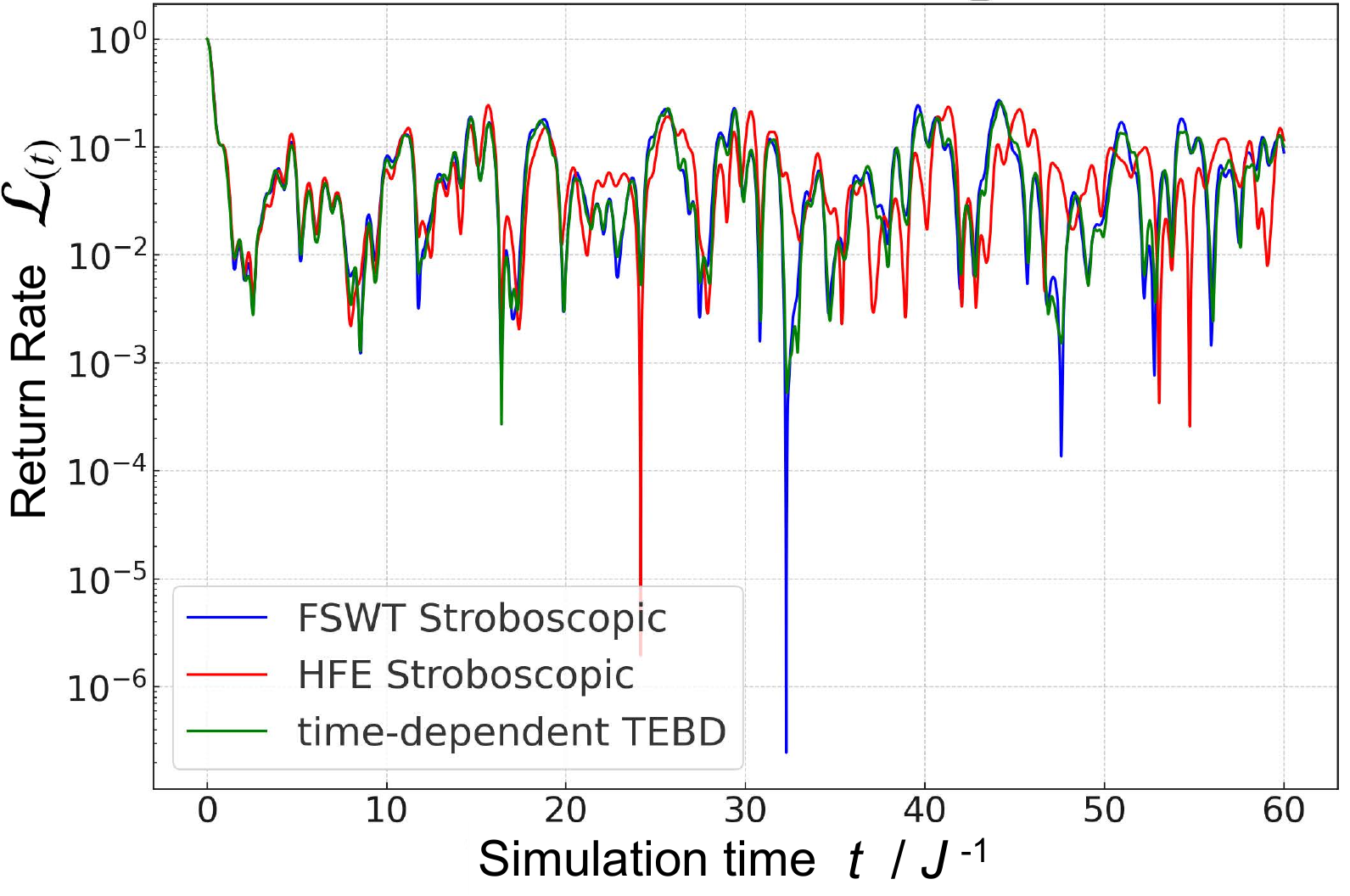}
    \caption{The return rate Eq.~(\ref{eq.Loschmidt}) for $\omega=8.5 J$, $U=3J$ and $g=\omL/4$. Noticeable deviations of FSWT dynamics appear at, e.g., time = 51 $J^{-1}$. }
    \label{fig:breakdown}
\end{figure}

Here in Fig.~\ref{fig:breakdown} we show the return rate dynamics in Fig.~\ref{fig:Evo}(c) for $\omega=8.5 J$, where the FSWT stroboscopic dynamics (which ignores the $\hat{y}_3$ correction) starts to noticeably deviate from the exact dynamics, showing a relative error $\mathcal{E}\sim0.2$ according to Eq.~(\ref{eq.NRMSE}). 

In addition, in Fig.~\ref{fig:Evo}(c), the relative error $\mathcal{E}$ of FSWT stroboscopic dynamics shows several local maxima around $\omL\approx 7J, 7.5J, 8J$ and $9J$. Since the Mott gap is $U=3 J$ for the Hubbard chain simulated in Fig.~\ref{fig:Evo}(c), these local maxima correspond to driving resonance above the Mott gap. Similar above-Mott-gap driving resonance has been reported previously in the linear absorption spectrum of the driven Hubbard clusters \cite{okamoto2021floquet}. For the parameters considered in Fig.~\ref{fig:Evo}(c), these resonances do not significantly affect the accuracy of our FSWT Hamiltonian (\ref{Floquet-H'}).

\section{Evaluation of the projection Eq.~(\ref{project_large_U_limit})}\label{appendix:project_large_U_limit}
In the derivation of Eq.~(\ref{project_large_U_limit}), we need to evaluate the projected Hamiltonian $\mathcal{P} \hat{h} \hat{H}'^{(2)} \mathcal{P}$, where $\mathcal{P} \equiv \Pi_{j=1}^{N} (1-\hat{n}_{j,\uparrow}\hat{n}_{j,\downarrow}) $ is the projector on the zero-doublon manifold. According to the definition of projector $\mathcal{P}$, we have 
\begin{equation}
\begin{split}
&\hat{c}_{j,s}^{\dag} \hat{c}_{j+1,s} \hat{n}_{j+1,\bar{s}} \mathcal{P} = \hat{c}_{j,s}^{\dag} \hat{c}_{j+1,s} \hat{n}_{j+1,\bar{s}} (1- \hat{n}_{j+1,s} \hat{n}_{j+1,\bar{s}}) \mathcal{P} =0, 
\end{split}
\end{equation}
and similarly $\hat{c}_{j,s}^{\dag} \hat{c}_{j+1,s} \hat{n}_{j,\bar{s}} \hat{n}_{j+1,\bar{s}} \mathcal{P} =0 $. %Meanwhile, 
Furthermore, since we consider the half-filling case, we have $\hat{n}_{j,\bar{s}} \to 1 - \hat{n}_{j,s} $, which results in
\begin{equation}
\begin{split}
\hat{c}_{j,s}^{\dag} \hat{c}_{j+1,s} \hat{n}_{j,\bar{s}} \mathcal{P} &\to \hat{c}_{j,s}^{\dag} \hat{c}_{j+1,s} (1 - \hat{n}_{j,s}) \mathcal{P} \\
&= \hat{c}_{j,s}^{\dag} \hat{c}_{j+1,s} \mathcal{P}.
\end{split}
\end{equation}
Thus, when the projector $\mathcal{P} $ is applied to the %$\mathcal{O}[g^2]$ order 
Floquet Hamiltonian~(\ref{Floquet-H'}), correct to $\mathcal{O}(J)$, we have
 
\begin{equation}
\begin{split}
\hat{H}'^{(2)} \mathcal{P} &= \frac{g^2 J}{\omL^2}
\sum\limits_{s} \sum\limits_{j=1}^{L-1} ( \hat{c}_{j,s}^{\dag} \hat{c}_{j+1,s} + \hat{c}_{j+1,s}^{\dag} \hat{c}_{j,s}   ) 
\left( 1 + \frac{\beta' + \gamma'}{2} (\hat{n}_{j,\bar{s}} +  \hat{n}_{j+1,\bar{s}} ) +  \delta' \hat{n}_{j,\bar{s}} \hat{n}_{j+1,\bar{s}} \right) \mathcal{P} \\
&= \frac{g^2 J}{\omL^2} \sum\limits_{s} \sum\limits_{j=1}^{L-1}
\hat{c}_{j,s}^{\dag} \hat{c}_{j+1,s}  \left( 1 + \frac{\beta' + \gamma'}{2} (\hat{n}_{j,\bar{s}} +  \hat{n}_{j+1,\bar{s}} ) +  \delta' \hat{n}_{j,\bar{s}} \hat{n}_{j+1,\bar{s}} \right) \mathcal{P} \\
&~~~~~~~~~~~~~~~~~   + \hat{c}_{j+1,s}^{\dag} \hat{c}_{j,s}   \left( 1 + \frac{\beta' + \gamma'}{2} (\hat{n}_{j,\bar{s}} +  \hat{n}_{j+1,\bar{s}} ) +  \delta' \hat{n}_{j,\bar{s}} \hat{n}_{j+1,\bar{s}} \right) \mathcal{P} \\
&= \frac{g^2 J}{\omL^2} \sum\limits_{s} \sum\limits_{j=1}^{L-1} (\hat{c}_{j,s}^{\dag} \hat{c}_{j+1,s} + \hat{c}_{j+1,s}^{\dag} \hat{c}_{j,s})
\left( 1 + \frac{\beta' + \gamma'}{2}  \right) \mathcal{P} \\
&= -\frac{g^2 }{\omL^2} \left( 1 + \frac{\beta' + \gamma'}{2}  \right) ~~ \times ~~ \hat{h}
 \mathcal{P}, \\
\end{split}
\end{equation}
and therefore
\begin{equation}\label{appendix-project_large_U_limit}
\begin{split}
&\mathcal{P} (\hat{h}+\hat{H}'^{(2)}) \frac{1}{-U} (\hat{h}+\hat{H}'^{(2)}) \mathcal{P} \\
&\approx \frac{1}{-U} \mathcal{P} \hat{h}^2 \mathcal{P} + \frac{1}{-U} \mathcal{P} \hat{h} \hat{H}'^{(2)} \mathcal{P} + \frac{1}{-U} \mathcal{P}  \hat{H}'^{(2)} \hat{h} \mathcal{P}  \\
&\approx \frac{1}{-U} \mathcal{P} \hat{h}^2 \mathcal{P} 
\times \left( 1 -2 \frac{g^2 }{\omL^2} \left( 1 + \frac{\beta' + \gamma'}{2}  \right)  \right)  \\
&= \bigg(  \frac{4 J^2}{U} (1-2\frac{g^2}{\omL^2}) + 4  \frac{g^2 J^2 }{\omL^2} \left( \frac{1}{U-\omL} + \frac{1}{\omL+U} \right)  \bigg) 
\times \sum\limits_{j=1}^{L-1} {\bf S}_j \cdot {\bf S}_{j+1}. 
\end{split} 
\end{equation}

\section{FSWT in strong-driving frame}\label{appendix:StrongDriveFSWT}
We start from the following driven Hamiltonian
\begin{equation}
\begin{split}
\hat{H}_t^{\text{lab}} &= -J \sum\limits_{\langle{\bf i},{\bf j}\rangle} \sum_s r_{{\bf j},{\bf i}} \hat{c}_{{\bf j},s}^{\dag} \hat{c}_{{\bf i},s} + U \sum_{{\bf j}} \hat{n}_{{\bf j},\uparrow} \hat{n}_{{\bf j},\downarrow} \\
&+ \sum_{{\bf j},s} \ep_{\bf j} \hat{n}_{{\bf j},s}  + g \sum_{{\bf j},s} (\phi_{\bf j} e^{i \omL t} + \phi_{\bf j}^* e^{-i \omL t}) \hat{n}_{{\bf j},s}  .
\end{split}
\end{equation}
The coefficient $r_{{\bf j},{\bf i}}$ allows the hopping to be non-uniform. The function $\phi_{\bf j}$ maps the position vector $\bf j$ into a complex value according to the laser drive in length gauge. For example, $\phi_{\bf j} =  ( j_x + i~j_y) / \sqrt{2} $ describes the coupling to a circularly polarised laser, while $\phi_{\bf j} =  j_x $ describes a linearly polarised laser.

To obtain driving effects at higher orders of driving strength $g$, we will apply FSWT (constructed in Chapter \ref{Chapter5}) in the strong-driving rotating frame. To work in this strong-driving frame, we apply the following PZW-type transform \cite{doi:10.1126/science.1119678,PhysRevLett.125.195301,10.21468/SciPostPhys.5.2.017}, $\vert \psi \rangle_t = \hat{U}^{\text{r}}_t \vert \psi \rangle_t^{\text{lab}} 
 $, where
\begin{equation}
\hat{U}^{\text{r}}_t = \exp \bigg( \frac{g}{\omL} \sum_{{\bf j},s} \hat{n}_{{\bf j}s}  (\phi_{\bf j} e^{i \omL t} - \phi_{\bf j}^* e^{-i \omL t})  \bigg).
\end{equation}
The resulting rotating-frame Hamiltonian, according to $\hat{H}_t = \hat{U}^{\text{r}}_t \hat{H}_t^{\text{lab}} (\hat{U}^{\text{r}}_t)^{\dag} + i (\partial_t \hat{U}^{\text{r}}_t) (\hat{U}^{\text{r}}_t)^{\dag}$, reads
\begin{equation}
\begin{split}
\hat{H}_t &= -J \sum\limits_{\langle{\bf i},{\bf j}\rangle} \sum_s e^{i A_{{\bf j},{\bf i}}  \sin (\omL t + B_{{\bf j},{\bf i}}) }  r_{{\bf j},{\bf i}}  \hat{c}_{{\bf j},s}^{\dag} \hat{c}_{{\bf i},s} \\
&+ \sum_{{\bf j},s} \ep_{\bf j} \hat{n}_{{\bf j},s} + U \sum_{{\bf j}} \hat{n}_{{\bf j},\uparrow} \hat{n}_{{\bf j},\downarrow}
\end{split}
\end{equation}
where $ A_{{\bf j},{\bf i}} \equiv \frac{2 g}{\omL} \vert \phi_{\bf j} - \phi_{\bf i} \vert$ and $B_{{\bf j},{\bf i}} \equiv \arg(\phi_{\bf j} - \phi_{\bf i})$. This is a multi-frequency driving Hamiltonian whose \textit{driving strength becomes the hopping parameter} $J$, which can be written as
\begin{equation}\label{H_t-strong-drive}
\begin{split}
\hat{H}_t = \hat{H}^{(0)} + \sum_{j=-\infty}^{\infty} \hat{H}^{(1)}_j e^{i j \omL t},
\end{split}
\end{equation}
where the superscript represents the orders of $J$. In Eq.~(\ref{H_t-strong-drive}), the undriven part no longer contains hopping, as given by
\begin{equation}
\hat{H}^{(0)} = \sum_{{\bf j},s} \ep_{\bf j} \hat{n}_{{\bf j},s} + U \sum_{{\bf j}} \hat{n}_{{\bf j},\uparrow} \hat{n}_{{\bf j},\downarrow},
\end{equation}
and the driving term oscillating at $e^{i j \omL t}$ reads
\begin{equation}
\hat{H}^{(1)}_j = -J  \sum\limits_{\langle{\bf i},{\bf j}\rangle} \sum_s \alpha^{[j]}_{{\bf j},{\bf i}}  \hat{c}_{{\bf j},s}^{\dag} \hat{c}_{{\bf i},s},
\end{equation}
where we define $ \alpha^{[j]}_{{\bf j},{\bf i}} \equiv  e^{ij B_{{\bf j},{\bf i}}}  \mathcal{J}_j ( A_{{\bf j},{\bf i}} ) r_{{\bf j},{\bf i}} $, which satisfies $ \alpha^{[-j]}_{{\bf j},{\bf i}} = \left( \alpha^{[j]}_{{\bf i},{\bf j}} \right)^*$ since the driven Hamiltonian is Hermite. Here $\mathcal{J}_j$ is the $j$th kind Bessel function.
In this rotating frame, the driving term has a static component $\hat{H}^{(1)}_0$. 

FSWT in this strong-driving frame finds a time-periodic unitary transform, i.e., the micro-motion operator $\hat{U}_t = \exp\left( \sum_{k=1}^{\infty}\sum_{j=1}^{\infty} e^{ij\omL t} \hat{f}^{(k)}_j - H.c. \right)$, which eliminates the oscillation in $\hat{H}_t$ perturbatively in orders of driving strength $J$. The resulting time-independent Floquet Hamiltonian $\hat{H}'$ gives the evolution operator $\hat{\mathcal{U}}'_{t,t_0}$. The whole process decouples the lab-frame evolution operator, such that $\hat{\mathcal{U}}_{t,t_0}^{\text{lab}} = (\hat{U}^{\text{r}}_t)^{\dag} \hat{U}_t^{\dag} \hat{\mathcal{U}}'_{t,t_0} \hat{U}_{t_0} \hat{U}^{\text{r}}_{t_0}$.

The $\mathcal{O}(J)$ lowest order Sylvester equation is, for all $j\neq0$,
\begin{equation}\label{FSWT-f^1_j}
\hat{H}^{(1)}_{j} + [\hat{f}^{(1)}_j,\hat{H}^{(0)}] - j \omL \hat{f}^{(1)}_j = 0,
\end{equation}
whose solution is
\begin{equation}
\hat{f}^{(1)}_j = -J \sum\limits_{\langle{\bf i},{\bf j}\rangle} \sum_s   \frac{ \alpha^{[j]}_{{\bf j},{\bf i}} } { \omL_{{\bf j},{\bf i}} } \hat{c}_{{\bf j},s}^{\dag} \hat{c}_{{\bf i},s} \hat{\theta}^{{\bar s},{\bar s}}_{ {\bf j},{\bf i} }
\end{equation}
where we have defined the parameter $\omL_{ {\bf j},{\bf i} } \equiv j\omL + \ep_{\bf j} - \ep_{\bf i}$ and the operator
\begin{equation}
\hat{\theta}^{s,s'}_{ {\bf j},{\bf i} } \equiv 1+\beta_{ {\bf j},{\bf i} } \hat{n}_{{\bf j} s} + \gamma_{ {\bf j},{\bf i} } \hat{n}_{{\bf i} s'} + \delta_{ {\bf j},{\bf i} } \hat{n}_{{\bf j} s} \hat{n}_{{\bf i} s'}
\end{equation}
where the coefficients
\footnote{For compactness, the dependence on the Fourier index $j$ is abbreviated in the notation of $\omL_{ {\bf j},{\bf i} }, \beta_{ {\bf j},{\bf i} }, \gamma_{ {\bf j},{\bf i} } \text{ and } \delta_{ {\bf j},{\bf i} } $.}
are defined as 
\begin{equation}\label{parameters'}
\begin{split}
\beta_{ {\bf j},{\bf i} } &= \frac{-U}{\omL_{ {\bf j},{\bf i} } +U} ~~~~~ 
\gamma_{ {\bf j},{\bf i} } = \frac{U}{\omL_{ {\bf j},{\bf i} } -U} \\ 
\delta_{ {\bf j},{\bf i} } &=  - \beta_{ {\bf j},{\bf i} } - \gamma_{ {\bf j},{\bf i} } .  
\end{split}
\end{equation}

then the Floquet Hamiltonian, correct to $\mathcal{O}(J^2)$, reads

\begin{equation}
\begin{split}
\hat{H}' &= \hat{H}'^{(0)} + \hat{H}'^{(1)} + \hat{H}'^{(2)} + \mathcal{O}(J^3) \\
&= \hat{H}^{(0)} + \hat{H}^{(1)}_0 + \frac{1}{2} \sum_{j=1}^{\infty}\left( [\hat{f}^{(1)}_j,\hat{H}^{(1)}_{-j}] + H.c. \right) \\
&= -J \sum\limits_{\langle{\bf i},{\bf j}\rangle} \sum_s \mathcal{J}_0 ( A_{{\bf j},{\bf i}} ) ~ r_{{\bf j},{\bf i}} \hat{c}_{{\bf j},s}^{\dag} \hat{c}_{{\bf i},s} + \sum_{{\bf j},s} \ep_{\bf j} \hat{n}_{{\bf j},s} + U \sum_{{\bf j}} \hat{n}_{{\bf j},\uparrow} \hat{n}_{{\bf j},\downarrow} \\
&~~~ + \frac{J^2}{2} \sum_{j=1}^{\infty} 
\left( \left[
\sum\limits_{\langle{\bf i},{\bf j}\rangle} \sum_s   \frac{ \alpha^{[j]}_{{\bf j},{\bf i}} } { \omL_{{\bf j},{\bf i}} } \hat{c}_{{\bf j},s}^{\dag} \hat{c}_{{\bf i},s} \hat{\theta}^{{\bar s},{\bar s}}_{ {\bf j},{\bf i} } 
,  \sum\limits_{\langle{\bf i'},{\bf j'}\rangle} \sum_{s'}  \alpha^{[-j]}_{{\bf j'},{\bf i'}} \hat{c}_{{\bf j'},s'}^{\dag} \hat{c}_{{\bf i'},s'}
\right] ~+~ H.c.\right) 
\end{split}
\end{equation}
In this result, the first line comes from $\hat{H}^{(0)} + \hat{H}^{(1)}_0$, where the hopping is renormalised by the dynamical localisation effect, with the intrinsic interactions in the undriven system unchanged. The second line contains the $\mathcal{O}(J^2)$ driving induced interaction, which evaluates to
\begin{equation}
\begin{split}
\hat{H}'^{(2)} &= J^2 \sum_{j=1}^{\infty} \sum\limits_{ \{ {\bf i - j} \} } \sum_s ( \hat{n}_{{\bf j}s} - \hat{n}_{{\bf i}s} ) \left(  \frac{ \big\vert \alpha^{[j]}_{{\bf j},{\bf i}} \big\vert^2  } { \omL_{{\bf j},{\bf i}} }  \hat{\theta}^{{\bar s},{\bar s}}_{ {\bf j},{\bf i} }  -  \frac{ \big\vert \alpha^{[j]}_{{\bf i},{\bf j}} \big\vert^2  } { \omL_{{\bf i},{\bf j}} }  \hat{\theta}^{{\bar s},{\bar s}}_{ {\bf i},{\bf j} }    \right)  \\
&+ \frac{J^2}{2} \sum_{j=1}^{\infty} \sum\limits_{ \{ {\bf i - j} \} } \sum_s \hat{c}_{{\bf j},s}^{\dag} \hat{c}_{{\bf j},{\bar s}}^{\dag} \hat{c}_{{\bf i},{\bar s}}  \hat{c}_{{\bf i},s} ~\alpha^{[j]}_{{\bf j},{\bf i}} \alpha^{[-j]}_{{\bf j},{\bf i}}  \bigg( \frac{ \beta_{{\bf j},{\bf i}} - \gamma_{{\bf j},{\bf i}} }{ \omL_{{\bf j},{\bf i}} } + \frac{ \beta_{{\bf i},{\bf j}} - \gamma_{{\bf i},{\bf j}} }{ \omL_{{\bf i},{\bf j}} } \bigg)    + H.c.\\
&+ \frac{J^2}{2} \sum_{j=1}^{\infty} \sum\limits_{ \{ {\bf i - j} \} } \sum_s \hat{c}_{{\bf j},s}^{\dag} \hat{c}_{{\bf i},{\bar s}}^{\dag} \hat{c}_{{\bf j},{\bar s}}  \hat{c}_{{\bf i},s} \bigg( - \frac{ \big\vert \alpha^{[j]}_{{\bf j},{\bf i}} \big\vert^2 }{ \omL_{{\bf j},{\bf i}} } ( \beta_{{\bf j},{\bf i}} - \gamma_{{\bf j},{\bf i}} ) - \frac{ \big\vert \alpha^{[j]}_{{\bf i},{\bf j}} \big\vert^2 }{ \omL_{{\bf i},{\bf j}} } ( \beta_{{\bf i},{\bf j}} - \gamma_{{\bf i},{\bf j}} ) \bigg)   + H.c.\\
&+ \frac{J^2}{2} \sum_{j=1}^{\infty} \sum\limits_{ \{ {\bf i - j - k} \} } \sum_s \hat{c}_{{\bf k},s}^{\dag} \hat{c}_{{\bf i},s} \bigg( - \alpha^{[j]}_{{\bf j},{\bf i}} \alpha^{[-j]}_{{\bf k},{\bf j}} \big( \frac{ \hat{\theta}^{{\bar s},{\bar s}}_{ {\bf j},{\bf i} } }{ \omL_{ {\bf j},{\bf i} } } + \frac{ \hat{\theta}^{{\bar s},{\bar s}}_{ {\bf j},{\bf k} } }{ \omL_{ {\bf j},{\bf k} } }  \big) + 
\alpha^{[j]}_{{\bf k},{\bf j}} \alpha^{[-j]}_{{\bf j},{\bf i}} \big( \frac{ \hat{\theta}^{{\bar s},{\bar s}}_{ {\bf k},{\bf j} } }{ \omL_{ {\bf k},{\bf j} } } + \frac{ \hat{\theta}^{{\bar s},{\bar s}}_{ {\bf i},{\bf j} } }{ \omL_{ {\bf i},{\bf j} } }  \big) \bigg) + H.c.\\
&+ \frac{J^2}{2} \sum_{j=1}^{\infty} \sum\limits_{ \{ {\bf i - j - k} \} } \sum_s \hat{c}_{{\bf j},s}^{\dag} \hat{c}_{{\bf k},{\bar s}}^{\dag} \hat{c}_{{\bf j},{\bar s}} \hat{c}_{{\bf i},s} \bigg( - \alpha^{[j]}_{{\bf j},{\bf i}} \alpha^{[-j]}_{{\bf k},{\bf j}} \big( \frac{ \hat{\mu}^{\bar s}_{ {\bf j},{\bf i} } }{ \omL_{ {\bf j},{\bf i} } } + \frac{ \hat{\mu}^{s}_{ {\bf j},{\bf k} } }{ \omL_{ {\bf j},{\bf k} } }  \big) + 
\alpha^{[j]}_{{\bf k},{\bf j}} \alpha^{[-j]}_{{\bf j},{\bf i}} \big( \frac{ \hat{\nu}^{s}_{ {\bf k},{\bf j} } }{ \omL_{ {\bf k},{\bf j} } } + \frac{ \hat{\nu}^{\bar s}_{ {\bf i},{\bf j} } }{ \omL_{ {\bf i},{\bf j} } }  \big) \bigg) + H.c.\\
&+ \frac{J^2}{2} \sum_{j=1}^{\infty} \sum\limits_{ \{ {\bf i - j - k} \} } \sum_s \hat{c}_{{\bf j},s}^{\dag} \hat{c}_{{\bf j},{\bar s}}^{\dag} \hat{c}_{{\bf k},{\bar s}} \hat{c}_{{\bf i},s} \bigg(  \alpha^{[j]}_{{\bf j},{\bf i}} \alpha^{[-j]}_{{\bf j},{\bf k}} \big( \frac{ \hat{\mu}^{\bar s}_{ {\bf j},{\bf i} } }{ \omL_{ {\bf j},{\bf i} } } - \frac{ \hat{\nu}^{s}_{ {\bf k},{\bf j} } }{ \omL_{ {\bf k},{\bf j} } }  \big) + 
\alpha^{[j]}_{{\bf j},{\bf k}} \alpha^{[-j]}_{{\bf j},{\bf i}} \big( \frac{ \hat{\mu}^{s}_{ {\bf j},{\bf k} } }{ \omL_{ {\bf j},{\bf k} } } - \frac{ \hat{\nu}^{\bar s}_{ {\bf i},{\bf j} } }{ \omL_{ {\bf i},{\bf j} } }  \big) \bigg) + H.c.\\
%note that this Floquet Hamiltonian is constructed in the strong driving frame, thus it can show difference comapred with the previous FSWT Hamiltonian constructed in lab frame
\end{split}
\end{equation}
Unlike $\langle {\bf i},{\bf j}\rangle$ which is nonequivalent to $\langle {\bf j},{\bf i}\rangle$, the summation $\{ {\bf i - j} \}$ here runs over all 2-site-connected bonds, thus $\{ {\bf i - j} \}$ and $\{ {\bf j - i} \}$ are equivalent which only need to be included once. Likewise, the summation $\{ {\bf i - j - k} \}$ runs over all 3-site-connected bonds where $\bf j$ is the middle site, thus $\{ {\bf i - j - k} \}$ and $\{ {\bf k - j - i} \}$ are equivalent and should be included only once.
To shorten the forula, we have defined $\hat{\mu}^{s}_{ {\bf j},{\bf i} } \equiv \beta_{ {\bf j},{\bf i} } + \delta_{ {\bf j},{\bf i} } \hat{n}_{{\bf i}s}$ and $\hat{\nu}^{s}_{ {\bf j},{\bf i} } \equiv \gamma_{ {\bf j},{\bf i} } + \delta_{ {\bf j},{\bf i} } \hat{n}_{{\bf j}s}$.

In the absence of the on-site potential $\ep_{\bf j}=0$, the above result is reduced according to $\omL_{{\bf j},{\bf i}} \to \omL$, and $\beta, \gamma, \delta$ become spatial independent.